\documentclass[twocolumn]{aastex63}

\pdfoutput=1 

\usepackage[T1]{fontenc}

\usepackage{url}
\usepackage{graphicx}
\usepackage[caption=false]{subfig}

\usepackage{booktabs}
\usepackage{dcolumn}

\usepackage[graphicx]{realboxes}

\usepackage{multirow}

\usepackage{longtable}

\usepackage{amsmath}

\usepackage{rotating}

\usepackage{lipsum}

\usepackage{mwe}

\usepackage{amsmath,amssymb}

\makeatletter
\DeclareRobustCommand{\coverline}[1]{
  \binrel@{#1}\binrel@@{\overline{#1}}%
}
\makeatother

\usepackage[utf8]{inputenc}
\usepackage{graphicx}
\usepackage{amsmath}
\usepackage{listings}
\usepackage{float}

\usepackage{subfig}

\usepackage{alltt}
\usepackage{amssymb}
\usepackage{amsfonts}
\usepackage{mathtools}

\usepackage[capitalise]{cleveref}

\crefdefaultlabelformat{\textbf{#2#1#3}}
\crefname{figure}{\textbf{Fig.}}{\textbf{Figures}} 

\usepackage{hyperref}

\usepackage[utf8]{inputenc}
\usepackage[T1]{fontenc}

\usepackage{lipsum}
\usepackage{graphicx}

\newcommand{\tablefootnote}

\shorttitle{Fundamental Properties of M dwarfs and M subdwarfs}

\submitjournal{ApJ}

\shorttitle{Sample article}
\shortauthors{Hejazi et al.}

\begin{document}

\title{Chemical Properties of the Local Disk and Halo. II.  Abundances of  3745 M dwarfs and Subdwarfs from Improved Model Fitting of Low-Resolution Spectra}

\author{Neda Hejazi}
\affil{Department of Physics and Astronomy, Georgia State University, Atlanta, GA 30303, USA}

\author{S\'ebastien L\'epine}
\affiliation{Department of Physics and Astronomy, Georgia State University, Atlanta, GA 30303, USA}

\author{Thomas Nordlander}
\affiliation{Research School of Astronomy \& Astrophysics, Australian National University, Canberra, ACT 2611, Australia}
\affiliation{The ARC Centre of Excellence for All Sky Astrophysics in 3 Dimensions, Canberra, ACT 2611, Australia}


\begin{abstract}
We present a  model-fit pipeline to determine the stellar parameters of M-type dwarfs, which is an improvement upon our previous work described in Hejazi et al. 2020.  We apply this pipeline to analyze the low-resolution (R$\sim$2000) spectra of 3745 M dwarfs/subdwarfs, collected at the MDM Observatory, Lick Observatory, Kitt Peak National Observatory, and Cerro Tololo Interamerican Observatory. We examine the variation of the inferred parameter values in the HR diagram constructed from their Gaia Early Data Release 3 (EDR3) parallaxes and optical magnitudes. We also study the distribution of our stars in the abundance diagram of [$\alpha$/Fe] versus [M/H]  and inspect the variation of their metallicity class, effective temperature, and surface gravity as well as their Galactic velocity components U, V, and W in this diagram. In addition, the analyses of the stars’ projected motions in the two-dimensional UV, VW, and UW planes and the variation of their chemical parameters in these planes, and also their distribution in the abundance-velocity diagrams are important parts of this study. The precision of our model-fit pipeline is confirmed by the clear stratification of effective temperature and chemical parameters in the HR diagram, the similarity of  the stars’ distribution in the  [$\alpha$/Fe] versus [M/H] diagram and in the metallicity-velocity planes with those from other studies, revealing substructure in the abundance-velocity diagrams, and  chemical homogeneity between the components of a set of binary systems. 

\end{abstract}


\section{Introduction} 
M-type dwarfs dominate the stellar population of our Galaxy, including $\sim$ 70$\%$ of all stars by number (Reid \& Gizis 1997; Bochanski et al. 2010) and hold enormous potential for being used as tools to probe  the structure and chemo-dynamical evolution of the Milky Way, provided kinematics and chemical composition can be reliably measured for very large numbers of them. Metal-deficient, M subdwarfs, in particular, are kinematically associated with the old Galactic halo and thick disk populations and carry the nucleosynthetic signatures of the early evolution of the Galaxy.  M dwarfs/subdwarf are more reliable probes of chemical composition in old populations because, due to the complexity of  the evolutionary sequence of more massive stars,  various enrichment processes may change their chemical pattern. For example, ``atomic diffusion'' can perturb the atmospheric elemental abundances of of more massive dwarfs with shallow convection zones, i.e.,  the higher the mass, the larger the effect (Gao et al. 2018). In comparison to more massive FGK dwarfs, atomic diffusion has a negligible impact on M-type dwarfs  because of their lower mass and deeper convection zone. The Mixing (``dredge-up'') process can also vary the surface abundance values to some extent in  giants, causing an offset in detailed chemical abundances  between dwarfs and giants  even with a common birthplace. Being unevolved main-sequence stars, there are therefore no ambiguities in the surface composition  of M dwarfs caused by the dredge-up process. Consequently,  the atmospheric abundances of M-type dwarfs  are pristine indicators of  the chemical properties of the progenitor molecular clouds where these stars formed (unless they merge or exchange mass with exoplanets, comets, or their companion in binary systems). 

M dwarfs have also been shown to be excellent targets to measure the isotopic abundance ratios of titanium (an $\alpha$ element) as a robust test of Galactic chemical evolution (GCE) models (Chavez \& Lambert 2009).  The accurate measurement of Ti  isotopic ratios in stars and the interstellar medium (ISM)  provides important information about the Galaxy's enrichment history, and  in particular, the production of Ti in different types of supernovae (that yield different isotopic ratios). However, because the thermal and macroscopic broadening of atomic lines in stars is comparable to the isotopic splitting, the isotopic effects can essentially  be observed in molecular bands, rather than in atomic lines (although the small  broadening of atomic lines in the ISM makes it possible to measure isotopic effects even in atomic lines in such environments),  and for this reason, isotopic analyses can be  conceived using the spectra of cool stars, which are dominated by molecular lines. Particularly, the prevailing TiO molecular bands in M dwarf spectra allow the measurement of relative abundances for the most stable isotopes of Ti, resulting in valuable clues on the nucleosynthesis of $\alpha$ elements.

However, M-type dwarfs have not been used to their fullest potential in Galactic archaeological studies within the last decade, as compared to numerous  analyses using FGK dwarfs and  giants. Due to their intrinsic faintness, the acquisition of high-resolution, high signal-to-noise spectra of M dwarfs and M subdwarfs demands large telescopes with long exposure times. High-resolution spectroscopy of M dwarfs has therefore been limited to small samples of very nearby stars, mostly from the Galactic disk (e.g., Mann et al. 2013; Rajpurohit et al. 2014, 2018; Veyette et al. 2017; Woolf 2020). On the other hand,  low-resolution spectra of M dwarfs/subdwarfs  have been acquired  in large numbers using more economically efficient observations (e.g., West et al. 2011; Lepine et al. 2013; Zhang 2021). To exploit such a wealth of data, we have attempted to develop an automated pipeline to infer stellar parameters, i.e., effective temperature T$_\textrm{\footnotesize{eff}}$, overall metallicity [M/H], alpha-element enhancement [$\alpha$/Fe], and surface gravity log \emph{g} for  M dwarfs/subdwarfs by fitting low-resolution spectra to the BT-Settl model grid calculated from the PHOENIX model atmospheres (see Hejazi et al. 2020, hereafter Paper I, for a detailed description). The resulting  values  enable us to analyze the correlation between these parameters, in particular between [M/H] and [$\alpha$/Fe] that has proven powerful in detecting and identifying distinct stellar populations (e.g., Adibekyan, 2012, 2013; Hayden et al. 2015 and references therein). The relationship between the stellar chemical parameters and  photometric and kinematic properties of stars is  also of great importance in distinguishing different  populations (e.g., Lepine et al. 2007;  2013; Hawkins et al. 2015; Yan et al. 2019, also see Section 9 in Paper I for more useful references in this regard).

 As it turns out,  different spectroscopic  studies utilize different methods using spectra with different  resolutions and over different wavelength regimes, and  a comparison of the results unfortunately shows significant systematic offsets in  abundance measurements. To assemble a precise  abundance dataset, it is thus critical to obtain  chemical abundances  of all stars in  a sample using the same technique over the same wavelength region with observed spectra of  comparable  resolutions. Although our low-resolution spectroscopic analyses  cannot result in  highly accurate values of   chemical parameters,  since we apply the same technique over a certain spectral range to all stars, the resulting  measurements are precise enough to separate  different stellar population using photometric, chemical and kinematic diagrams.

The work reported here  is an improvement upon the method described in Paper I with a more extensive investigation of the model-fit process and its complications. We revisit the spectra from the sample of Paper I, which focused on high proper-motion stars ($\mu{\geq}0.4{\arcsec}{\textrm{yr}}^{-1}$),  but also expand the analysis to an additional 2045  spectra from nearby (d$<$25 pc) stars with lower proper motions, down to  $\mu=0.03{\arcsec}{\textrm{yr}}^{-1}$. In Section 2, we  describe the spectroscopic sample used in this study. The selected model grid is detailed in section 3. The correlation and degeneracy effect between physical parameters that may cause systematic uncertainties in the  best-fit parameters values are also analyzed in Sections 3. We outline our revised model-fit technique in Section 4 and the error analysis in Section 5.  The  resulting values from our pipeline and their distributions in several parametric and photometric diagrams are presented in Sections 6 and 7. The effect of surface gravity variation and the residual correlation between wavelength datapoints on the best-fit parameter estimates are also addressed in these sections. We analyze the precision of the model-fit chemical parameter values  using a set of common proper-motion pairs in Section 8. The distributions of our stars in the kinematic planes as well as in the abundance-velocity diagrams are examined in Section 9. Finally, we summarize our work in Section 10.

\section{Spectroscopic Sample}
Our dataset consists of spectra from 3745 M dwarfs/subdwarfs collected between 2002 and 2012 in multiple observing runs at the MDM observatory, the Kitt Peak National Observatory (KPNO),  the Lick Observatory, and the Cerro Tololo Interamerican Observatory (CTIO). A summary of the observations and data reduction is presented in Paper I (Section 2). For the present analysis, spectral variances were re-calculated from the original spectra using  the read noise and Poisson statistics of the integrated photon count at each spectral pixel, evaluated before sky subtraction and spectrophotometric calibration.

All stars are classified using the template-fit method outlined in Paper I (section 3)  into twelve metallicity classes (from 1 for very metal-rich stars to 12 for very metal-poor stars) and into standard spectral subtypes. The sample consists of stars with spectral subtype between M0.0 and M6.5  and with 2MASS magnitudes \textit{K}$\leq$15, excluding  objects with unreliable parallaxes and/or photometric magnitude measurements from Gaia DR2/EDR3 (Gaia collaboration et al. 2018; 2021). All the objects used here are either stars that were originally selected for follow-up observations because of their high proper motions ($\mu{\geq}0.4{\arcsec}{\textrm{yr}}^{-1}$) as measured in the SUPERBLINK proper motion survey (L\'epine \& Shara 2005), or because their pre-Gaia photometric distances suggested they were within 25 parsecs of the Sun (L\'epine 2005). All the  brighter objects  with \textit{J}$<$9  (1564 stars) were previously presented and analyzed in L\'epine et al. (2013), and the subset of stars with large proper motions ($\mu{\geq}0.4{\arcsec}{\textrm{yr}}^{-1}$, 1544 stars) was presented and analyzed in Paper I.

Data for the 3745 stars are presented in Tables 1 and 2. Table 1 lists right ascensions, declinations, parallaxes, and proper motions obtained from Gaia EDR3 for 3545 of the stars; entries could not be found in EDR3 for the remaining 200 objects, and for those we are listing their data from the Gaia DR2 catalog. The Gaia flag in the last column of Table 1  shows the  Data Release from which the data is drawn. Table 2 presents the optical and infrared magnitudes of the stars as listed in the Gaia catalogs (DR2 or EDR3 as appropriate) and from  the 2MASS survey.

\section{Synthetic Spectra}

\subsection{Grid Selection}
We utilize the latest version of BT-Settl synthetic spectra generated by the PHOENIX model atmospheres, as described in Paper I (Section 5.1). Our grid selection is almost the same as explained in Paper I (Section 5.2), limited to T$_\textrm{\footnotesize{eff}}$ from 2700 to 4000 K (consistent with our  selected M0.0-M6.5 spectral type range) in steps of 100 K, log \emph{g} from 4.5 to 5.5 dex in steps of 0.5 dex, [M/H] from $-$2.5 to +0.5 dex in steps of 0.5 dex, and with a range of [$\alpha$/Fe] values: from $-$0.2 to +0.4 dex for [M/H]$\geqslant$$-$0.5 and from 0 to +0.6 dex 
for [M/H]<$-$0.5 dex, in steps of 0.2 dex\footnote{Micro-turbulence velocity, $\xi$, is another parameter that can affect spectral absorption lines, but is assumed to be important  in the radiative part of the atmosphere when calculating  BT-Settl synthetic spectra. The turbulent convective velocity as determined by the mixing length theory is added to micro-turbulence velocity in quadrature, but in practice, only increases it by a few percent in M dwarfs, and mainly in deeper layers that do not significantly contribute  to the line core formation. This parameter  is estimated for M dwarfs to a first approximation assuming a simple   function of  effective temperature: $\xi$=(T$_\textrm{\footnotesize{eff}}$ - 1250 K)/7200 (km/s).}. 
  
We perform  a four-dimension spline interpolation at every step of 50 K in T$_\textrm{\footnotesize{eff}}$, 0.05 dex in [M/H], 0.025 dex in  [$\alpha$/Fe], and 0.05 in log \emph{g}, giving  a total of 869,211 grid points in the 4D parameter space. In the present work, we introduce an expanded  set of grid points in surface gravity, i.e., log \emph{g} from 4.5 to 5.5 dex in steps of  0.05 dex, whereas Paper I examined a more limited set of gravity grid points  from 4.8 to 5.2 dex with a step size of 0.1 dex. Moreover, to reduce the computational time of   model-fit runs,  we first apply a Gaussian broadening kernel and then resample all original and interpolated synthetic spectra across  a wavelength step of 0.1 {\AA}. Our analysis shows that the resulting model-fit parameters of our observed spectra (which have a spectral resolution of $\simeq$2-3 {\AA}) do not change when using these down-sampled synthetic spectra\footnote{We randomly selected 300 stars from our sample and applied the pipeline to this subset using both  original and down-sampled synthetic spectra and found no significant difference in the two sets of results.}, as compared to parameters inferred from model spectra sampled on the original, high-resolution wavelength grid (0.02 {\AA}).

\subsection{Spectral Structure and Synthetic Fitting}

In the presence of low temperatures and high pressures,  molecules such as  TiO, VO, CO, CaH, FeH, MgH, H$_\textrm{\footnotesize{2}}$, and H$_\textrm{\footnotesize{2}}$O are the main opacity source in  the atmospheres of M dwarfs (Allard \& Hauschildt 1995). The modeling of stellar atmospheres therefore requires  accurate opacity data  for millions of molecular transitions and spectral lines. These molecules have significant impacts on  the detailed structure of M dwarf spectra, influencing  both individual lines and  the overall spectral shape through the cumulative effect of large numbers of overlapping lines at various wavelength ranges (Valenti et al. 1998). 

The forests of molecular lines hinder the identification of  continuum levels even at high spectral resolution, which  complicates the accurate determination of atomic line strengths and also introduce obstacles in the techniques that are readily applied to spectra of more massive dwarfs, i.e., F, G, and K  dwarfs, for deriving their physical parameters. The most accurate approach to determine the stellar parameters  of M dwarfs is to normalize both the observed spectrum  of interest and  all the synthetic spectra relative to their corresponding continuum, and compare the resulting rectified spectra to find the best-fit model. Although the continuum of synthetic spectra  of M dwarfs can, in principle, be calculated using the number density of key species contributing to the continuous opacity such as H, H$^\textrm{\footnotesize{$-$}}$, and H$_\textrm{\footnotesize{2}}$, the dominant molecular bands make it difficult to properly define the continuum level of an observed M dwarf spectrum, particularly in the optical regime where the TiO molecular lines cause substantial flux depressions against the continuum background. This is a significant issue because the comparison between a continuum-normalized observed spectrum and a set of continuum-normalized synthetic spectra,   with regard to the real depth of spectral lines and features (measured from the continuum) that  indeed depend on  atmospheric parameters, is expected to be the most reliable way to search for the best-fit model. Unfortunately, this approach is not applicable  to M dwarfs, which poses challenges in estimating their stellar parameters. 

There are two common methods used to perform  synthetic fitting over broad spectral ranges when the continuum cannot be correctly determined.  In one method, both the observed spectrum and  the synthetic spectra are normalized to a pseudo-continuum profile separately, and these pseudo-rectified spectra are then used  in the minimization procedure to obtain the best-fit model (e.g., Lee et al. 2011a; Kuznetsov et al. 2019).  In another approach, each synthetic spectrum in the  model grid, or alternatively the observed spectrum, is flux-renormalized by a polynomial that, in either case, can be determined from the ratio of the observed spectrum to each of the synthetic spectra (e.g., L\'epine et al. 2013; Hejazi et al. 2020), or  equivalently, from the loss function as described in  Zhang et al. (2021). The flux-renormalized  synthetic spectra (or observed spectrum) will then be compared to the unchanged observed spectrum (or  synthetic spectra)  to find the best matched model. In both methods, the observed spectrum is compared to the synthetic models in terms of  ``apparent line depths'',  that are measured from  pseudo-continuum levels,  and  ``apparent spectral features'', effectively overlooking potentially important physical information embedded in the real continuum.

\subsection{Spectral Sensitivities}

Prior to describing our model-fit technique, we  aim to investigate the sensitivity of synthetic spectra to their physical parameters and examine  how their spectral morphology changes when each parameter is varied. This  can help one understand how two synthetic  spectra with different parameters may be  degenerate, which can lead to possible systematic errors in resulting model-fit parameters.  To this end, we illustrate  the change in the  apparent spectral shape  of  synthetic spectra that are degraded to the typical resolution of our observed spectra.  For ease of comparison, we normalize these spectra relative to the flux at an arbitrary wavelength, e.g., the longest wavelength in the plots mentioned below (8725 {\AA}), setting the flux at this wavelength to unity.

The shape, depth, and broadness of prominent spectral lines are found to vary  with physical parameters in a complicated way. Figures 1-4 show how the spectral shape changes as one physical parameter (effective temperature, metallicity, $\alpha$-element enhancement, or surface gravity) is varied while the other parameters are kept fixed.  The position of some key elements and the regions over which certain molecular bands have an important effect are also presented in these figures. In Figure 1, synthetic spectra are shown for different values of temperature,  i.e., T$_\textrm{\footnotesize{eff}}$=3000, 3200, 3400, 3600, 3800, and 4000 K; all models have  [M/H]=0.0 dex,  [$\alpha/$Fe]=+0.2 dex, and  log \emph{g}=5.0 dex. As T$_\textrm{\footnotesize{eff}}$ decreases,  the star becomes significantly redder and the overall flux drops. In cooler stars,  there is less  thermal energy to populate higher atomic states, so only low-excitation atomic lines remain prominent. Furthermore, with decreasing effective temperature,  more atoms contribute to  form molecules and grains, which increases the effect of molecular bands on individual atomic spectral lines. While some atomic lines such as Ca I ($\sim$6104, 6124, 6164, 6441, 6463, and 6496 {\AA}) Ca II ($\sim$8500, 8544, and 8664 {\AA}), Li I ($\sim$6710 {\AA}), Fe I ($\sim$8329 and 8390 {\AA}), Ti I ($\sim$8438 {\AA}), and V I ($\sim$8692 {\AA}) are weaker at low temperatures, the alkali atomic lines, i.e.,  the doublet  K I ($\sim$7667 and 7701 {\AA}) and Na I ($\sim$8186 and  8197 {\AA}), are  prominent even in the spectra of very cool M dwarfs. These alkali atomic lines are therefore of importance in the determination of atmospheric parameters, especially at low spectral resolution. 

The dependence of spectral morphology on metallicity is more complex. Figure 2 depicts the model spectra with [M/H]=$-$2.0, $-$1.5, $-$1.0, $-$0.5, 0.0, and +0.5 dex; all these models are associated with T$_\textrm{\footnotesize{eff}}$=3400 K, [$\alpha/$Fe]=+0.2 dex, log \emph{g}=5.0 dex. As metallicity increases, the molecular bands become progressively stronger, increasingly depressing the flux level.  It is important to note that with increasing metallicity, the TiO bands (and more likely other oxide metals) grow stronger, but the CaH (and also other hydrides) bands are not affected. This is because  when metallicity increases, the gas pressure decreases and thus the concentration of hydride molecules remains nearly the same (Bessell 1991). The effect of metallicity on  real and apparent line strengths differs from one line to another, which is  due to  the correlation of various factors with stellar metal content. In general, the spectral shape dramatically changes at low metallicities;  the  weaker the molecular bands are, the more prominent individual atomic lines become. Further, due to different levels of flux depression, the spectral features in metal-poor spectra significantly differ from those of metal-rich ones, resulting in significant  changes in broadband colors even for stars with similar effective temperatures. This is the reason why our metal-poor stars ([M/H]$\leq$$-$1.0 dex) are normally well separated from the metal-rich ones ([M/H]$\geq$$-$0.5 dex)  using the inferred metallicities from the synthetic fitting (see section 6).

\begin{figure}[h]
\centering
\subfloat
        {\includegraphics[ height=3.5cm, width=9cm]{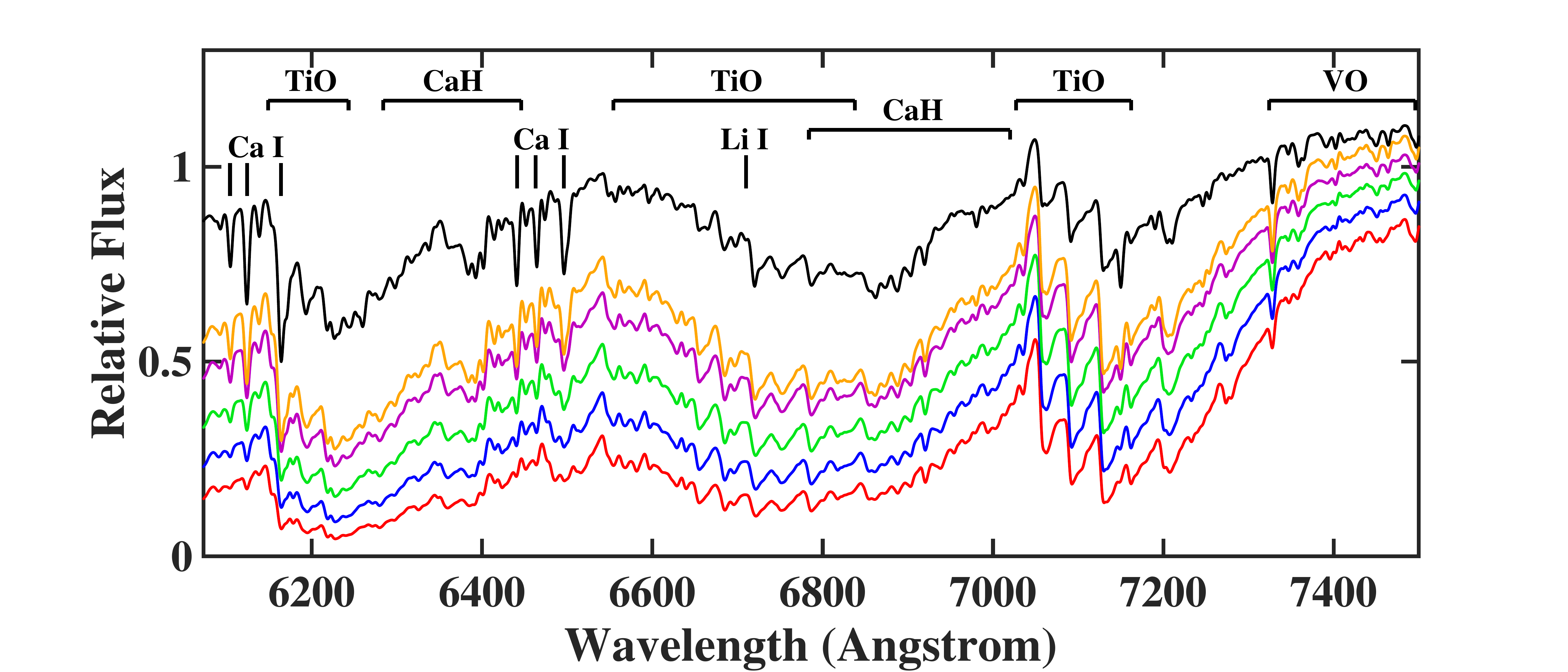}}
  \vspace{-0.38cm}
   \subfloat
         {\includegraphics[ height=3.5cm, width=9cm]{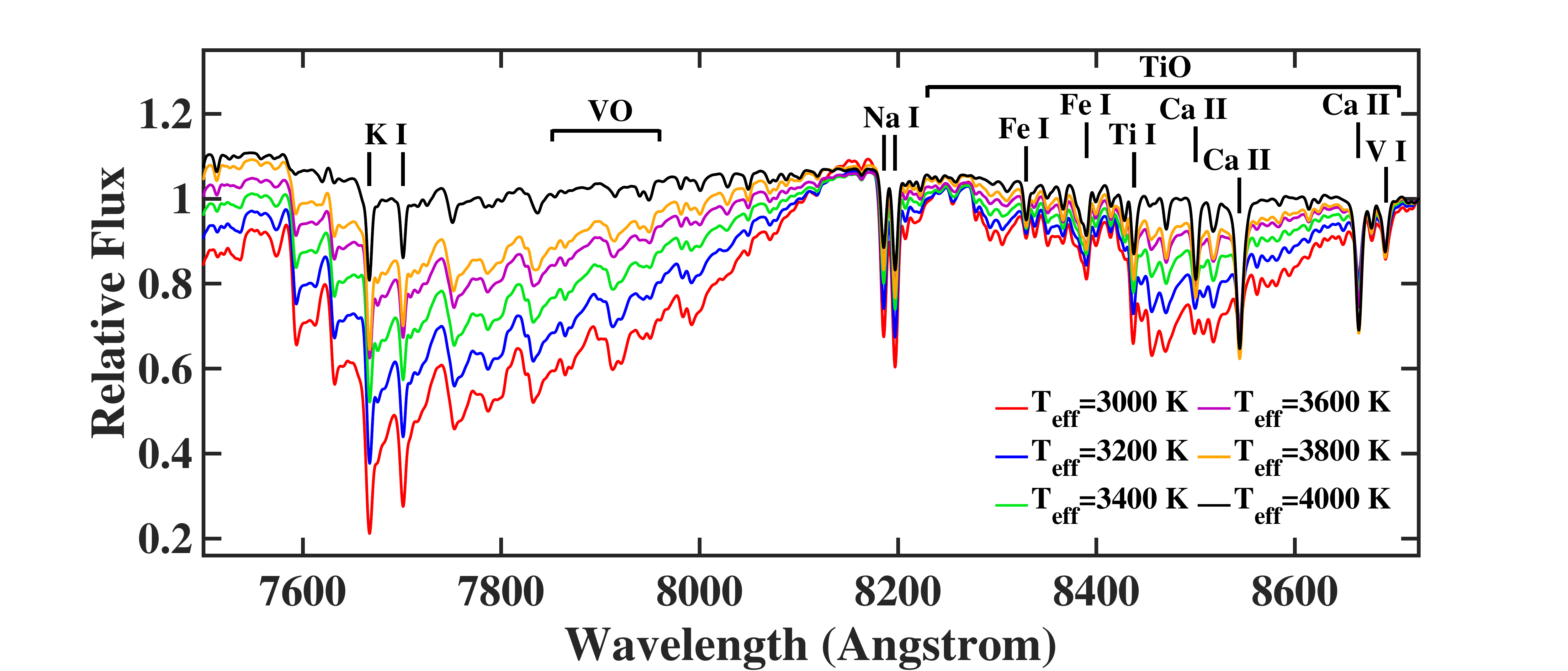}}   
\caption
        {\footnotesize{Spectral sensitivity to effective temperature: synthetic spectra with [M/H]=0.0 dex,  [$\alpha/$Fe]=+0.2 dex, and  log \emph{g}=5.0 dex, but different  values of T$_\textrm{\footnotesize{eff}}$= 3000, 3200, 3400, 3600, 3800, and 4000 K. The spectra are normalized at 8725 {\AA}.}}
    \end{figure}

\begin{figure}[h]
\centering
\subfloat
        {\includegraphics[  height=3.5cm, width=9cm]{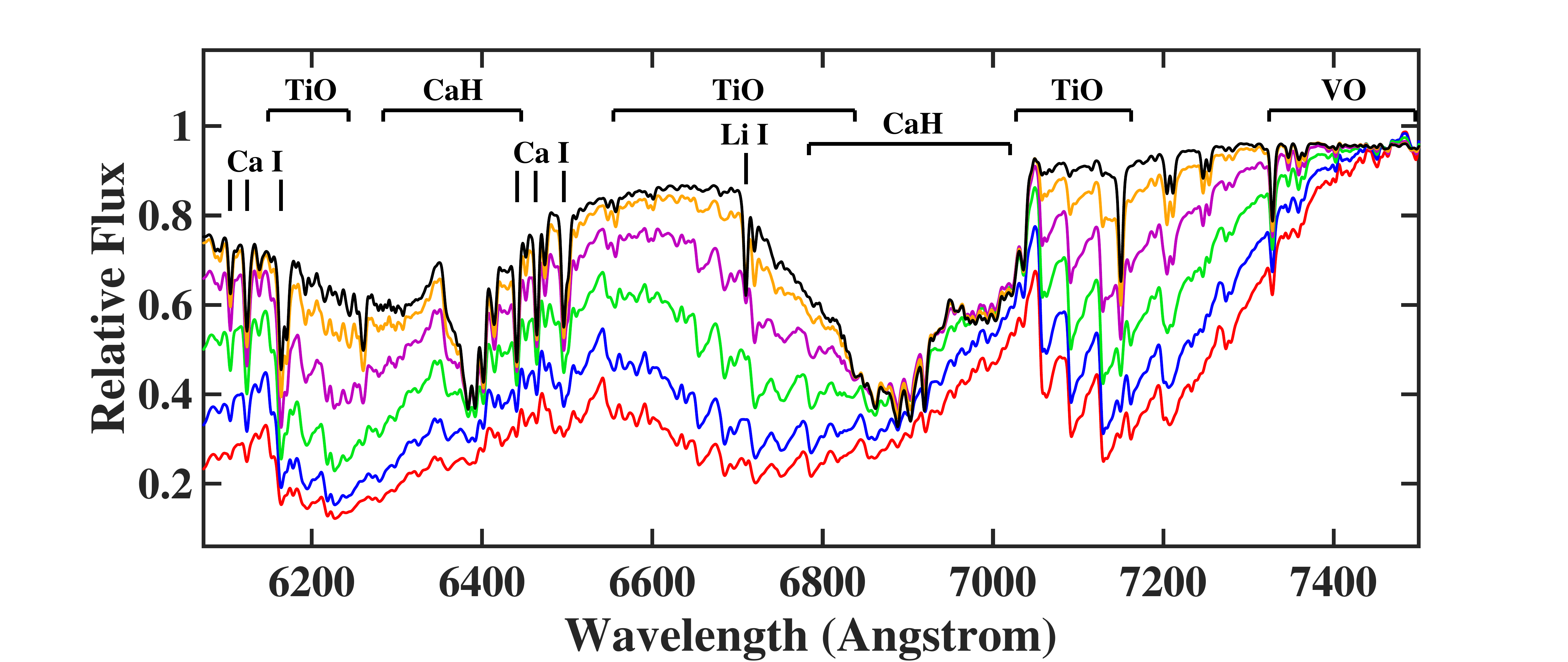}}
 \vspace{-0.38cm}
\subfloat
         {\includegraphics[  height=3.5cm, width=9cm]{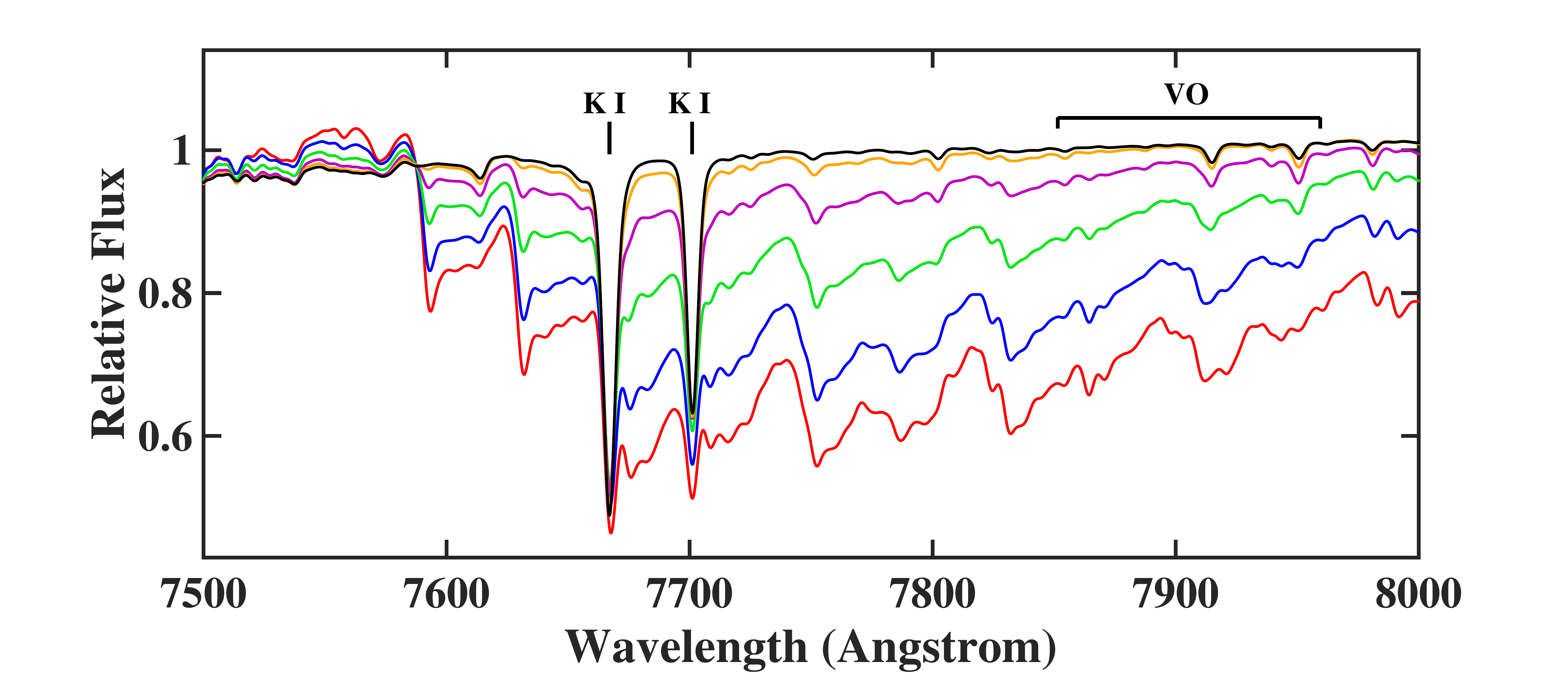}}
\vspace{-0.38cm}
\subfloat
        {\includegraphics[  height=3.5cm, width=9cm]{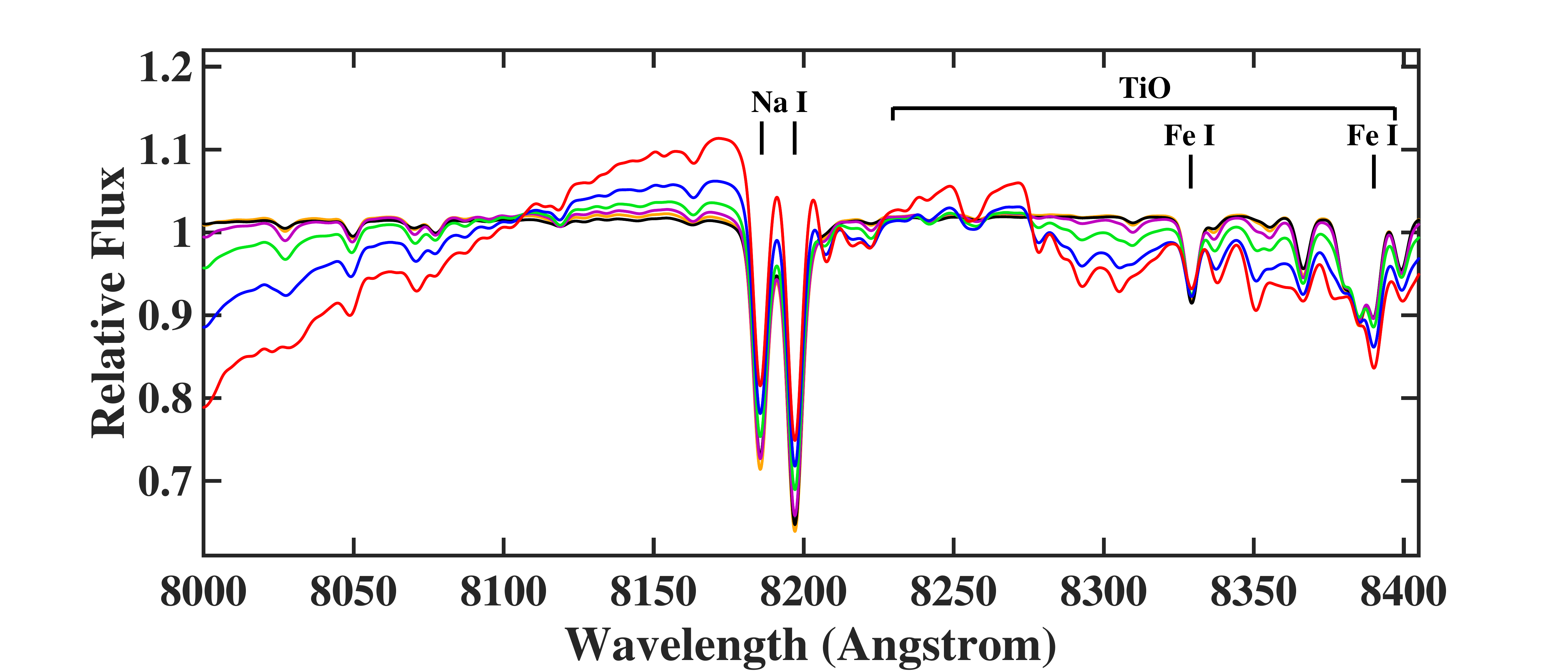}}
   \vspace{-0.38cm}
\subfloat
         {\includegraphics[  height=3.5cm, width=9cm]{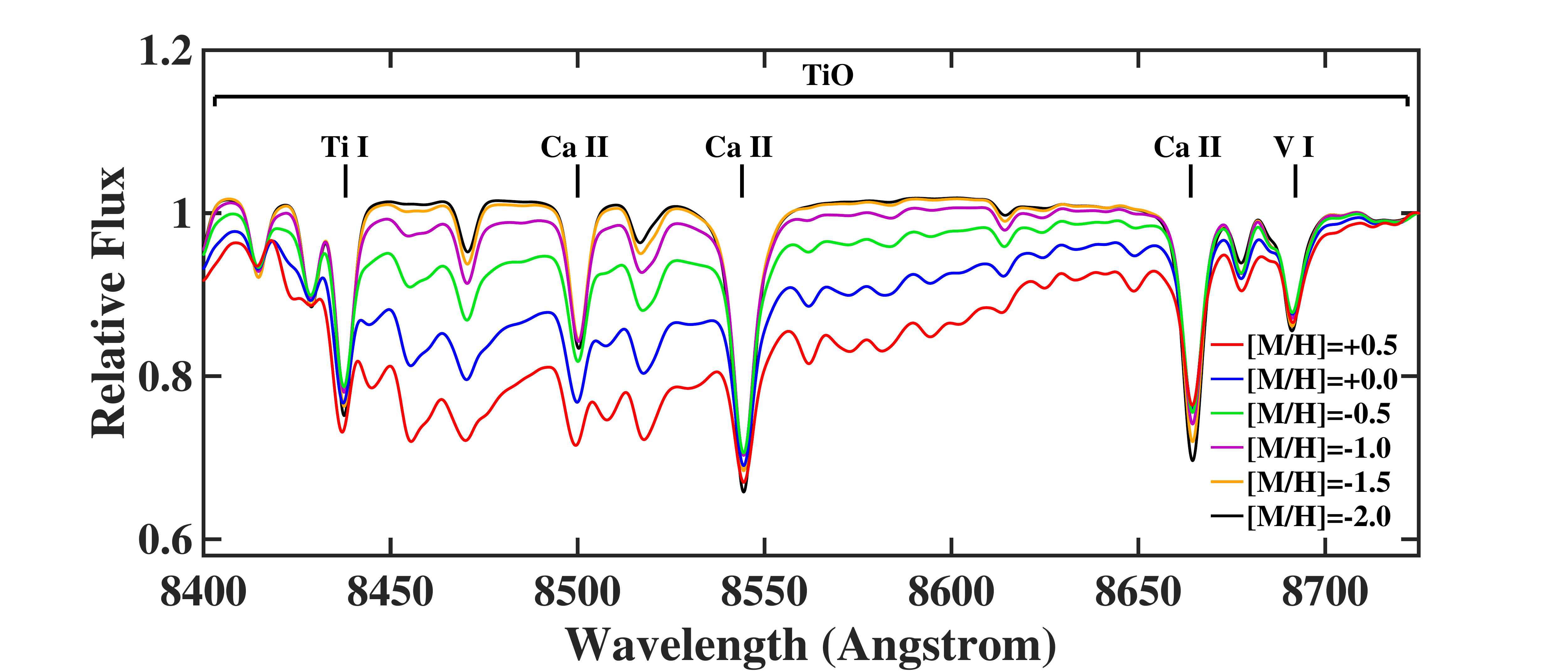}}
\caption
        {\footnotesize{Spectral sensitivity to metallicity: synthetic spectra with T$_\textrm{\footnotesize{eff}}$=3400 K, [$\alpha/$Fe]=+0.2 dex, log \emph{g}=5.0 dex, and [M/H]=$-$2.0, $-$1.5, $-$1.0, $-$0.5, 0.0, and +0.5 dex. The spectra are normalized at 8725 {\AA}.}}
\end{figure}

\begin{figure}[ht]
\centering
\subfloat
        {\includegraphics[ height=3.5cm, width=9cm]{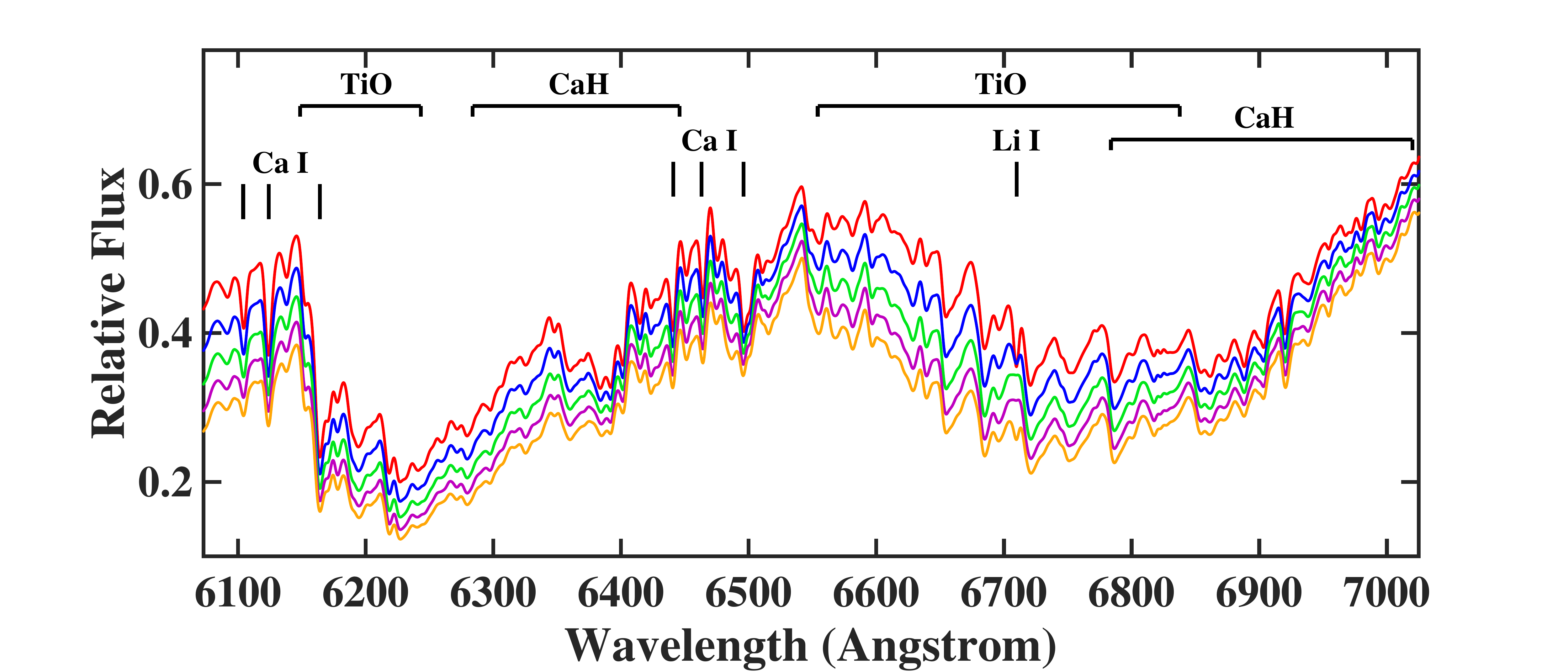}}
 \vspace{-0.38cm}
\subfloat
         {\includegraphics[height=3.5cm, width=9cm]{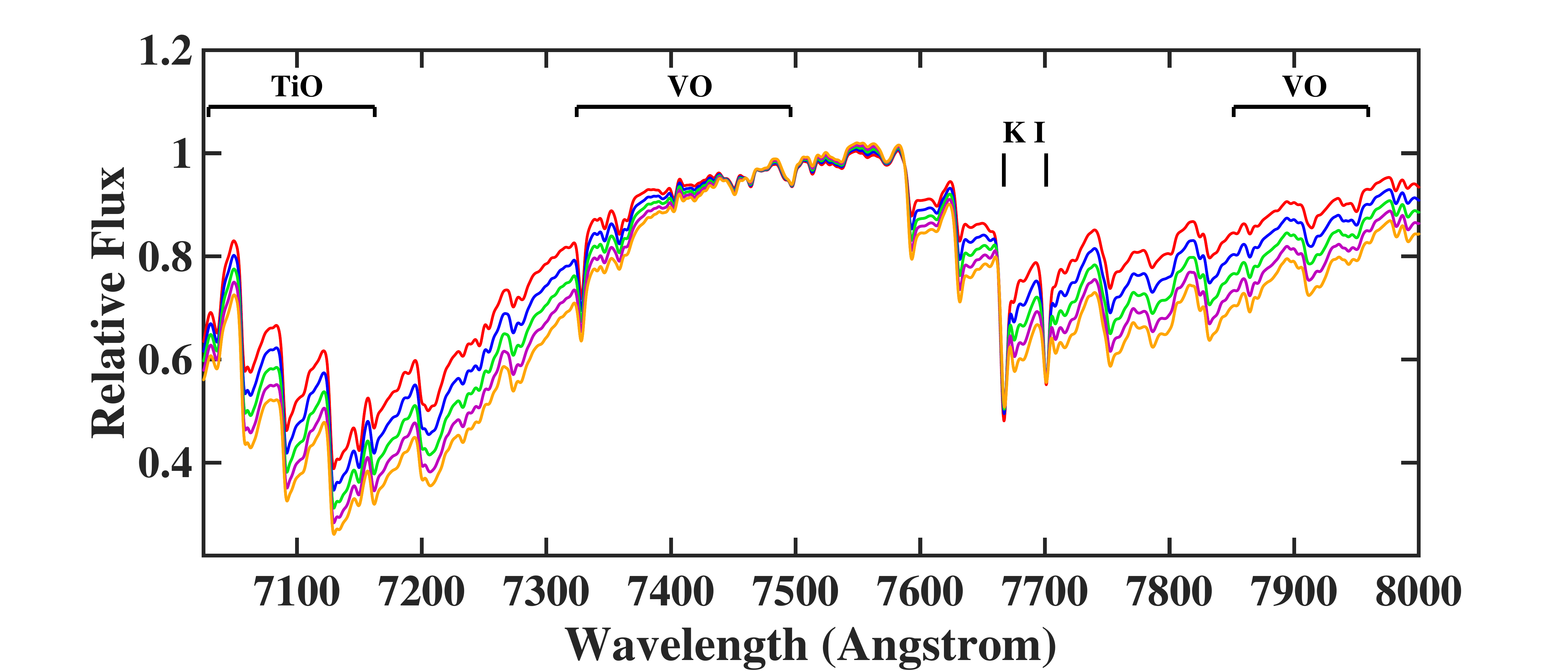}}
\vspace{-0.38cm}
\subfloat
        {\includegraphics[ height=3.5cm, width=9cm]{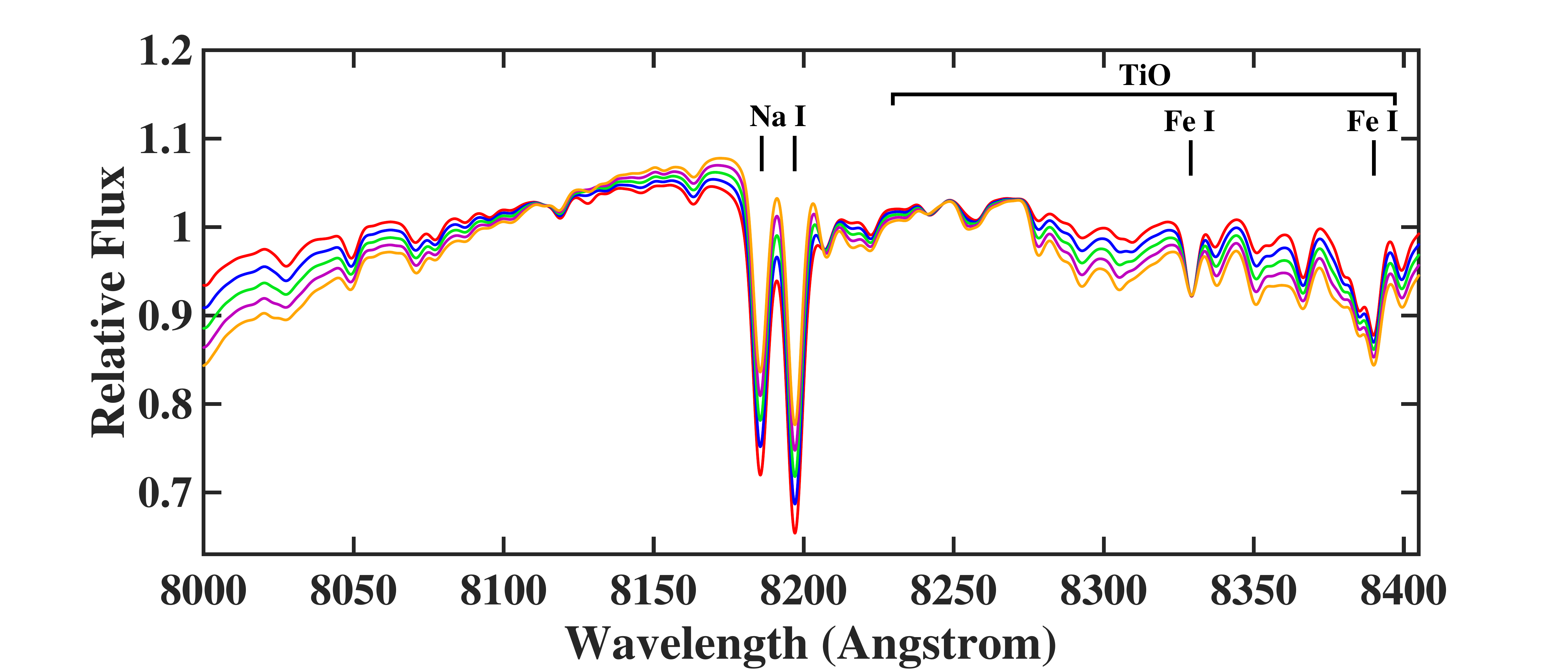}}
   \vspace{-0.38cm}
\subfloat
         {\includegraphics[ height=3.5cm, width=9cm]{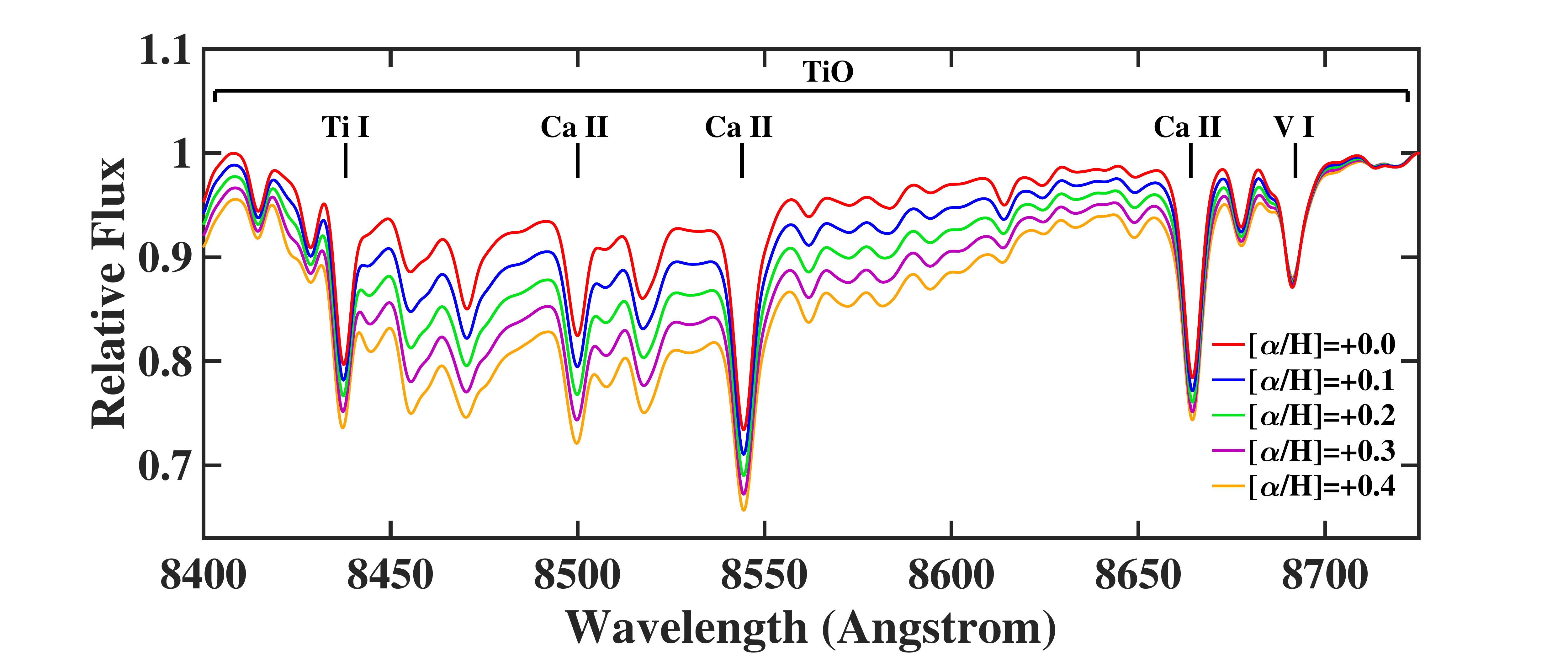}}
\caption
        {\footnotesize{Spectral sensitivity to $\alpha$-element enhancement: synthetic spectra with T$_\textrm{\footnotesize{eff}}$=3400 K, [M/H]=0.0 dex,  log \emph{g}=5.0 dex, and [$\alpha/$Fe]=0.0, +0.1, +0.2, +0.3, and +0.4 dex. The spectra are normalized at 8725 {\AA}.}}
\end{figure}

\begin{figure}[ht]
\centering
\subfloat
        {\includegraphics[ height=3.5cm, width=9cm]{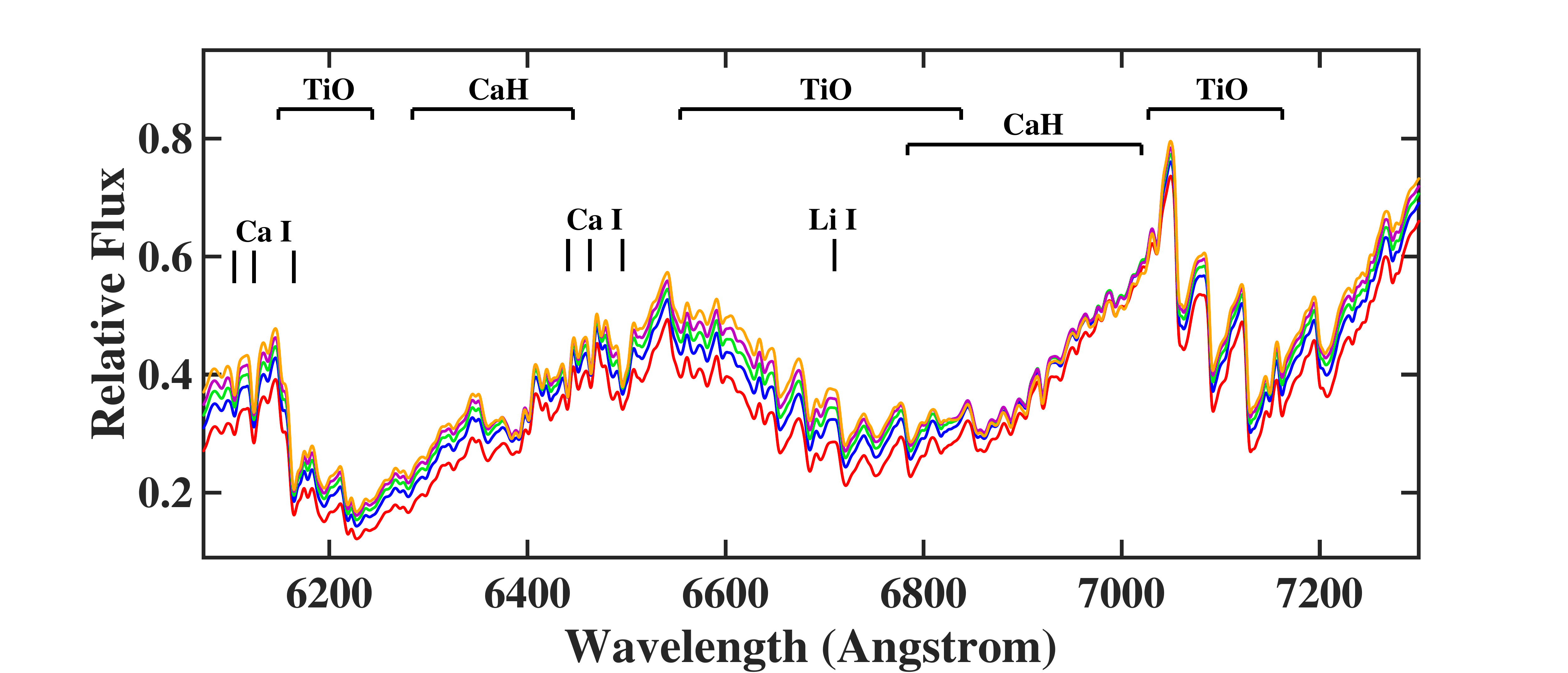}}
 \vspace{-0.38cm}
\subfloat
         {\includegraphics[height=3.5cm, width=9cm]{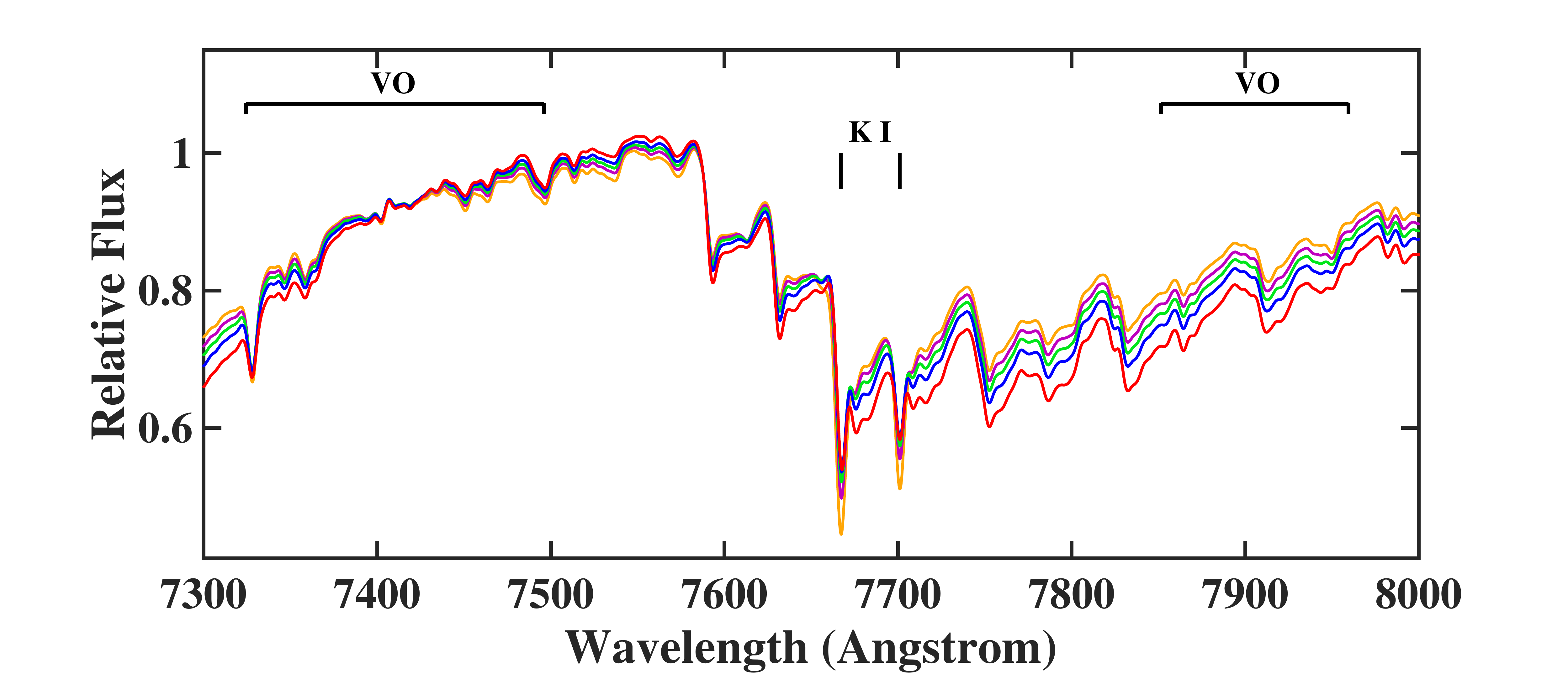}}
\vspace{-0.38cm}
\subfloat
        {\includegraphics[ height=3.5cm, width=9cm]{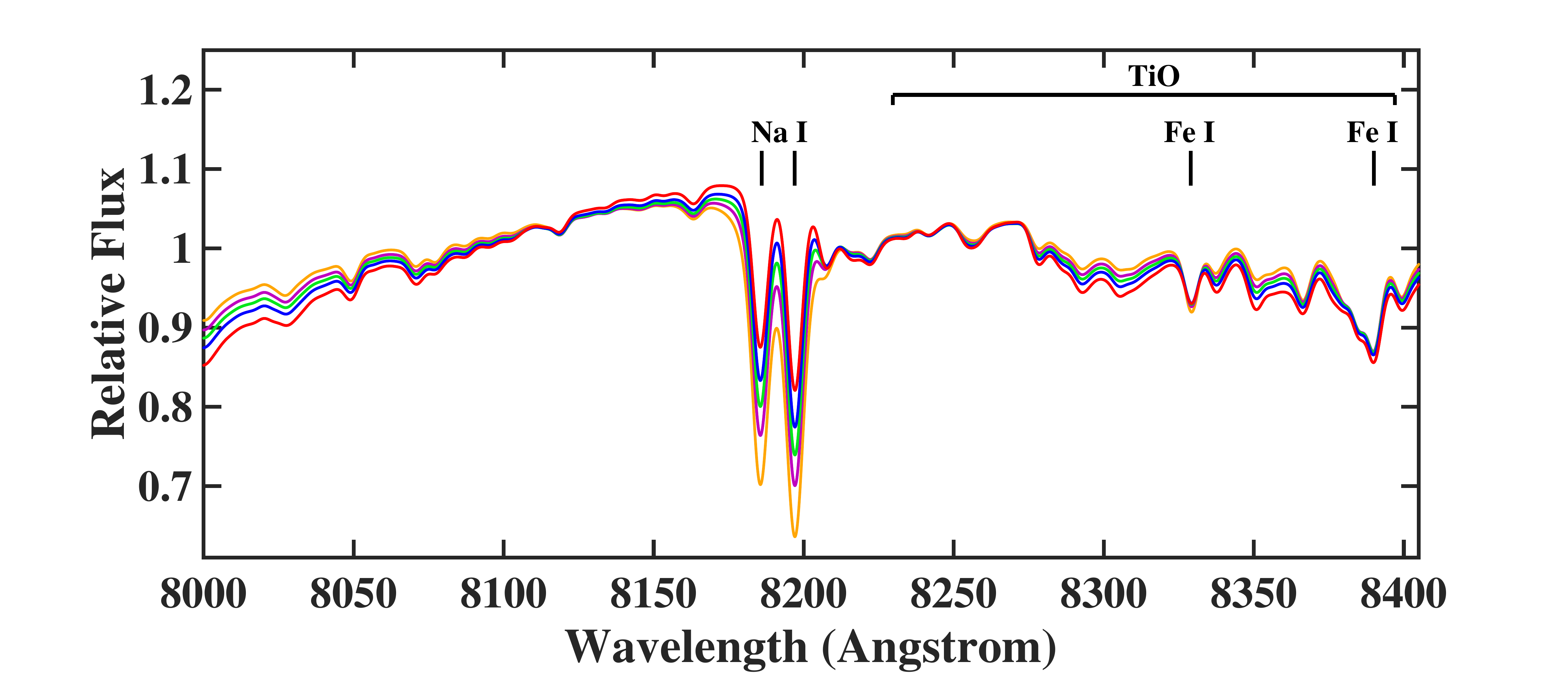}}
   \vspace{-0.38cm}
\subfloat
         {\includegraphics[ height=3.5cm, width=9cm]{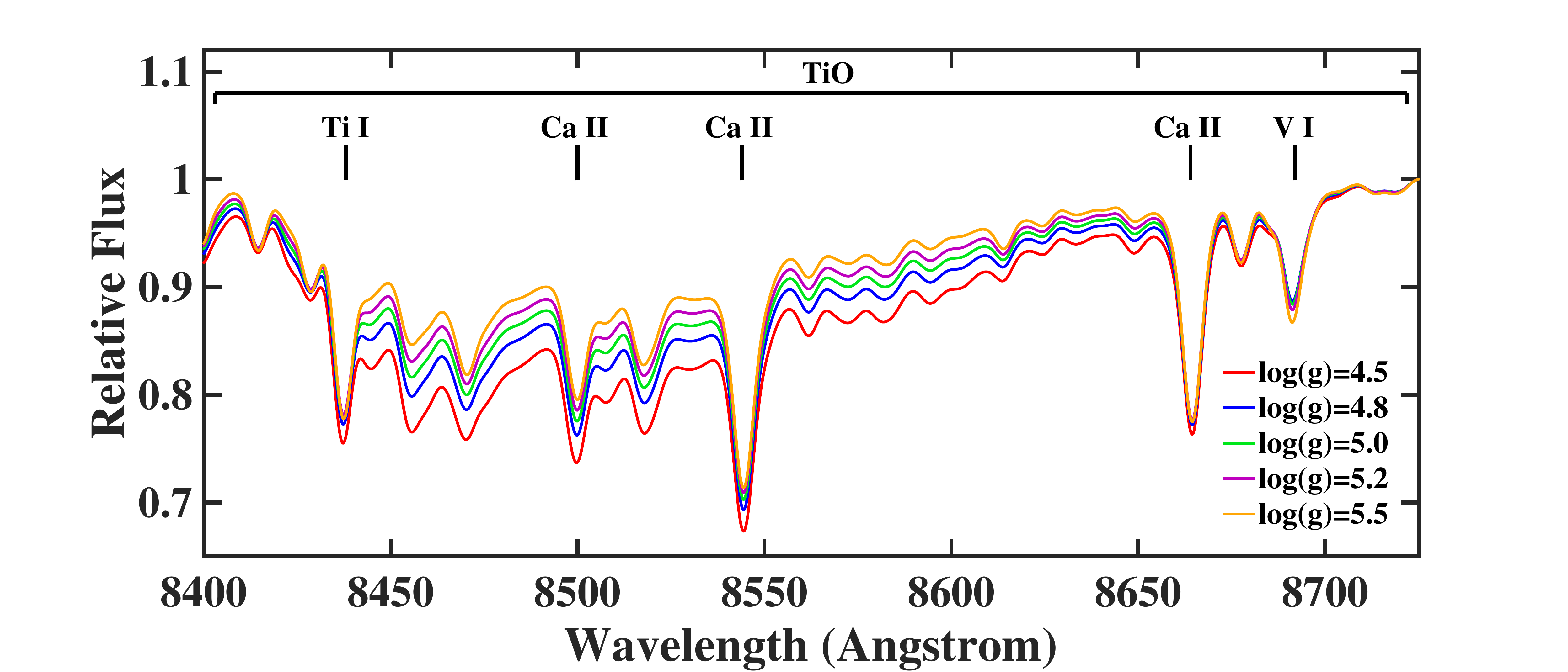}}
\caption
        {\footnotesize{Spectral sensitivity to surface gravity: synthetic spectra with T$_\textrm{\footnotesize{eff}}$=3400 K, [M/H]=0.0 dex, [$\alpha/$Fe]=+0.2 dex, and log \emph{g}=4.5, 4.8, 5.0, 5.2, and 5.5 dex. The spectra are normalized at 8725 {\AA}.}}
\end{figure}

\begin{figure}[ht]
\centering
\subfloat
        {\includegraphics[ height=3.5cm, width=9cm]{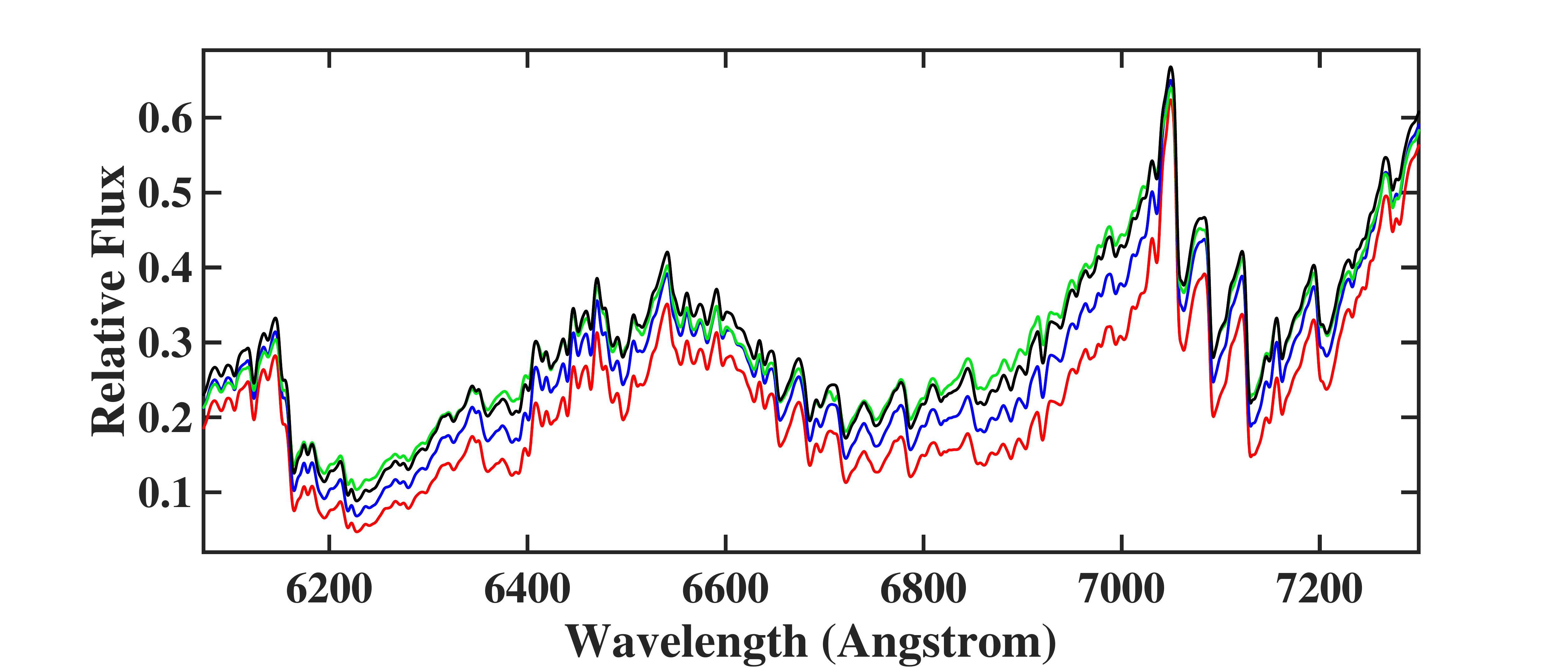}}
   \vspace{-0.38cm}

   \subfloat
         {\includegraphics[ height=3.5cm, width=9cm]{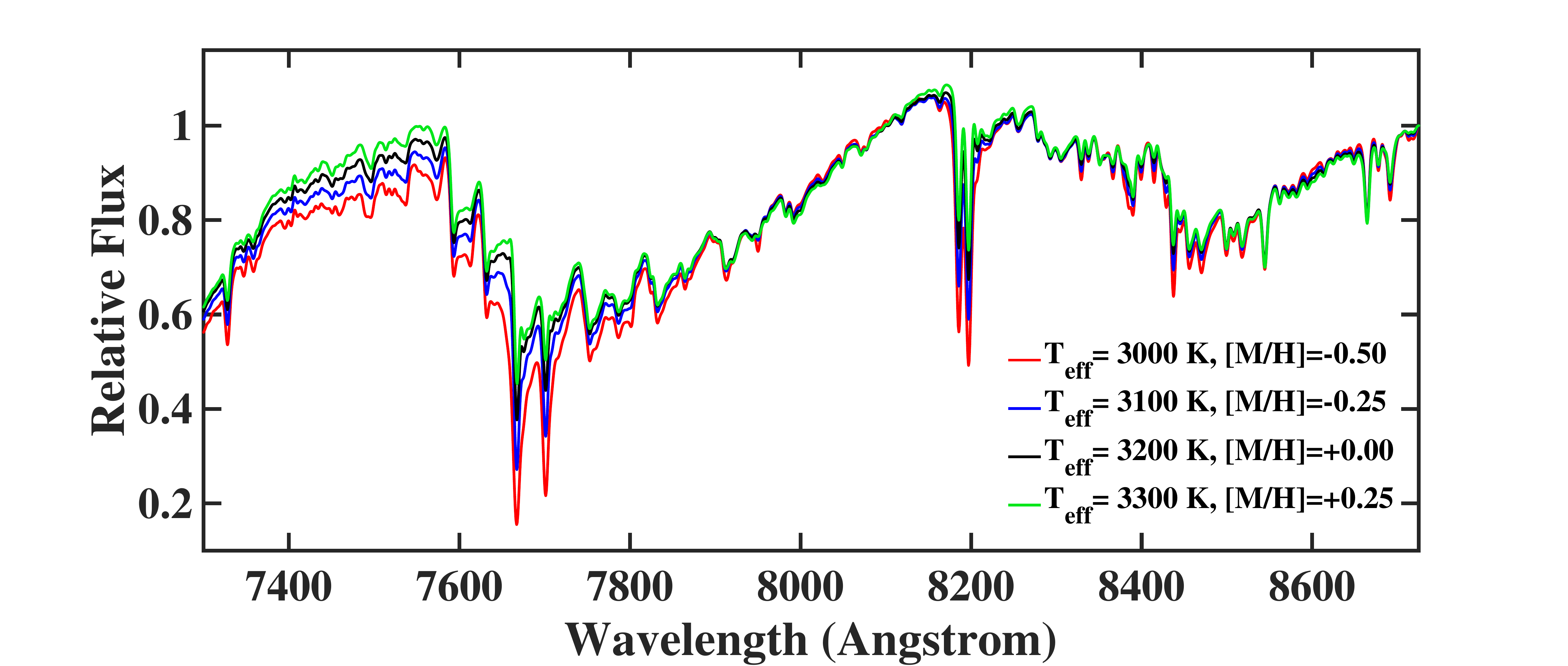}}   
 \caption
        {\footnotesize{Spectral degeneracy from effective temperature and metallicity: synthetic spectra with [$\alpha/$Fe]=0.2 dex, log \emph{g}=5.0, and  T$_\textrm{\footnotesize{eff}}$=3000, 3100, 3200, and 3300 K and [M/H]=$-$0.5, $-$0.25, 0.0, and +0.25 dex, respectively. The spectra are normalized at 8725 {\AA}.}}
    \end{figure}

\begin{figure}[ht]
\centering
\subfloat
        {\includegraphics[ height=3.5cm, width=9cm]{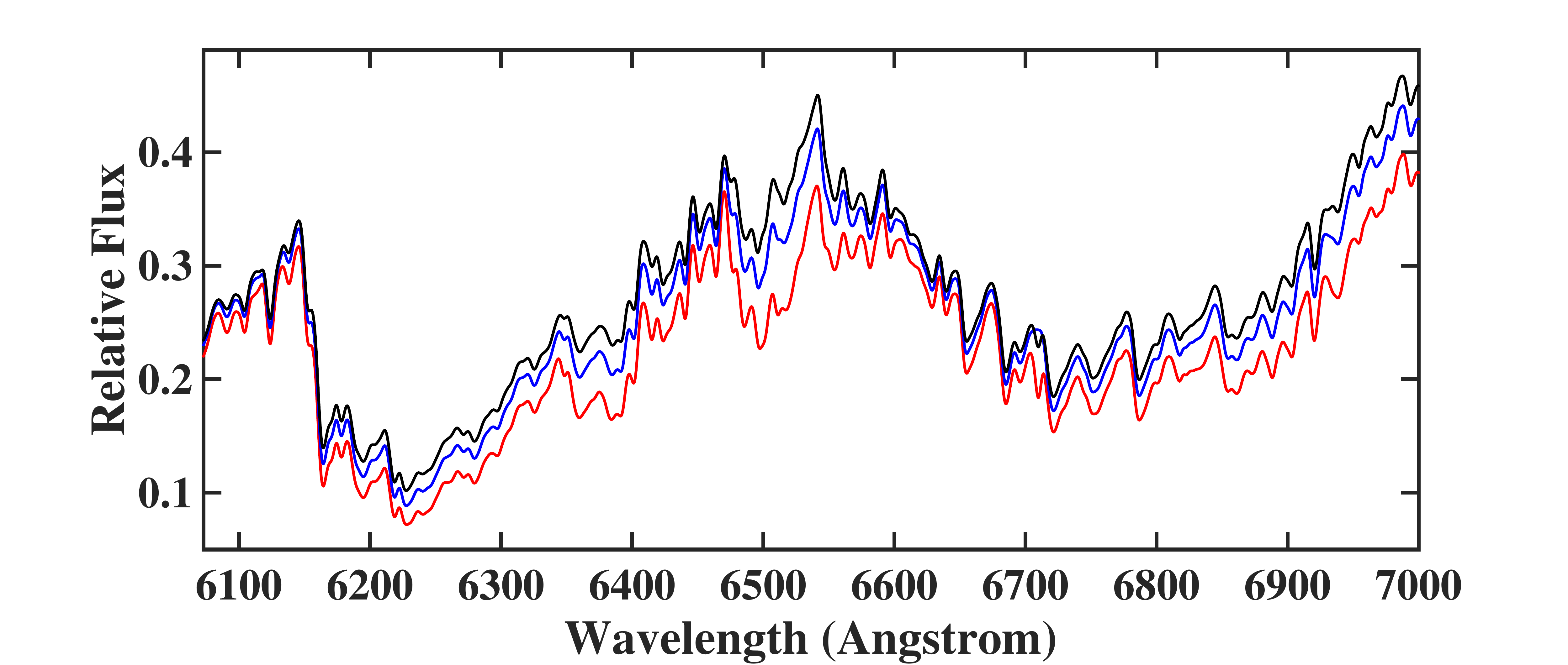}}
 \vspace{-0.38cm}

\subfloat
         {\includegraphics[ height=3.5cm, width=9cm]{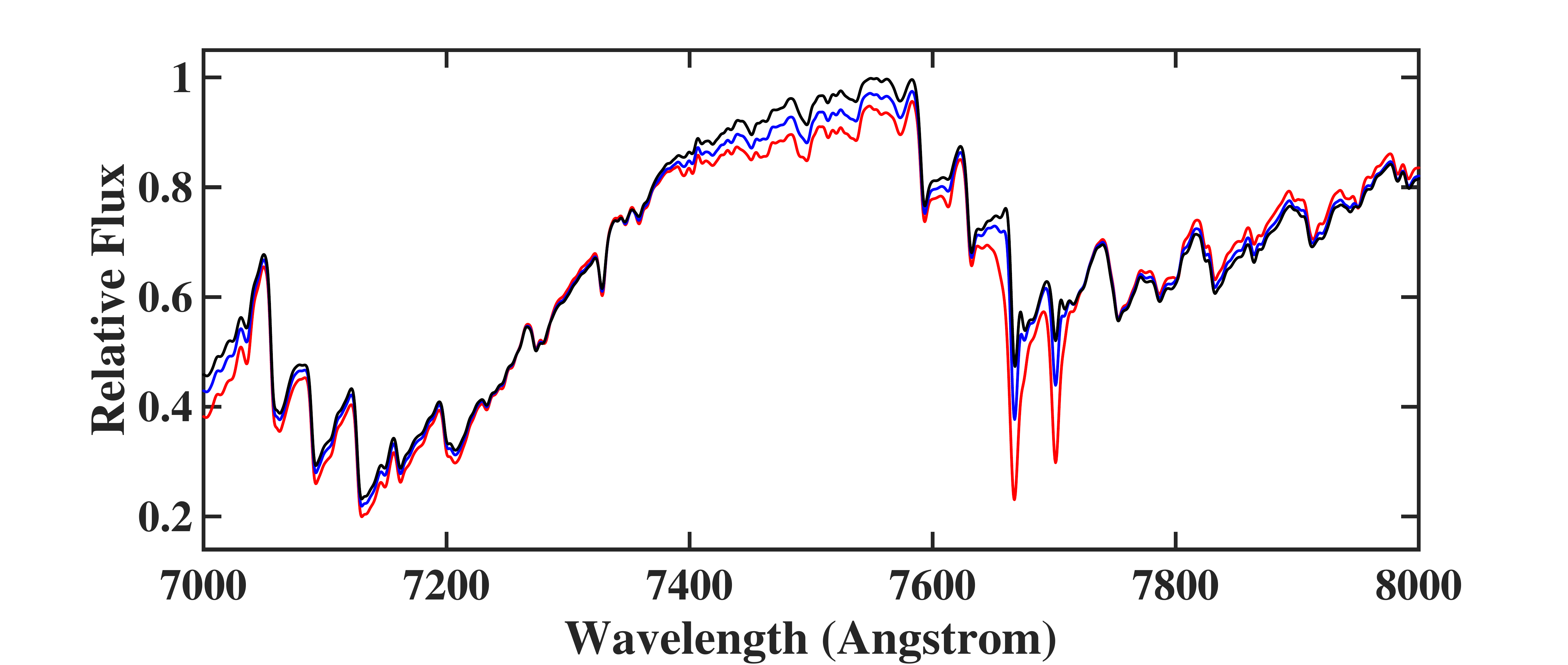}}
\vspace{-0.38cm}

\subfloat
        {\includegraphics[ height=3.5cm, width=9cm]{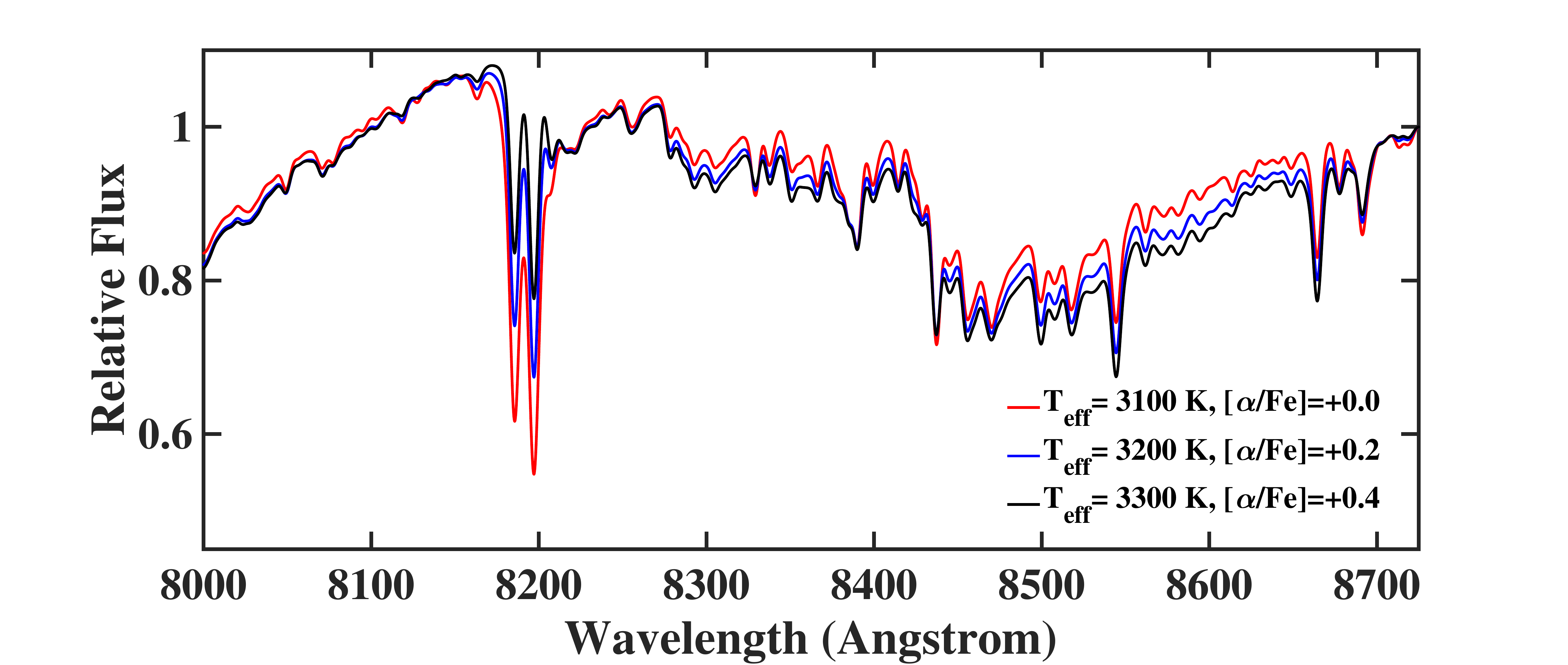}}
\caption
        {\footnotesize{Spectral degeneracy from effective temperature and $\alpha$-element enhancement: synthetic spectra with [M/H]=0.0 dex, log \emph{g}=5.0,  T$_\textrm{\footnotesize{eff}}$=3100, 3200, and 3300 K, and [$\alpha$/Fe]= 0.0, +0.2, and +0.4 dex, respectively. The spectra are normalized at 8725 {\AA}.}}
\end{figure}

\begin{figure}[ht]
\centering
\subfloat
        {\includegraphics[ height=3.5cm, width=9cm]{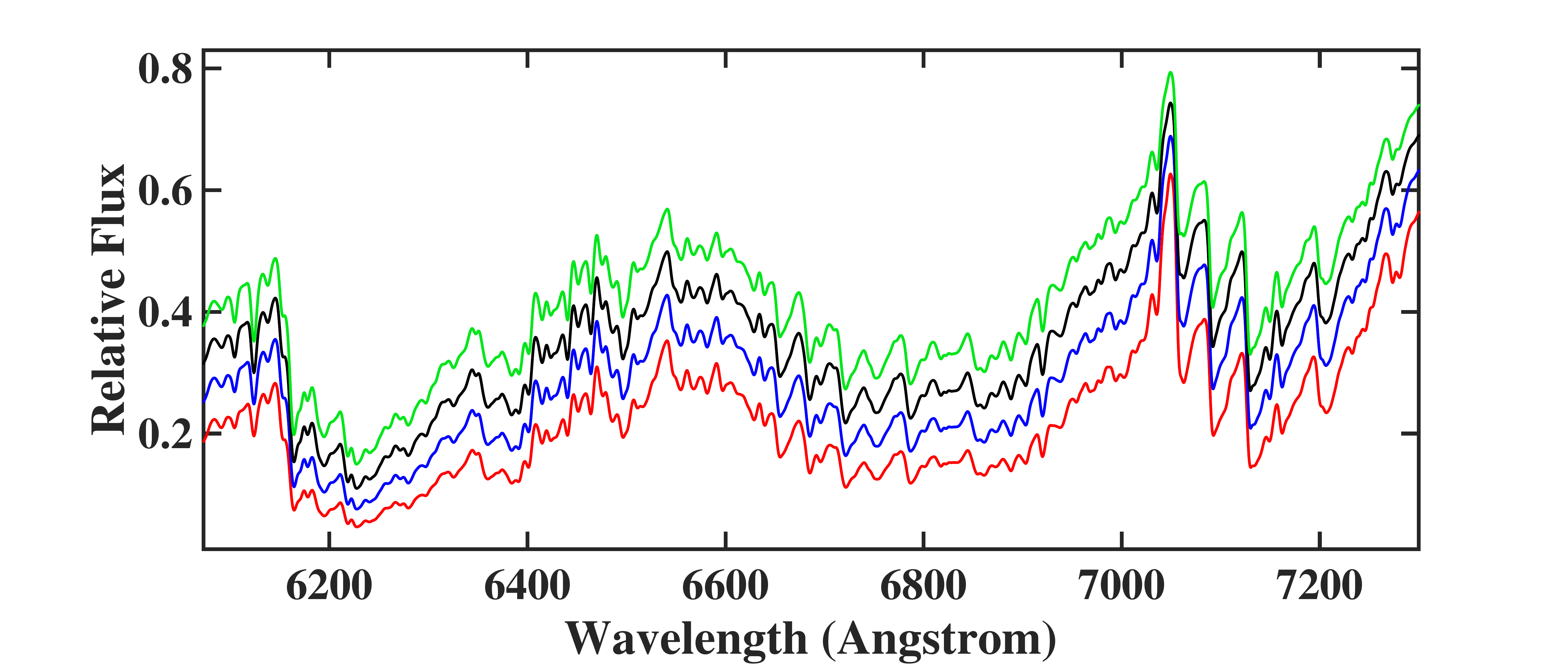}}
 \vspace{-0.38cm}

\subfloat
         {\includegraphics[ height=3.5cm, width=9cm]{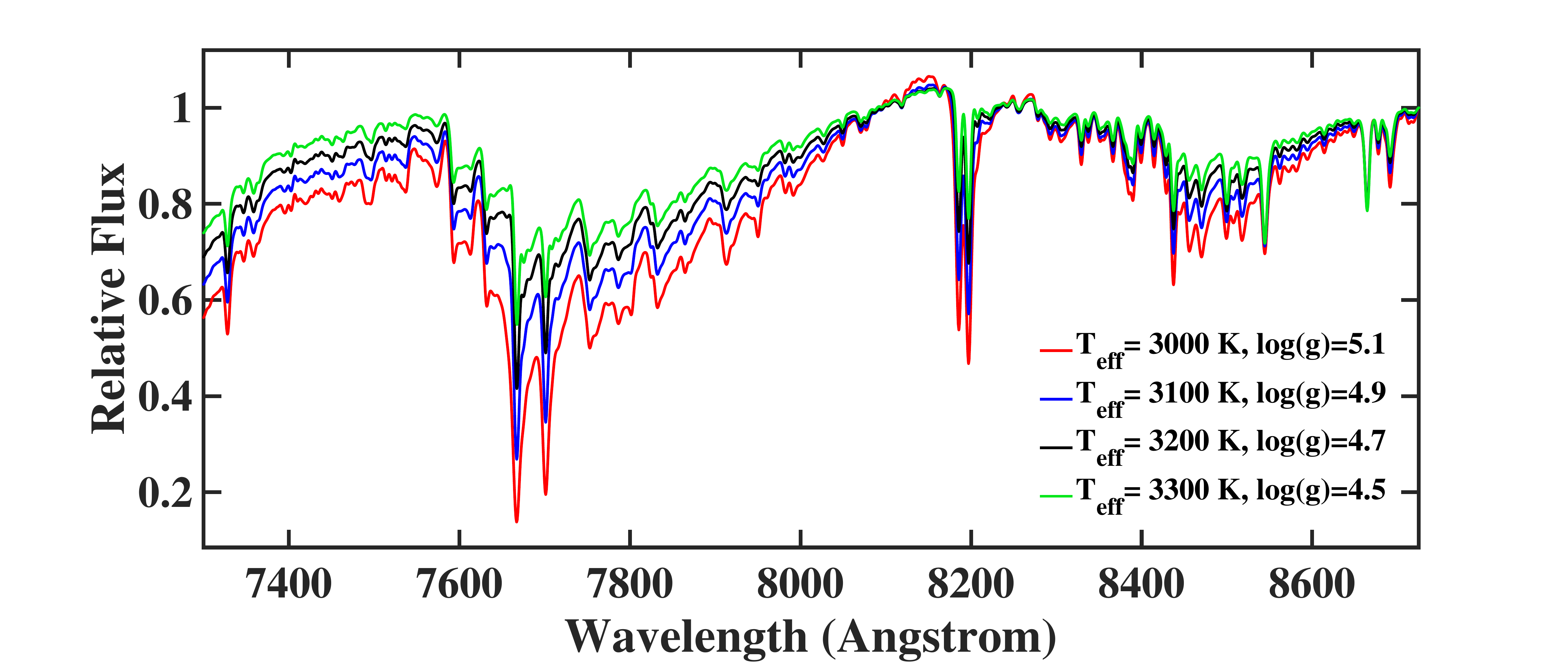}}
 \caption
        {\footnotesize{Spectral degeneracy from effective temperature and surface gravity: synthetic spectra with [M/H]=$-$0.5 dex, [$\alpha$/Fe]=+0.2 dex,    T$_\textrm{\footnotesize{eff}}$=3000, 3100, 3200,  and 3300 K, and log \emph{g}=5.1, 4.9, 4.7, and 4.5 dex, respectively. The spectra are normalized at 8725 {\AA}.}}
\end{figure}

\begin{figure}[ht]
\centering
\subfloat
        {\includegraphics[ height=3.5cm, width=9cm]{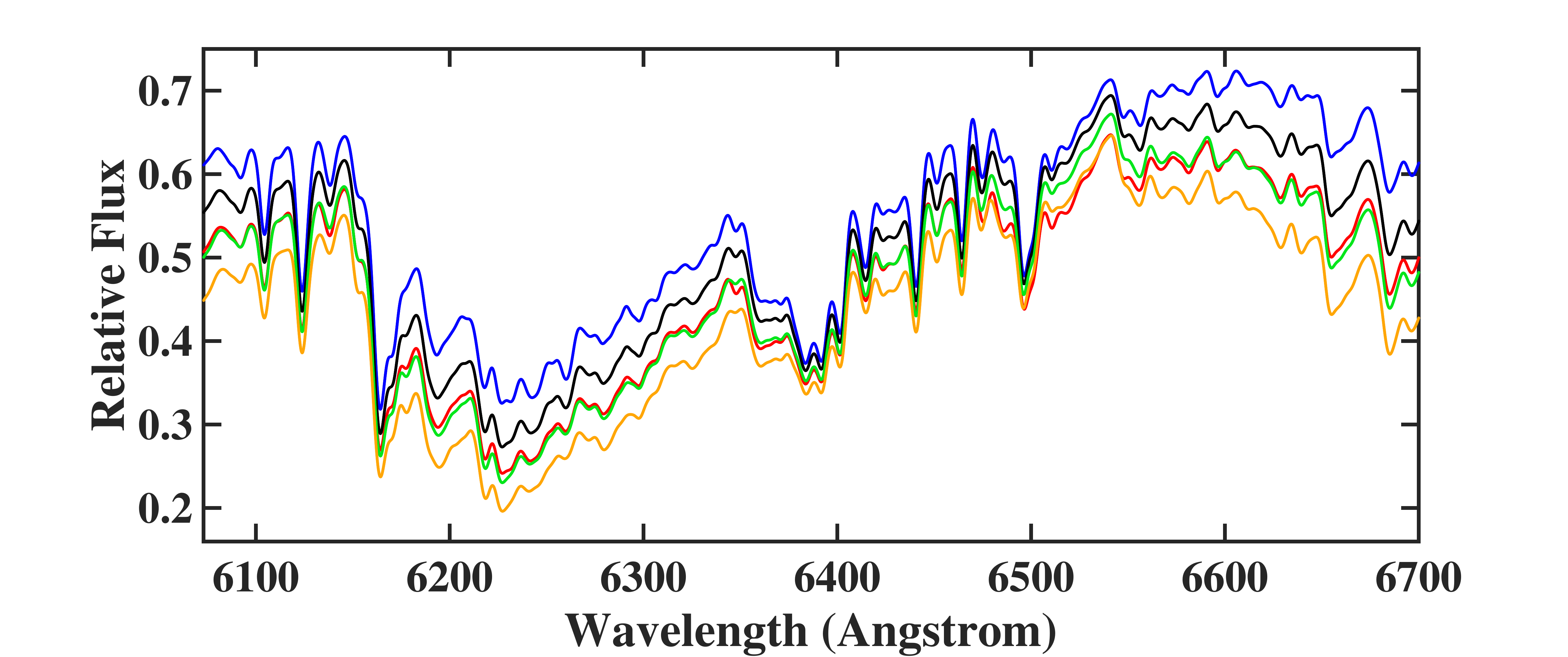}}
 \vspace{-0.38cm}
\subfloat
         {\includegraphics[ height=3.5cm, width=9cm]{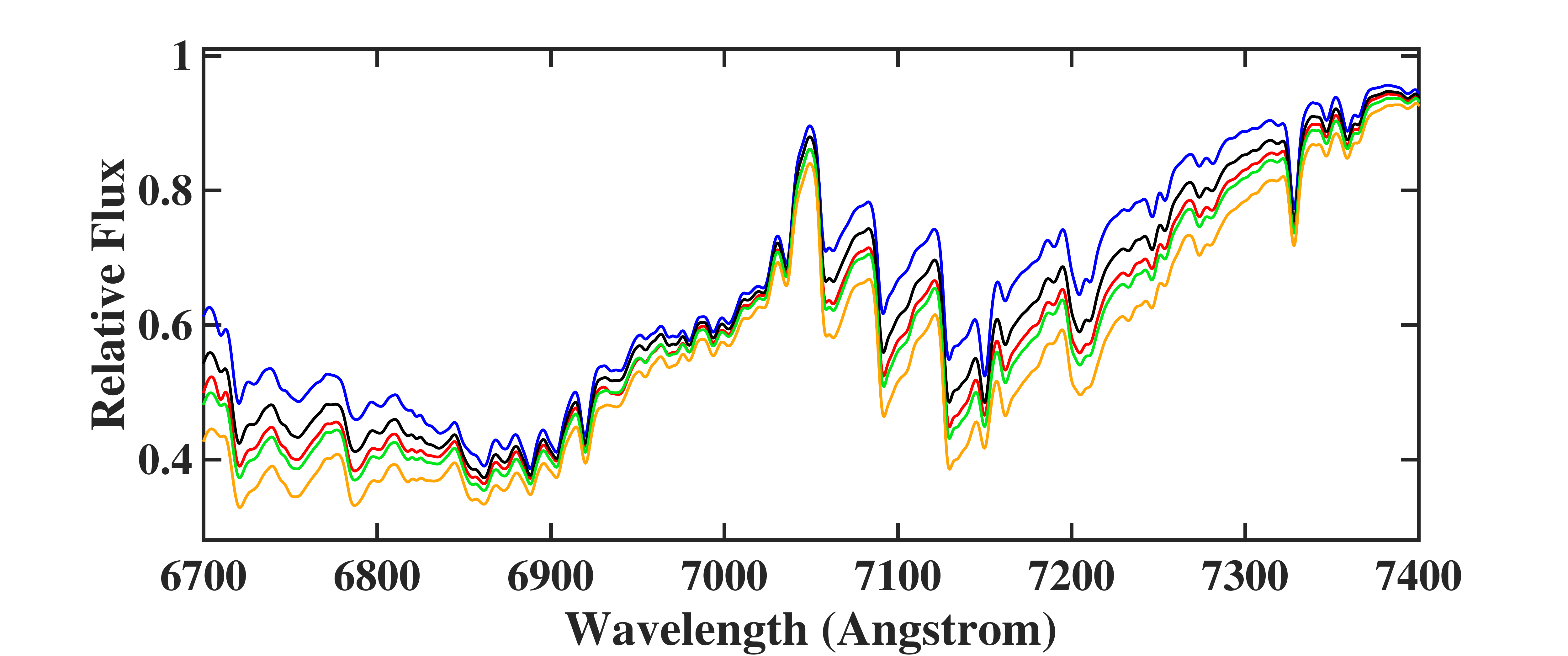}}
\vspace{-0.38cm}
\subfloat
        {\includegraphics[ height=3.5cm, width=9cm]{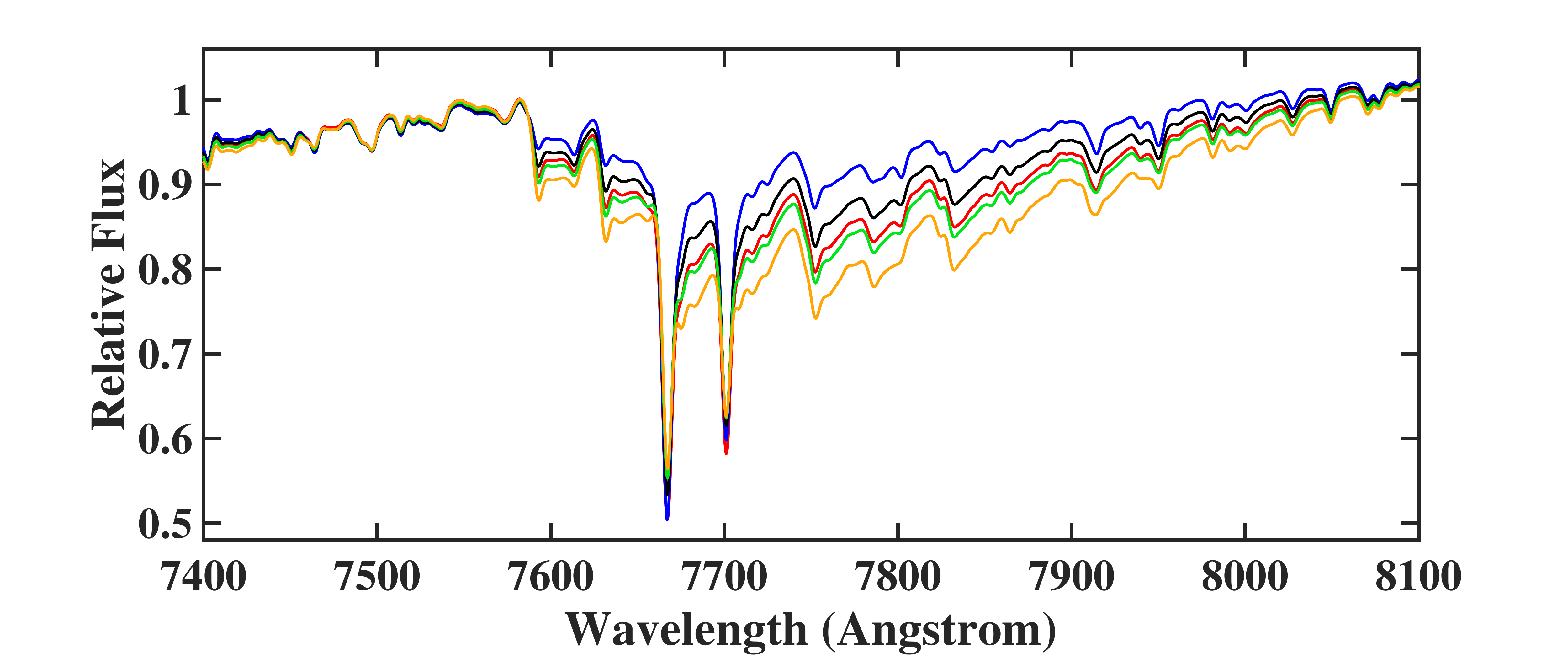}}
   \vspace{-0.38cm}
\subfloat
         {\includegraphics[height=3.5cm, width=9cm]{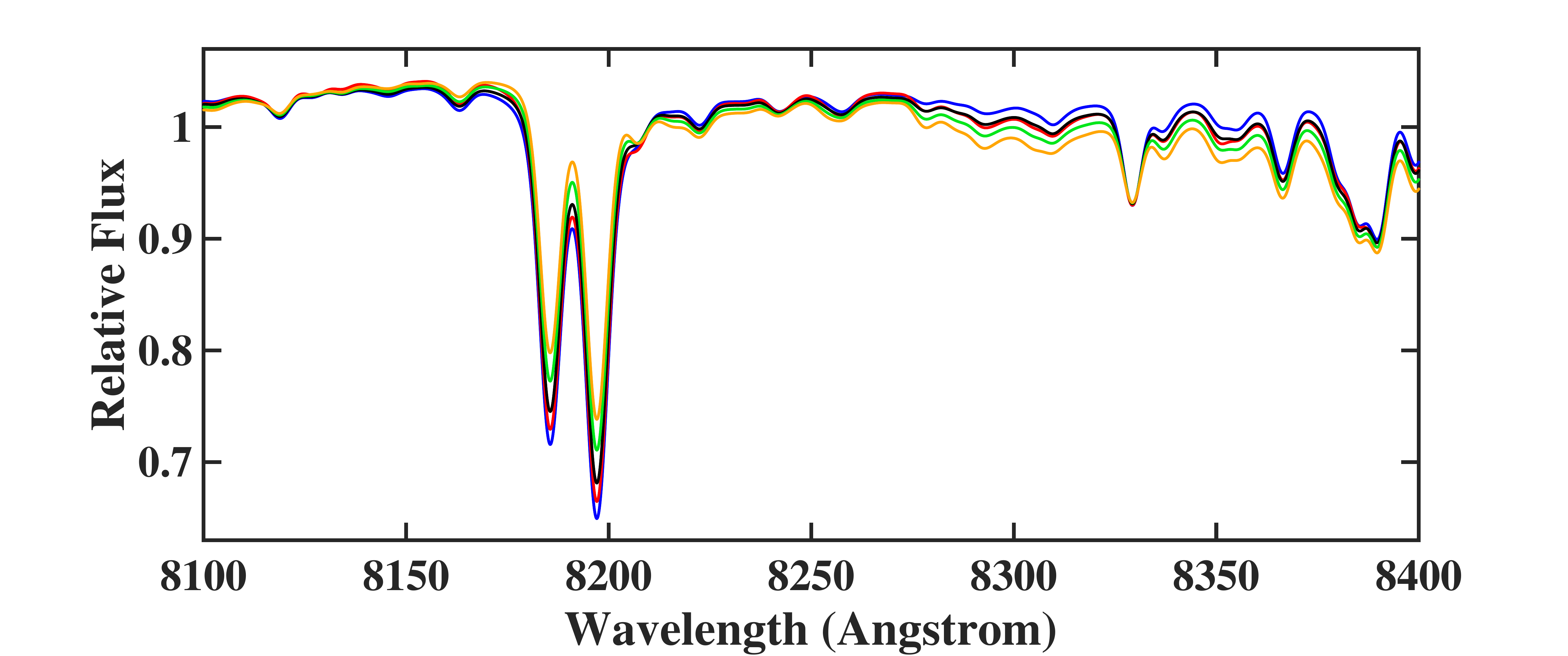}}
    \vspace{-0.38cm}
\subfloat
         {\includegraphics[ height=3.5cm, width=9cm]{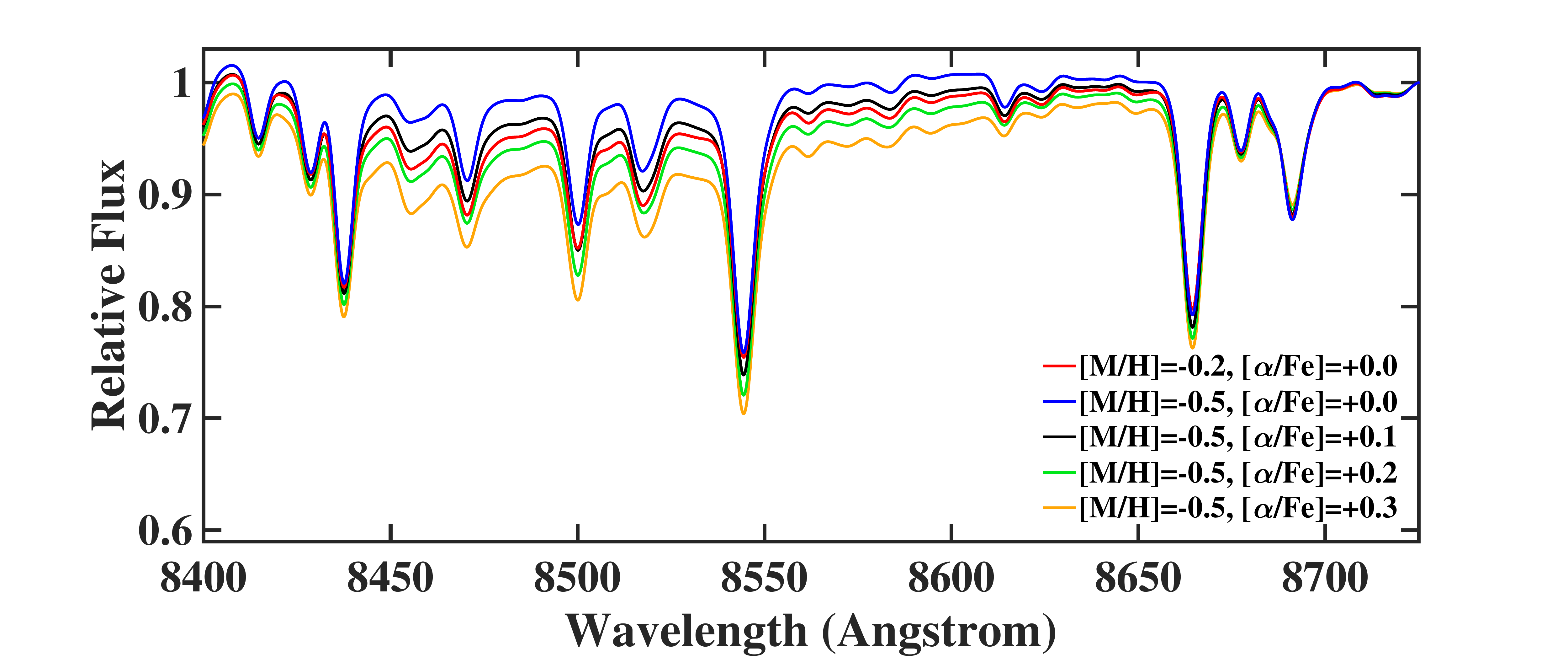}}
\caption
        {\footnotesize{Spectral degeneracy from metallicity and $\alpha$-element enhancement: synthetic spectra with T$_\textrm{\footnotesize{eff}}$=3400 K and log \emph{g}=5.0 dex, but five different combinations of [M/H] and [$\alpha$/Fe]: ([M/H]=$-$0.2 dex, [$\alpha$/Fe]=+0.0 dex), ([M/H]=$-$0.5 dex, [$\alpha$/Fe]=+0.0 dex), ([M/H]=$-$0.5 dex, [$\alpha$/Fe]=+0.1 dex), ([M/H]=$-$0.5 dex, [$\alpha$/Fe]=+0.2 dex), and ([M/H]=$-$0.5 dex, [$\alpha$/Fe]=+0.3 dex). The spectra are normalized at 8725 {\AA}.}}
\end{figure}

\begin{figure}[ht]
\centering
\subfloat
        {\includegraphics[ height=3.5cm, width=9cm]{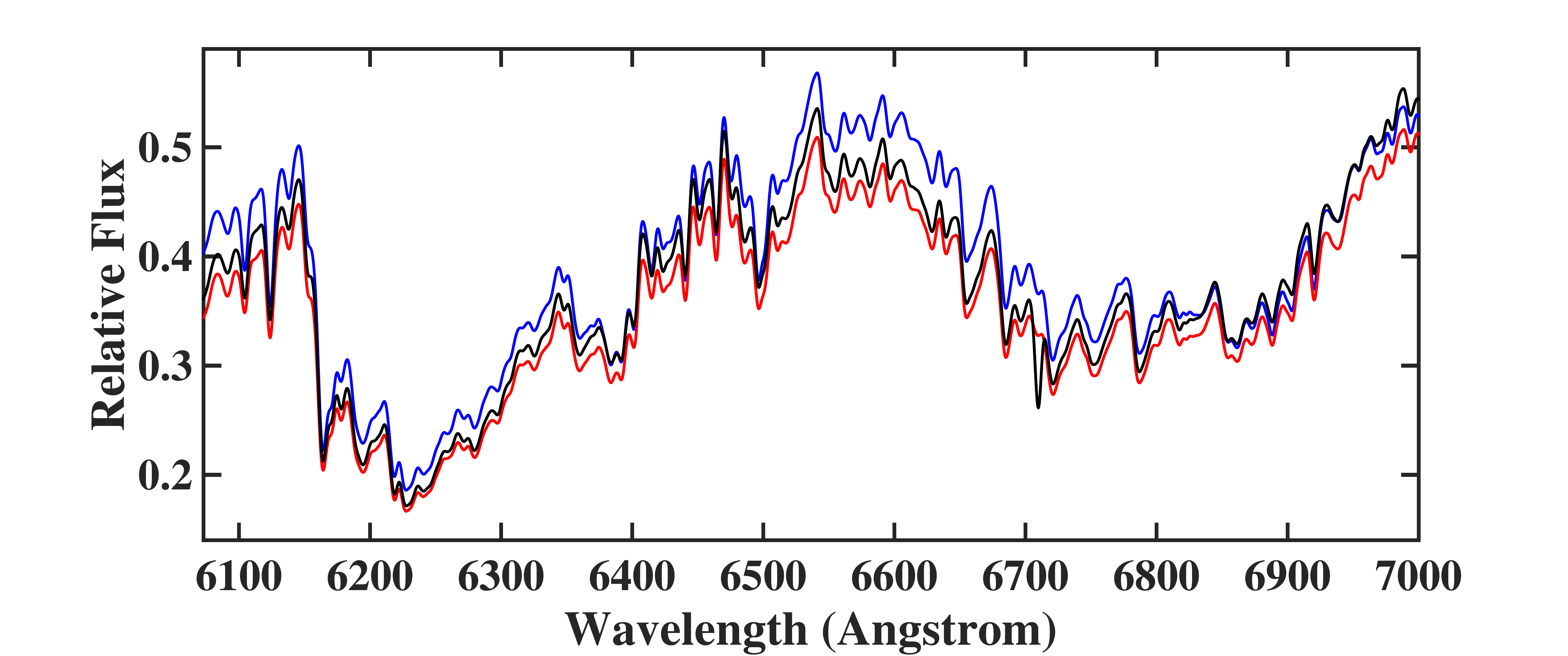}}
 \vspace{-0.38cm}

\subfloat
         {\includegraphics[ height=3.5cm, width=9cm]{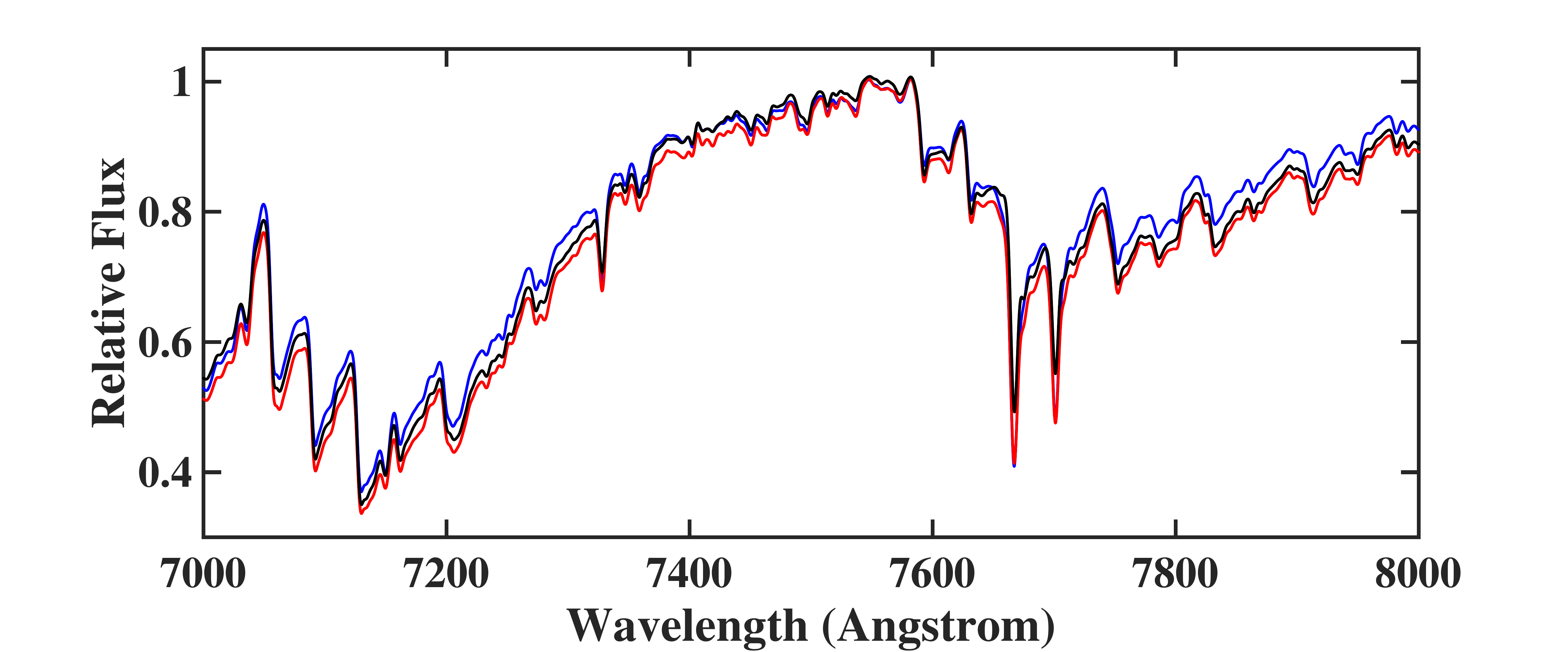}}
 \vspace{-0.38cm}

\subfloat
        {\includegraphics[ height=3.5cm, width=9cm]{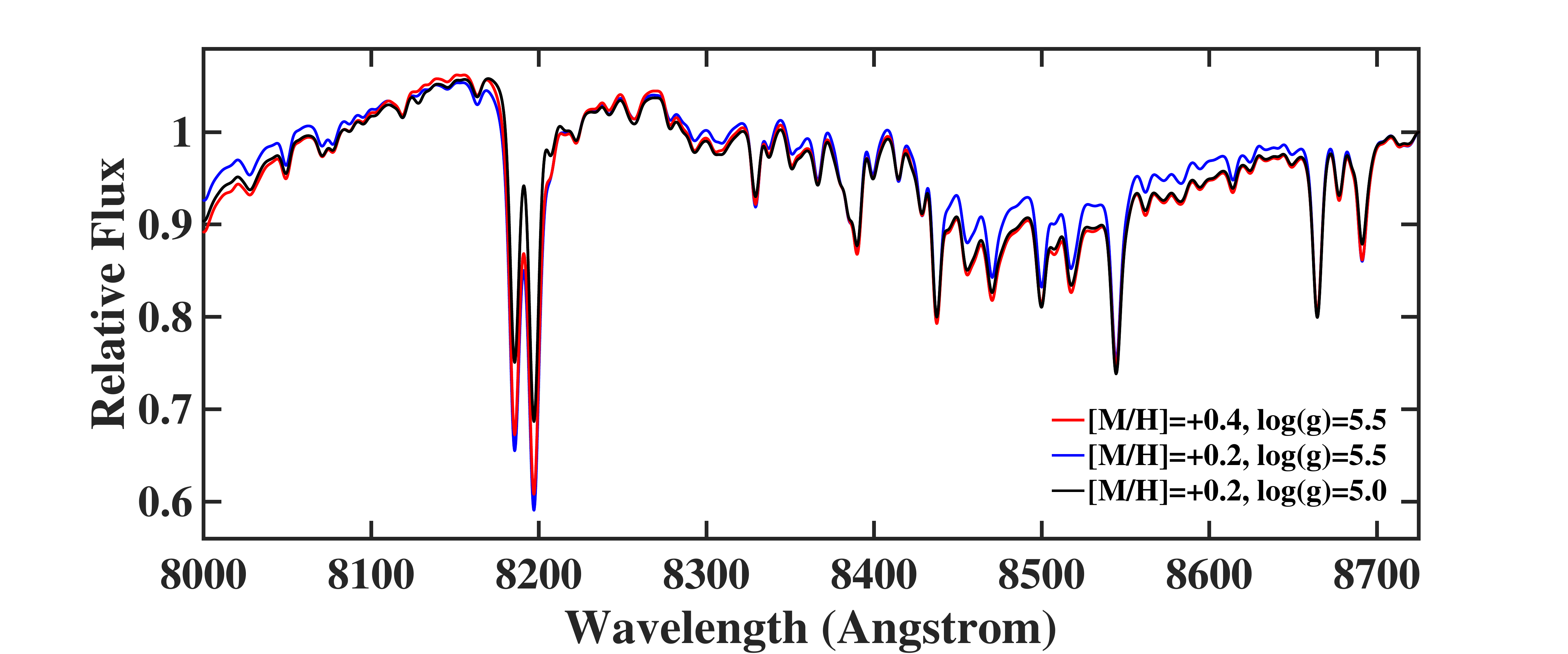}}
\caption
        {\footnotesize{Spectral degeneracy from metallicity and surface gravity: synthetic spectra with  T$_\textrm{\footnotesize{eff}}$=3400 K,  [$\alpha$/Fe]=0.0 dex, [M/H]=+0.2, +0.2, and +0.4 dex, and log \emph{g}=5.0, 5.5, and 5.5 dex, respectively. The spectra are normalized at 8725 {\AA}.}}
\end{figure}

\begin{figure}[ht]
\centering
\subfloat
        {\includegraphics[height=3.5cm, width=9cm]{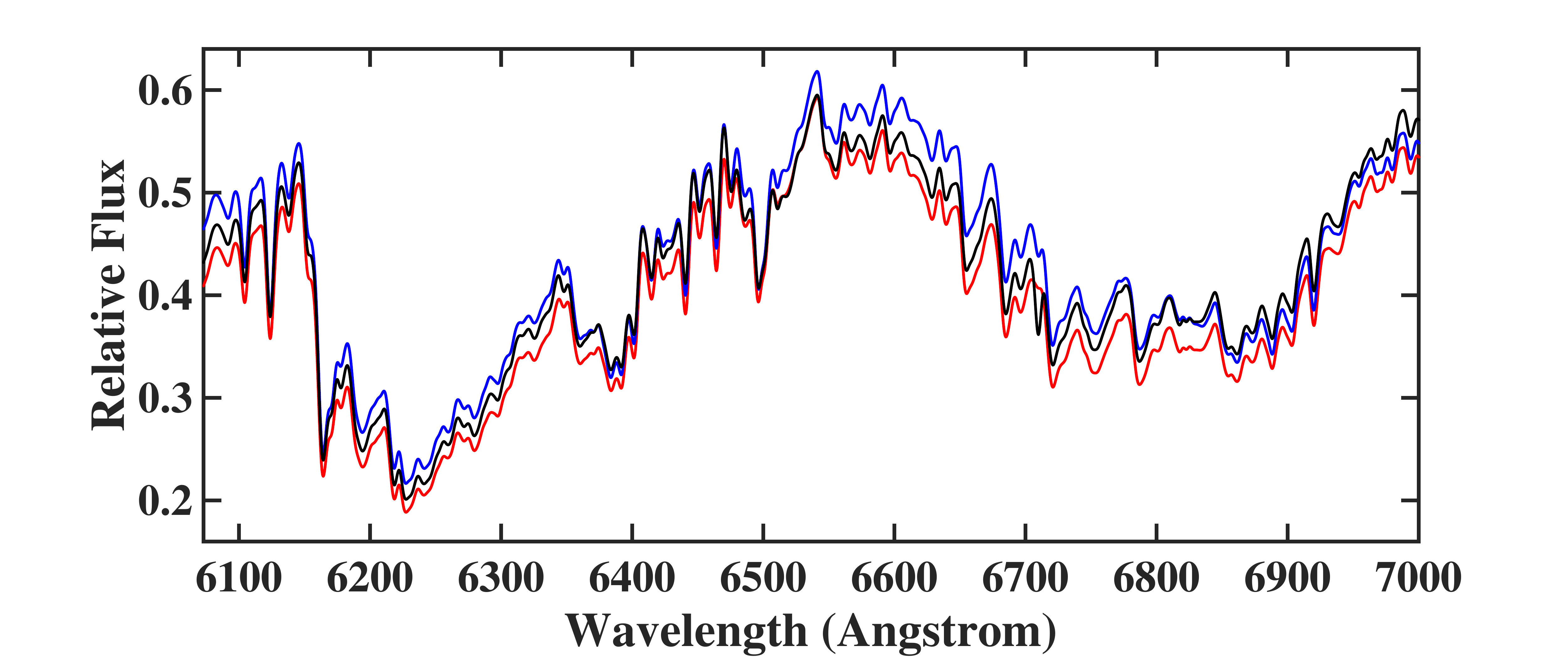}}
\vspace{-0.38cm}
\subfloat
         {\includegraphics[height=3.5cm, width=9cm]{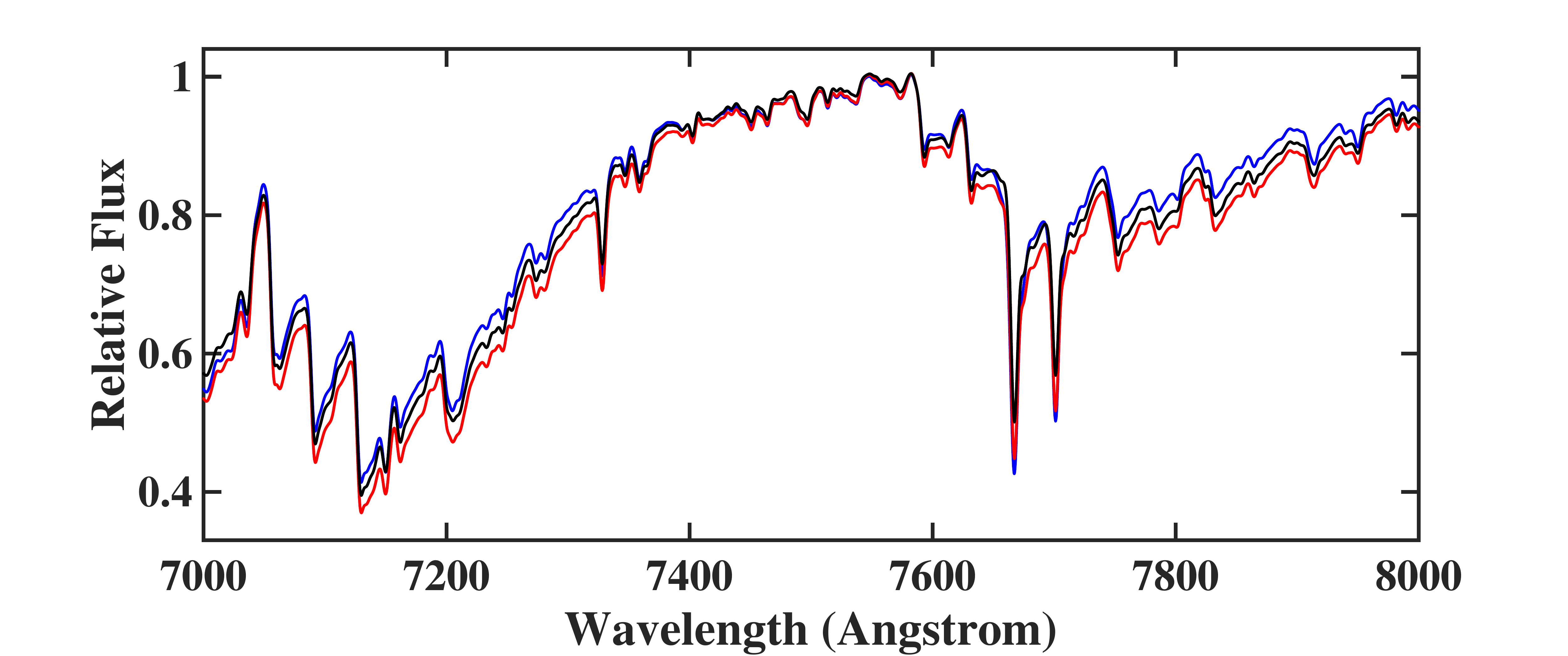}}
\vspace{-0.38cm}
\subfloat
        {\includegraphics[height=3.5cm, width=9cm]{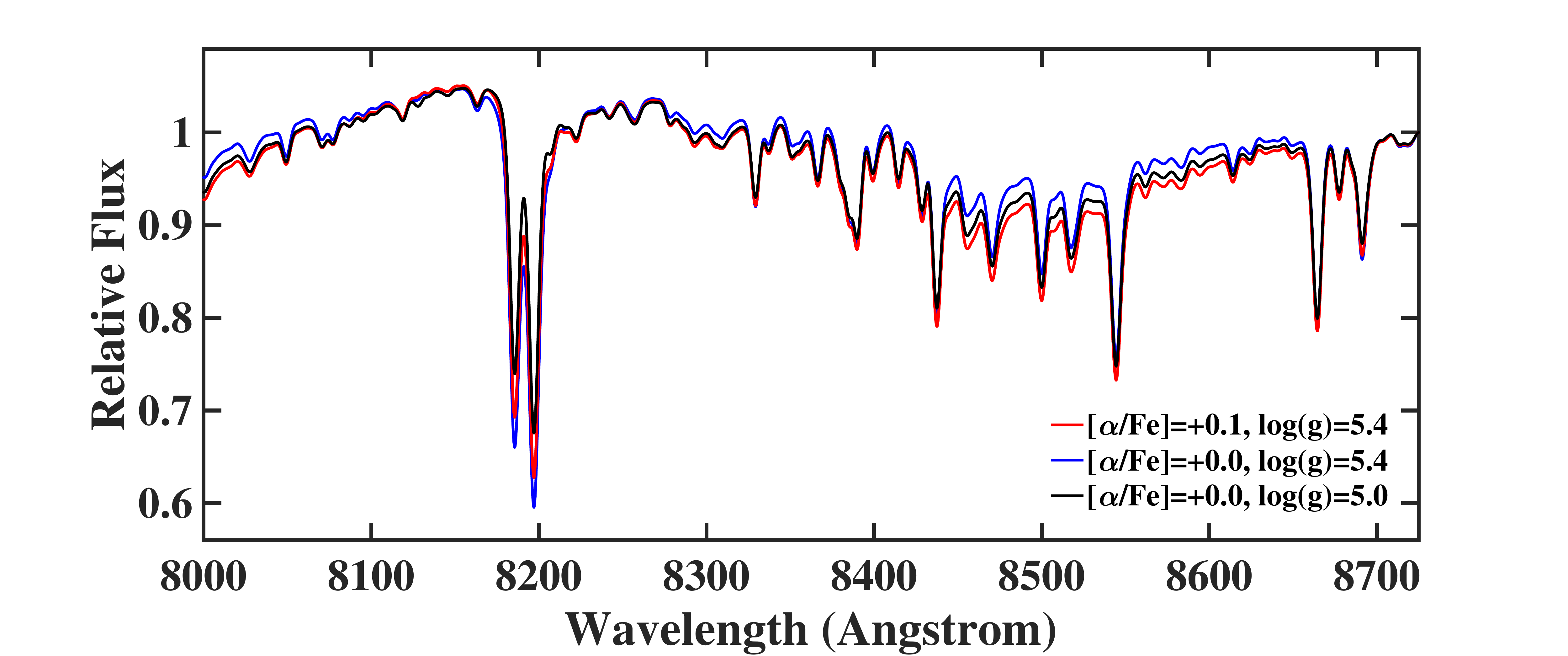}}
\caption
        {\footnotesize{Spectral degeneracy from $\alpha$-element enhancement and surface gravity: synthetic spectra with  T$_\textrm{\footnotesize{eff}}$=3400 K,  [MH]=0.0 dex,  [$\alpha$/Fe]=+0.0, +0.0, and +0.1 dex, and log \emph{g}=5.0, 5.4, and 5.4 dex, respectively. The spectra are normalized at 8725 {\AA}.}}
\end{figure}

The influence of $\alpha$-element enhancement on spectral shape is shown in Figure 3, which displays the spectra with [$\alpha/$Fe]=0.0, +0.1, +0.2, +0.3, and +0.4 dex but all having the same T$_\textrm{\footnotesize{eff}}$=3400 K, [M/H]=0.0 dex, log \emph{g}=5.0 dex.  As can be seen from the figure, higher values of [$\alpha/$Fe] generally cause a higher degree of flux depression from molecular bands. The atomic lines are remarkably influenced by the background molecular opacities, but in different ways. For example, the $\alpha$-element lines, i.e., Ca I, Ca II,  and Ti, are significantly blended with the ambient molecular bands at higher [$\alpha/$Fe] values. On the other hand, there is an opposite trend for the other atomic lines such as the doublet Na I that grow stronger when [$\alpha/$Fe] decreases. Decreasing the abundances of  $\alpha$ elements may lead to  increasing the abundances of some other elements  to keep  the overall metallicity constant for the plotted model spectra, though many factors have to be taken into account to completely explain the effect of  [$\alpha/$Fe] on spectral lines and features.

Figure 4 demonstrates the change of spectral structure due to the variation of surface gravity, showing the spectra with different values of log \emph{g}=4.5, 4.8, 5.0, 5.2, and 5.5 dex but the same T$_\textrm{\footnotesize{eff}}$=3400 K, [M/H]=0.0 dex, and [$\alpha/$Fe]=+0.2 dex. Clearly, surface gravity has a weaker effect on overall flux and spectral shape, even over a wide  range of 1.0 dex, as compared to the other parameters.  Nevertheless, this parameter plays an important role in the determination of the other stellar parameters using  synthetic model fitting (Section 6). The sensitivity of spectral lines to  log \emph{g} varies from one element to another.  For instance, the doublet K I and Na I become stronger as log \emph{g} increases, which can be explained by the variation of  the population of atomic states with pressure.  Based on numerical tests, we find that for a typical  model with a low value of  log \emph{g}, the majority of  Na  and K atoms are  ionized while Ca and Ti are mainly  neutral.  Increasing log \emph{g} thus increases the population of neutral Na I and K I atoms, but has no effect on the population of neutral Ca I and Ti I atoms.

All these examinations imply that each parameter has a unique effect on the flux level and the full structure of synthesized spectra over a wide spectral range. Some combinations of the parameters
shown in Figures 1-4 may not, however, happen in real stars. For example, since  M dwarfs don't age,  a change in log \emph{g} for a fixed mass (or T$_\textrm{\footnotesize{eff}}$) must be due to changes in the abundance scale. In other words, stars with the same T$_\textrm{\footnotesize{eff}}$ and the same chemical parameters will have nearly the same log \emph{g}. This means while other parameters are constant, the variation of log \emph{g} will be extremely narrow in reality, as compared to the broad range  (1.0 dex)  presented  in Figure 4.

In effect,  the spectral sensitivity to a particular physical parameter  depends  on the value of the other parameters.  The shape of spectral lines and features changes in a complex  way when more than one parameter vary, and for this reason,  to fully understand the exact dependence of spectral structure on stellar parameters, extensive spectroscopic analyses  combined  with high-resolution optical and infrared spectra (which is beyond the scope of the present investigation) may ultimately be required.

\subsection{Spectral Degeneracies and Systematics}

 Apparent similarities and degeneracies between spectra  have been shown to be a serious problem in  M dwarf synthetic fitting   (e.g., Passegger et al. 2016 and 2018; Rains et al. 2021). Despite the complicated dependence of spectral shapes on physical parameters, there are  
 clear similarities in the overall morphology of model spectra generated using different
 parameter values, which are expected to cause  significant  levels of  degeneracy in model fitting, 
 in particular when the observed/synthetic spectrum is flux-renormalized to deal with the absence of a clearly defined continuum, as is usually the case. The perplexing correlations between parameters makes a full degeneracy analysis difficult, and in this section, we will  illustrate the cases arising from the simultaneous variation of two parameters.  A complete exploration  of  spectral degeneracy when more than two parameters are involved remains very challenging and is beyond the scope of the present paper. Our model-fit results indicate systematic relationships  between physical parameters  that likely have physical origins (for example, a reduced metallicity is expected to naturally yield a higher  $\alpha$-element enhancement), other correlations are likely due to issues in model atmospheres or  systematic errors in inferred parameter values resulting from degeneracies in synthetic spectra, or even a combination of all these factors. Our purpose here is to attempt  to find  the underlying reason for these  trends, however, one should keep in mind that the broad conclusions below  are only valid for modeling of low-resolution spectra, because at high resolutions, the more detailed spectral structures can break the degeneracies that are plaguing  spectral modeling at low resolutions.

\textbf{Effective Temperature and Metallicity:} The correlation between effective temperature and metallicity is essential in  synthetic fitting and resulting parameter values.  As shown in Section 6, our observed spectra resist being fit with high-temperature (T$_\textrm{\footnotesize{eff}}$$\gtrsim$3600 K), high-metallicity  ([M/H]$\gtrsim$0.2 dex) model spectra, no matter which initial values, minimization formalism, or  iterative process are chosen in the model-fit pipeline.  And this is despite the fact that many of our lower temperature stars are fitted with high-metallicity model spectra, which means that at least some of our hotter stars should be high-metallicity objects as well. This indicates that  high-temperature, high-metallicity synthetic models may not be  in agreement with observations, which makes a perceptible systematic trend  in our estimated parameters. In the atmosphere of M dwarfs, the number of molecules decreases with increasing  effective temperature, which in turn reduces the influence of molecular bands on the spectral flux. On the other hand, as metallicity increases, the abundance of heavy elements rises, which makes the effect of molecular lines stronger. Apparently, model-fits of low-resolution spectra are not able to correctly disentangle these two effects for  high-temperature, high metallicity M dwarfs, and most likely tend to fit high-metallicity stars with models of slightly lower effective temperatures as a response to seeing stronger molecular bands. This problem is compounded by the fact that high-temperature M dwarfs have weak molecular bands (more like K dwarfs) which provides reduced leverage in the model-fitting compared with cooler M dwarfs whose molecular bands are very strong. 

An examination of synthetic spectra also shows some sources of  degeneracies that arise from the coincident variation of T$_\textrm{\footnotesize{eff}}$ and [M/H]. As [M/H] rises,  molecular band opacities increase, which depress the continuum and, in turn, decrease apparent line depths (as measured from the local pseudo-continuum).  These changes can be somewhat nullified by increasing T$_\textrm{\footnotesize{eff}}$, which has the opposing effect of reducing molecular opacities and increasing apparent line depths. This effect is illustrated in Figure 5. There is, for example, a clear resemblance between any two models with a difference of 100 K in T$_\textrm{\footnotesize{eff}}$ and  0.25 dex in [M/H] over many wavelength regions. However, the significant discrepancies  in some line strengths and features  between models with more than 100 K apart in T$_\textrm{\footnotesize{eff}}$ break this degeneracy.  We find a variation of T$_\textrm{\footnotesize{eff}}$ within $\sim$100 K with a simultaneous  variation of [M/H] within $\sim$0.25-0.30 dex may lead to degeneracy,  at both  high and low metallicities. Such degeneracies cannot, however, explain the above-described systematic trend that only occurs for the high-temperature, high-metallicity  stars.

\textbf{Effective Temperature and $\alpha$-element Enhancement:}  As detailed in Section 6, our model-fit  results  also show a systematic trend between T$_\textrm{\footnotesize{eff}}$ and  [$\alpha$/Fe] mainly for  the near-solar-metallicity stars ($-$0.5$\leq$[M/H]$\leq$+0.5 dex), again regardless the initial values, the  $\chi$$^\textrm{\footnotesize{2}}$ formalism, and the extent of the examination model grid used in the  minimization processes.  Our high-temperature stars tend to fit with low values of [$\alpha$/Fe], while  low-temperature stars are better fitted with high values  of [$\alpha$/Fe], which is very unlikely to represent a real correlation since our stars are selected from a volume-limited sample, i.e., they should all have similar distributions of [$\alpha$/Fe] regardless of temperature. Instead, this effect likely points to issues with the synthetic spectra. Most of the optical region in M dwarf spectra is strongly influenced by the  molecular lines of  TiO that is composed of two $\alpha$ elements (i.e., Ti and O). As effective temperature increases, the number of TiO molecules decreases, which should not be confused with a decrease in $\alpha$-element abundances. However, as one can see from Figure 3, the decrease in [$\alpha$/Fe] also reduces the effect of molecular bands on the overall flux, and this can be the reason why higher-temperature stars are better matched with models of lower  [$\alpha$/Fe] values. This trend therefore introduces systematic errors in modeling the atmosphere of M dwarfs which cannot properly differentiate  between the effect of the TiO molecular bands and $\alpha$-element abundances. 

The two parameters T$_\textrm{\footnotesize{eff}}$ and [$\alpha$/Fe] can also cause spectral degeneracies to some extent. An increase in [$\alpha$/Fe] does not affect the resulting spectrum in the exactly  opposite way as an increase in T$_\textrm{\footnotesize{eff}}$. Despite this fact, there are still similarities in low-resolution spectra over some important wavelength regions between any two  spectra with a difference of 100 K in T$_\textrm{\footnotesize{eff}}$ and 0.2 dex in  [$\alpha$/Fe],   as shown in Figure 6. However, the significant differences in spectral morphology between models with effective temperatures that differ by more than 100 K restrict the degeneracy to a relatively narrow range of T$_\textrm{\footnotesize{eff}}$ values. All in all, a variation of T$_\textrm{\footnotesize{eff}}$ within $\sim$100 K along with a variation  of  [$\alpha$/Fe] within $\sim$0.2 dex   can produce relatively degenerate synthetic spectra, in both high- and low-metallicity regimes, but such a temperature change  cannot be responsible for the above-described systematic trend observed in our resulting model-fit parameters.

\textbf{Effective Temperature and Surface Gravity:} The relation between effective temperature and surface gravity is of importance in characterizing stellar atmospheres. As presented in Section 6, our synthetic fitting shows a tendency for the higher-temperature stars to be better matched with lower values of log \emph{g}, and inversely, the lower-temperature stars are better matched with higher values of log \emph{g}, primarily for stars of  near-solar metallicities, which is  consistent with the prediction of stellar  evolutionary theory (Section 4).  In addition, these two  parameters can induce degeneracies when varied together simultaneously, as presented in Figure 7. For instance, any two spectra with a difference of 100 K in T$_\textrm{\footnotesize{eff}}$ and 0.2 dex in log \emph{g} are quite similar, giving rise to   degeneracy effects in the synthetic model fitting. However, the considerable discrepancies in some atomic lines and features between the models with more than 100 K apart  in T$_\textrm{\footnotesize{eff}}$ do not result in systematic errors in the inferred model-fit parameters.   We note a variation of T$_\textrm{\footnotesize{eff}}$  within $\sim$100 K together with  a variation of log \emph{g} within $\sim$0.2 dex  may yield degenerate spectra,  at both high and low metallicities,   although such degeneracies may not be extensive enough to cause the  observed relation between these two parameters in our results.

\textbf{Metallicity and $\alpha$-element Enhancement:}  There is a general correlation between the values of metallicity and $\alpha$-element enhancement derived from our model-fit pipeline: [$\alpha$/Fe] increases as  one moves from metal-rich M dwarfs to metal-poor M subdwarfs, which is  in line with the trends predicted by  Galactic chemical evolutionary  models (Section 6). But because $\alpha$ elements have a dominant effect on M dwarf spectra, these two chemical parameters affect the spectral shape in roughly the same way and an increase in [M/H] may be counteracted by a decrease in [$\alpha$/Fe] and vice versa. However, the change of spectral shape due to  a large variation in [M/H] (e.g., by  $\gtrsim$ 0.5 dex) cannot be fully compensated by adjusting [$\alpha$/Fe]  over our selected grid range. As a result, the degeneracy can only happen when these two parameters vary over small ranges, as shown in Figure 8. Considering the model with  [M/H]=$-$0.2  and [$\alpha$/Fe]=0.0 dex (red) as the reference, we begin with the model having  [M/H]=$-$0.5  and [$\alpha$/Fe]=0.0 dex (blue) and increase  [$\alpha$/Fe] by 0.1 (black) and then by 0.2 dex (green). One can see how the spectral shape becomes more similar   to the reference model by varying [$\alpha$/Fe] over many spectral regions. However, increasing [$\alpha$/Fe] by $\gtrsim$ 0.3 dex (yellow)  makes the spectral shape more  dissimilar to  the reference model, and this is where the degeneracy breaks. On closer examination, we find that a variation of [M/H] within $\sim$0.3-0.35 dex and a variation of  [$\alpha$/Fe] within  $\sim$0.2 dex at the same time may create degenerate synthetic spectra in both low- and high-metallicity regimes. While such degeneracies are too limited to be responsible to the  global  trends in the [M/H]-[$\alpha$/Fe] diagram, some local trends associated with other parameters may be attributed to these spectral degeneracies (Section 6). Apart from this degeneracy, these two parameters are tightly correlated together over larger ranges, as shown in Section 5.

\textbf{Metallicity and Surface Gravity:}  The correlation between metallicity and surface gravity is critical  in  synthetic-model fitting. Our model-fit results show a systematic relation between these two parameters when surface gravity is allowed to vary, in spite of the selected initial values, the minimization procedure, or the extent of the examination grid, as described more at length in Section 6.  There appears to be some degeneracies between spectra with a small difference in metallicity  but with a  relatively large difference in surface gravity (as compared to the parameter ranges in our model gird), as illustrated  in Figure 9.  A  decrease in metallicity by 0.2 dex  normally raises the flux level and increases the apparent line depths, which may be counteracted with a decrease in  surface gravity by 0.5 dex to some extent over a significant part of the selected wavelength coverage. However, the spectral morphology begins to become  inconsistent as the difference in  metallicity and surface gravity increases, which breaks the degeneracy. We find a metallicity change  within $\sim$0.2 dex and a gravity change  within $\sim$0.5 dex can create degenerate synthetic spectra, resulting in systematic trends in the best-fit parameter values, mostly for stars of  near-solar metallicities. This can contribute to the observed relationship between [M/H] and log \emph{g} in our results for  [M/H]$\geq$$-$0.5 dex, when log \emph{g} is treated as a free parameter; i.e., the higher-metallicity stars tend to have higher values of surface gravity and the lower-metallicity stars tend to have lower values of surface gravity. However, the effect is not as important for metal-poor spectra, which implies that degeneracies from these two parameters are less significant  in the low-metallicity regime.  This is the reason why  our results show no considerable systematic relation between metallicity and surface gravity for metal-poor M subdwarfs.

\textbf{$\alpha$-element Enhancement and Surface Gravity:} 
As will be shown in section 6 below, there is   a  trend between our [$\alpha$/Fe] best-fit estimates and the inferred values of log \emph{g} from empirical photometric relations (which are kept fixed in the model fitting), particularly for  high-metallicity stars.  This trend follows  in very much the same way as the trend between [$\alpha$/Fe]  and T$_\textrm{\footnotesize{eff}}$, which indeed reflects the tight correlation between T$_\textrm{\footnotesize{eff}}$ and log \emph{g}. On the other hand,  there is a weak degeneracy effect between [$\alpha$/Fe] and log \emph{g} when both are varied together. Figure 10 compares  the  spectral shapes of three models with distinct values of  log \emph{g} and [$\alpha$/Fe].  There is some level of similarity in the spectral morphology of two   spectra with a difference of 0.1 dex in [$\alpha$/Fe] and 0.4 dex in log \emph{g}. The effect is, however, not as strong as the degeneracy between [M/H] and log \emph{g}, but can be  responsible for the systematic shifting of some metal-rich stars to higher values of   [$\alpha$/Fe]  and log \emph{g} (Section 6).  We also find that this degeneracy begins to  break once the difference in  [$\alpha$/Fe] and  log \emph{g} increases. Our examination shows a weaker degeneracy effect for metal-poor spectra, and as a consequence, there is no noticeable trend between the model-fit values of these two parameters in the low-metallicity regime.

As mentioned above, a thorough investigation of the correlation between parameters needs more careful analyses and examinations.  Detailed spectral shapes are influenced by all stellar parameters together, making the understanding of degeneracy effects difficult. Overall,  when  observed spectra provide limited information  because of low spectral resolutions, instrumental noise, and renormalization of the continuum, these spectra can be easily fitted to a range of synthetic spectra with different stellar parameter values, which introduces significant uncertainties in the spectral modeling. Ideally, one would want to know the effective ranges over which the different physical parameters may produce degenerate synthetic spectra, and thus determine the level of uncertainty for each physical parameter for any specific star. This is a challenging proposition in general, but the rough estimates provided above in a few specific cases provide a general guideline to what one should expect. 

The executive summary is  effective temperature has the largest effect on the shape and flux level of spectra, and for this reason, it cannot make significant degeneracies along with other parameters beyond a variation of $\sim$100 K, and this is the reason why this parameter is usually well determined by our spectroscopic model fitting. Any trend or relationship  related to T$_\textrm{\footnotesize{eff}}$  is most likely due to a physical origin or deficiencies in the model atmospheres.  On the other hand, the three other parameters, i.e.,  [M/H], [$\alpha$/Fe], and  log \emph{g}, produce variations in the synthetic spectra that are strongly correlated, and this causes significant degeneracies at low spectral resolutions. Most notably, there may be a critical degeneracy effect in the model fitting (Section 6.3) when the change of spectral morphology due to a relatively large variation of a parameter (e.g.,  log \emph{g}) can be compensated  by
 a relatively small variation of another parameter (e.g., [M/H] or [$\alpha$/Fe]).

\begin{figure}\centering
       {\includegraphics[ height=5.8cm, width=9.5cm]{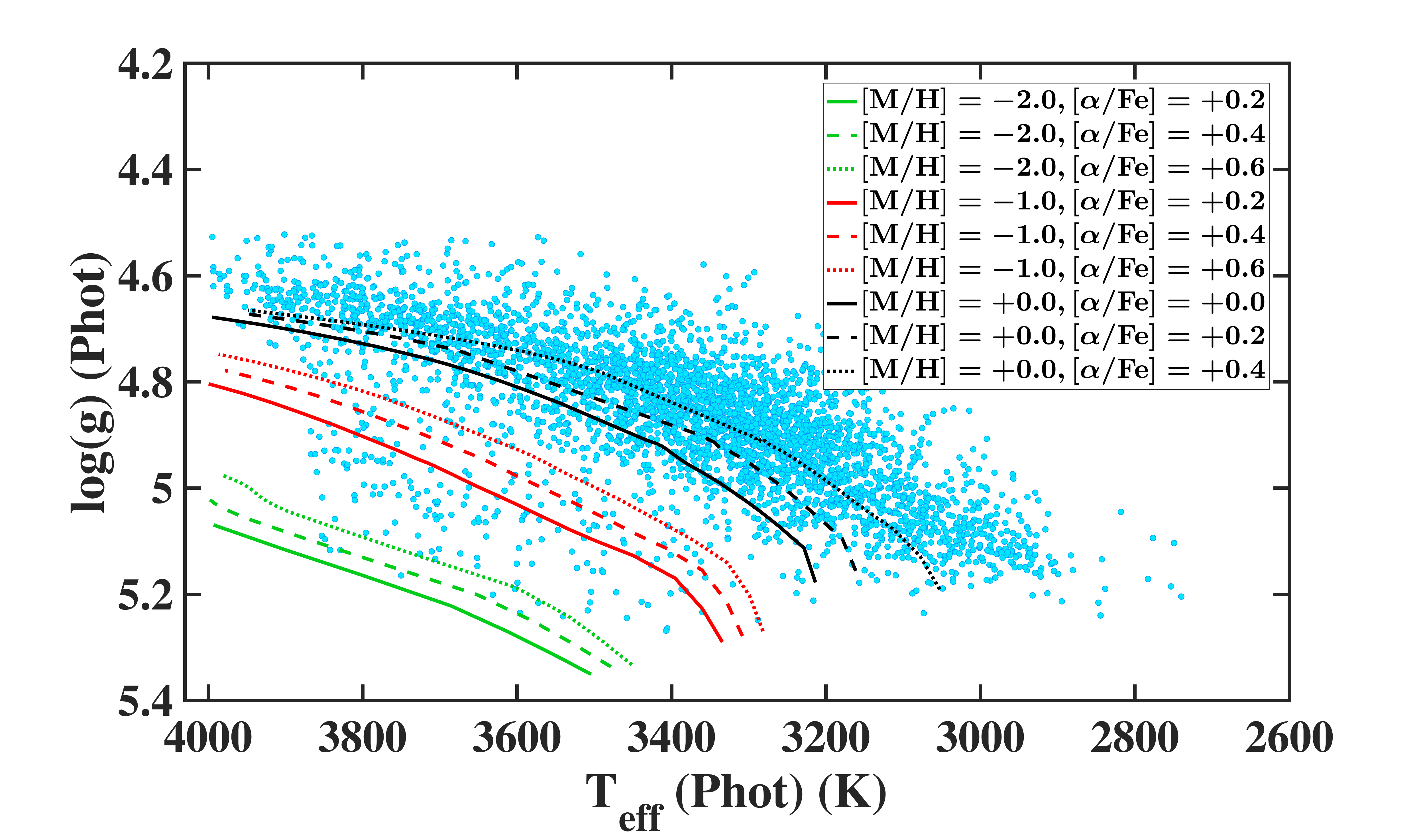}}
 \caption
        { \footnotesize{Photometric surface gravity versus photometric effective temperature for the 3745 stars (blue dots)  along with the isochrones for a  stellar age of 10 Gyr, and  chemical parameters [M/H]=+0.0 dex and [$\alpha$/Fe]=+0.0, +0.2, and +0.4 dex (solid, dashed and dotted black lines, respectively),  [M/H]=$-$1.0 dex and [$\alpha$/Fe]=+0.2, +0.4, and +0.6 dex (solid, dashed and dotted red lines, respectively), and [M/H]=$-$2.0 dex and [$\alpha$/Fe]=+0.2, +0.4, and +0.6 dex (solid, dashed and dotted green lines, respectively) from the Dartmouth stellar evolutionary models.}}
\end{figure}

\begin{figure}\centering
       {\includegraphics[ height=5.4cm, width=9.5cm]{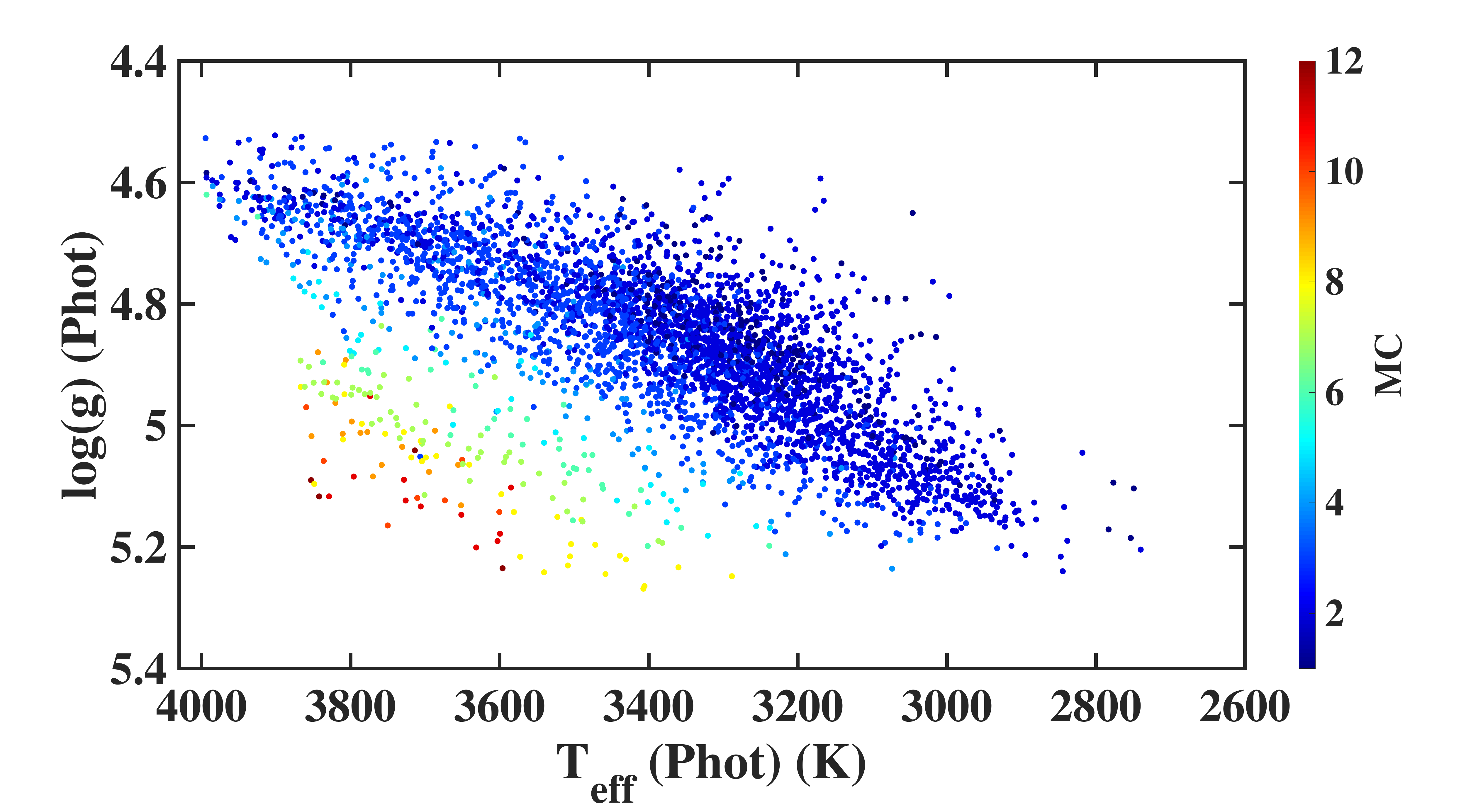}}
 \caption
        {\footnotesize{Photometric surface gravity versus photometric effective temperature for the 3745 stars (the same as shown in Figure 11), color-mapped according to their metallicity classes.}}
\end{figure}

\section{Synthetic-Fit Method}

In comparison to our previous work, we extent the fitting spectral region to 6090-8700 {\AA}, which is covered by the majority of our observed spectra. However, there is a  number of  stars whose spectra do not fully encompass this wavelength region and may have a starting wavelength  longer than 6090 {\AA} or an ending wavelength  shorter than 8700 {\AA}.  Nevertheless, these stars all minimally cover the spectral range between 6300 {\AA}  and 8600 {\AA}, which includes the most important features sensitive to stellar parameters  (see Paper I). The effect of missing spectral portions  on the resulting best-fit stellar parameters is therefore negligible. We exclude the  region 6550-6580  {\AA}, consisting of the H$_\textrm{\footnotesize{$\alpha$}}$ emission line of hydrogen that is  detected  in some of  our stars, but cannot be reproduced by model atmospheres. The atomic line Li I ($\sim$ 6710 {\AA}) is not well-modeled for a number of synthetic spectra, particularly in the  metal-poor regime, and consequently, we remove the region 6690-6726 {\AA} from the fitting region. In addition,  the wavelength ranges contaminated by telluric absorption bands, i.e., 6855-6885 {\AA}, 7590-7655 {\AA}, 8125-8170 {\AA}, and 8216-8270 {\AA} are also masked. It should be noted that these regions are consistent with the most prominent parts in the transmission spectrum (R$\sim$4000) of the Earth's atmosphere over the KPNO  (Hinkle, Wallace \&  Livingston 2003). We also develop a routine to automatically reject highly noisy and problematic regions as well as occasional spectral artifacts from each observed spectrum.

\subsection{Minimization Process} 
  
We carry out  a number of $\chi$$^\textrm{\footnotesize{2}}$ minimizations, each of which compares one observed spectrum with a number of synthetic spectra associated with  known values of the parameters T$_\textrm{\footnotesize{eff}}$, [M/H], [$\alpha$/Fe], and  log \emph{g}; these four parameters are referred to as the "primary" parameters. In each minimization, one or more of the primary parameters are allowed to vary (``free''), while the others are kept fixed (``constant''). Radial velocity shift, $v_{r}$, and  convolution factor, C\footnote{Synthetic spectra  are convolved by a Gaussian profile to match the spectral resolution of observed spectra.}, both referred to as "secondary" parameters, may also be varied or held fixed.  All minimizations follow the same sequence, as summarized below:

\begin{enumerate}
\item  Free parameters  and their corresponding ranges  are determined, and a  specified value is assigned to each constant parameter.
\item  Synthetic spectra are convolved based on the convolution factor C.
\item Observed wavelengths are shifted according to the radial velocity shift $v_{r}$ and the synthetic spectra are interpolated at the shifted wavelengths.  
\item  The observed spectrum is divided by each interpolated synthetic spectrum and the ratio is fitted by a polynomial of  high order  (order of 8) with three successive rejection runs of 10$\sigma_{r}$, 8$\sigma_{r}$, and 6$\sigma_{r}$, where $\sigma_{r}$  is the root mean square error (RMSE) between the quotient and the fit.
\item  The  interpolated synthetic spectra are multiplied by their corresponding polynomials evaluated at each shifted observed wavelength to renormalize synthetic spectra relative to the observed spectrum.  

\item The flux-renormalized, interpolated model spectra at the shifted wavelengths, F$_{\textrm{mod}}$($\lambda_{\textrm{i,shift}}$), are compared to the observed spectrum at the unshifted wavelengths, F$_{\textrm{obs}}$($\lambda_{\textrm{i,unshift}}$), by a $\chi$$^\textrm{\footnotesize{2}}$  as defined by

\begin{equation}
    \chi^\textrm{\footnotesize{2}} =   {
                \sum \limits_{i} \frac{(\mathrm{F}_{\mathrm{obs}}(\lambda_{\mathrm{i,unshift}})  - \mathrm{F}_{\mathrm{mod}}(\lambda_{\mathrm{i,shift}}))^{2}} {\mathrm{E}_{\mathrm{obs}}(\lambda_{\mathrm{i,unshift}})^{2}}} 
\end{equation}
where E$_{\textrm{obs}}$($\lambda_{\textrm{i,unshift}}$) is the empirical uncertainty (statistical and instrumental error) of the observed flux at the unshifted wavelengths.
\item The best-fit model spectrum minimizes the $\chi$$^\textrm{\footnotesize{2}}$ value.
\end{enumerate}
\noindent

Our technique comprises several sequential minimizations that  follow the above-described routine, but with different sets of free and constant  parameters and different ranges within which the selected free parameters vary. Resulting parameters from each minimization are used in the next minimization, progressively improving  fit solution.

\subsection{Initial Parameter Values} 
Prior to running the pipeline, we provide an  initial estimate for each primary parameter [M/H], T$_\textrm{\footnotesize{eff}}$, and   log \emph{g} based on the star's estimated  metallicity class (MC, 1<MC<12 from standard spectral classification) and  available photometric calibrations for M dwarfs. We also provide an initial value for [$\alpha$/Fe] by applying a simple relation between [M/H] and [$\alpha$/Fe] as follows:  we obtain  the  initial  metallicity estimates  based on  the measured metallicity class of our  stars  and the relationship between metallicity class and [M/H] derived in Paper I (Figure 34). Motivated by our previous results and other studies of nearby FGK dwarfs (e.g., Adibekyan et al. 2012, 2013; Recio-Blanco et al. 2014), we define a first-order relationship between [M/H] and [$\alpha$/Fe] to estimate the initial values of [$\alpha$/Fe]: 

\begin{equation}
\begin{aligned}
& \mathrm{[\alpha/Fe]}=+0.5 \text{ for }   \mathrm{-2.5 \leq [M/H] \leq -2.0}\\
& \mathrm{[\alpha/Fe]}=+0.4 \text{ for }   \mathrm{-1.9 \leq [M/H] \leq -1.5} \\
& \mathrm{[\alpha/Fe]}=+0.3 \text{ for }   \mathrm{-1.4 \leq [M/H] \leq -1.0} \\
& \mathrm{[\alpha/Fe]}=+0.2 \text{ for }   \mathrm{-0.9 \leq [M/H] \leq -0.5} \\
& \mathrm{[\alpha/Fe]}=+0.1 \text{ for }   \mathrm{-0.4 \leq [M/H] \leq -0.1} \\
& \mathrm{[\alpha/Fe]}=+0.0 \text{ for }   \mathrm{+0.0 \leq [M/H] \leq +0.5}.
\end{aligned}
\end{equation}
\noindent

We calculate the initial values of T$_\textrm{\footnotesize{eff}}$  using the photometric calibrations  of Mann et al. (2015) who developed a number of metallicity-independent relationships between  T$_\textrm{\footnotesize{eff}}$ and colors associated with the SDSS (\textit{r,z}), Johnson-Cousins (\textit{V,I$_\textrm{\footnotesize{c}}$}), 2MASS (\textit{J,H}), and Gaia (\textit{G}$_\textrm{\footnotesize{BP}}$,\textit{G}$_\textrm{\footnotesize{RP}}$)  bandpasses. We choose the relation with the least scatter ($\sigma$=49 K), which is based on the 2MASS and Gaia colors \textit{J$-$H} and \textit{G}$_\textrm{\footnotesize{BP}}$$-$\textit{G}$_\textrm{\footnotesize{RP}}$:

 \begin{equation}
\mathrm{T_{\footnotesize{eff}}} = 3500{\{}\sum \limits_{i=0}^{{i=4}}  a_{i}(\mathit{J-H})^{i} + \sum \limits_{i=5}^{{i=6}}  a_{i}(\mathit{G_{\footnotesize{BP}}}- \mathit{G_{\footnotesize{RP}}})^{i}{\}}
\end{equation}
\noindent where ${a}_{0}$=3.1720, ${a}_{1}$=$-$2.4750, ${a}_{2}$=1.0820, ${a}_{3}$=$-$0.2231, ${a}_{4}$=0.0174, ${a}_{5}$=0.0878, and ${a}_{6}$=$-$0.0436. We also obtain an initial value for the  radius ($\sigma$=2.89{\%}) and the mass  ($\sigma$=2.0{\%})  of the stars using the  metallicity-independent, photometric relations from Mann et al. (2015) and Mann et al. (2019), respectively, as follows:

\begin{equation}
log(\frac{\mathrm{R_{*}}}{\mathrm{R_{\sun}}}) = b_{1}+b_{2}\mathit{M_{K}}+b_{3}\mathit{M_{K}}^2
\end{equation}

\begin{equation}
log(\frac{\mathrm{M_{*}}}{\mathrm{M_{\sun}}}) =   
                \sum \limits_{i=0}^{i=5}  c_{i}(\mathit{M_{K}} - zp)^{i} 
\end{equation}
\noindent 
where \textit{M$_{\footnotesize{K}}$} is the absolute magnitude in 2MASS \textit{K} bandpass,  ${b}_{1}$=1.9515, ${b}_{2}$=$-$0.3520, ${b}_{3}$=0.0168, ${c}_{0}$=$-$0.6420, ${c}_{1}$=$-$0.2080, ${c}_{2}$=$-$8.43$\times$10$^{-4}$, ${c}_{3}$=7.87$\times$10$^{-3}$, ${c}_{4}$=1.42$\times$10$^{-4}$, ${c}_{5}$=$-$2.13$\times$10$^{-4}$,  and $zp$=7.5. We then determine the  surface gravity of the stars  by the equation (Prsa et al. 2016):

\begin{equation}
{log(\emph{g}})=4.438+log(\frac{\mathrm{M_{*}}}{\mathrm{M_{\sun}}}) -2log(\frac{\mathrm{R_{*}}}{\mathrm{R_{\sun}}}).
\end{equation}
 \noindent
 The photometric calibration of T$_\textrm{\footnotesize{eff}}$ (Eq. 3)  was established before the launch of the Gaia mission; the pre-launched Gaia bandpasses were somewhat off from their later derived values, and the estimated temperatures from this relation are unlikely to be reliable. However, since we use these estimates only  as initial values to start the iterative model-fit pipeline, the photometric values derived above should be sufficiently accurate for our purposes. As shown in Section 6, our determined effective temperatures show strong correlation with color and absolute magnitude in the HR diagram, indicating their high precision, even as only approximate initial values of T$_\textrm{\footnotesize{eff}}$ are used.

Figure 11 shows the photometric surface gravity versus photometric effective temperature of the 3745 stars in our sample. Overplotted are isochrones for a stellar age of 10 Gyr  and metallicities [M/H]=$-$2.0, $-$1.0, and +0.0 dex,  each with three different values of [$\alpha$/Fe], all from the Dartmouth stellar evolutionary model grid (Dotter et al. 2008). The solar-metallicity isochrone with [$\alpha$/Fe]=0 appears to be offset towards higher values of  T$_\textrm{\footnotesize{eff}}$ and log \emph{g}, while the solar-metallicity isochrones with higher values of $\alpha$-element enhancement, i.e.,  [$\alpha$/Fe]=+0.2 and +0.4, are better matched with the average near-solar metallicity, disk stars (see Figure 31). However, it is difficult to properly interpret  the location of isochrones relative to the stars' distribution  because  the photometric temperatures and surface gravities, and also isochrones  may be inconsistent to some extent with the true parameter values. Figure 12 displays the same plot as shown in Figure 11, but color-mapped based on the metallicity class of the stars. Although the empirical photometric relations have not been tested for  low-metallicity stars ([M/H]$<$$-$0.6 dex) that make up about 6$\%$ of our sample, the  metal-poor stars (MC$\geq$6) are appropriately located in the low-metallicity region of the diagram. Therefore, we believe the photometric estimates provide acceptable initial values for the low-metallicity stars to start the pipeline.

\subsection{Fitting Pipelines} 
Our main goal is to develop a synthetic-fit pipeline that can be accomplished  in a reasonable amount of time. The performing of  a minimization process over 860,000 grid points is extremely time-consuming and exploring all the grid points is a very inefficient process. To this end, our pipeline carries out multiple grid searches  over limited ranges of parameters, beginning with the derived initial values as described above. This procedure significantly reduces the required number of computing processors (CPU or central processing unit) and  the operation time, providing an efficient technique for large, low-resolution surveys.

We examine two different methods and analyze the resulting best-fit parameter values and their distributions in parameter spaces as well as in photometric and kinematic diagrams. In the first method, that we name the ``normal method'',  the residual correlation is not taken into account and we normally include all  the wavelength data points in the fitting.  In the second, that we call the ``reduced-correlation method'', the residual correlation between neighboring data points is considered and the minimizations  are performed over a subset of wavelength points that are distanced from each other to reduce the effect of this correlation. Each method is  also  investigated in two ways;  one approach keeps  surface gravity constant  and one treats surface gravity as a free parameter. The results from these methods are then compared to each other and  the influence of the residual correlation and surface gravity on derived best-fit parameter values is addressed separately.

\subsubsection{Normal Method} 
Beginning with the above-described initial parameter values  and rounding each value to the closest grid point of the corresponding parameter, if necessary, and also setting an initial convolution factor C equal to the average value of our previous results (Paper I), the model-fit pipeline carries out  ten  successive minimization steps each of which  follows the same routine as outlined in Section 4.1. Best-fit parameter values inferred at each step are used in the next step as revised initial parameters. The first five steps examine different values of the parameters on a coarser model grid, while the last five steps examine the values on a finer model grid. This pipeline, when surface gravity is allowed to vary, is outlined as below:

\begin{enumerate}
\item Vary the radial velocity shift $v_{r}$ from $-$500 to 500 km/s, in steps of 10 km/s.
\item Vary  the convolution factor C from 0.1 to 8, in steps of 0.1.
\item Vary T$_\textrm{\footnotesize{eff}}$ within {$\pm$}200 K, in steps of  100 K, around its corresponding initial value.
\item Simultaneously vary  [M/H] within {$\pm$}0.5 dex, in steps of 0.1 dex,   [$\alpha$/Fe] within {$\pm$}0.3 dex, in steps of 0.05 dex,  and log \emph{g} within {$\pm$}0.2 dex, in steps of 0.1 dex, around their respective initial values.
\item Setting all parameters derived from previous steps as new initial values, iteratively repeat all the steps from 1 to 4  until the parameter values converge.
\item Setting all parameters derived from the step 5 as new initial values, vary $v_{r}$ from $-$500 to 500 km/s, in steps of 5 km/s.
\item Vary  C from 0.05 to 8, in steps of 0.05.
\item Vary T$_\textrm{\footnotesize{eff}}$ within {$\pm$}100 K, in steps of  50 K, around its corresponding initial value.
\item Simultaneously vary  [M/H] within {$\pm$}0.25 dex, in steps of 0.05 dex,   [$\alpha$/Fe] within {$\pm$}0.1 dex, in steps of 0.025 dex,  and log \emph{g} within {$\pm$}0.1 dex, in steps of 0.05 dex, around their respective initial values.
\item Setting all parameters derived from the steps 6-9 as new initial values, iteratively repeat these steps  until the parameter values converge.
\end{enumerate}

In the case when surface gravity is regarded as a fixed parameter, the pipeline executes the same steps as above, except for the steps 4 and 9, in which  log \emph{g} is not allowed to change and its values is kept constant and equal to its initial value.

As it happens, surface gravity plays an important role in low-resolution spectral fitting, although it has the least effect on the spectral shape as compared to the other parameters, notably effective temperature and metallicity. Generally, the accurate determination of  surface gravity, even in high-resolution analyses, has proved to be challenging. Based on our careful grid search,  when this parameter is allowed to vary over a large range of values, it tends to diverge and find best-fit values near the limits of the range for some stars, which do not represent physically realistic solutions knowing that our objects are dwarf stars. The other parameters are accordingly adjusted to this value of log \emph{g} and unlikely fitted to the correct models. For this reason, we minimize this effect by limiting the range of values, within which log \emph{g} is free to change in each minimization step, i.e., at most {$\pm$}0.2 dex.  Nevertheless, there is still a considerable  variation of surface gravity during the model fitting, which has a significant  influence on our final  results and the distributions of stars in different  diagrams (Section 6).  The best treatment for this problem, particularly in low-resolution spectroscopy,  is to use the acceptable values of surface gravity determined using independent methods (e.g., photometry) and keep this parameter fixed in the entire model-fit pipeline. Unfortunately, reliable estimates of  log \emph{g} are not available for our stars, and this effect remains as a source of uncertainty  for our low-resolution analysis.

\subsubsection{Reduced-Correlation Method} 
Due to the finite resolution of astronomical spectrographs, the instrumental line-spread function (LSF, that describes the distribution of a monochromatic light source at the local plane) is usually sampled over several pixels, resulting in a broadening of observed spectral features. Consequently, adjacent pixels or wavelength datapoints do not represent the independent samples of the true spectrum, and a discrepancy between a synthesized and observed spectral feature may yield a correlated residual that spreads over multiple datapoints (Czekala et al. 2015, hereafter C15). As emphasized in C15 (and pointed out by the first author through our private communication), the Poisson errors (or noises) from astrophysical or instrumental effects inserted on the observed spectrum are likely uncorrelated, but it is the models that do not have enough degrees of freedom to fit the data accurately. A minor mismatch between the model and observed spectra can create correlated fit residuals that propagate over a characteristic scale comparable to the instrumental broadening. 

To account for this effect, C15 have developed a formulation of $\chi$$^\textrm{\footnotesize{2}}$ minimization, including  a nontrivial covariance matrix with off-diagonal terms that measures the correlation between any pair of wavelength residuals, which is iterated using Markov Chain Monte
Carlo (MCMC) simulations (Starfish\footnote{http://iancze.github.io/Starfish}). Although this framework can perfectly apply 
for detailed analyses of a small number of stars at a time,  the large size of our spectroscopic dataset, combined with the time-consuming and computationally expensive MCMC methods, made this procedure too prohibitive to run on available computer resources. As an alternative, we  introduce a significantly quicker  approach to reduce the residual correlation between adjacent datapoints, as described below.
 
Figure 13 compares the observed spectra (red) with their corresponding  flux-renormalized best-fit models (blue), at unshifted wavelengths,  for four typical  stars  spanning a wide range of metallicities with parameters: T$_\textrm{\footnotesize{eff}}$=3300, 3500, 3500, and 3400 K, [M/H]=+0.3, 0.0, $-$0.6, and $-$1.1 dex, [$\alpha$/Fe]=+0.2, +0.05, +0.275, and +0.325 dex, and  log \emph{g}=5.35, 5.0, 5.3, and 5.35 dex, respectively. The respective residuals R$_{\textrm{i}}$  between these spectra and their best-fit models, as described by: 

\begin{equation}
   \mathrm{R}_{\mathrm{i}}=   {
                {\mathrm{F}_{\mathrm{obs}}(\lambda_{\mathrm{i,unshift}})  - \mathrm{F}_{\mathrm{mod}}(\lambda_{\mathrm{i,shift}})}}, 
\end{equation}
\noindent
which is evaluated at each  datapoint, are also shown in the lower panels. The autocorrelation length scale (ACL) of  the residual array  R$_{\textrm{i}}$ can be determined using the  autocorrelation function (ACF), i.e., from the trend in the autocorrelation value as a function of the pixel lag number. The ACF of the residual arrays from the four stars in Figure 13 are shown in Figure 14.  Here we calculate their residual ACF relative to their best-fit model from the ``normal method'' when surface gravity is a free parameter\footnote{In general, there is a smaller distance between the observed spectra and their best-fit models derived from the pipeline in which log \emph{g} is a free parameter, which leads to apparently lower values of $\chi$$^\textrm{\footnotesize{2}}$, as compared to those inferred from the pipeline in which log \emph{g} is constant. However,  there may be a strong degeneracy effect underlying the resulting parameter values from the former method for some stars, which is difficult to perceive just by comparing the observed spectrum and and its best-fit model. Nevertheless, we find no considerable difference in the  estimated ACL using the two approaches.}. Since the stars were observed by similar instrumental setups, all the plots show more and less similar decreasing trends with pixel lag number.  The  ACL  is defined by the lag numbers  where the ACF of the residuals are reduced by about a third ($\sim$20-40{\%}) of the value at lag number=0. For our observed spectra, we find that ACL$\sim$10 pixels in the vast majority of cases. In  other words,  the residual at each point is significantly propagated over $\sim$10 neighboring  datapoints, which means that the residuals of any 10 adjacent datapoints are highly correlated. There is a relationship between  the ACL and the spectral resolution of the related spectrograph: the higher resolution, the less instrumental broadening, and the shorter ACL. 

In the ``reduced-correlation method'', we therefore perform each minimization routine over every 10-th wavelength datapoint, i.e.,  selecting a subset of one tenth of the total datapoints that show little to no correlation and thus efficiently sample our low-resolution spectra. Starting from the first datapoint, we select every 10-th wavelength, and apply the model-fit pipeline to these selected wavelengths to derive the parameter values. We then repeat the same process, but starting from the second  datapoint and derive a second set of parameter values. This approach is repeated ten times up to the tenth datapoint and each time a new set of parameter values are inferred. We lastly take an average of the ten values obtained from each step and report this average as our final best-fit estimate.

Since we run this procedure on each spectrum after running the ``normal method'' described in Section 4.3.1, we use the final best-fit values from the ``normal method''  as initial values in this ``reduced-correlation method''.  We  again perform the ``reduced-correlation method'' for two cases of surface gravity, i.e., when it is a free or a constant parameter, and the parameter estimates  from the ``normal method'' using each  gravity-modeling approach are then used as initial values in the corresponding approach of the "reduced-correlation method''. We also treat  the   secondary parameters  as fixed parameters in the entire process, equal to the well-determined values obtained from the ``normal method''. The pipeline, in which surface gravity is a variable parameter, is thus shortened to only three steps, as follows:

\begin{enumerate}
\item Vary T$_\textrm{\footnotesize{eff}}$ within {$\pm$}100 K, in steps of  50 K, around its corresponding initial value.
\item Simultaneously vary  [M/H] within {$\pm$}0.25 dex, in steps of 0.05 dex,   [$\alpha$/Fe] within {$\pm$}0.1 dex, in steps of 0.025 dex,  and log \emph{g} within {$\pm$}0.1 dex, in steps of 0.05 dex, around their respective initial values.
\item Setting all parameters derived from  the steps 1 and 2 as new initial values, iteratively repeat these steps  until the parameter values converge.
\end{enumerate}

 In the case when surface gravity is a fixed parameter, the pipeline performs the three steps above, except for the steps 2 and 3 in which log \emph{g} is kept  constant equal to its initial value.

\begin{figure}\centering
\subfloat
        {\includegraphics[ height=5.2cm, width=9.3cm]{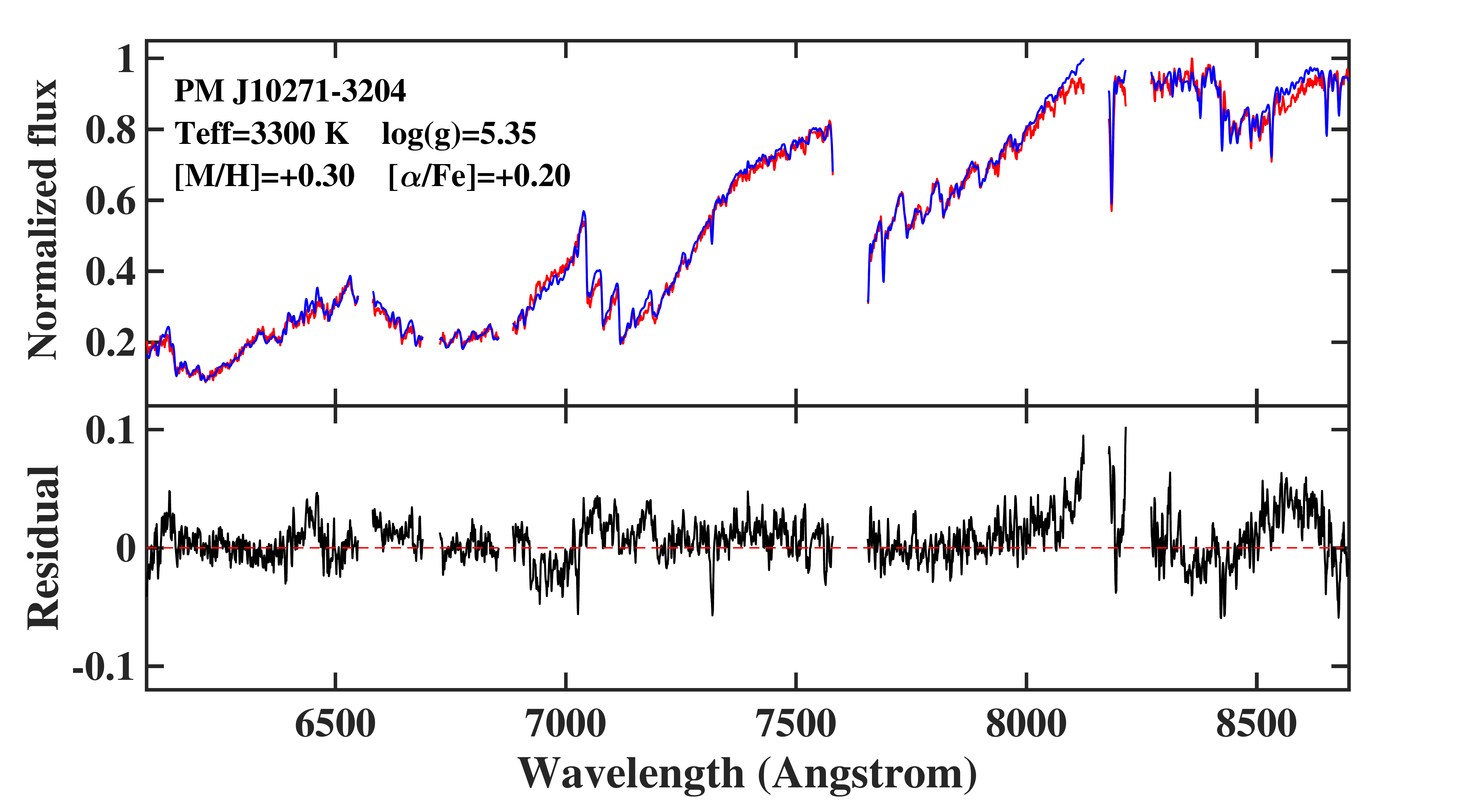}}       
\vspace{-0.37cm}
    \subfloat
        {\includegraphics[ height=5.2cm, width=9.3cm]{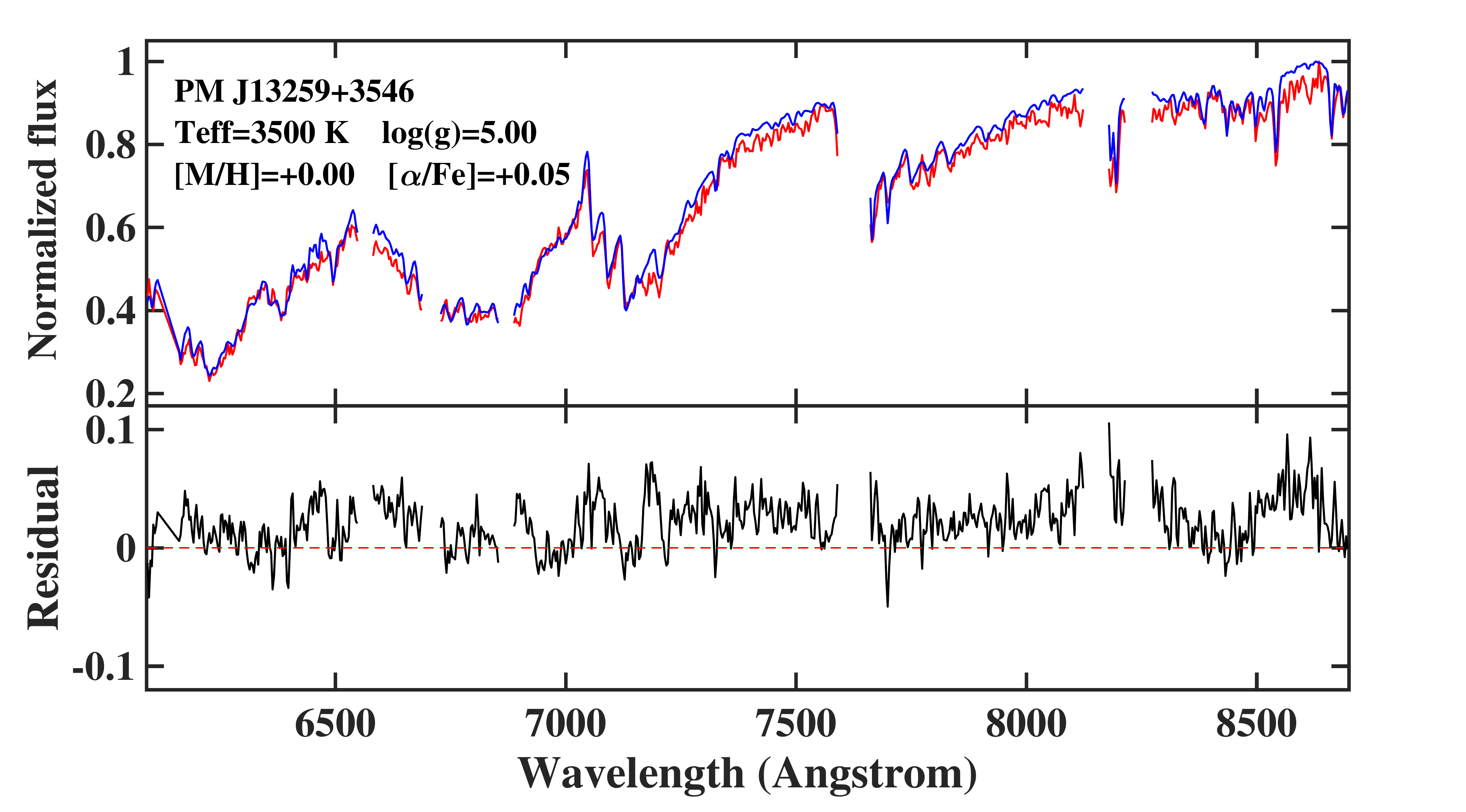}}      
\vspace{-0.37cm}
    \subfloat
        {\includegraphics[ height=5.2cm, width=9.3cm]{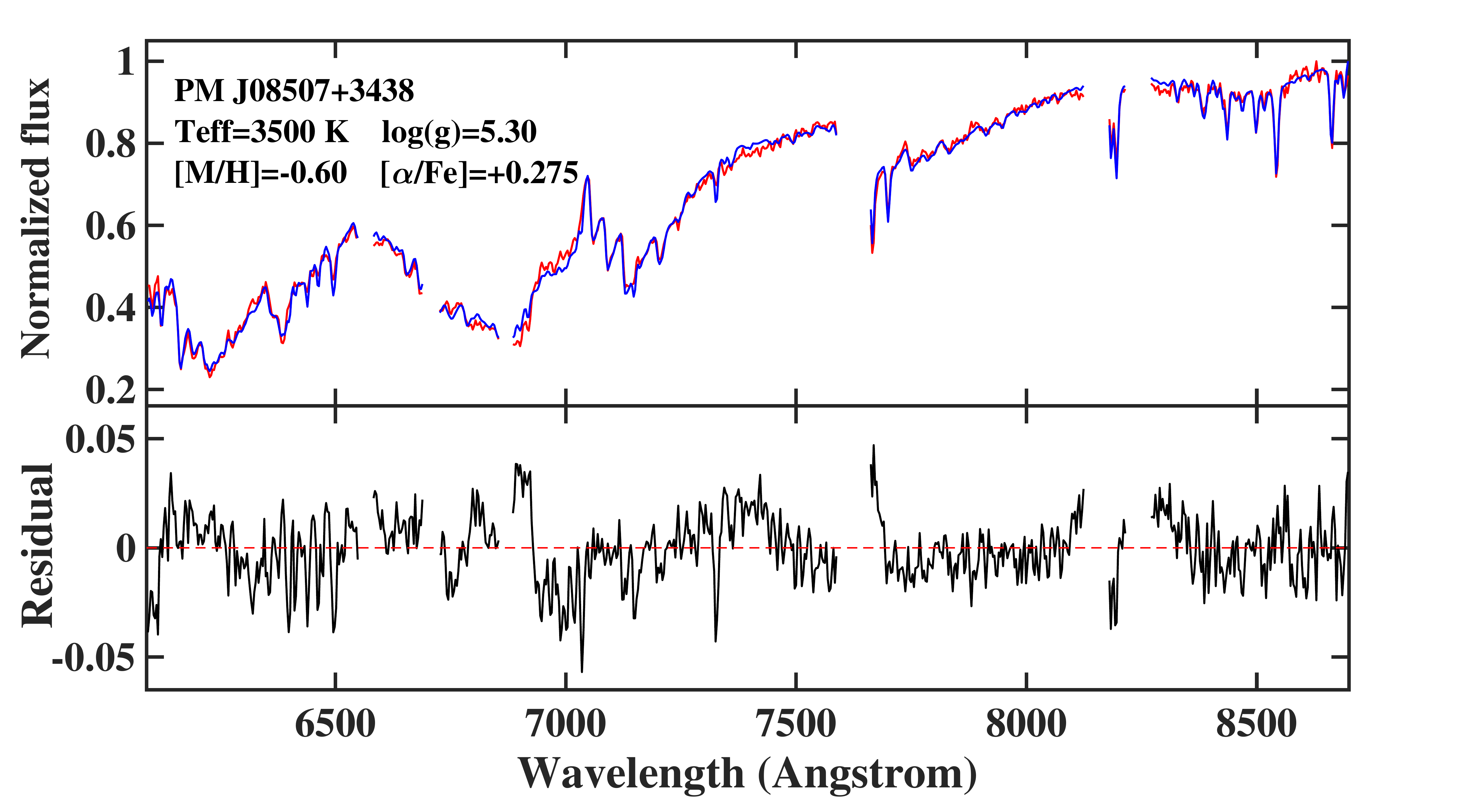}}   
\vspace{-0.37cm}
    \subfloat
        {\includegraphics[ height=5.2cm, width=9.3cm]{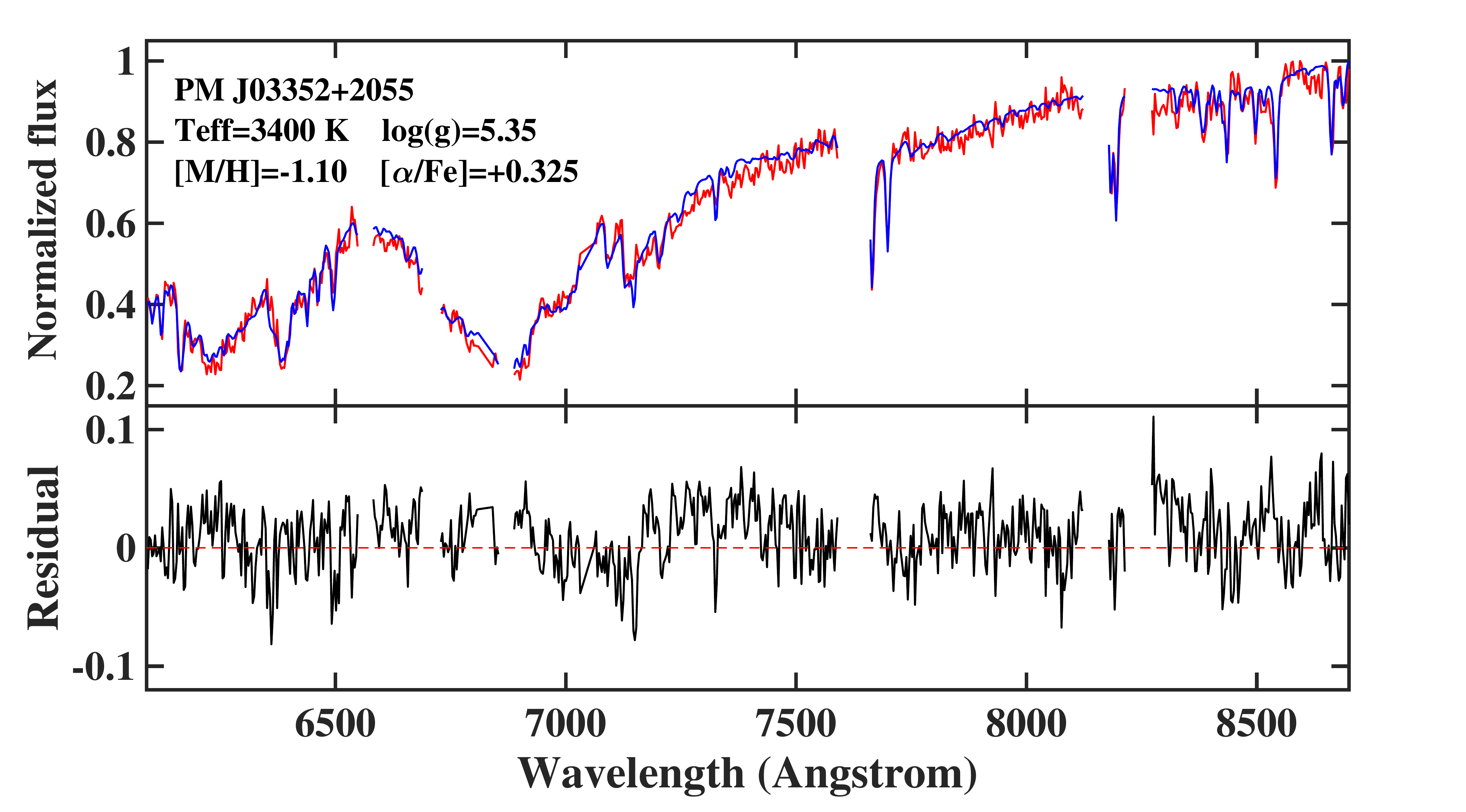}}                 
   \caption
        {\footnotesize{Comparison between the observed spectra (red) and their respective  flux-adjusted, best-fit models (blue), at unshifted wavelengths,  for four stars with metallicities  ranging from +0.3 dex to $-$1.10 dex. The associated residuals between the spectra and their best-fit models are also shown in the lower panels. The parameters of these stars are reported in the legends.}}
\end{figure}

\begin{figure}\centering
\subfloat
        {\includegraphics[ height=5cm, width=9cm]{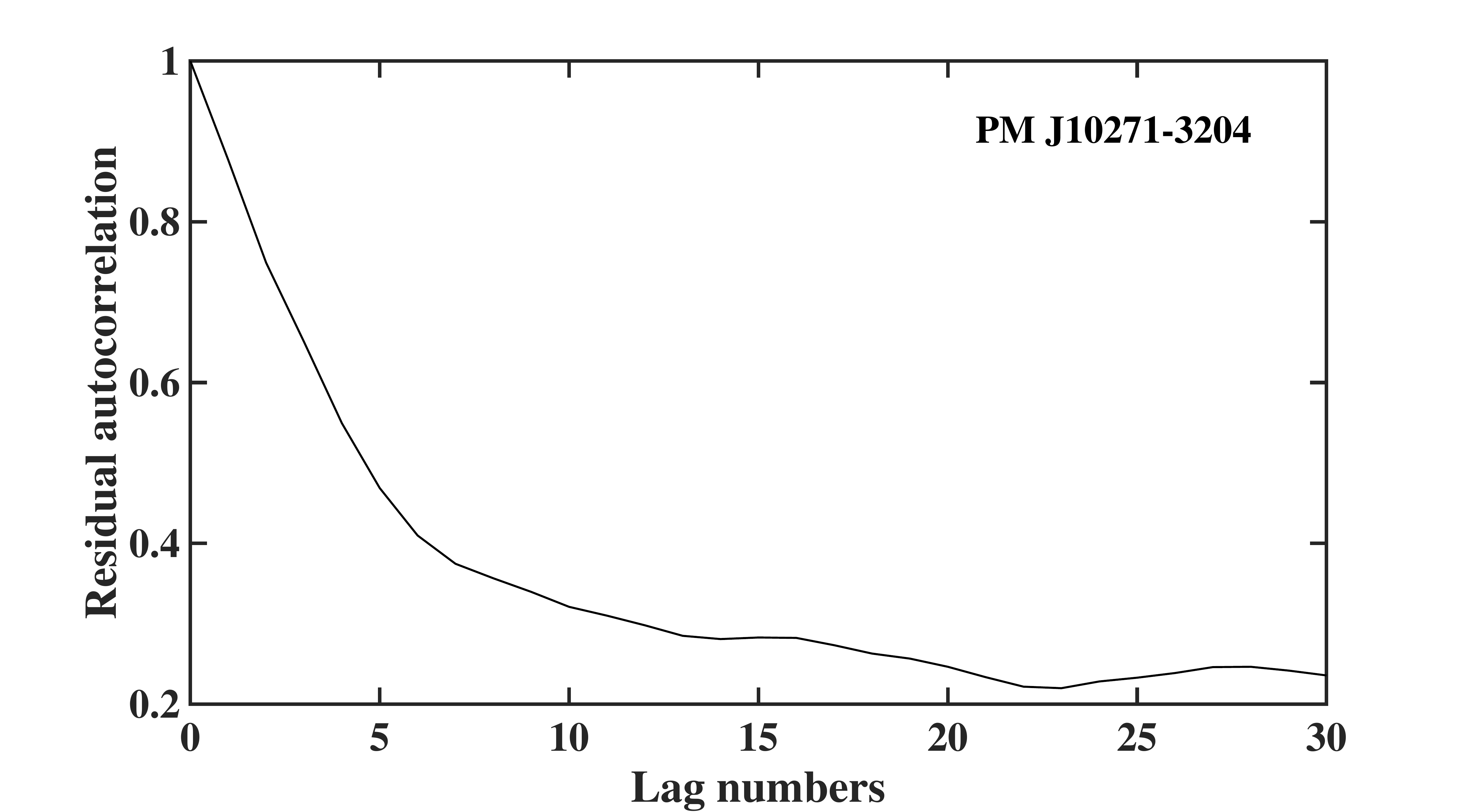}}       
\vspace{-0.1cm}

    \subfloat
        {\includegraphics[ height=5cm, width=9cm]{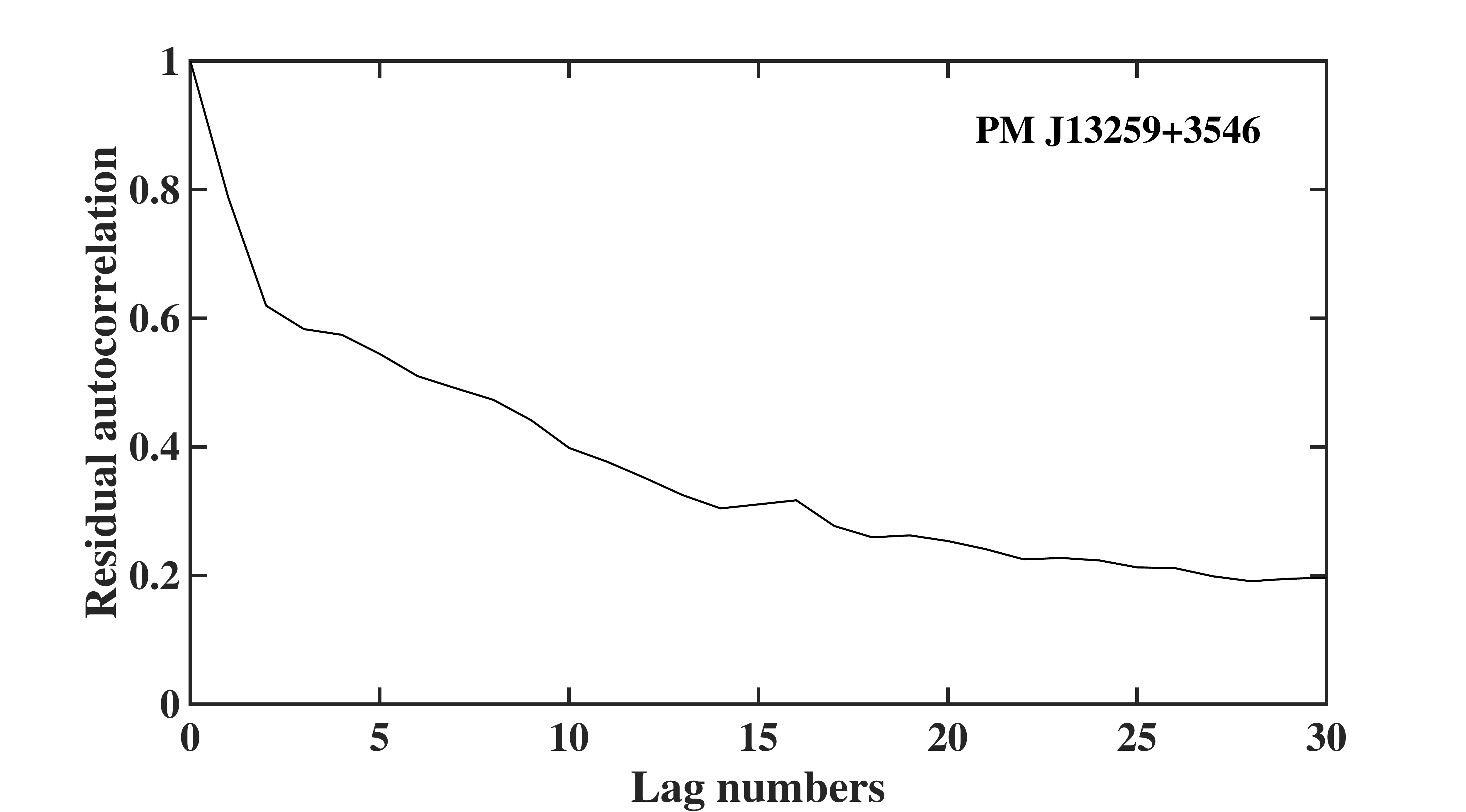}}      
\vspace{-0.1cm}

    \subfloat
        {\includegraphics[ height=5cm, width=9cm]{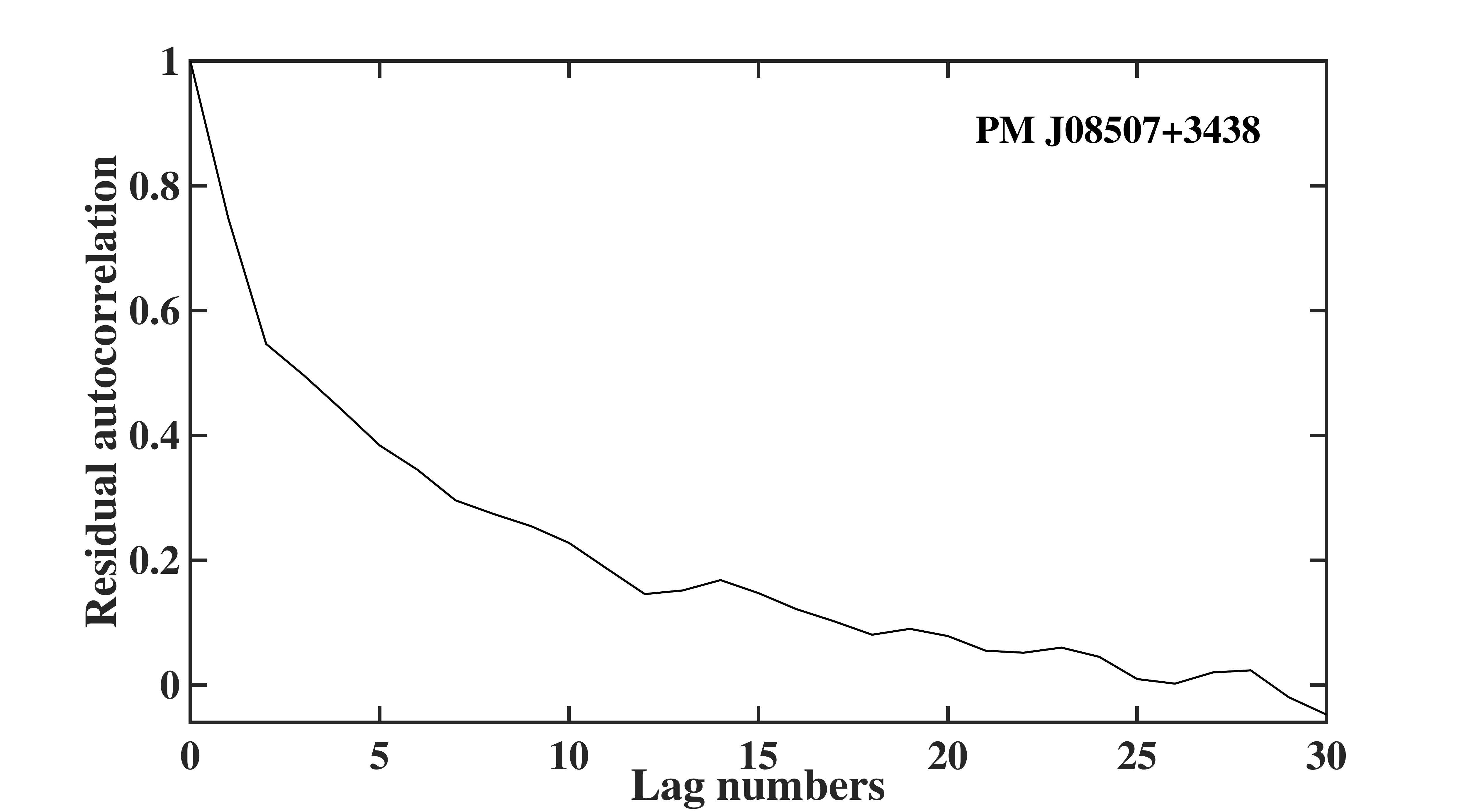}}   
\vspace{-0.1cm}

    \subfloat
        {\includegraphics[ height=5cm, width=9cm]{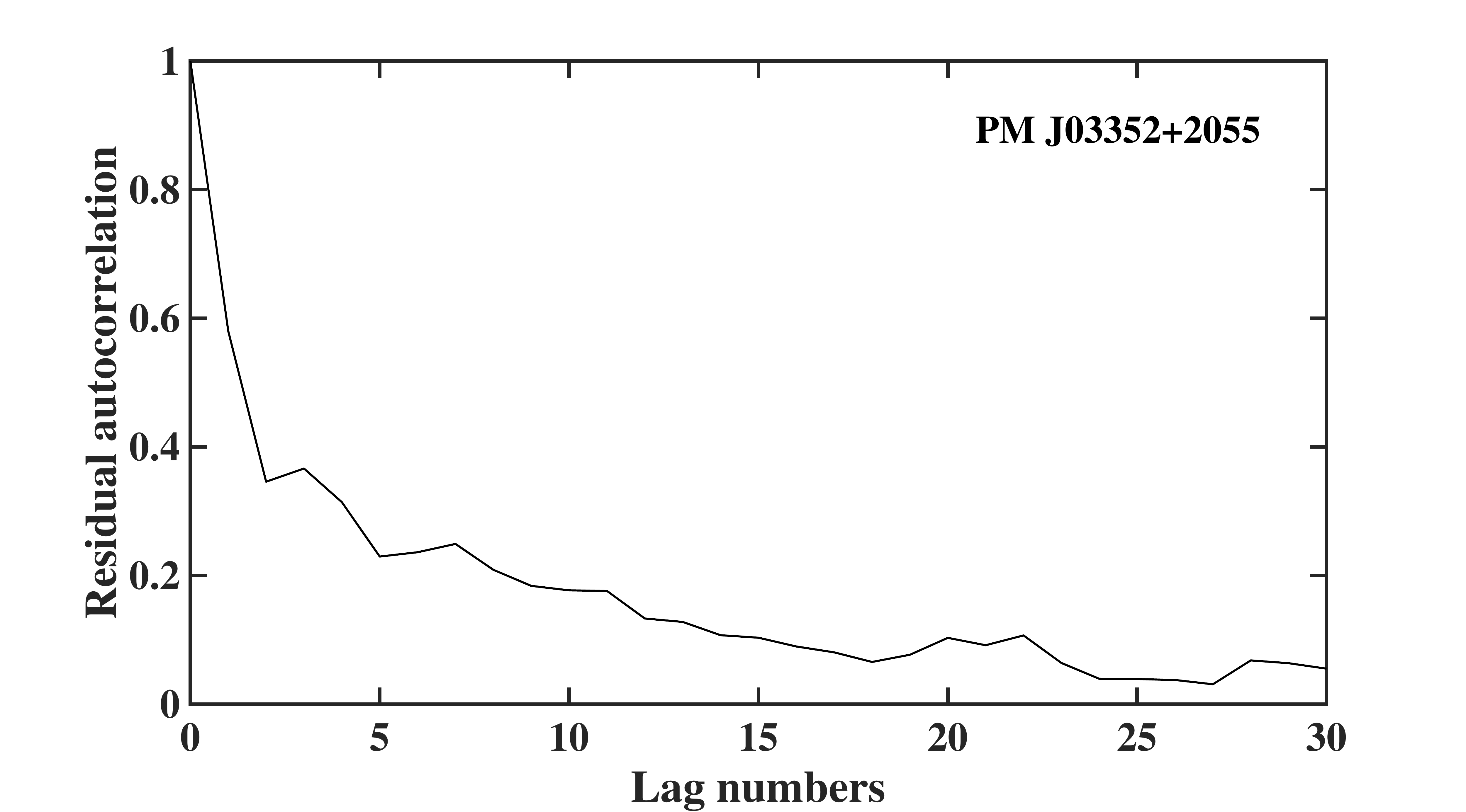}}                 
   \caption
        {\footnotesize{Residual Autocorrelation versus lag numbers for the same stars as those shown in Figure 13.}}
\end{figure}

\begin{figure*}\centering
\subfloat 
      {\includegraphics[ height=4.8cm, width=6cm]{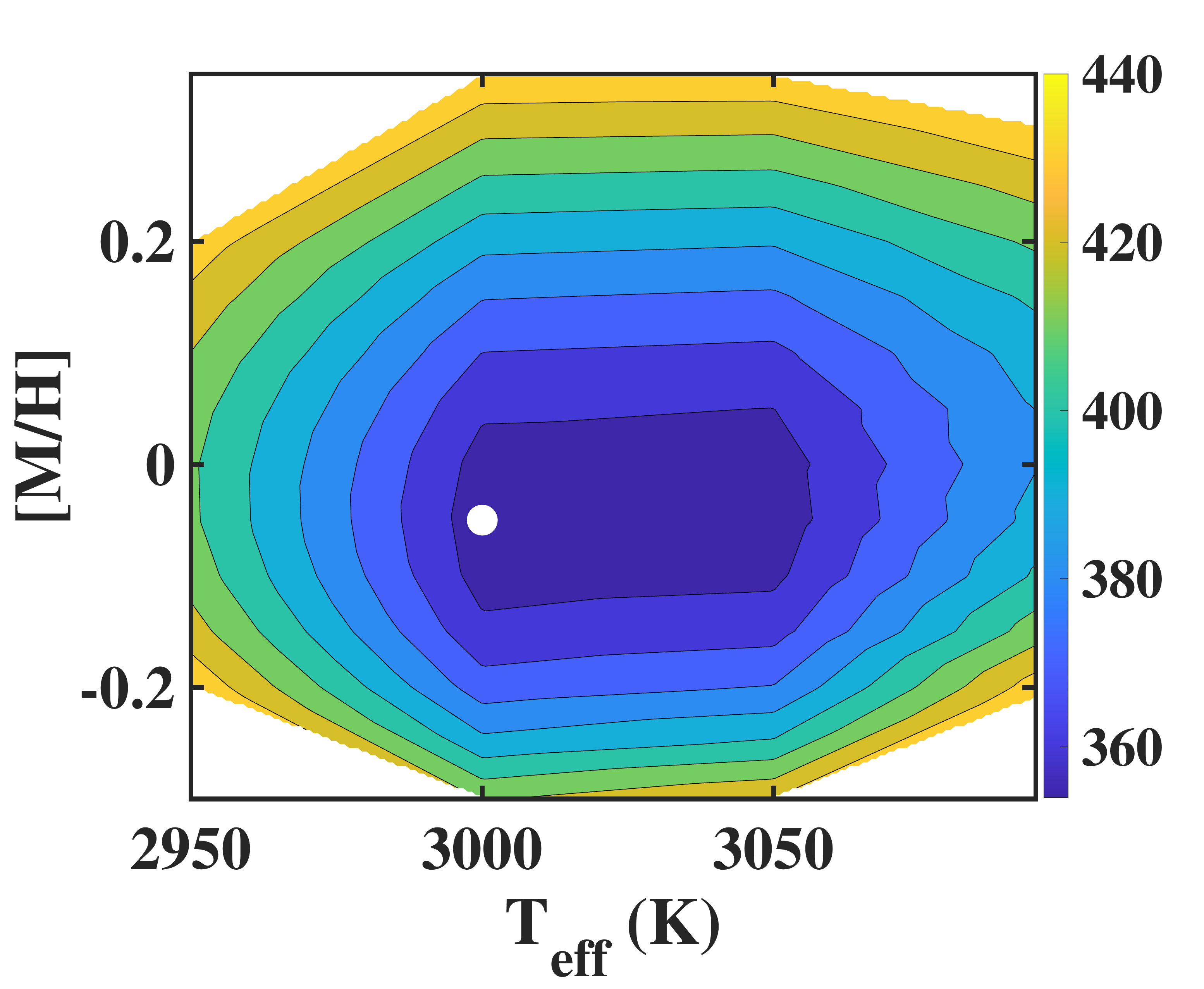}}               
\hspace{-0.1cm}
\subfloat 
      {\includegraphics[ height=4.8cm, width=6cm]{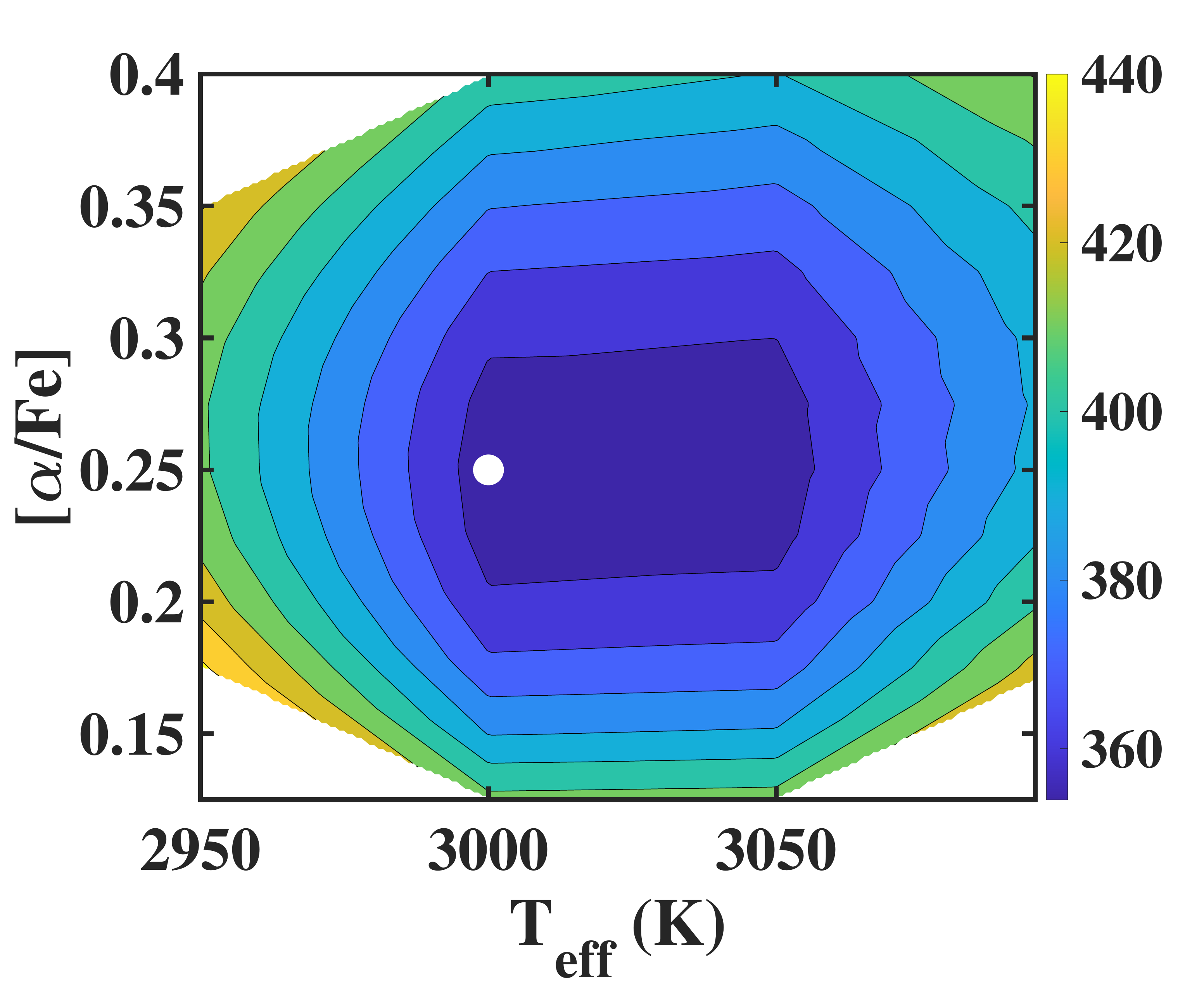}}               
\hspace{-0.1cm} 
 \subfloat      
         {\includegraphics[ height=4.8cm, width=6cm]{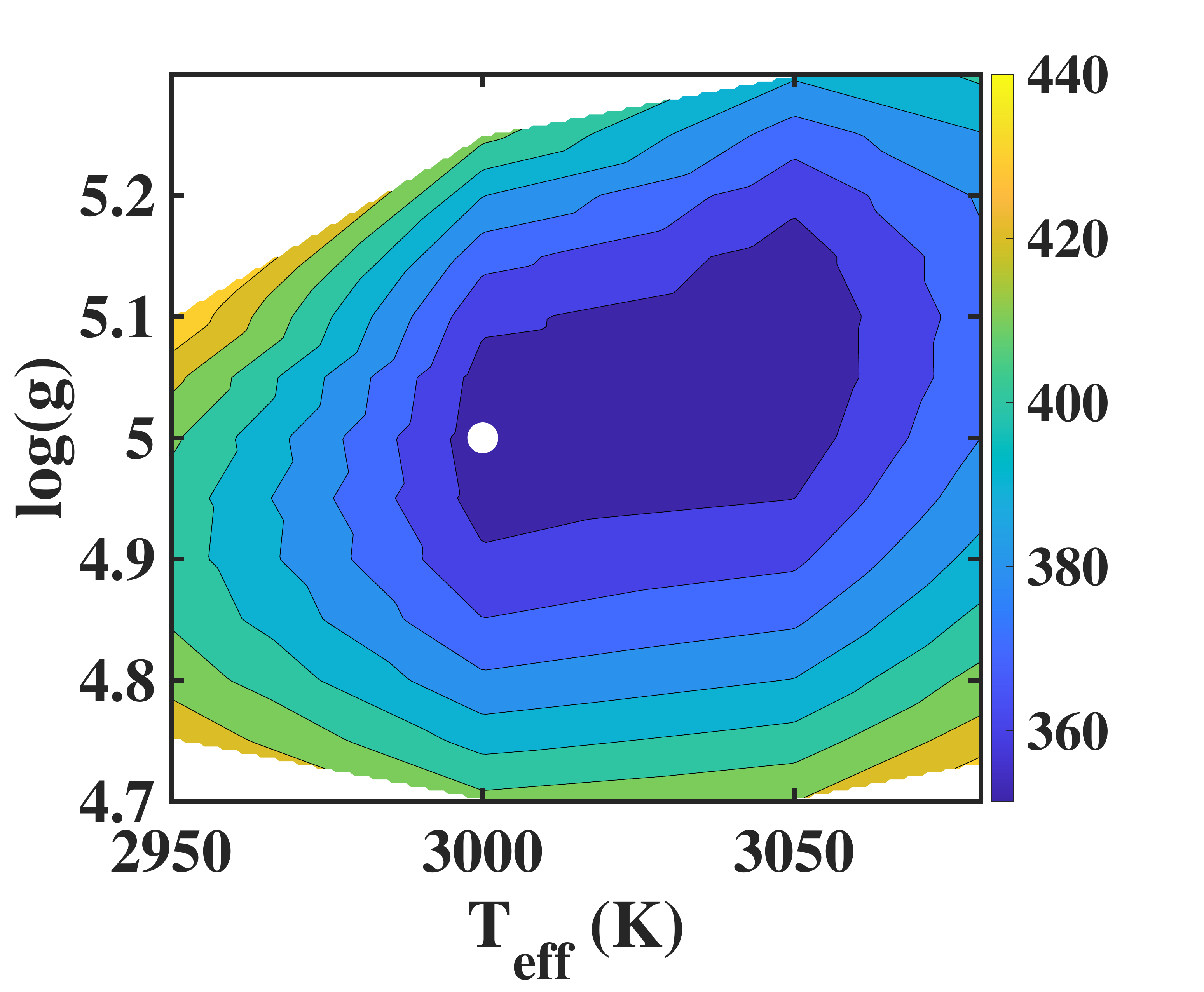}}
 \vspace{-0.35cm}

 \subfloat 
      {\includegraphics[ height=4.8cm, width=6cm]{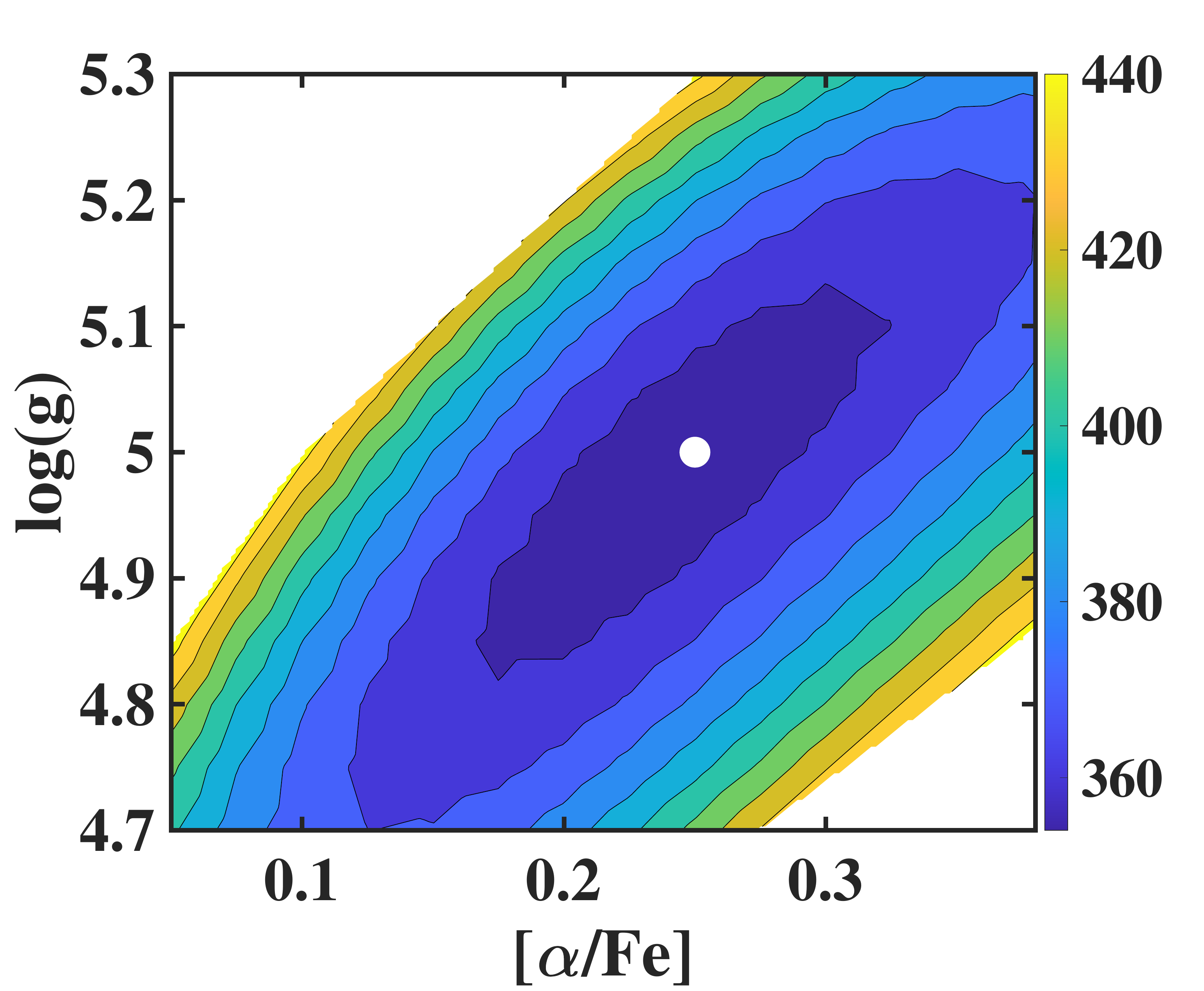}}               
\hspace{-0.1cm}
\subfloat 
      {\includegraphics[ height=4.8cm, width=6cm]{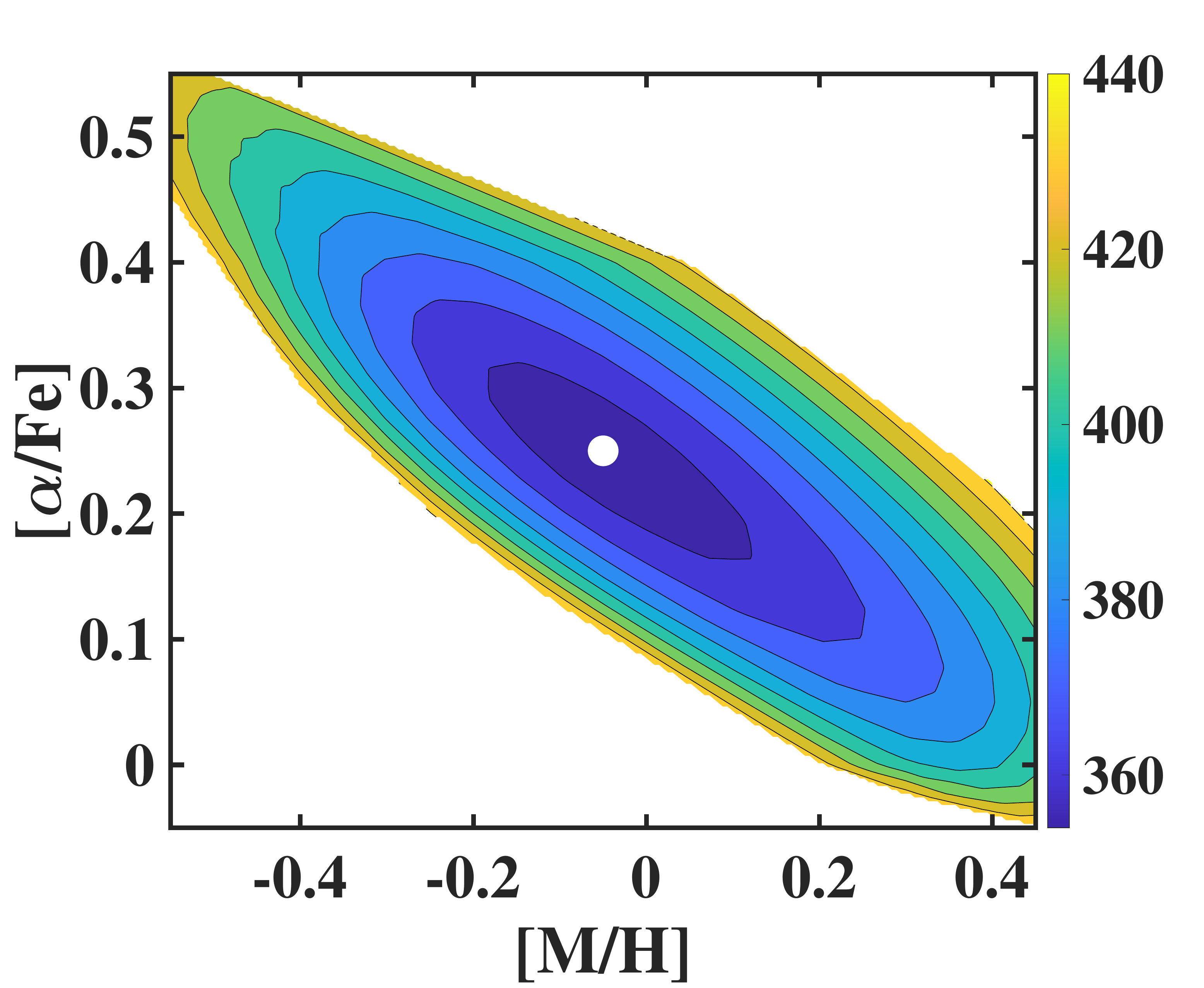}}               
\hspace{-0.1cm} 
 \subfloat      
         {\includegraphics[ height=4.8cm, width=6cm]{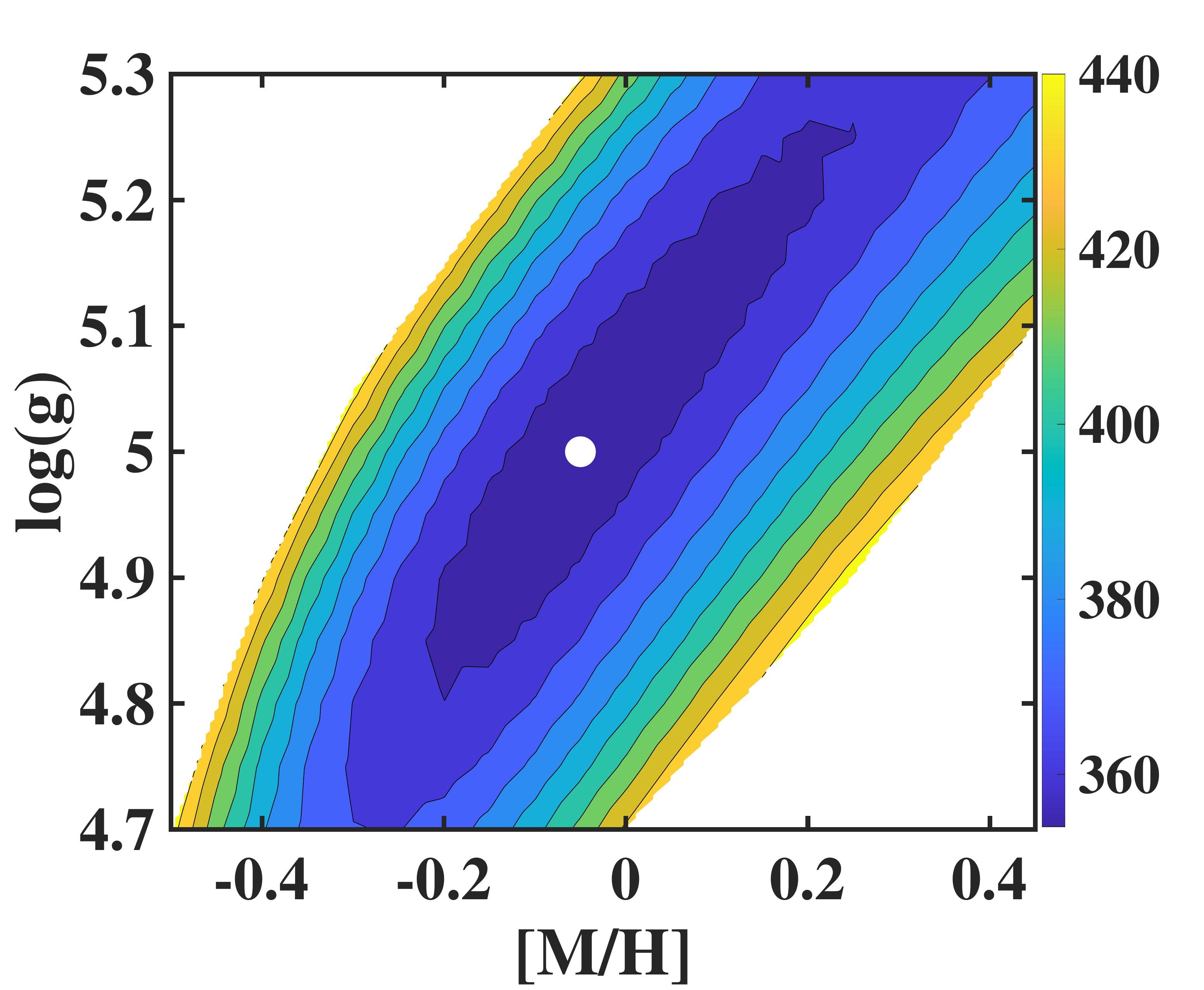}}        
 \caption
        {\footnotesize{$\chi$$^\textrm{\footnotesize{2}}$ maps in two-dimensional parameter planes for the star PM J01305+2008  with T$_\textrm{\footnotesize{eff}}$=3000 K, [M/H]=$-$0.05 dex, [$\alpha$/Fe]=+0.25 dex, and  log \emph{g}=5.0 dex. The best-fit grid point is shown by a white dot. The number of wavelength datapoints used in the $\chi$$^\textrm{\footnotesize{2}}$ calculation is 682.}}
\end{figure*}

\begin{figure*}\centering
\subfloat 
      {\includegraphics[ height=4.8cm, width=6cm]{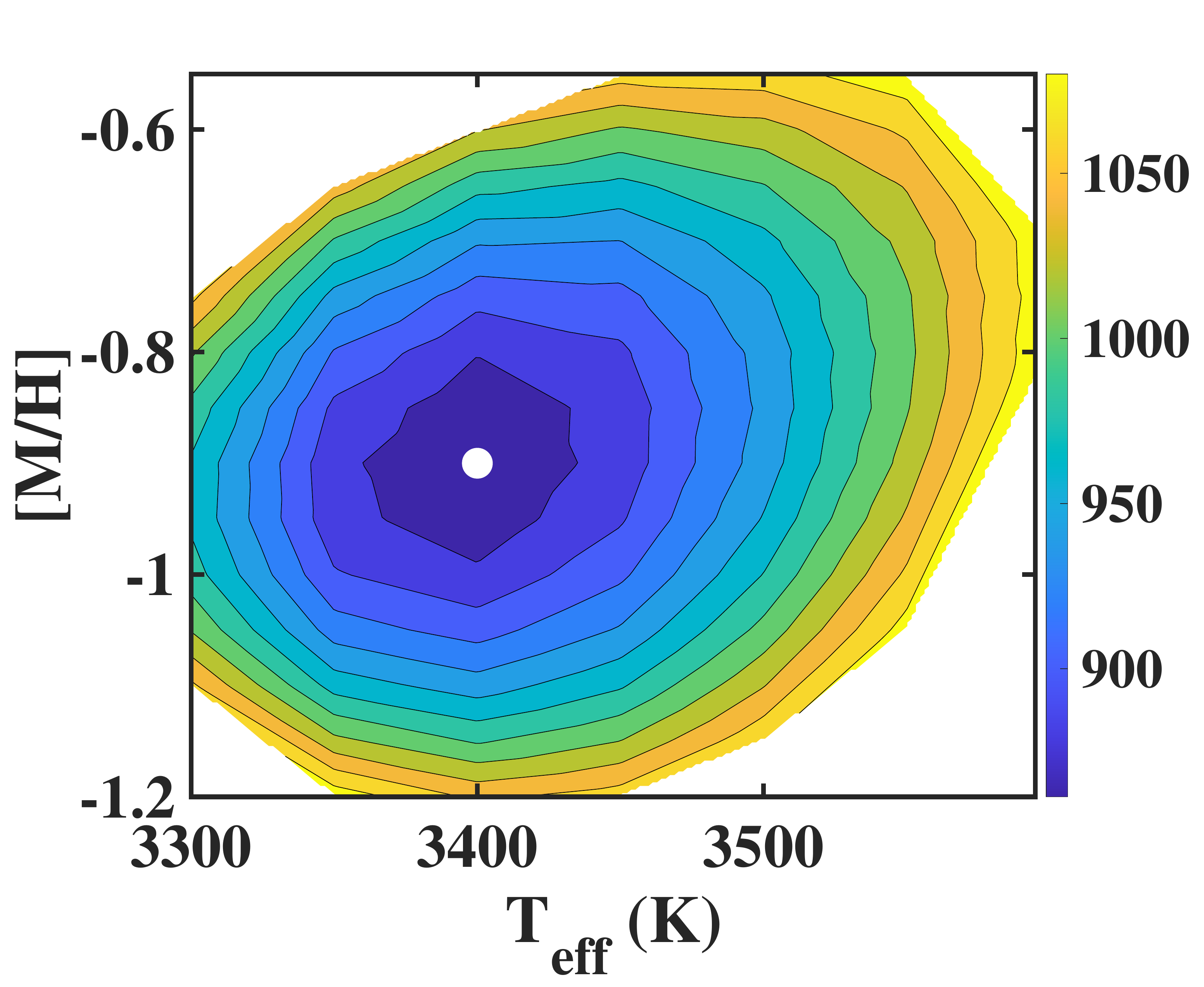}}               
\hspace{-0.1cm}
\subfloat 
      {\includegraphics[ height=4.8cm, width=6cm]{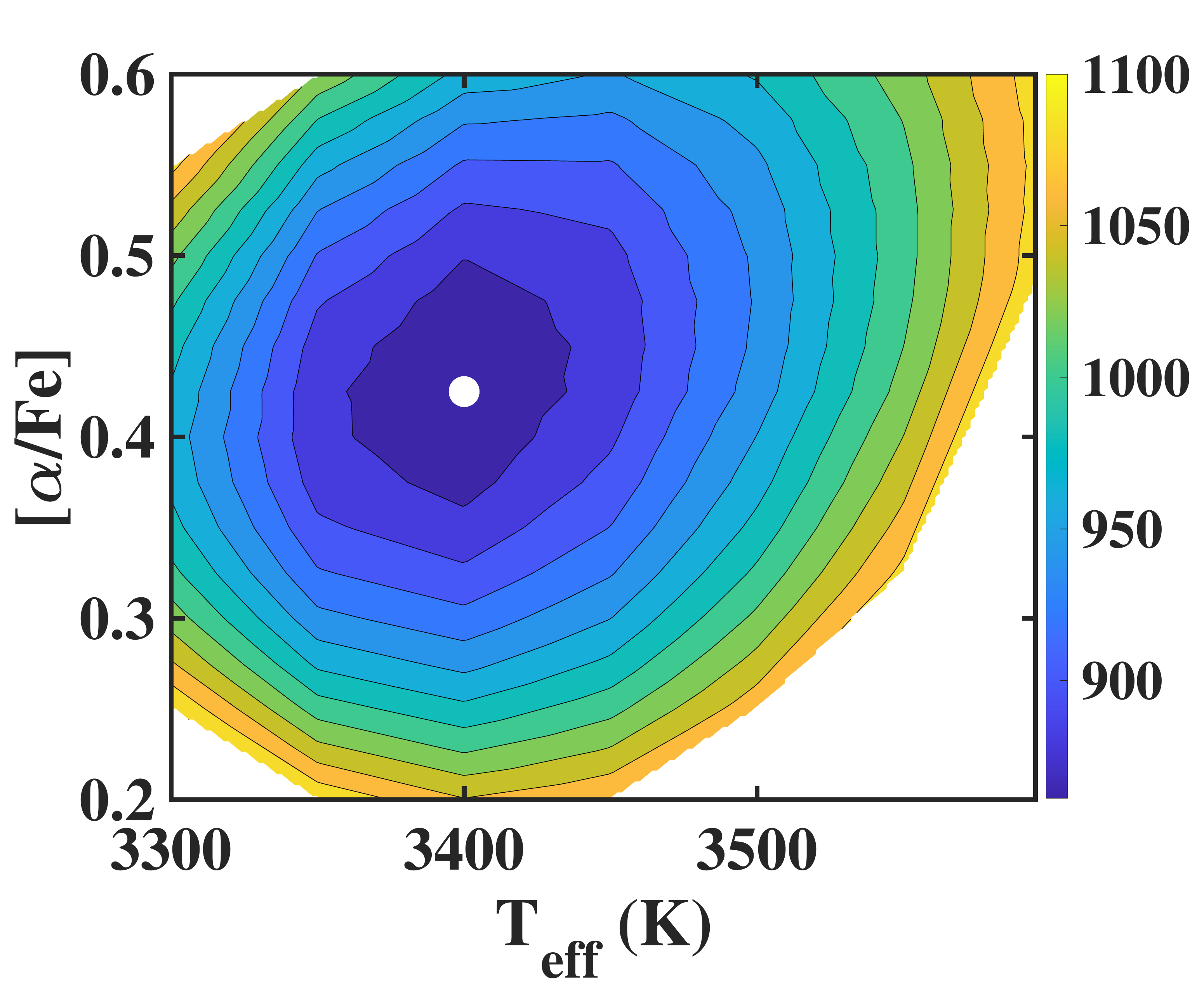}}               
\hspace{-0.1cm} 
 \subfloat      
         {\includegraphics[height=4.8cm, width=6cm]{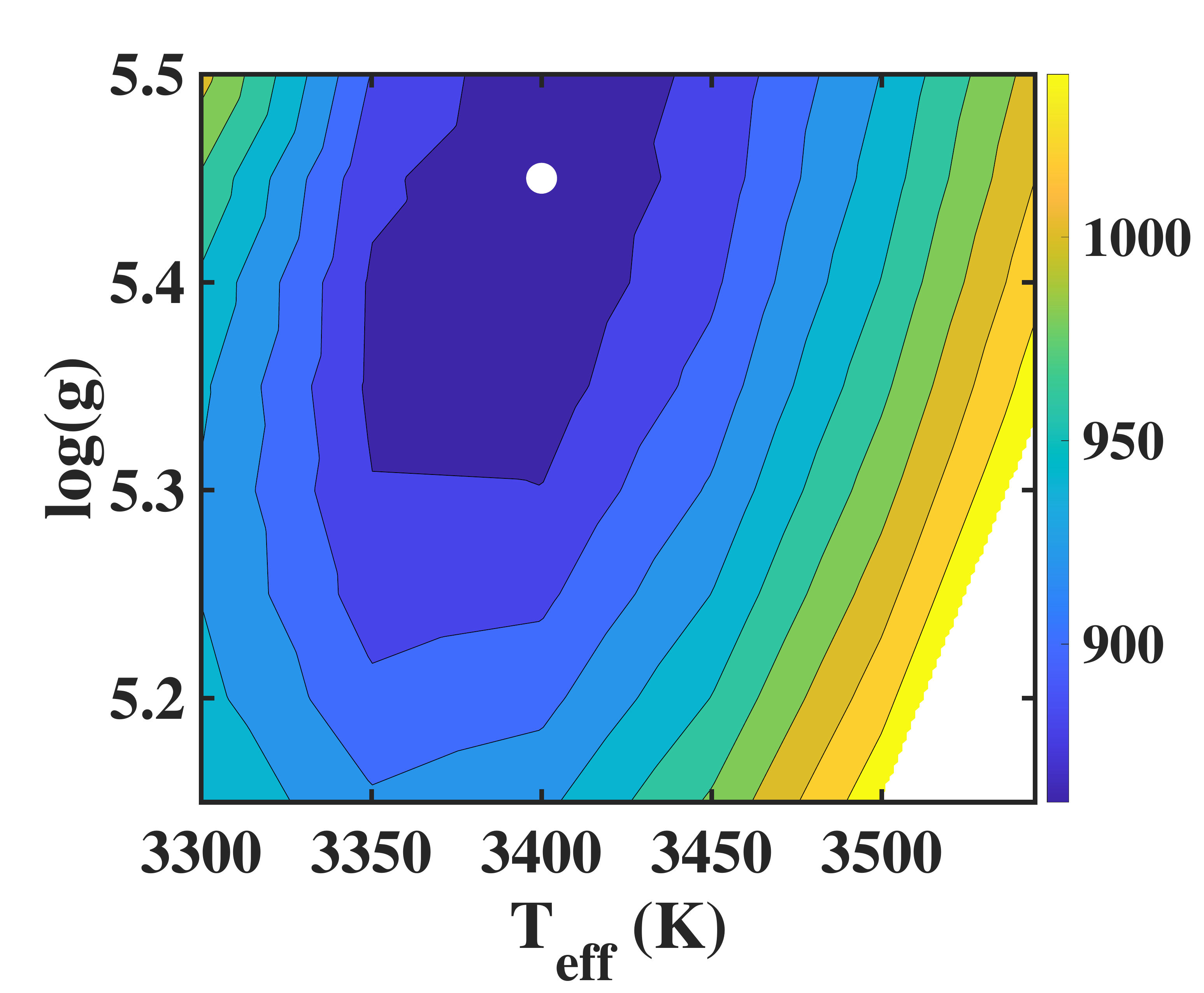}}
 \vspace{-0.35cm}

 \subfloat 
      {\includegraphics[ height=4.8cm, width=6cm]{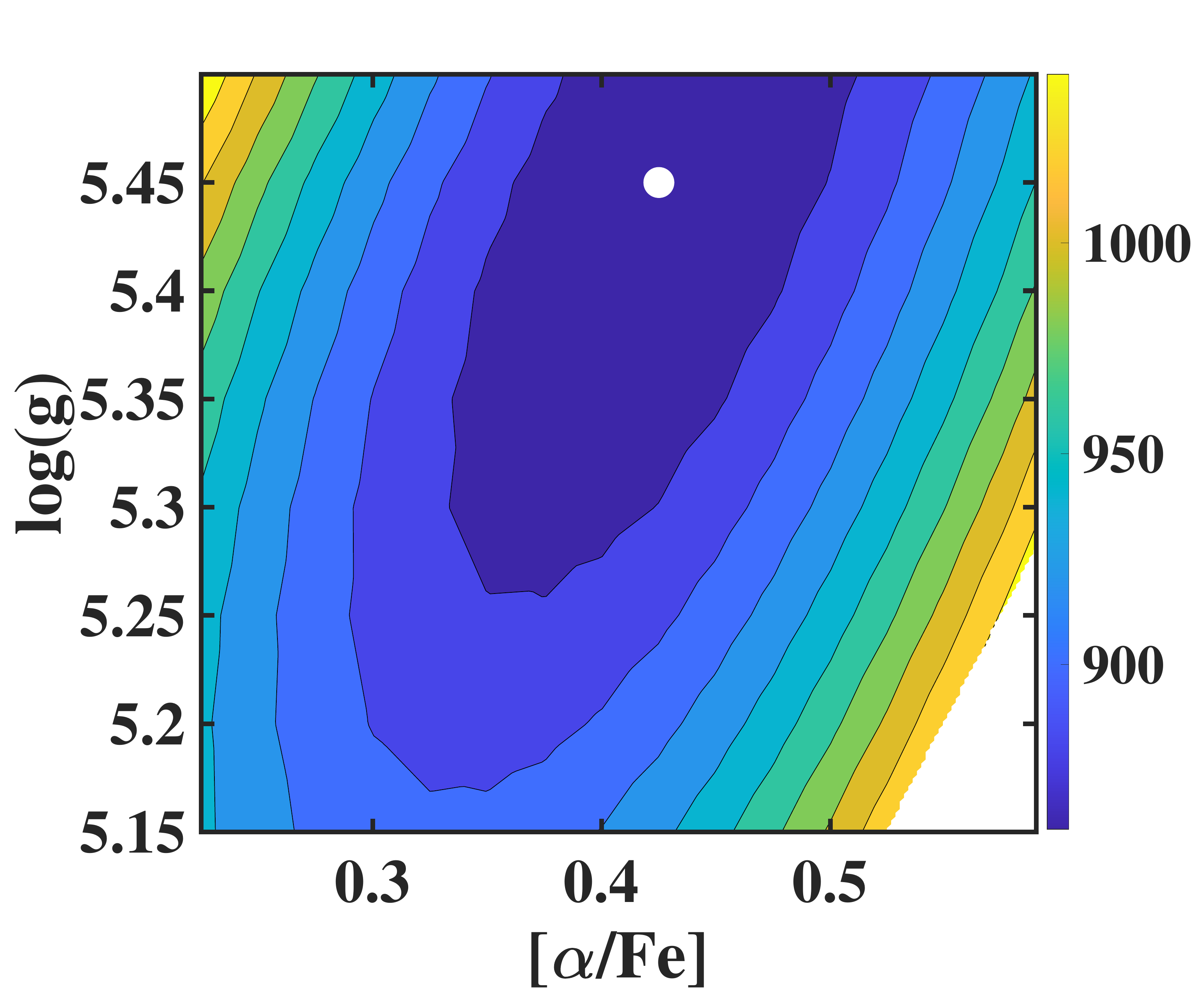}}               
\hspace{-0.1cm}
\subfloat 
      {\includegraphics[height=4.8cm, width=6cm]{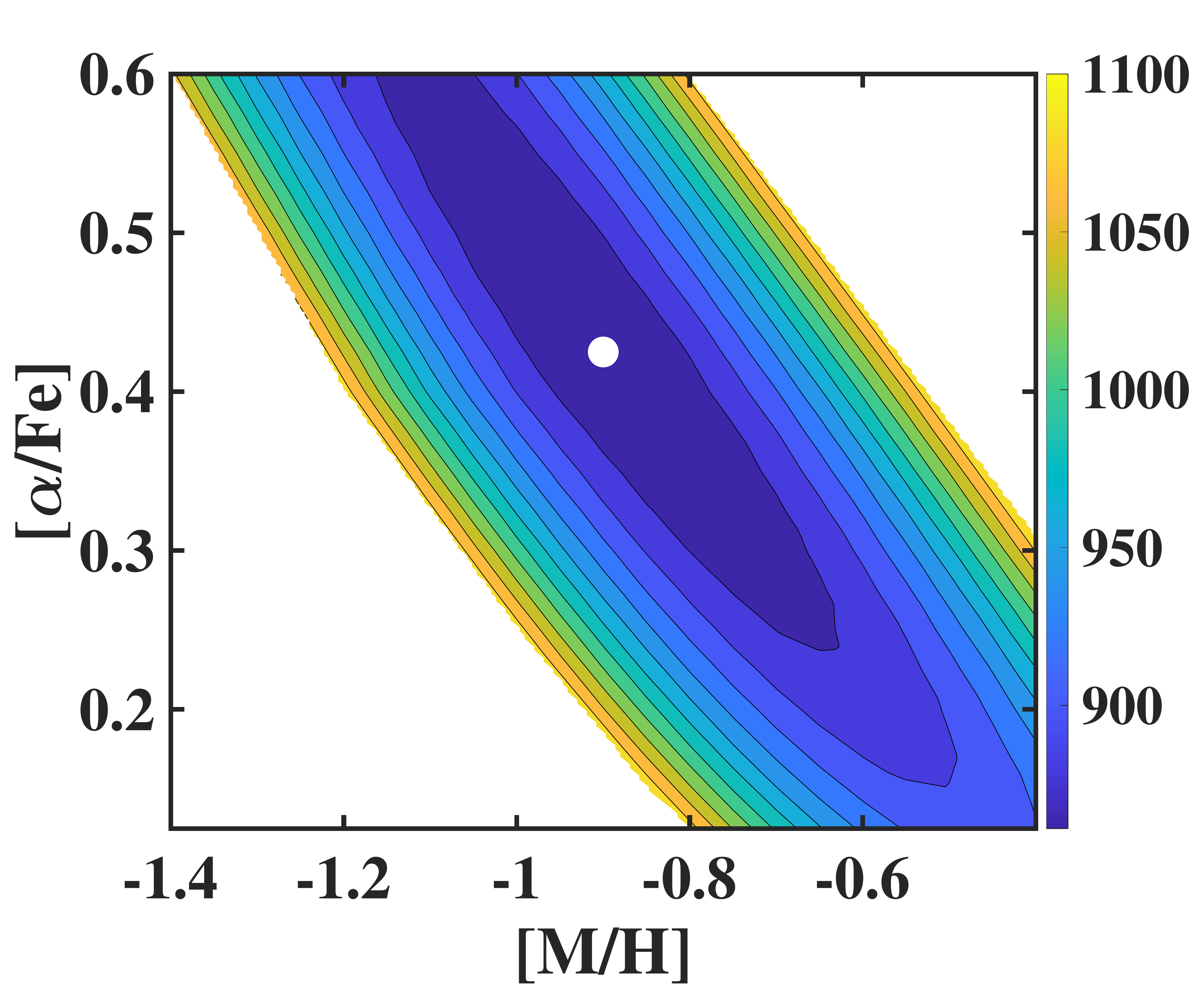}}               
\hspace{-0.1cm} 
 \subfloat      
         {\includegraphics[ height=4.8cm, width=6cm]{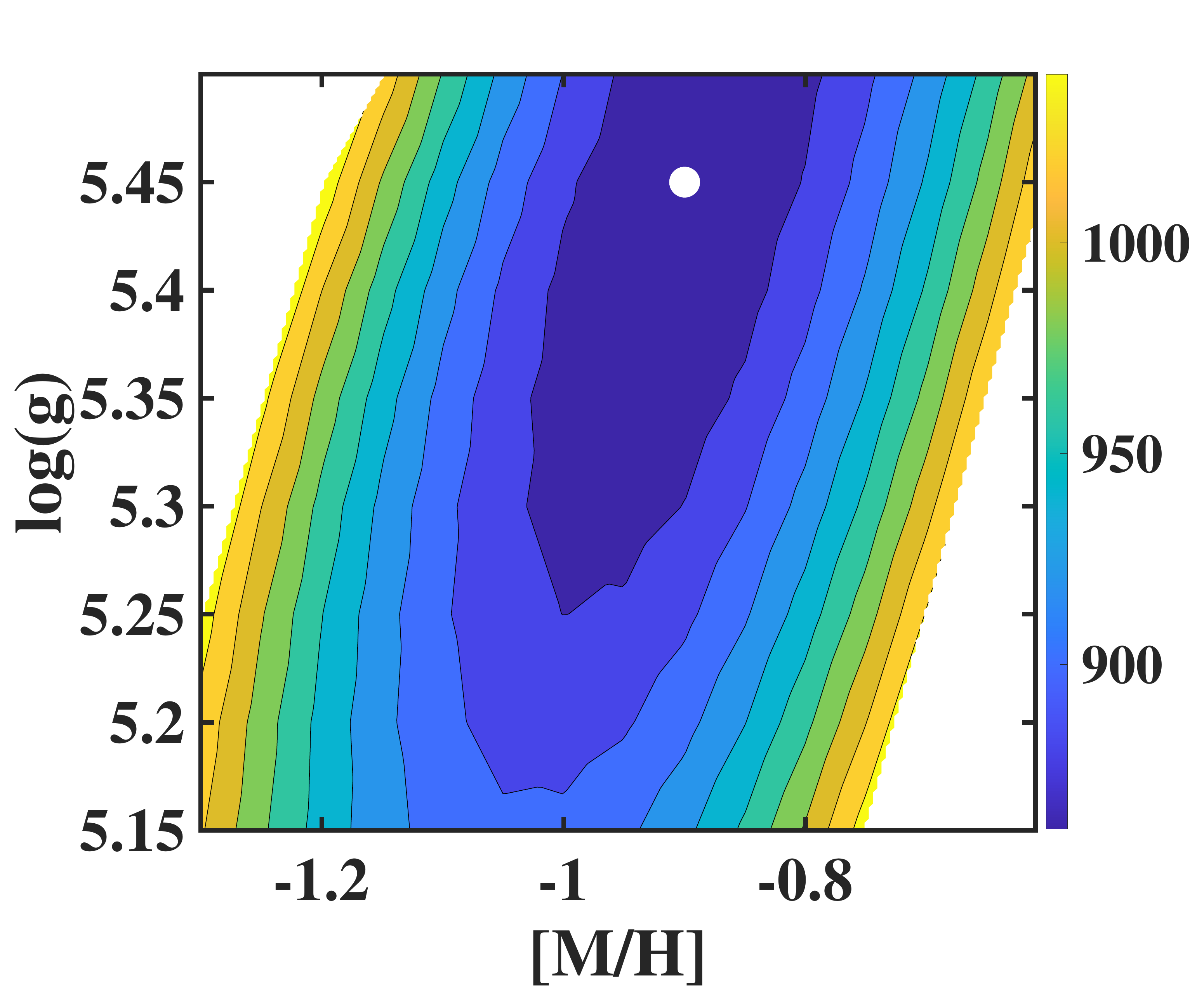}}        
 \caption
        {\footnotesize{$\chi$$^\textrm{\footnotesize{2}}$ maps in two-dimensional parameter planes for the star PM J03440+2730 with T$_\textrm{\footnotesize{eff}}$=3400 K, [M/H]=$-$0.9 dex, [$\alpha$/Fe]=+0.425 dex, and  log \emph{g}=5.45 dex. The best-fit grid point is shown by a white dot. The number of wavelength datapoints used in the $\chi$$^\textrm{\footnotesize{2}}$ calculation is 1063.}}
\end{figure*}

\begin{figure*}\centering
\subfloat 
      {\includegraphics[ height=4.8cm, width=6cm]{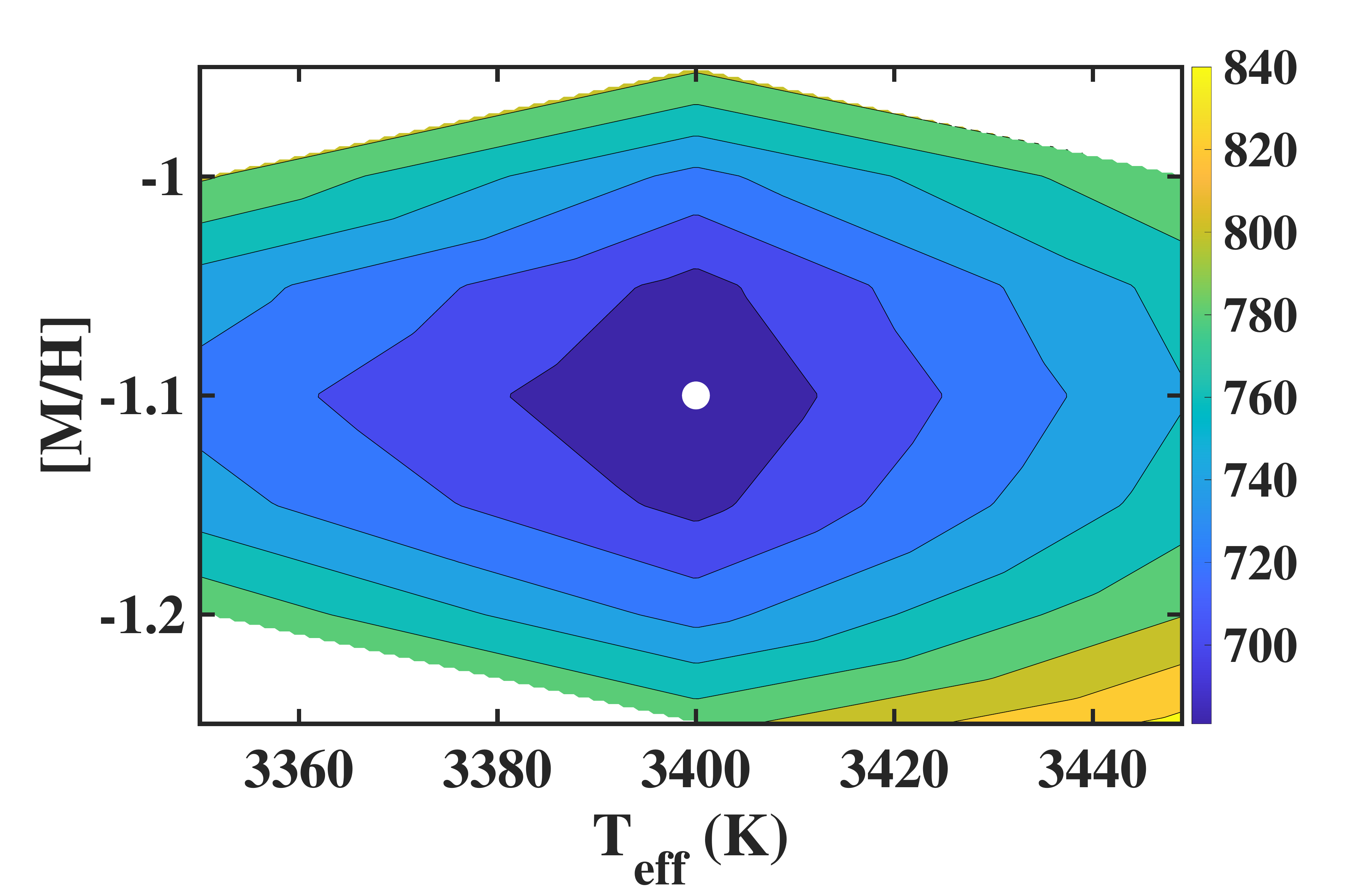}}               
\hspace{-0.1cm}
\subfloat 
      {\includegraphics[ height=4.8cm, width=6cm]{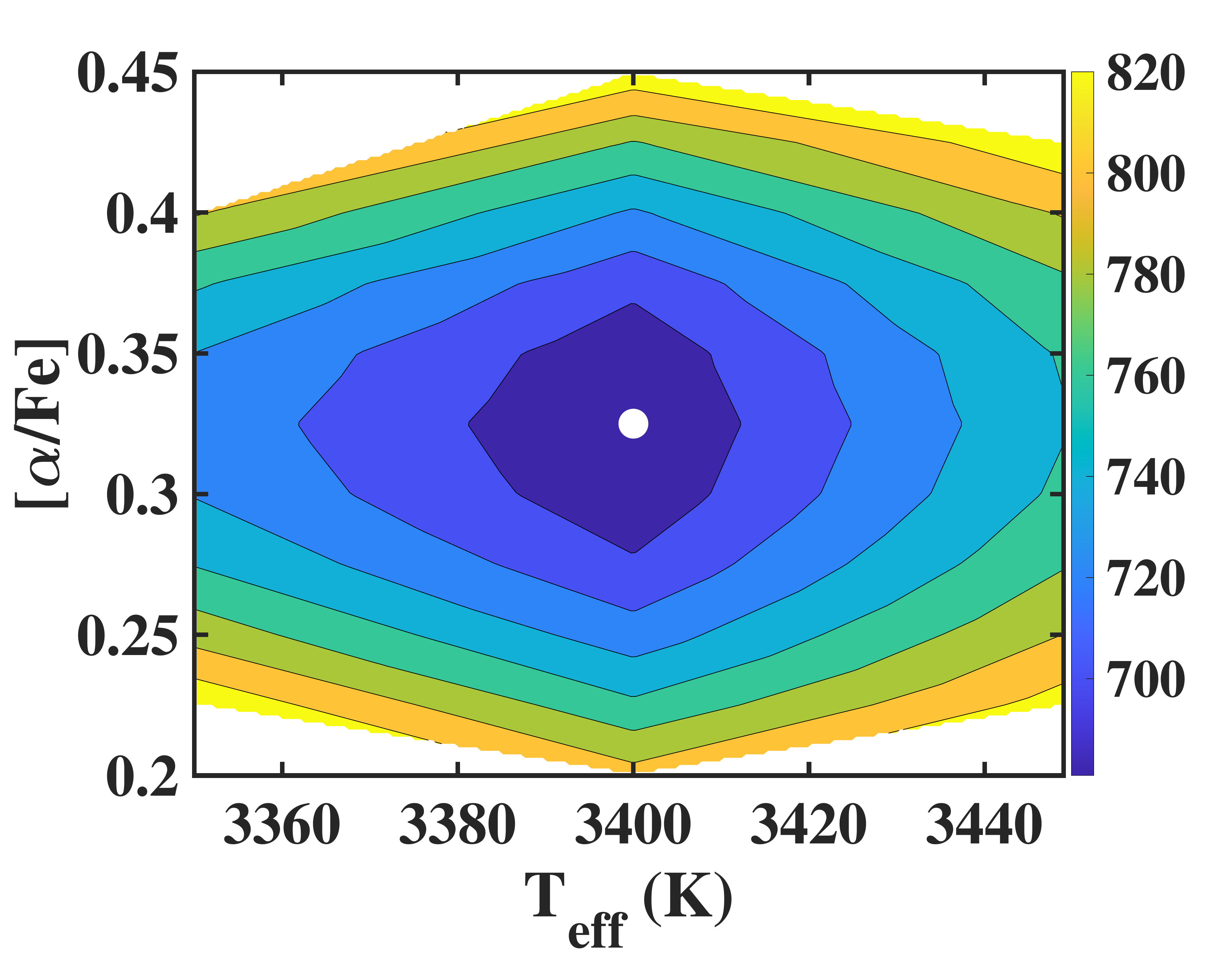}}               
\hspace{-0.1cm} 
 \subfloat      
         {\includegraphics[ height=4.8cm, width=6cm]{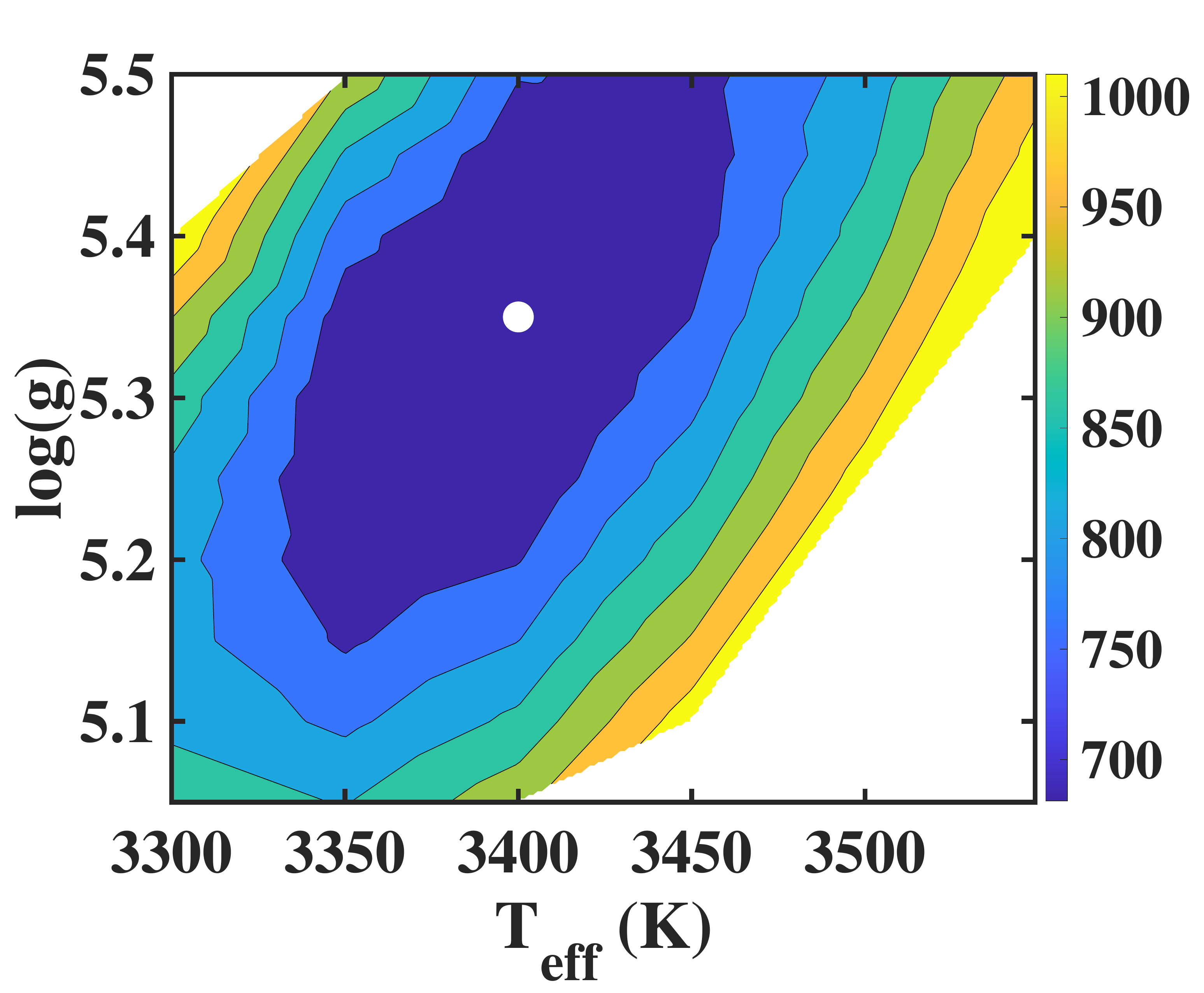}}
 \vspace{-0.35cm}

 \subfloat 
      {\includegraphics[height=4.8cm, width=6cm]{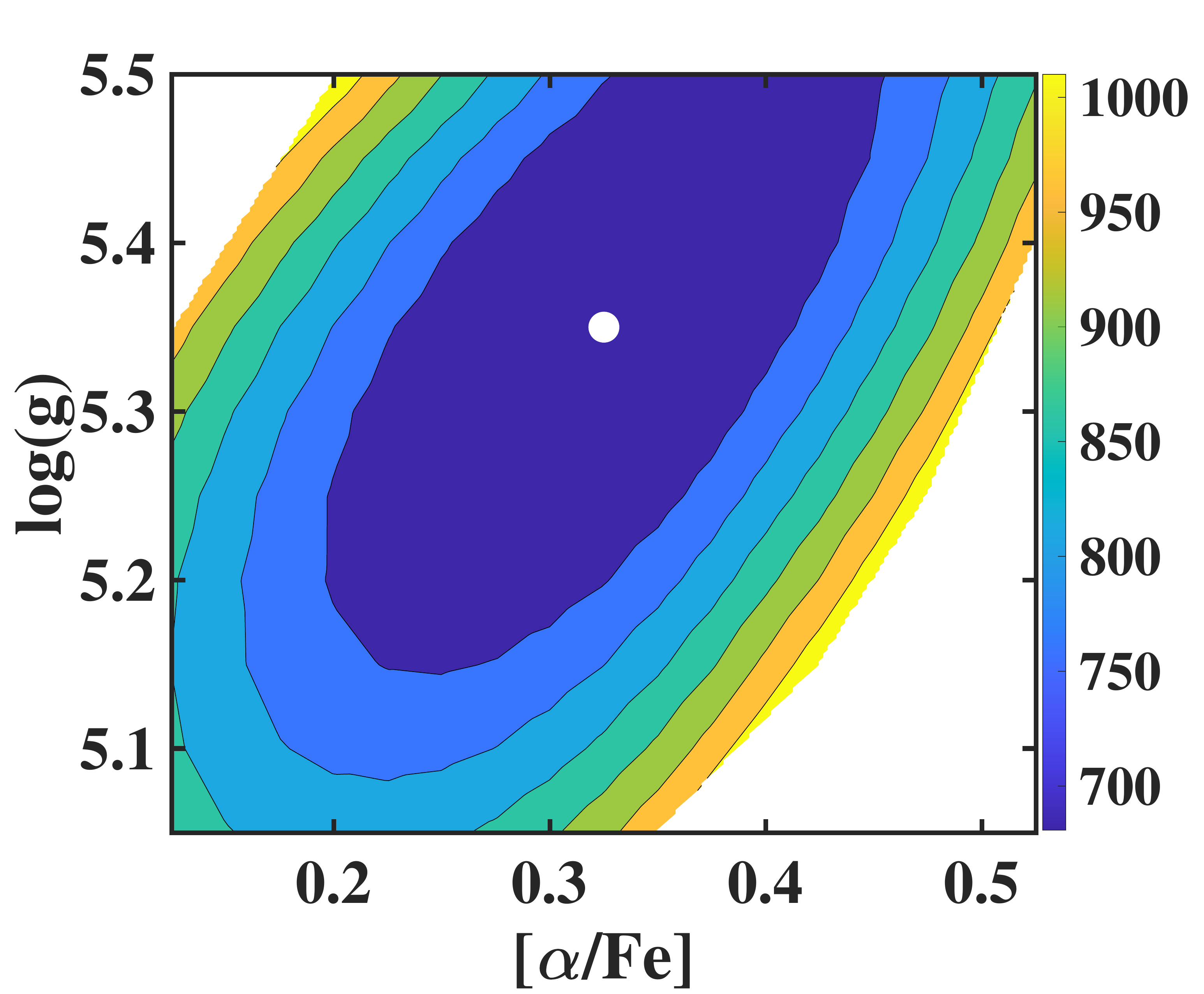}}               
\hspace{-0.1cm}
\subfloat 
      {\includegraphics[ height=4.8cm, width=6cm]{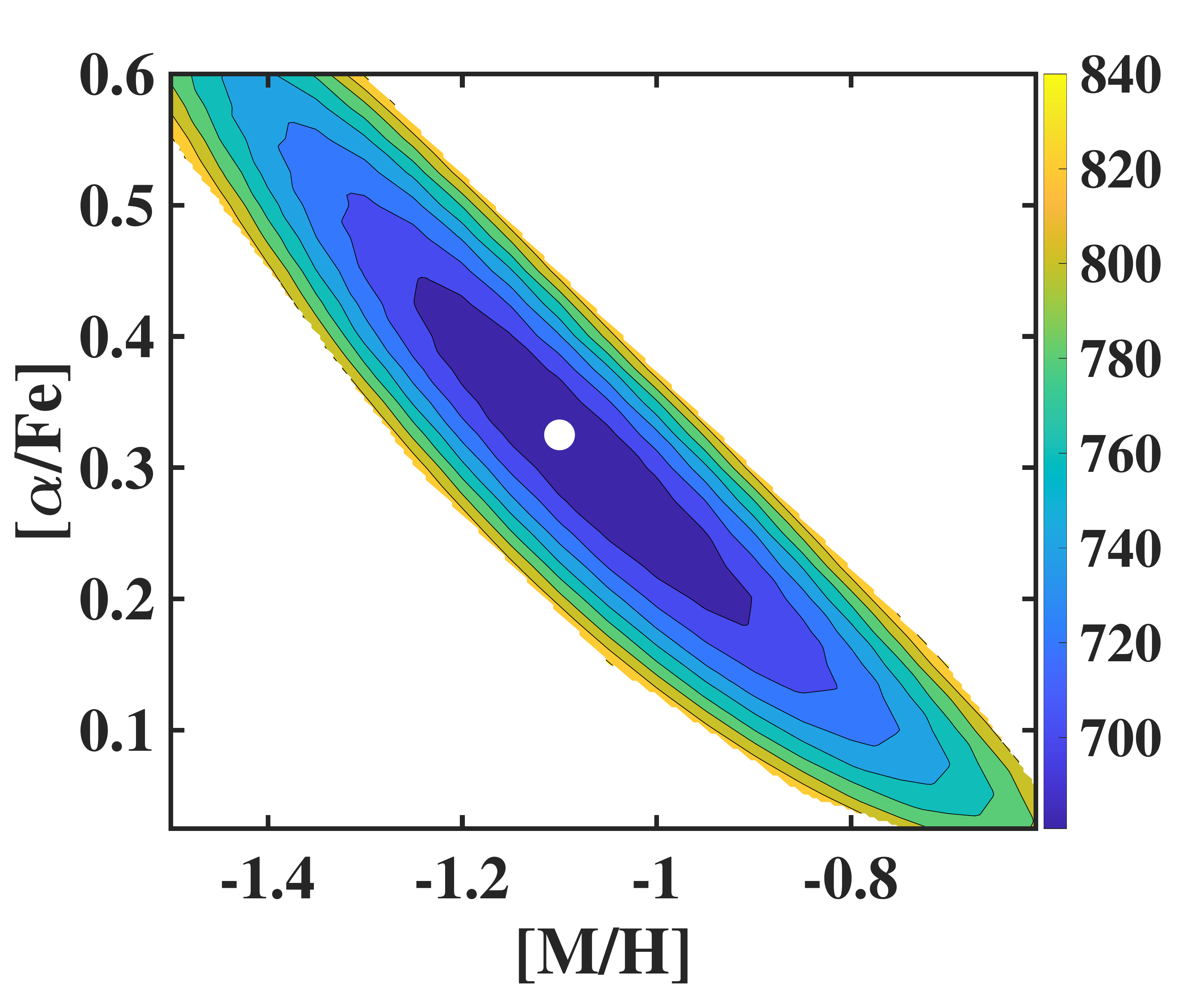}}               
\hspace{-0.1cm} 
 \subfloat      
         {\includegraphics[ height=4.8cm, width=6cm]{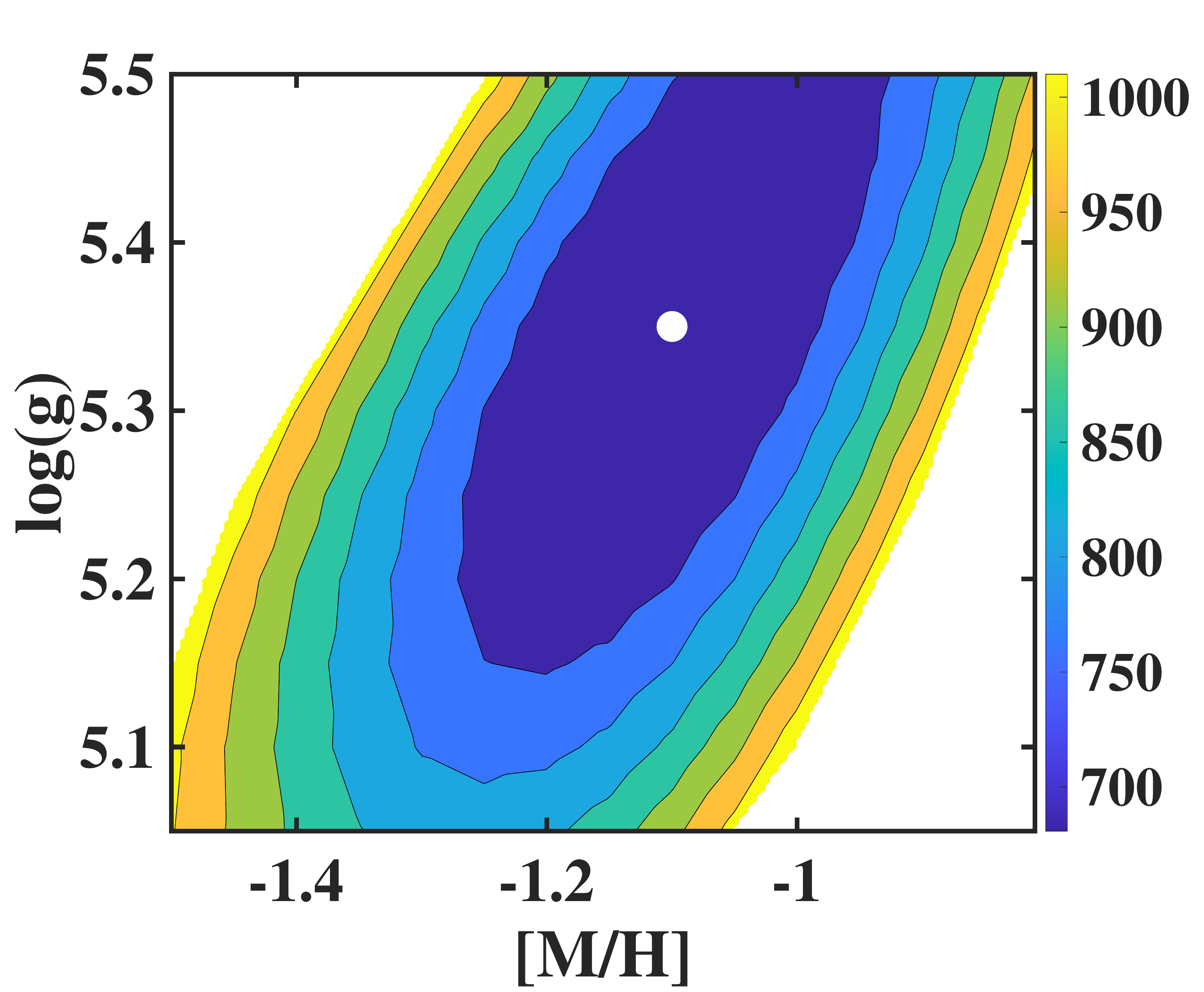}}        
 \caption
        {\footnotesize{$\chi$$^\textrm{\footnotesize{2}}$ maps in two-dimensional parameter planes for the star PM J03352+2055  with T$_\textrm{\footnotesize{eff}}$=3400 K, [M/H]=$-$1.1 dex, [$\alpha$/Fe]=+0.325 dex, and  log \emph{g}=5.35 dex. The best-fit grid point is shown by a white dot. The number of wavelength datapoints used in the $\chi$$^\textrm{\footnotesize{2}}$ calculation is 734.}}
\end{figure*}

\begin{figure*}\centering
\subfloat 
      {\includegraphics[height=4.8cm, width=6cm]{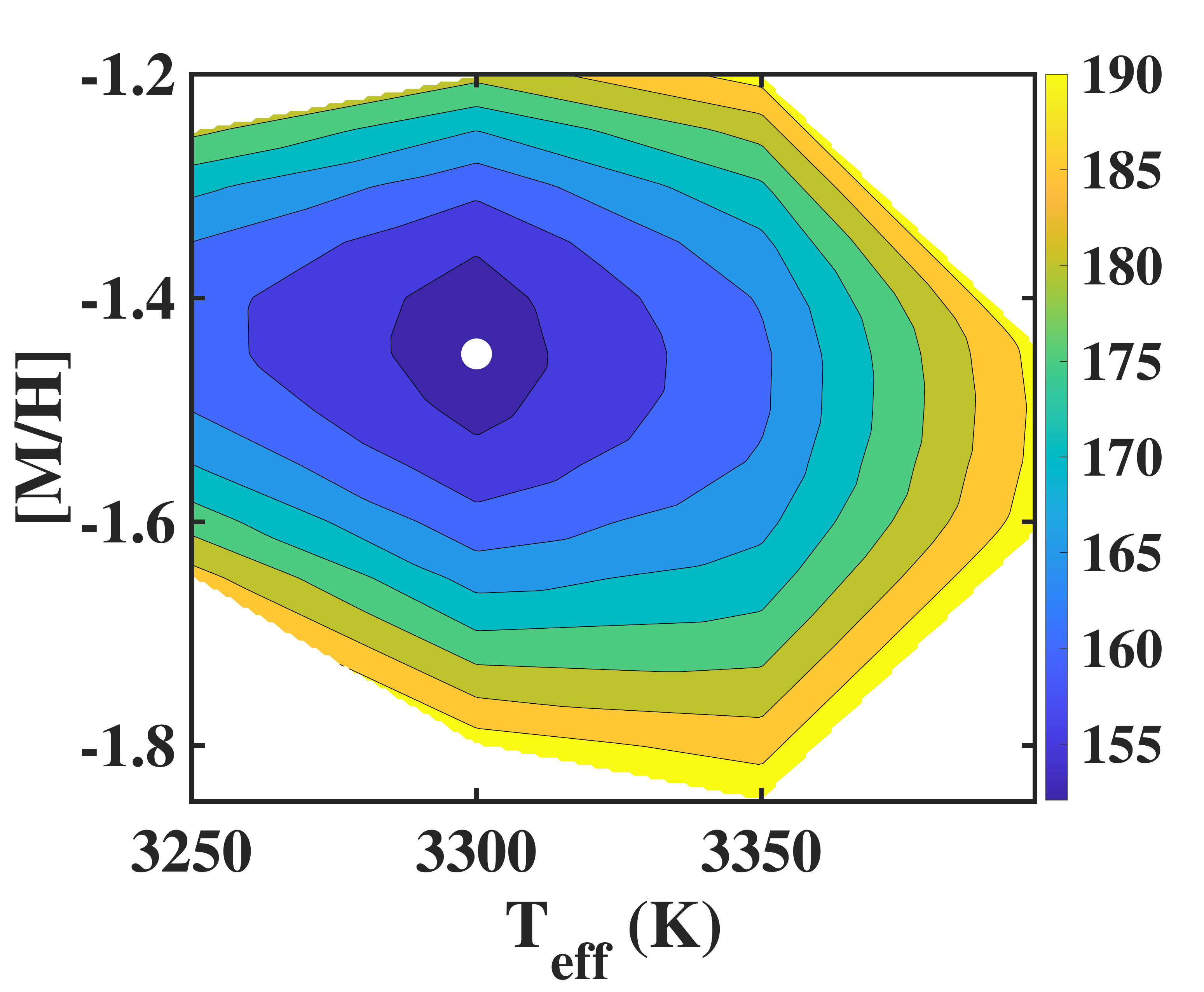}}               
\hspace{-0.1cm}
\subfloat 
      {\includegraphics[ height=4.8cm, width=6cm]{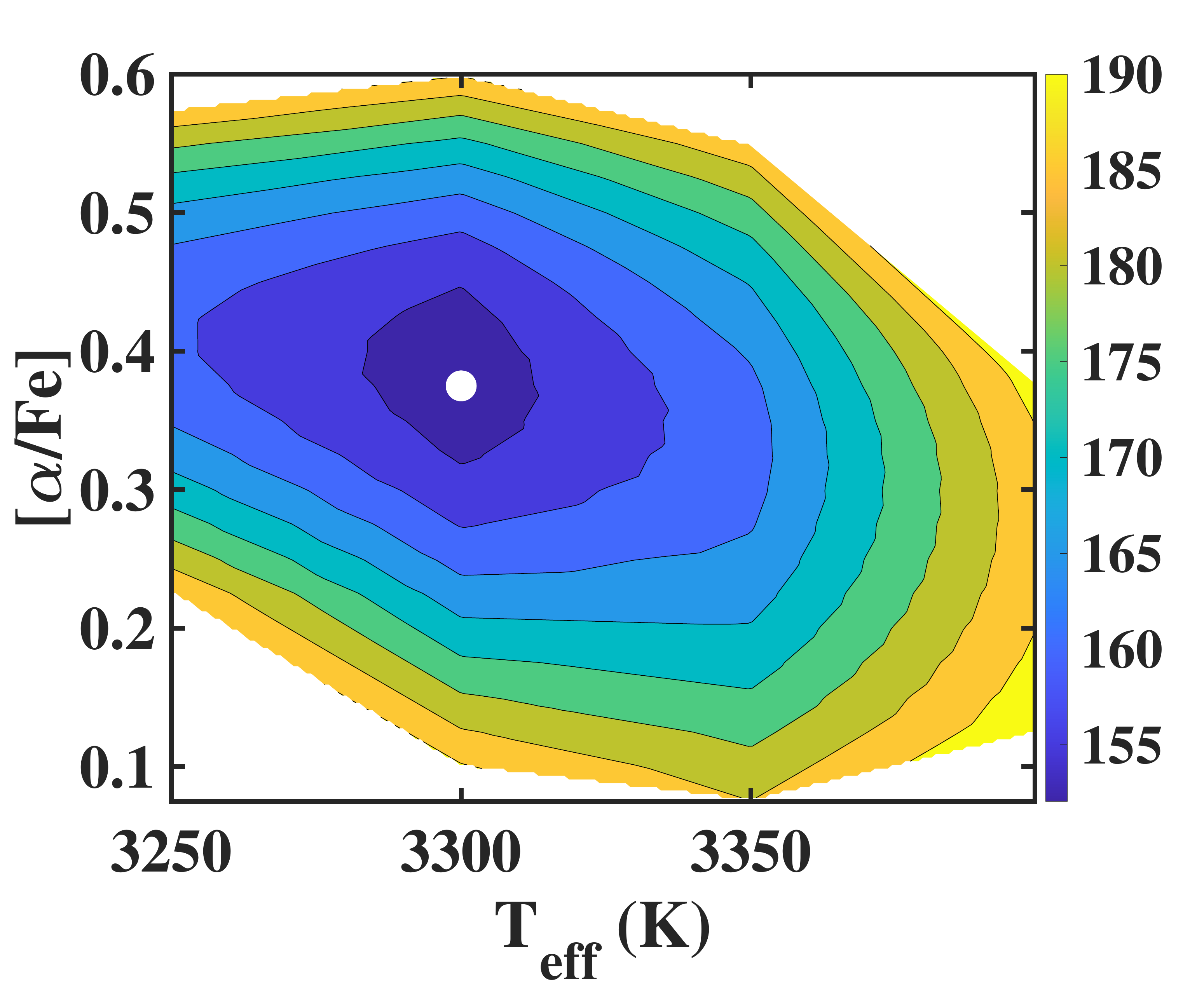}}               
\hspace{-0.1cm} 
 \subfloat      
         {\includegraphics[ height=4.8cm, width=6cm]{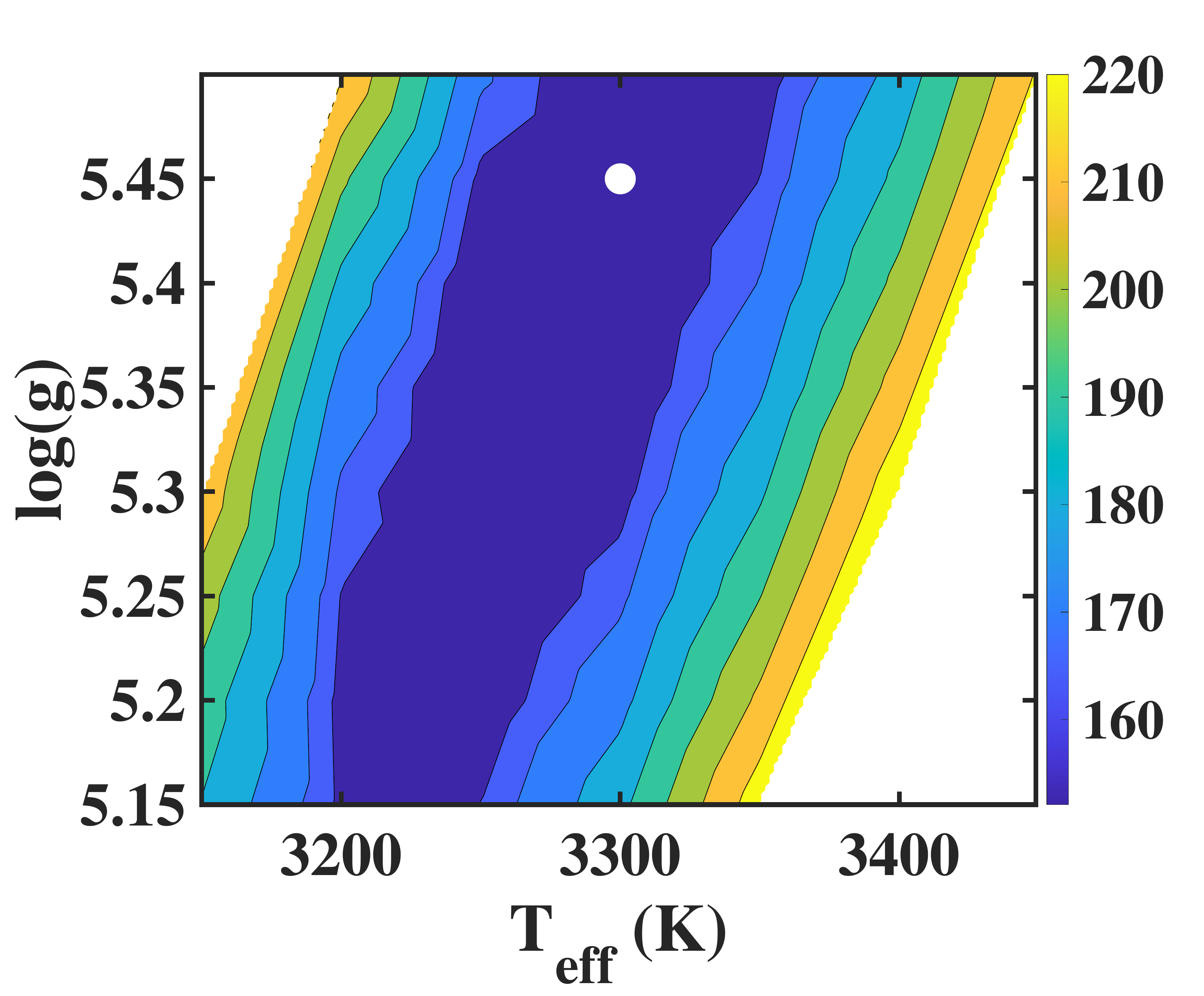}}
 \vspace{-0.35cm}

 \subfloat 
      {\includegraphics[ height=4.8cm, width=6cm]{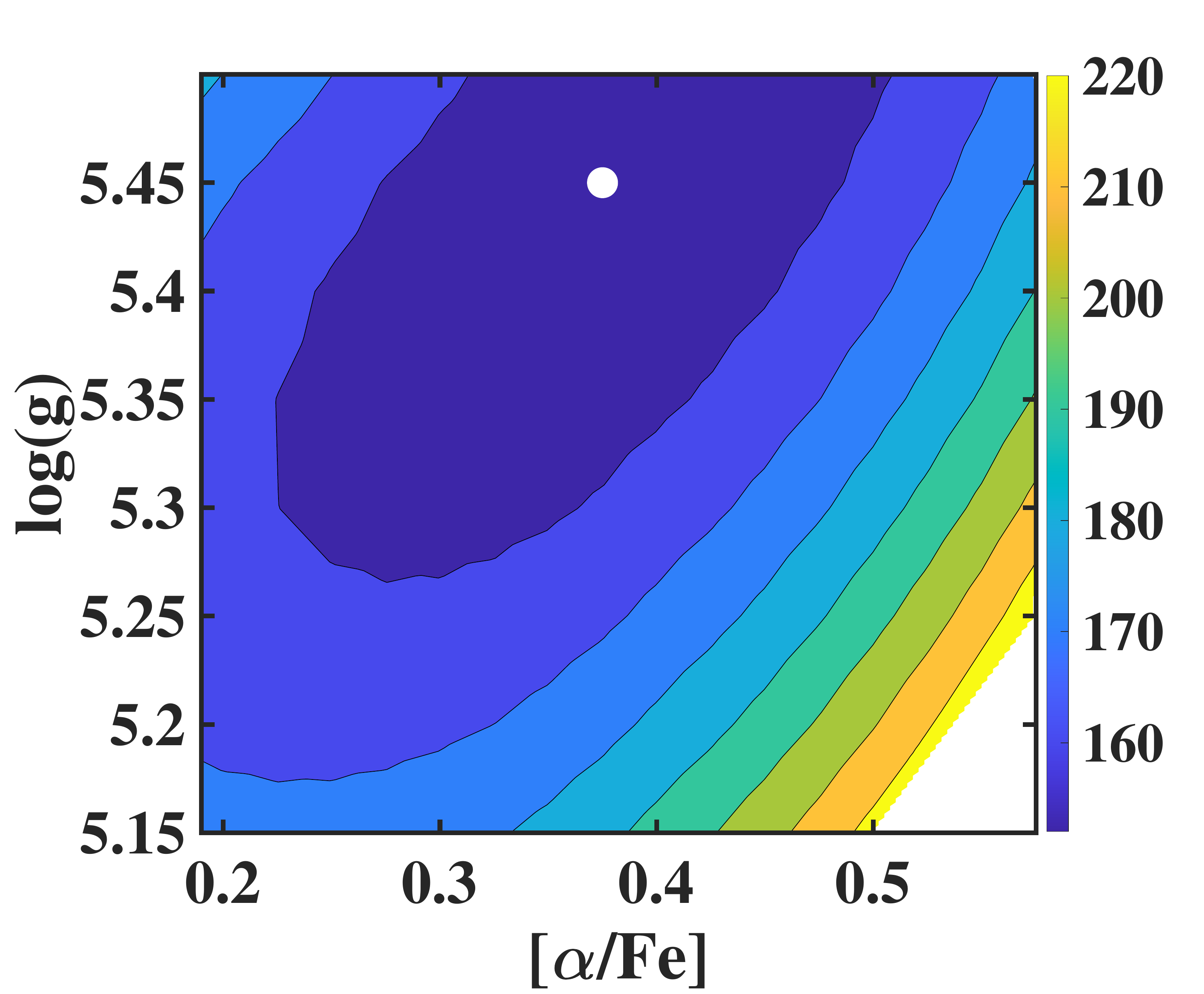}}               
\hspace{-0.1cm}
\subfloat 
      {\includegraphics[ height=4.8cm, width=6cm]{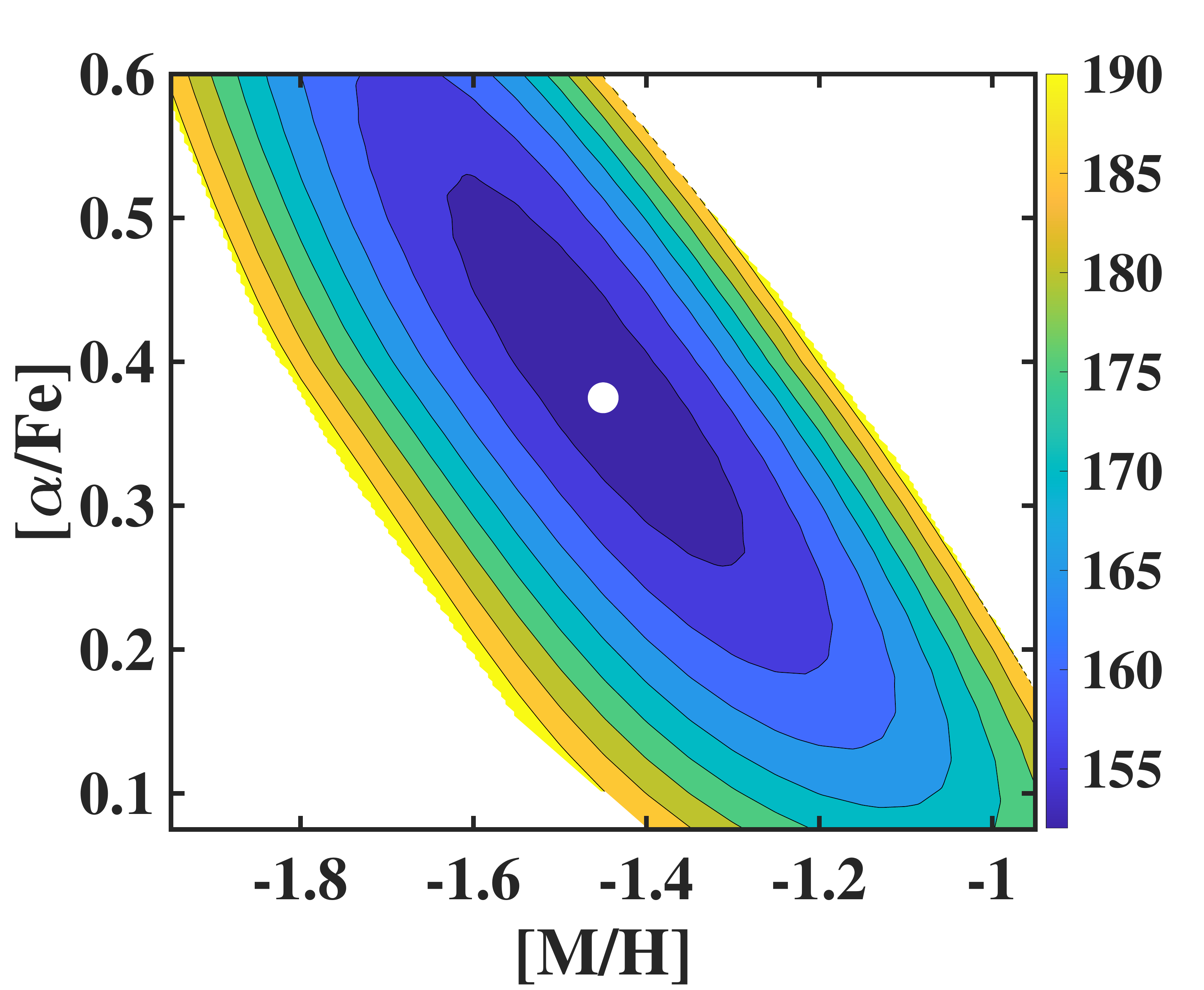}}               
\hspace{-0.1cm} 
 \subfloat      
         {\includegraphics[ height=4.8cm, width=6cm]{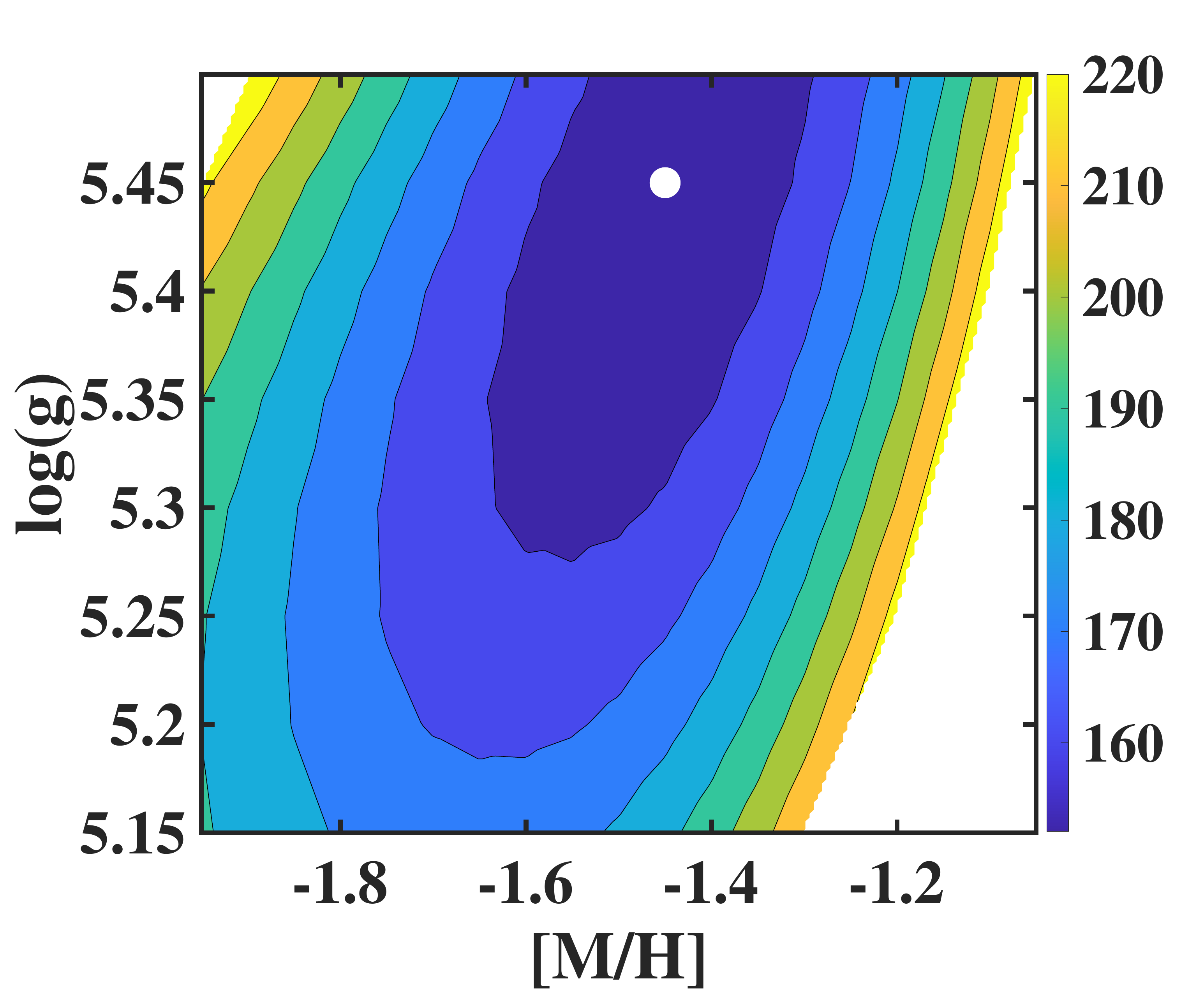}}        
 \caption
        {\footnotesize{$\chi$$^\textrm{\footnotesize{2}}$ maps in two-dimensional parameter planes for the star PM J06069+1706  with T$_\textrm{\footnotesize{eff}}$=3300 K, [M/H]=$-$1.45 dex, [$\alpha$/Fe]=+0.375 dex, and  log \emph{g}=5.45 dex. The best-fit grid point is shown by a white dot. The number of wavelength datapoints used in the $\chi$$^\textrm{\footnotesize{2}}$ calculation is 606.}}
\end{figure*}

\section{Error Analysis}
The uncertainty in our inferred parameters originates from distinct sources as follows:

\begin{itemize}
\item Uncertainties in observed spectra, e.g.,  instrumental errors and statistical noise from photon count. The observational uncertainties  are taken into account as an inverse weight in the $\chi$$^\textrm{\footnotesize{2}}$  formalism (Eq.1). 

\item Uncertainties in the interpolated synthetic spectra. Since the corresponding  spectra calculated from  the atmosphere code are not available yet (as their computations would be exceedingly time consuming), no comparison can be made to quantify the difference between the interpolated and calculated spectra. Nevertheless, we assume that these interpolated model spectra represent a fair approximation of the calculated ones.

\item Uncertainties in the flux-renormalization of synthetic spectra relative to the observed spectrum of interest. This is difficult to quantify using the present $\chi$$^\textrm{\footnotesize{2}}$ method. However, some other methods such as  Starfish (see above) provide a routine through MCMC simulations (C15) that can account for any uncertainties induced by the modification of model spectra. Unfortunately, such methods would be extremely computationally intensive for our large M dwarf/subdwarf sample, and  imperfections in the flux-renormalization of synthesized models thus remain as an underlying source of error in the inferred parameter. 

\item Errors from the residual correlation between wavelength datapoints. Such an error factor  can be significantly reduced by using the ``reduced-correlation method'' described in Section 4.3.2, although this does not make a considerable difference in the distribution of the derived parameters, which is likely due to the low spectral resolution of our spectra.

\item Errors  from  the degeneracy between physical parameters. As explained  in Section 3.4 (and also shown in Section 6), the degeneracy effect from log \emph{g} and each of chemical parameters, i.e., [M/H] and  [$\alpha$/Fe],  can cause significant errors in the resulting parameters. The best solution is to keep surface gravity fixed and equal to the values inferred from independent techniques. There is also a weaker degeneracy between [M/H] and  [$\alpha$/Fe], which may cause errors in the derived parameter values.

\item Uncertainties due to possibly undetected unresolved binaries. These binary systems have a composite spectrum, which can make significant errors in the derived stellar parameters.

\item Uncertainties associated with  limitations of models  to reproduce all observed spectral details. These are caused by the incompleteness of available opacity data, and by simplifying assumptions  in modeling stellar atmosphere, e.g.,  the local thermodynamic equilibrium (LTE) that overlooks  non-LTE (NLTE) effects, which may cause significant abundance errors (Bergemann \& Nordlander 2014; Nordlander et al 2017) or  simplifications used in the Mixing Length Theory (Kippenhahn \& Weigert 1990) connected with stellar convection. Such uncertainties are much more difficult to quantify and our error analysis does not include the systematic uncertainties due to the disagreement between models and observations. Nevertheless, we find systematic trends between parameters that are unlikely to have physical origins and may be attributed to the inadequacy of model atmospheres.

\end{itemize}

Figures 15-18 show the  $\chi$$^\textrm{\footnotesize{2}}$ maps in six two-dimensional parameter spaces for four typical stars in different metallicity regimes. The parameter values are estimated using the ``normal method'' in which surface gravity is considered as a free parameter. To create these two-dimensional maps, we vary two parameters that are shown in X and Y axes, while keeping the two other  parameters fixed and equal to their best-fit estimates (otherwise, the  $\chi$$^\textrm{\footnotesize{2}}$ would be a multivalued function). The best-fit grid point is presented by a white dot  in each plot. The number of wavelength  datapoints ($\sim$the degrees of freedom) used in the  $\chi$$^\textrm{\footnotesize{2}}$ calculation for each spectrum (i.e., after removing the regions including  the H$_\textrm{\footnotesize{$\alpha$}}$ emission line of hydrogen, telluric absorption lines, and  spectral artifacts as well as highly noisy ranges)  is shown in the respective captions. The best-fit grid points are located in the regions with lowest $\chi$$^\textrm{\footnotesize{2}}$ values, which shows the consistency of our model-fit pipeline, though in some cases the best-fit parameter is off-center from the $\chi$$^\textrm{\footnotesize{2}}$ distribution, which suggests small systematic errors in the minimization algorithm, likely due to the limited resolution of the model grid. We also note that the reduced $\chi$$^\textrm{\footnotesize{2}}$ ($\chi$$^\textrm{\footnotesize{2}}$/degrees of freedom) is often near or even less than 1, indicating a potential issue with the flux uncertainties (E$_{\textrm{obs}}$) that, in addition to Poisson noises,  incorporate terms propagated from uncertainties in the flux calibration. These errors are partially,  but not fully, canceled by flux-renormalization (Section 4.1), and the flux uncertainty values are therefore likely to be somewhat overestimated.

There is a relatively tight correlation between [M/H] and [$\alpha$/Fe], directed  from high-[M/H] and low-[$\alpha$/Fe] to low-[M/H] and high-[$\alpha$/Fe] values. As explained in Section 3, these two chemical parameters change  the spectral morphology in roughly the same way, but at different levels. This anti-correlation between [M/H] and [$\alpha$/Fe], however, suggests that the combined parameter [$\alpha$/Fe]+[M/H] should be significantly better constrained than either one individually. This anti-correlation  also appears as  tilted distributions  of stars with nearly  the same best-fit values of log \emph{g} (even with different effective temperatures) in the [$\alpha$/Fe] versus [M/H] diagram, no matter which gravity-modeling approach is used (Section 6.3). As can be seen from these figures, T$_\textrm{\footnotesize{eff}}$ is much less correlated with [M/H] and [$\alpha$/Fe], while the correlation between log \emph{g} and  the two chemical parameters is  prominent.

The joint error distribution of  T$_\textrm{\footnotesize{eff}}$, [M/H], [$\alpha$/Fe], and log \emph{g} can be estimated using the $\chi$$^\textrm{\footnotesize{2}}$ values of a model grid that spans over ranges  adequately wider  than the expected parameter errors\footnote{Critical values of the $\chi$$^\textrm{\footnotesize{2}}$ distribution with various degrees of freedom are tabulated and available in the literature. However, these values do not apply to our analysis, yielding too small parameter errors.}. Focusing on the ``normal method'',  we  find the ``joint confidence grid'',  that contains the  models with $\chi$$^\textrm{\footnotesize{2}}$ values within 20{\%} deviation from the minimum value, is large enough for calculating the errors of the derived parameters. The deviation from the minimum value of $\chi$$^\textrm{\footnotesize{2}}$ by 20{\%} typically spans parameter ranges around best-fit values within T$_\textrm{\footnotesize{eff}}$$\sim$$\pm$150-160 K, [M/H]$\sim$$\pm$0.4-0.45 dex, [$\alpha$/Fe]$\sim$$\pm$0.2-0.25 dex, and  log \emph{g}$\sim$$\pm$0.25 dex (if using the variable-gravity approach) that are considerably larger than the average errors of our results from Paper I. In addition, the variation of the inferred parameter values from the present work in the abundance diagram of [$\alpha$/Fe] versus [M/H] (Section 6.3) or abundance-velocity  spaces (Section 9) are quite comparable to those of other studies, which indicates that the expected parameter errors must be  less than the above values. In Section 9, we also identify substructure such as streams in abundance-velocity spaces that would not be possible if the resulting chemical parameters had large uncertainties. The 20{\%} deviation is a conservative choice,  in comparison with  the 5{\%}  departure chosen in the high-resolution analysis of Rajpurohit et al. (2014) and  the 10{\%}  departure chosen in our previous low-resolution analysis of Paper I.  The parameter errors  are then determined by the standard deviation of the individual confidence ranges that are  obtained by projecting the  four-dimensional joint confidence grid on  the corresponding one-dimensional  parameter spaces.

In the present study, the inferred parameter values from the ``reduced-correlation method'' are only used to compare with those from the ``normal method''  and  to investigate how the results change when using independent wavelength datapoints.  The standard deviation of the ten measurements from the mean value for each parameter is calculated to show the scatter of different estimates from different sets of nearly uncorrelated datapoints.

\section{Results: Normal Method}
We apply the ``normal method'' to the 3745 M dwarfs/subdwarfs for two approaches when surface gravity is kept constant (as specified by ``const-grav'') and when surface gravity is allowed to vary (as specified by ``var-grav''). In the following sections, we examine the distribution of the  inferred parameter values using different diagrams and address the effect of surface gravity variation on the model-fit values of the other parameters.

\subsection{Effective Temperature Distribution}

 Figure 19 compares the photometric temperatures with  the spectroscopic model-fit values inferred from the two gravity-modeling  approaches, which are randomized within  the step size of the temperature gird. Both plots present a  similar dispersion around the 1:1 line (black line), though there is a slightly better consistency between the  photometric temperatures and the model-fit values  from the constant-gravity approach (top panel) compared those from the variable-gravity approach  (bottom panel). These panels indicate  an average offset (downward shift relative to the 1:1 line) of $\sim$120-140 K for stars with T$_\textrm{\footnotesize{eff}}$$\gtrsim$3700 K and  $\sim$50 K  for stars with T$_\textrm{\footnotesize{eff}}$$\lesssim$3100-3200 K. The largest discrepancies occur in metal-poor stars with a typical offset (upward shift relative to the 1:1 line) of $\sim$170 K between two temperature measurements, which suggests that the photometric temperatures for low-metallicity stars may not be as reliable; this is likely because temperature-color relationships have been largely calibrated and tested using metal-rich stars. We thus assume that the spectroscopic model-fit values are more reliable in this case.

\subsection{Gaia HR Diagram}

We aim to examine the variation of parameter values in the  HR diagram using Gaia magnitudes and parallaxes. However, the position of stars in  such a diagram is affected by the precision of Gaia astrometric measurements which is quantified by a quality flag, i.e., the  Renormalized Unit Weight Error (RUWE)\footnote{The detailed description of RUWE can be found in the Gaia  DPAC public documentation: https://gea.esac.esa.int/archive/documentation/GDR2/Gaia\_ar \newline chive/chap\_datamodel/sec\_dm\_main\_tables/ssec\_dm\_ruwe.h \newline tml}. The RUWE is a measure of the astrometric noise and is expected to be  around 1.0 for a well-behaved measurement of a single star  where the single-star model provides a good fit to the astrometric observations, despite the primary color or magnitude. On the other hand, a RUWE value significantly greater than 1.0 (say, >1.4) indicates that the source is non-single or otherwise problematic for the astrometric solution. The RUWE is set to null for sources with only a two-parameter solution, since this value would be difficult to interpret in such cases. Some studies have used this parameter as an indication of binarity. For example, Ziegler et al. (2020) showed  that 86{\%} of their planet candidate sample with RUWE>1.4 had companions undetected by the Gaia survey, but observed in high-resolution follow-up. The effect and application of Gaia RUWE values can be found in a number of papers (e.g., Ziegler et al. 2020, Belokurov et al. 2020; Wood et al. 2021; Kervella et al. 2022).

Figure 20 shows the RUWE values of our 3745 stars versus their distances  (in  logarithmic scale). We find high values of RUWE for a sizable fraction of our sample. In total 1066 stars, i.e., 28{\%} of the sample, have  RUWE>1.4, of which 459 stars have 1.4<RUWE$\leq$2, 373 stars have 2<RUWE$\leq$10, and 234 stars have RUWE>10.  As seen from the figure, the high-RUWE stars are mostly  located at lower distances ($\lesssim$50 pc), presenting a high level of binarity. In general, unresolved binaries are easier to detect from astrometric deviations when the stars are relatively nearby. Our sample consists of a large number of nearby stars (within 25 pc) compared with typical  samples from Gaia that usually include stars at  larger distances, and consequently,  unresolved systems might not show up as well in the astrometric solution. Moreover, our spectroscopic survey selected nearby stars based on (pre-Gaia) photometric distances, and   binary stars  are more likely to be selected in a ``nearby'' sample based on a photometric distance cut because unresolved binaries are overluminous at a given color and thus appear to be closer. Figure 21 displays the Gaia HR diagram,  \textit{M$_{\footnotesize{G}}$} (absolute \textit{G} magnitude) versus  \textit{G}$_\textrm{\footnotesize{BP}}$-\textit{G}$_\textrm{\footnotesize{RP}}$ (top panel) and \textit{M$_{\footnotesize{G}}$} versus  \textit{G} -\textit{G}$_\textrm{\footnotesize{RP}}$ (bottom panel), of  two subsets: 2679 stars with RUWE<1.4 (as shown in blue) and 1066 stars with RUWE>1.4 (as shown in black). Clearly, the majority of high-RUWE stars reside towards the upper end of the distribution, which is likely due to being over-luminous as a result of their  binarity.

Figure 22 presents the Gaia HR diagram, \textit{M$_{\footnotesize{G}}$} versus  \textit{G}$_\textrm{\footnotesize{BP}}$-\textit{G}$_\textrm{\footnotesize{RP}}$,  of the full set of 3745 stars (top panels) and a cleaner subset of  2679 stars with RUWE<1.4  (bottom panels).  All these plots are color-mapped based on the  effective temperatures derived from both  procedures when  log \emph{g} is  fixed (left panels) and when  log \emph{g} is variable (right panels); in this case we find that there is no significant difference between the inferred temperatures from these two  gravity-modeling approaches.  Obviously, the effective temperature of stars is consistent  with their Gaia color and absolute magnitude, and this close agreement is a strong indication that our model-fit pipeline produces relatively high precision values for T$_\textrm{\footnotesize{eff}}$ in most cases. The absence of  high-temperature, low-metallicity stars  is a result of sample selection effect;  among metal-poor, low-mass stars, those with T$_\textrm{\footnotesize{eff}}$>3600 K look significantly bluer and tend to be classified as K dwarfs, and are thus excluded. In addition, low-temperature, metal-poor M subdwarfs are often  too faint to be observed using the selected telescopes and spectrographs (see Tables 1 and 2 in Paper I for more details), and for this reason,  are not present  in our sample. Both effects explain why metal-poor M subdwarfs are found over a much narrower range of T$_\textrm{\footnotesize{eff}}$ values (3000$\lesssim$T$_\textrm{\footnotesize{eff}}$$\lesssim$3600 K), while the range of T$_\textrm{\footnotesize{eff}}$ values for the more metal-rich M dwarfs is much broader  (2800$\lesssim$T$_\textrm{\footnotesize{eff}}$$\lesssim$4000 K). Figure 23 depicts the HR diagram, but with another form, i.e.,  \textit{M$_{\footnotesize{G}}$} versus  \textit{G}-\textit{G}$_\textrm{\footnotesize{RP}}$, of the same samples, color-mapped by the same effective temperatures, as shown in the respective panels of Figure 22. The impact of RUWE appears to be more important in the HR diagram with the color  \textit{G}-\textit{G}$_\textrm{\footnotesize{RP}}$ than the color  \textit{G}$_\textrm{\footnotesize{BP}}$-\textit{G}$_\textrm{\footnotesize{RP}}$.

\begin{figure}\centering
\subfloat
        [Fixed surface gravity]{\includegraphics[ height=4.8cm, width=6.8cm]{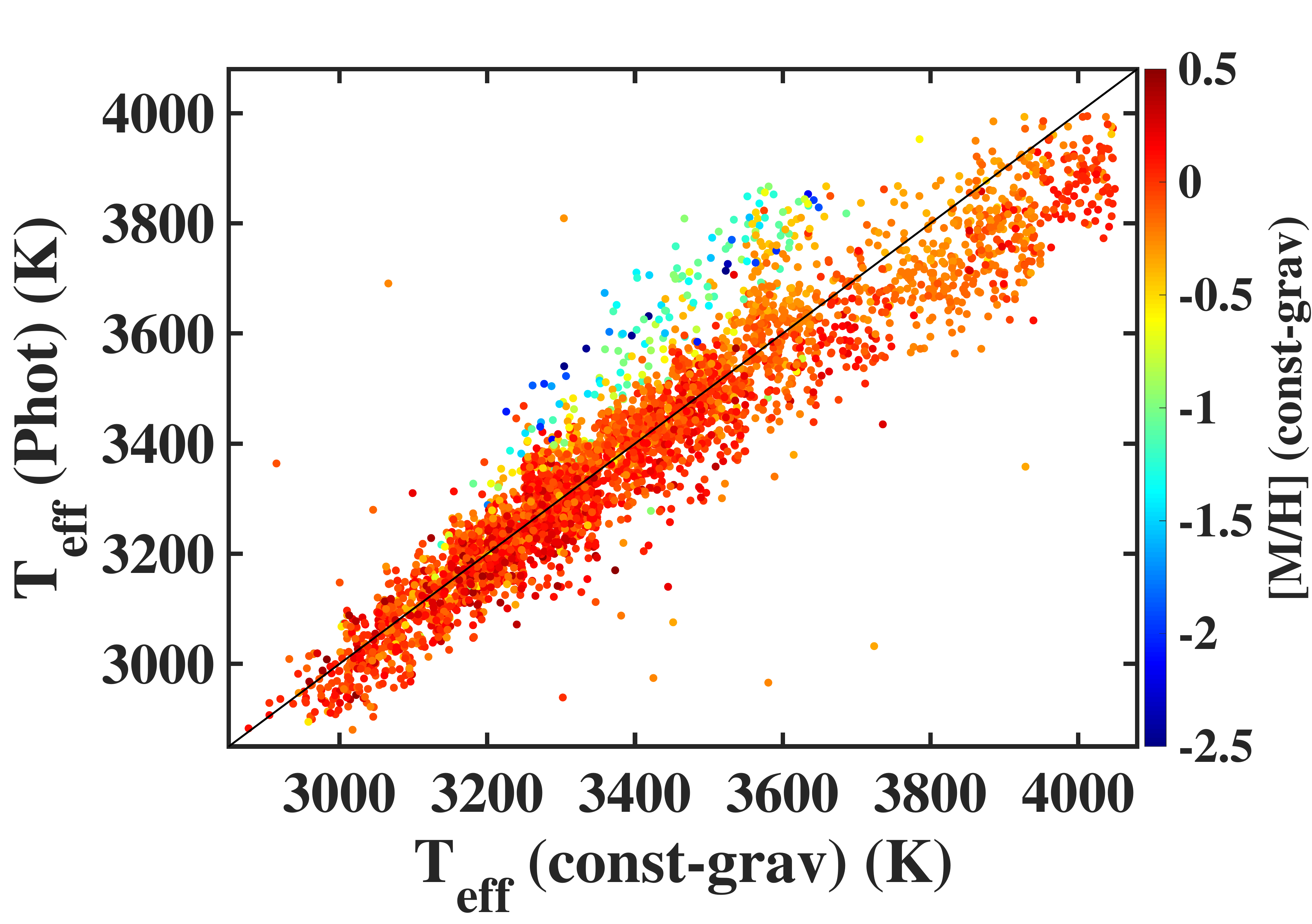}}
\vspace{-0.25cm}

\subfloat       
        [Variable surface gravity]{\includegraphics[ height=4.8cm, width=6.8cm]{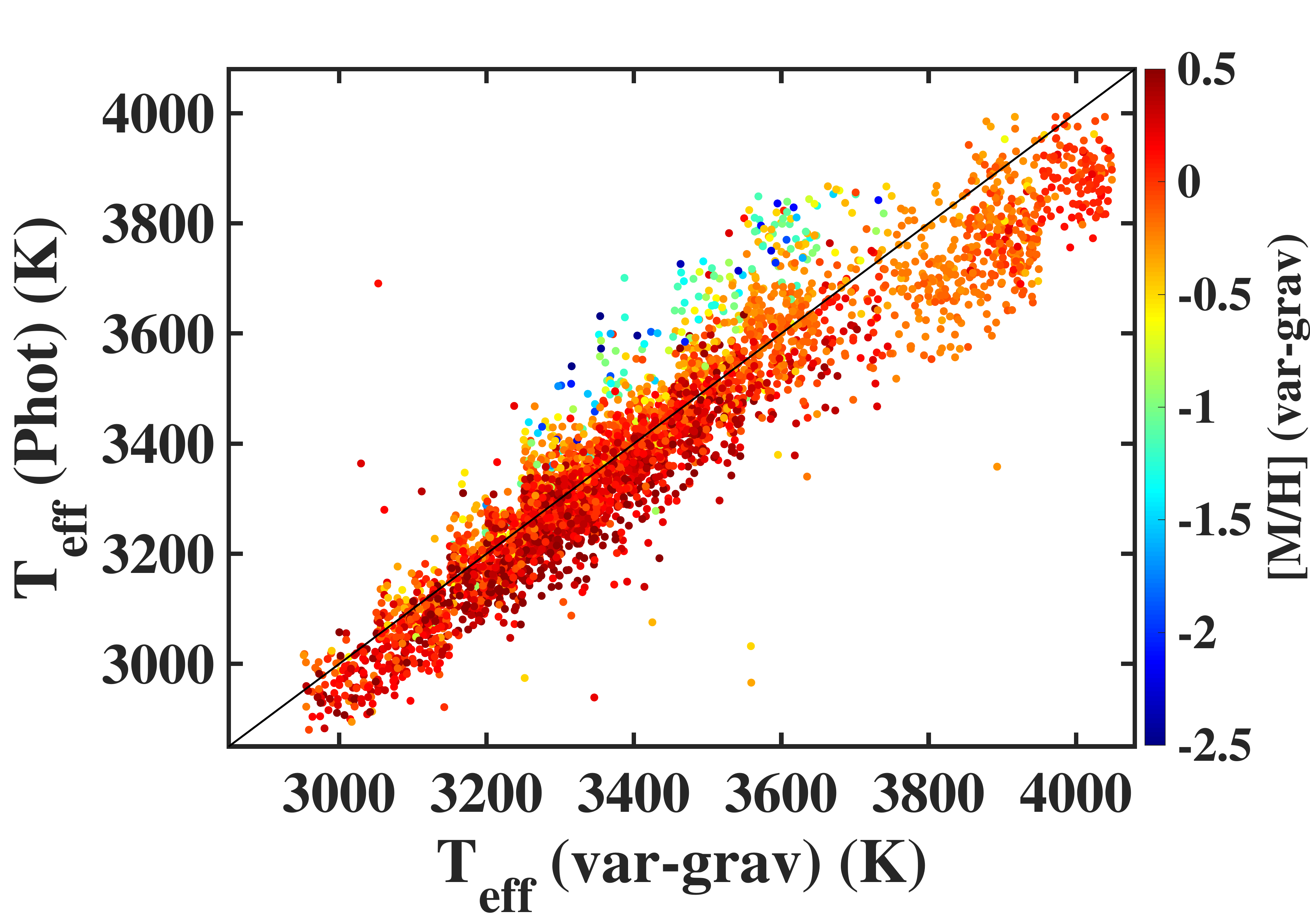}}
 \caption
        {\footnotesize{Comparison between the photometric temperatures and the inferred model-fit values for the 3745 stars, when surface gravity is considered as a fixed parameter  (top panel) and when surface gravity is considered as a free parameter (bottom panel), using the \textbf{normal method}.}}
\end{figure}

\begin{figure}\centering
       {\includegraphics[ height=5.5cm, width=8cm]{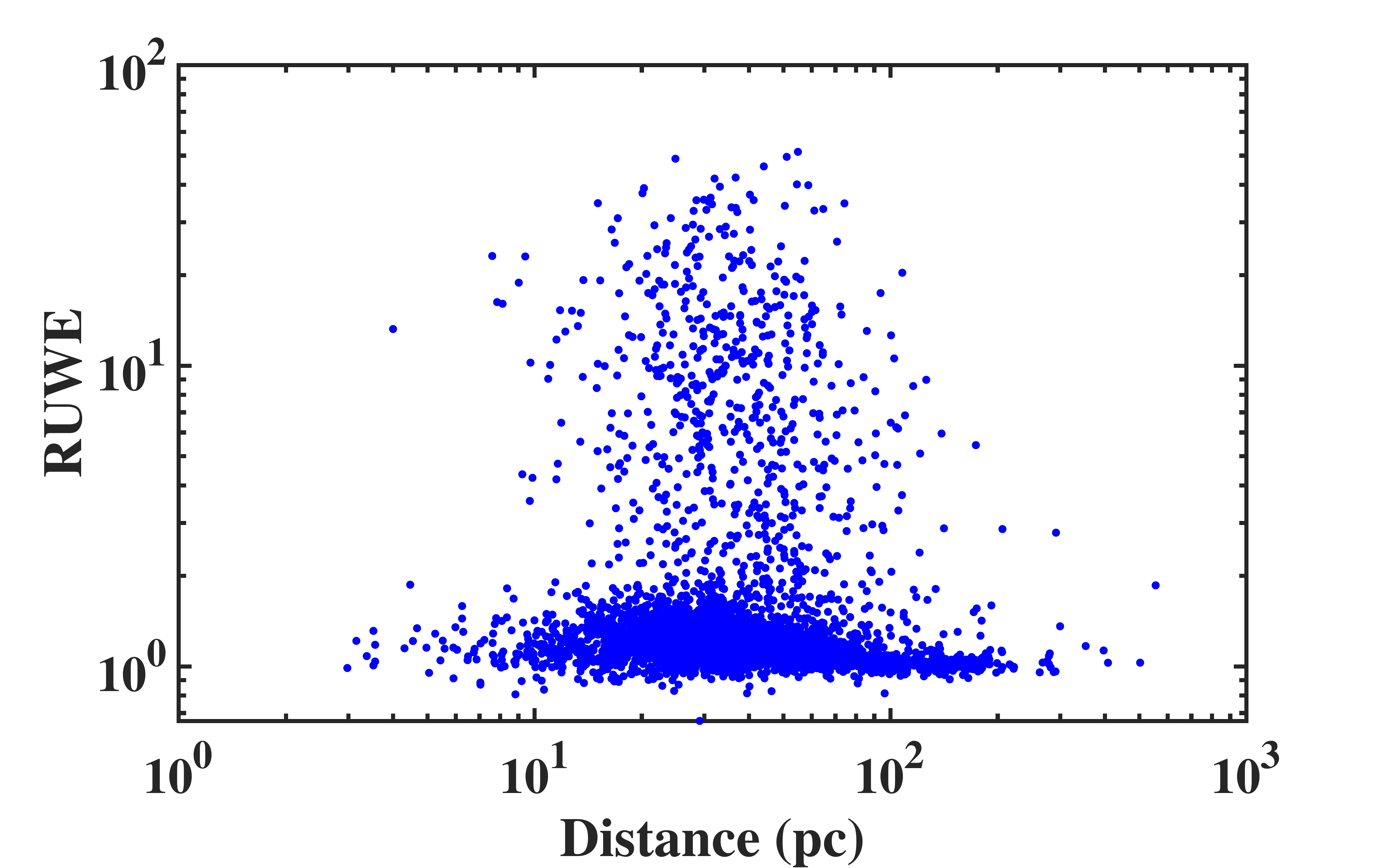}}
 \caption
        {\footnotesize{RUWE versus distance of the 3745 stars in  logarithmic scale.}}
\end{figure}

\begin{figure}\centering
\subfloat
       {\includegraphics[ height=4.8cm, width=7.5cm]{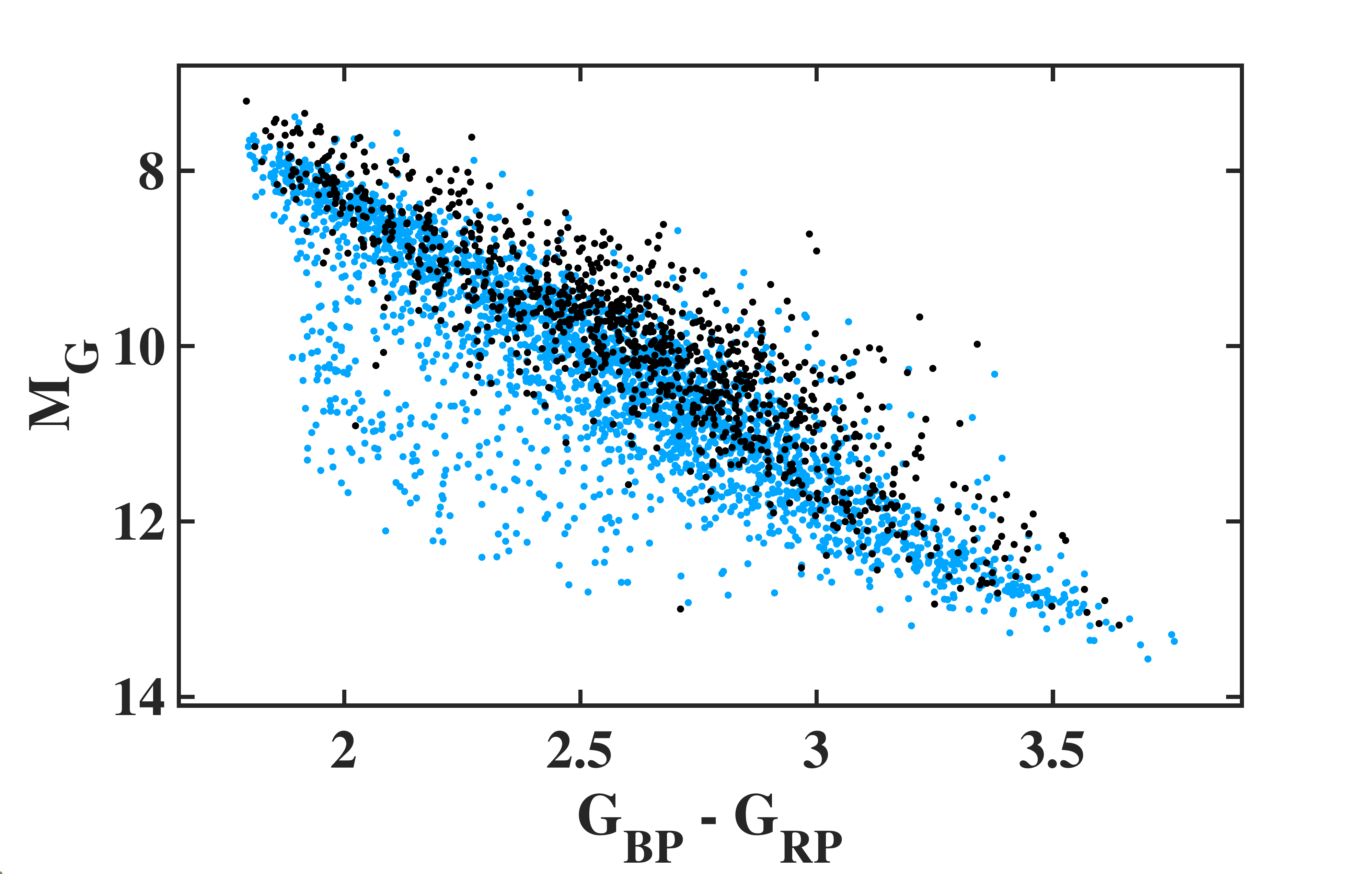}}
\vspace{-0.25cm}

 \subfloat
       {\includegraphics[ height=4.8cm, width=7.5cm]{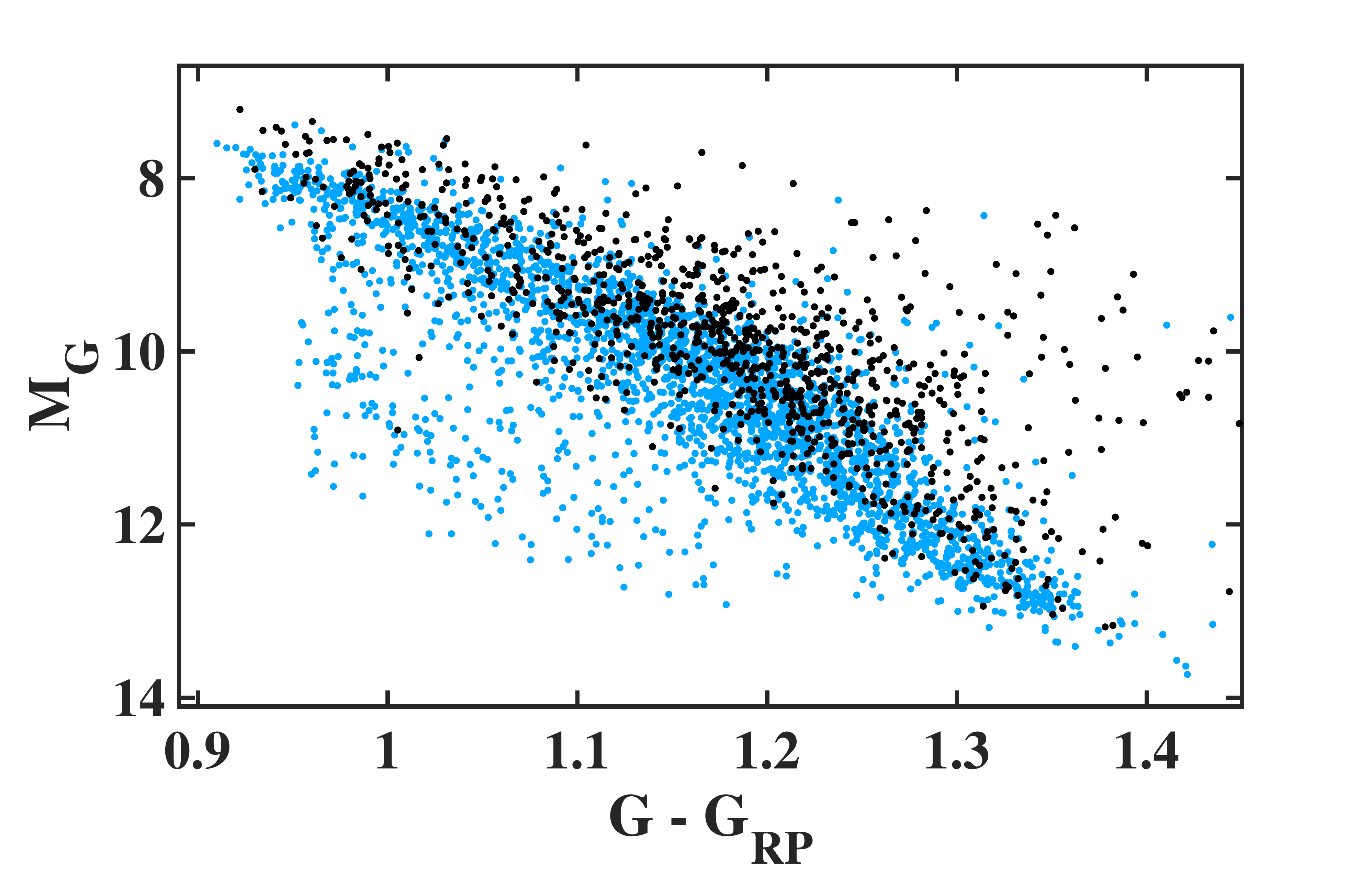}}
   \caption
        {\footnotesize{HR diagram,  \textit{M}$_\textrm{\footnotesize{G}}$ versus \textit{G}$_\textrm{\footnotesize{BP}}$ - \textit{G}$_\textrm{\footnotesize{RP}}$ (top panel) and \textit{M}$_\textrm{\footnotesize{G}}$ versus \textit{G} - \textit{G}$_\textrm{\footnotesize{RP}}$ (bottom panel), of the 3745 stars divided into two subsets: 2679 stars with RUWE<1.4 (blue) and 1066 stars with RUWE>1.4 (black).}}
\end{figure}

\begin{figure*}\centering
\subfloat
       [Fixed surface gravity]{\includegraphics[ height=5.1cm, width=8.2cm]{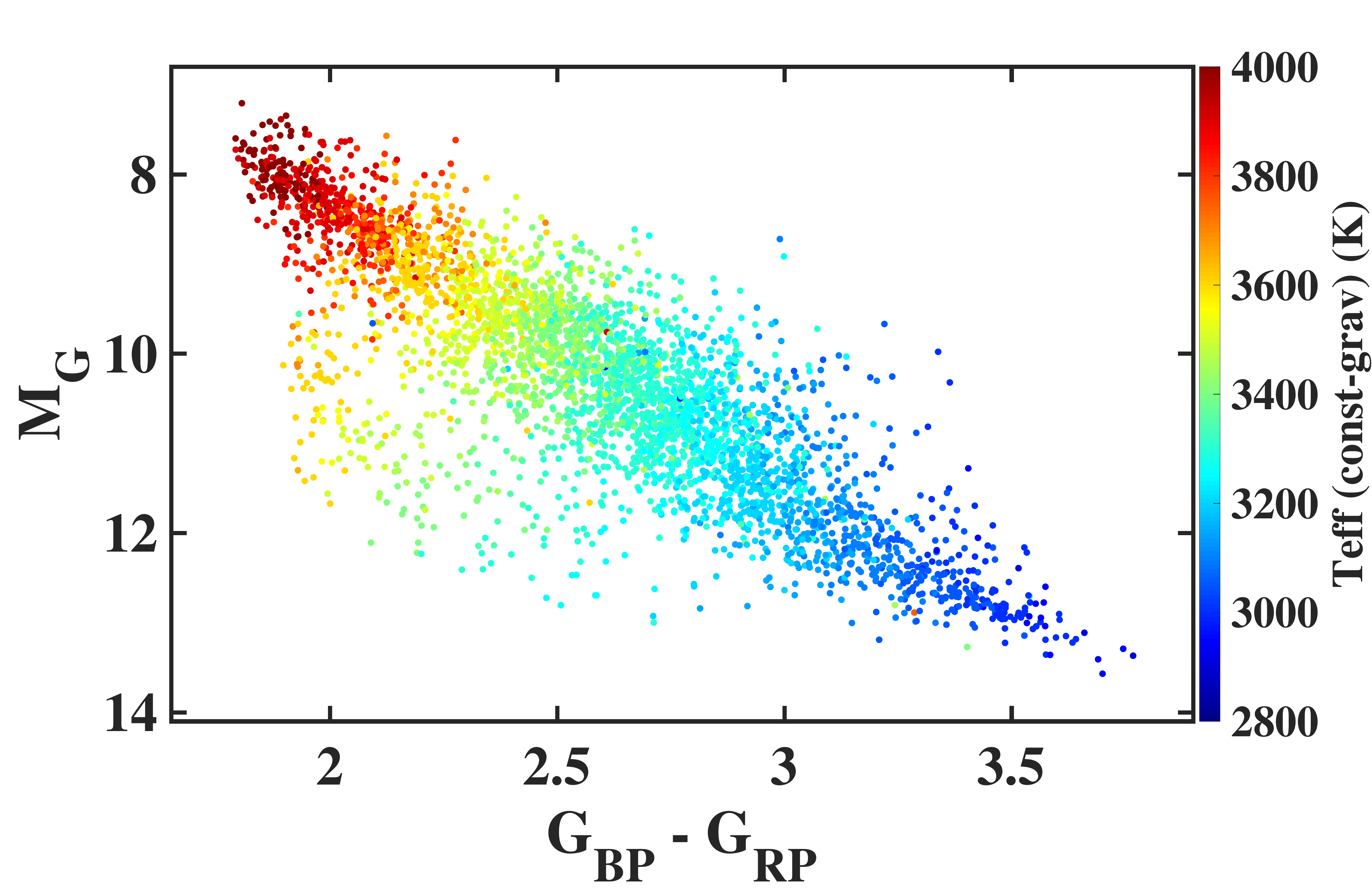}}
 \hspace{0.3cm}
 \subfloat      
        [Variable surface gravity]{\includegraphics[ height=5.1cm, width=8.2cm]{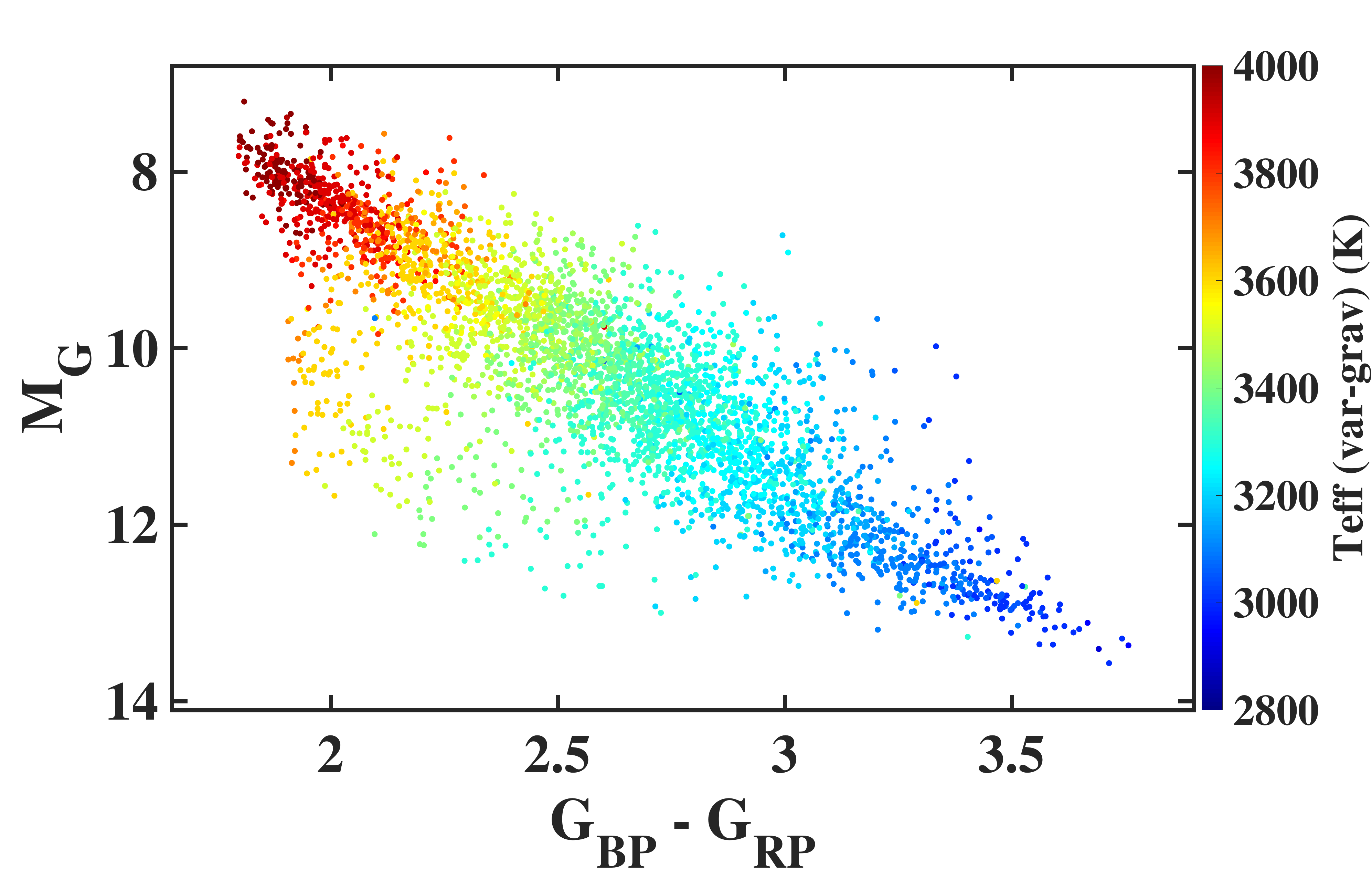}}
  \vspace{-0.4cm}

 \subfloat
         [Fixed surface gravity]{\includegraphics[ height=5.1cm, width=8.2cm]{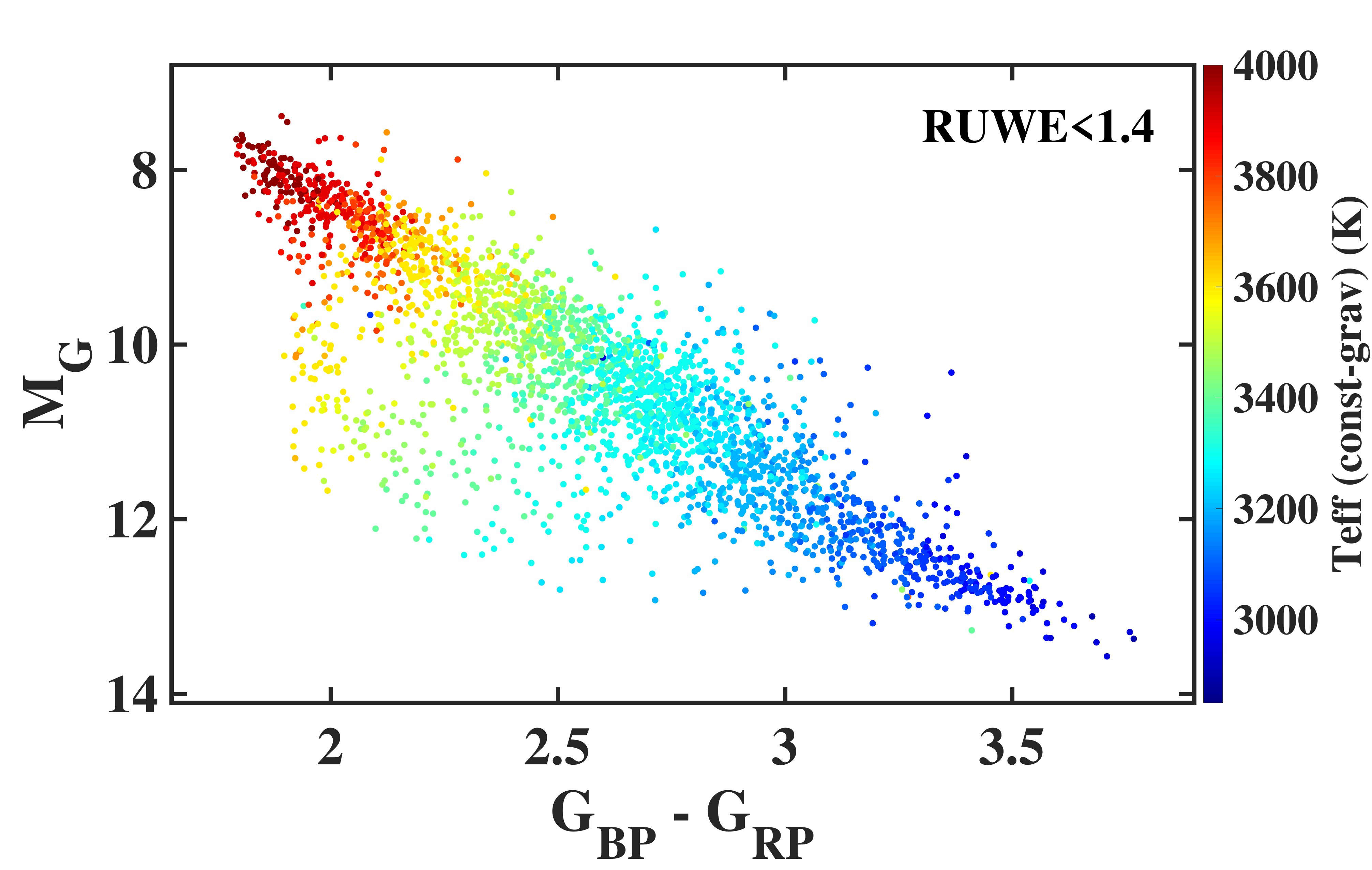}}   
 \hspace{0.3cm} 
 \subfloat 
         [Variable surface gravity]{\includegraphics[height=5.1cm, width=8.2cm]{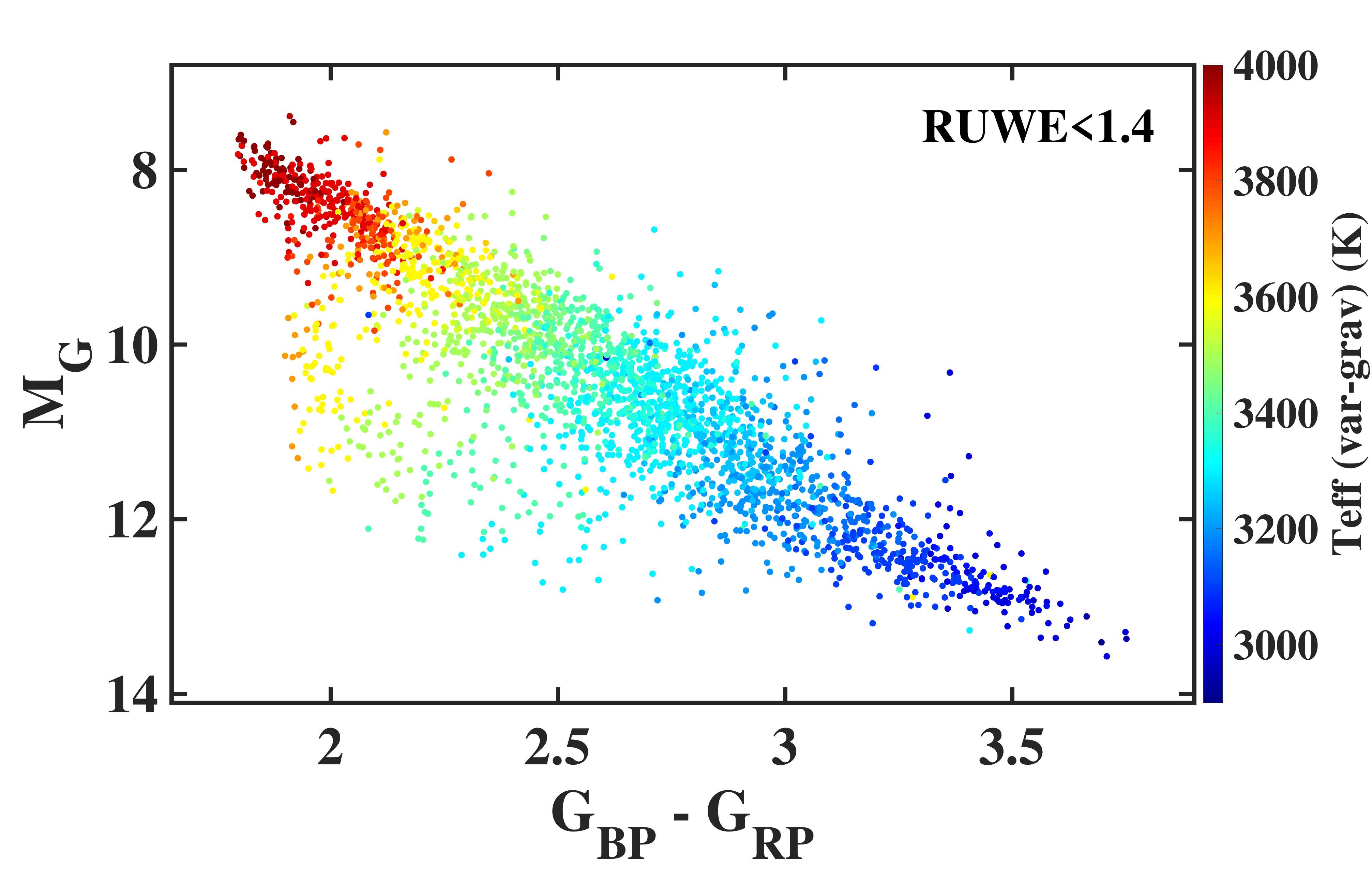}}  
\caption
        {\footnotesize{HR diagram,  \textit{M}$_\textrm{\footnotesize{G}}$ versus \textit{G}$_\textrm{\footnotesize{BP}}$ - \textit{G}$_\textrm{\footnotesize{RP}}$,  of the 3745 stars (top panels) and a subset of 2679 stars with RUWE<1.4 (bottom panels),  when surface gravity is set as a fixed parameter (left panels) and when surface gravity is set as a free parameter (right panels), color-mapped based on  effective temperatures, using the \textbf{normal method}.}} 
  \end{figure*}

\begin{figure*}\centering
\subfloat
       [Fixed surface gravity]{\includegraphics[ height=5.1cm, width=8.2cm]{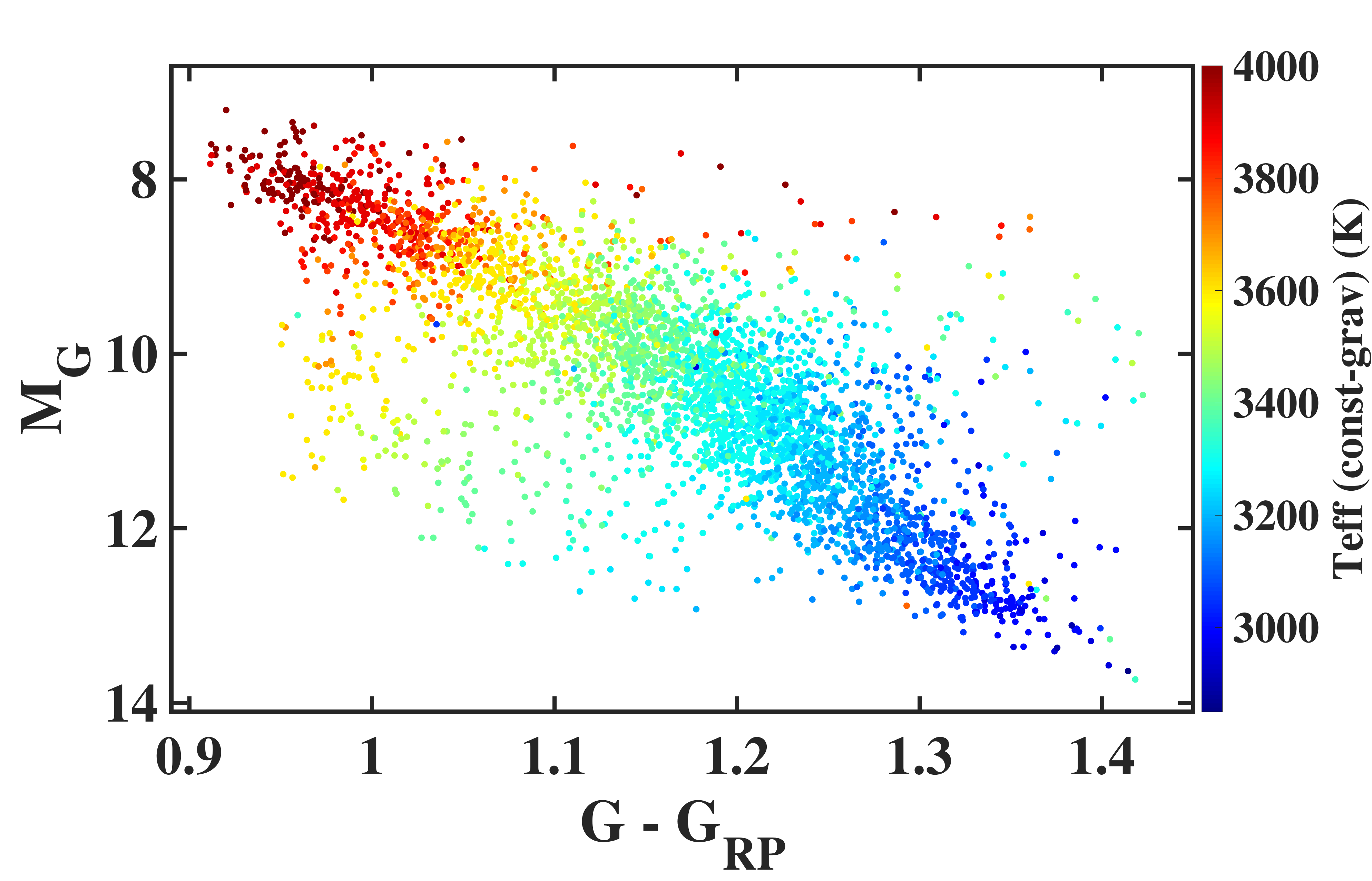}}
 \hspace{0.3cm}
 \subfloat      
        [Variable surface gravity]{\includegraphics[ height=5.1cm, width=8.2cm]{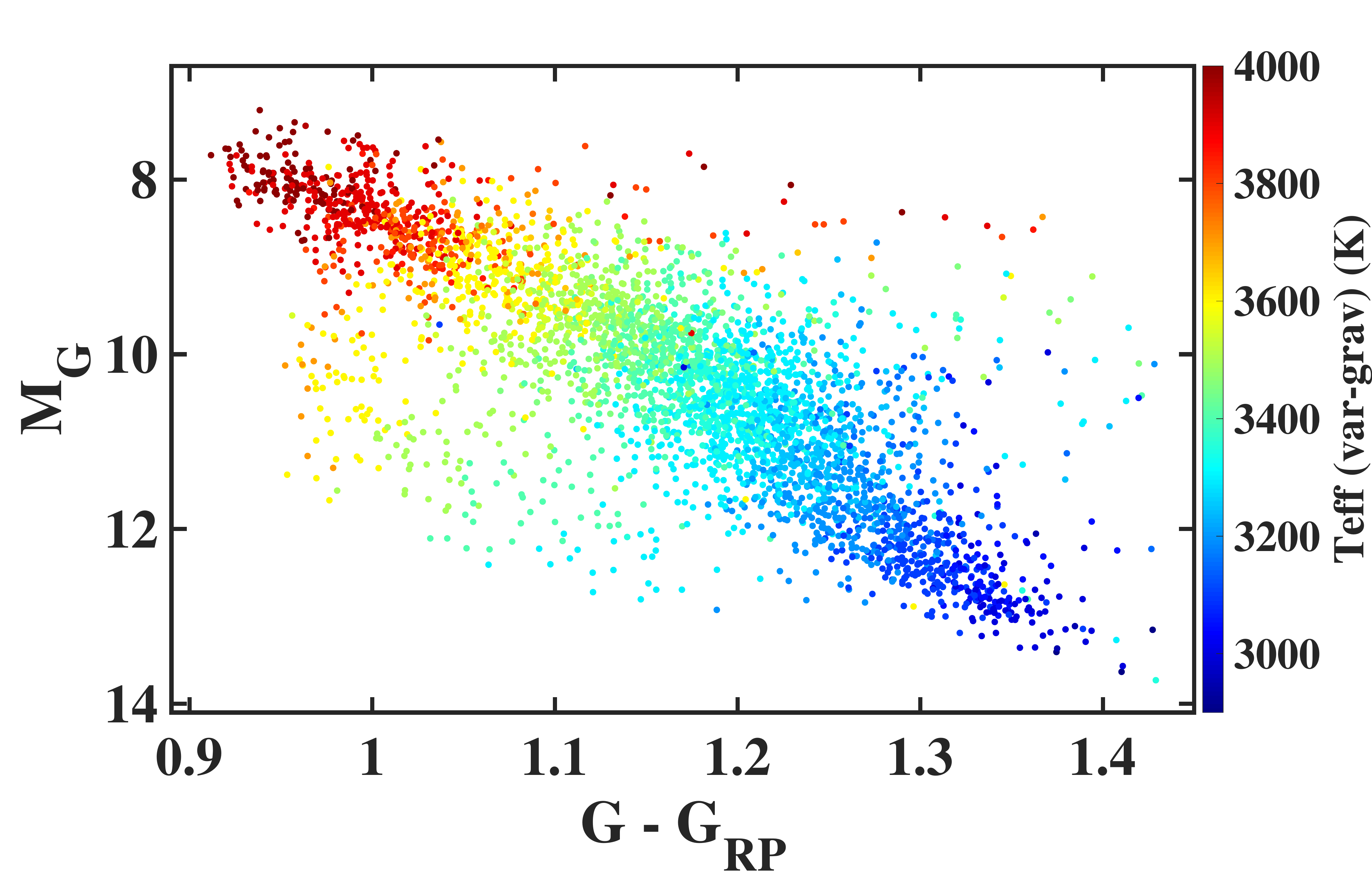}}
  \vspace{-0.4cm}

 \subfloat
         [Fixed surface gravity]{\includegraphics[ height=5.1cm, width=8.2cm]{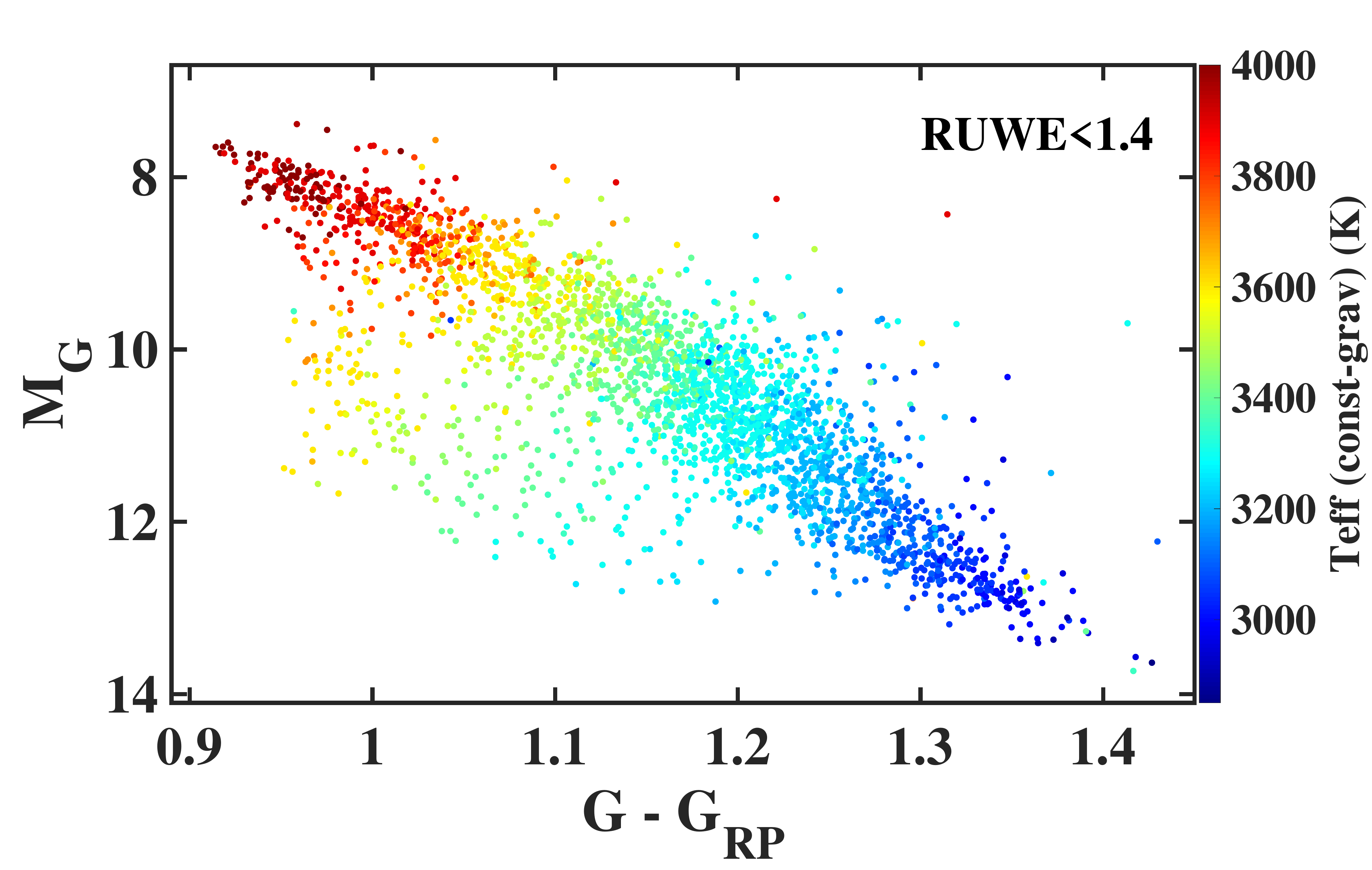}}   
 \hspace{0.3cm} 
 \subfloat 
         [Variable surface gravity]{\includegraphics[height=5.1cm, width=8.2cm]{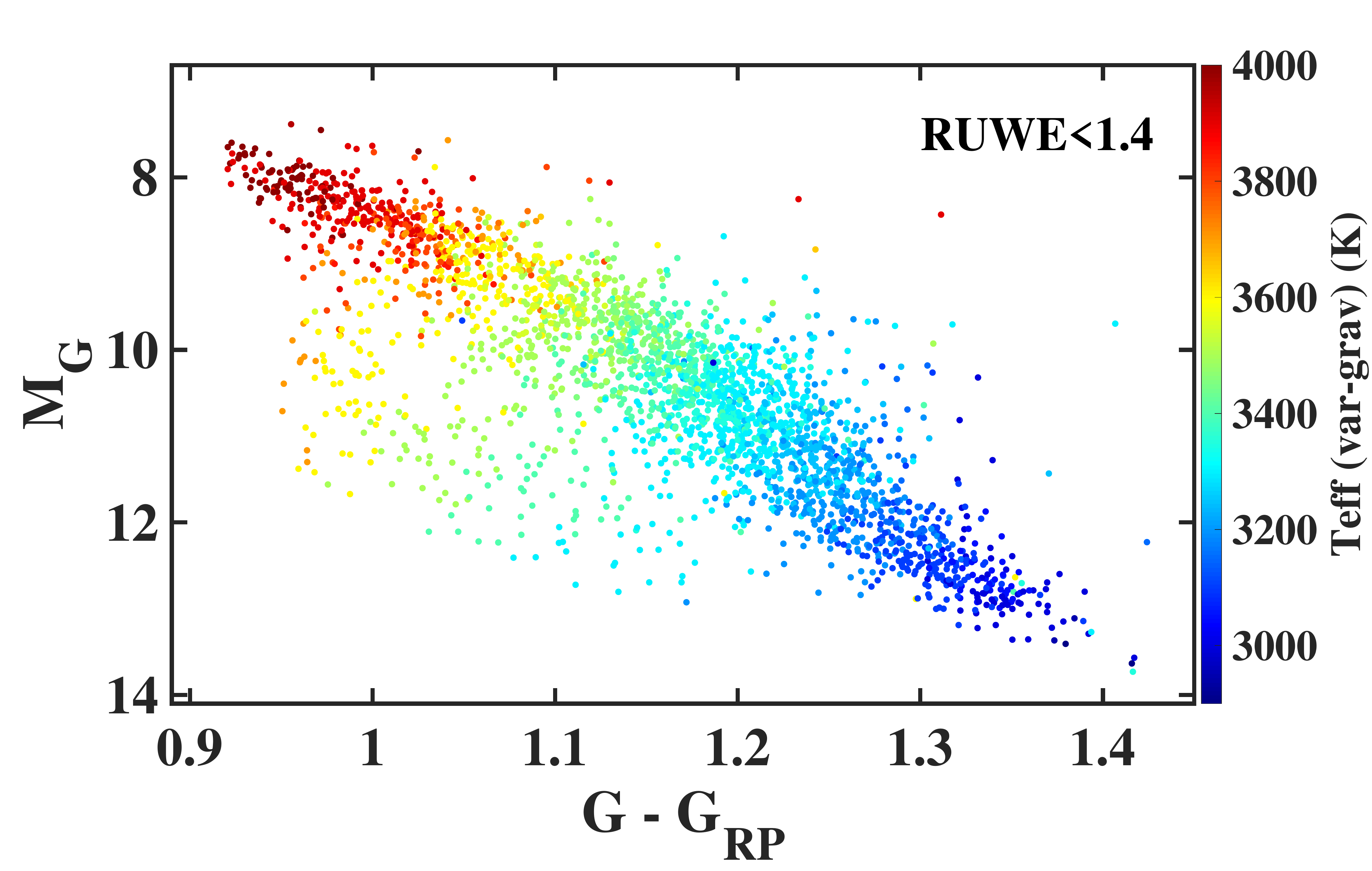}}  
\caption
        {\footnotesize{HR diagram,  \textit{M}$_\textrm{\footnotesize{G}}$ versus \textit{G} - \textit{G}$_\textrm{\footnotesize{RP}}$,  of the 3745 stars (top panels) and a subset of 2679 stars with RUWE<1.4 (bottom panels),  when surface gravity is set as a fixed parameter (left panels) and when surface gravity is set as a free parameter (right panels), color-mapped based on  effective temperatures, using the \textbf{normal method}.}} 
  \end{figure*}

\begin{figure*}\centering
\subfloat
       [Fixed surface gravity]{\includegraphics[ height=5.7cm, width=9.1cm]{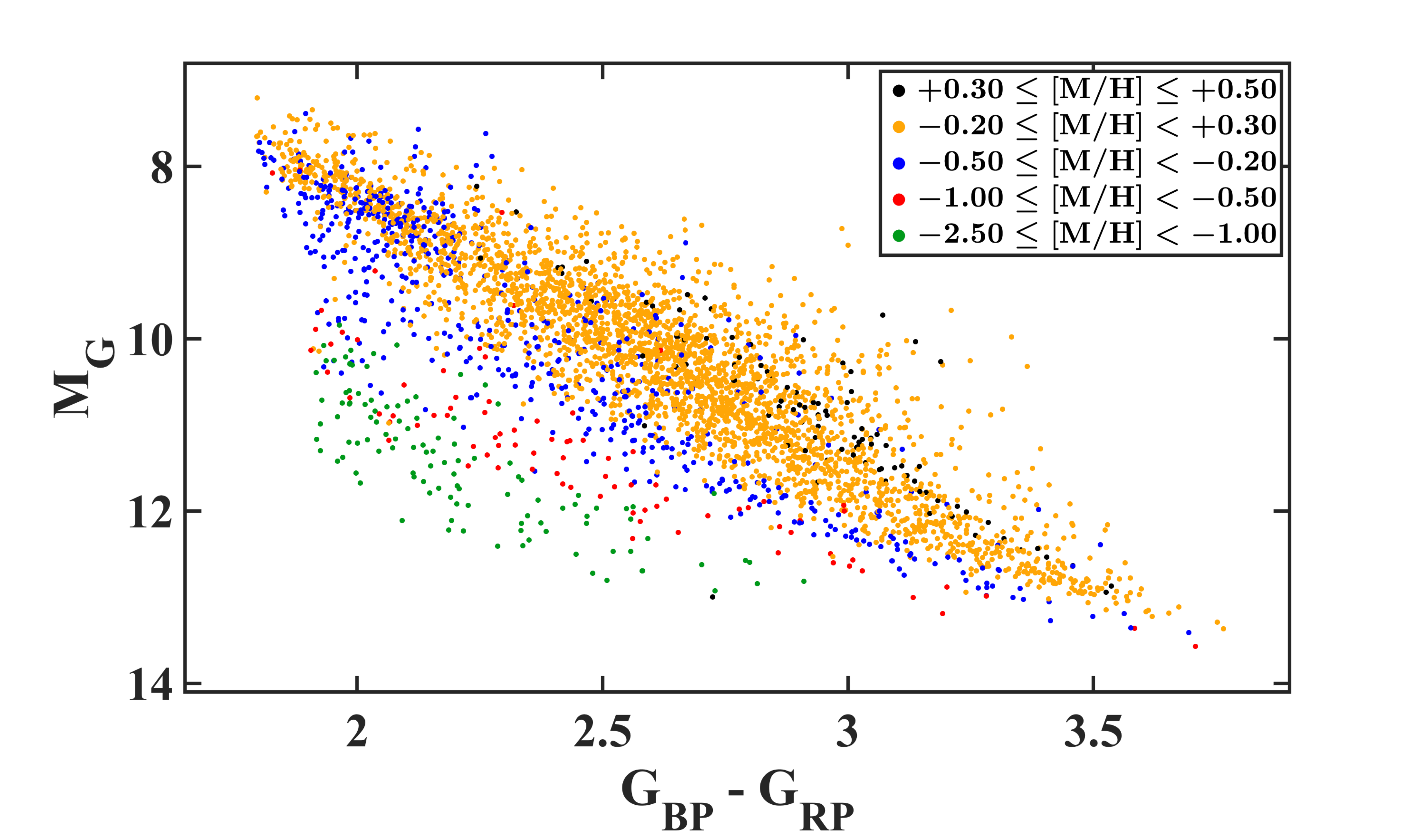}}
 \hspace{-0.35cm} 
 \subfloat      
        [Variable surface gravity]{\includegraphics[ height=5.7cm, width=9.1cm]{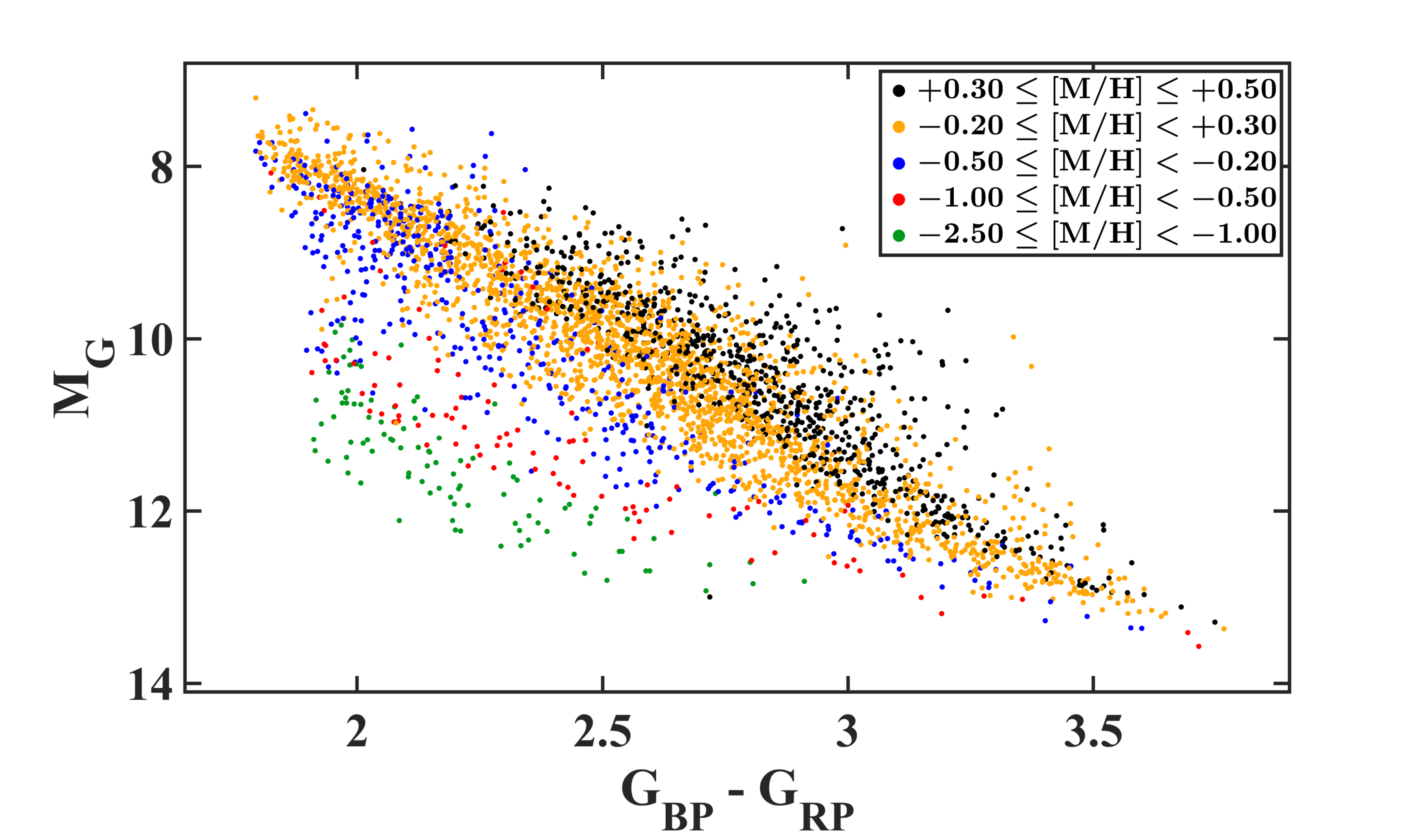}}
  \vspace{-0.35cm}

 \subfloat
         [Fixed surface gravity]{\includegraphics[ height=5.7cm, width=9.1cm]{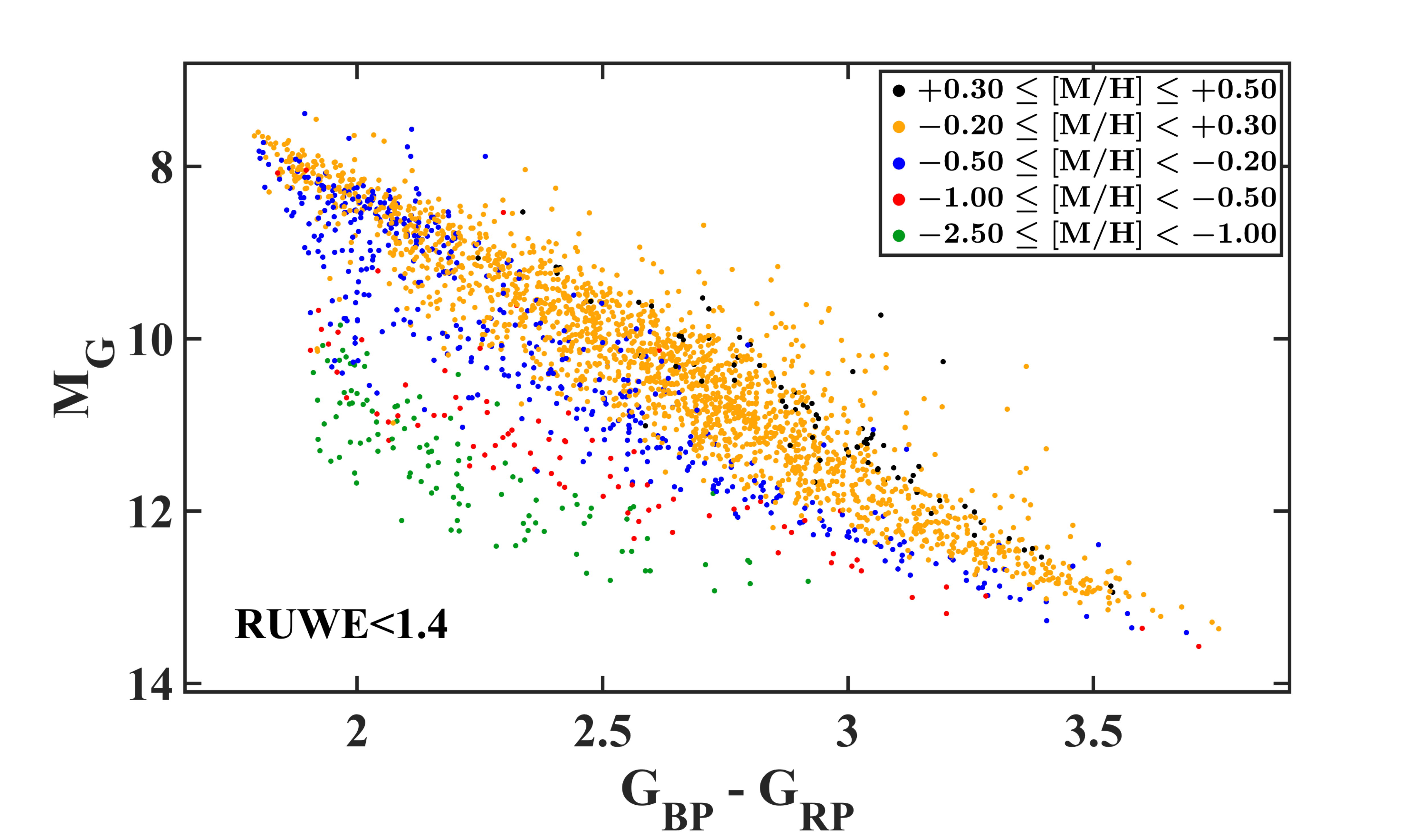}}   
 \hspace{-0.35cm} 
 \subfloat 
         [Variable surface gravity]{\includegraphics[height=5.7cm, width=9.1cm]{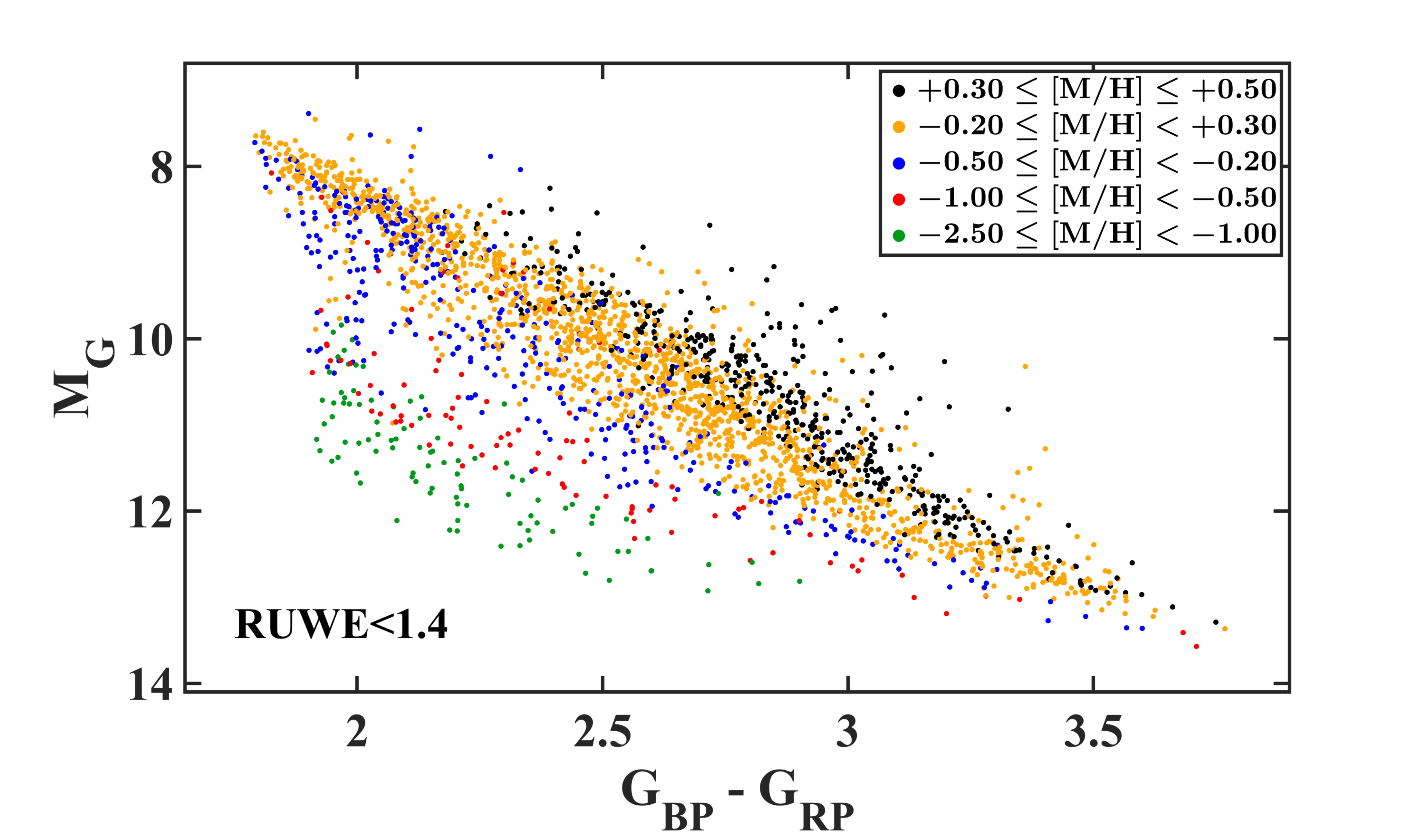}}  
\caption
         {\footnotesize{HR diagram,  \textit{M}$_\textrm{\footnotesize{G}}$ versus \textit{G}$_\textrm{\footnotesize{BP}}$ - \textit{G}$_\textrm{\footnotesize{RP}}$, of the 3745 stars (top panels)  and a subset of 2679 stars with RUWE<1.4 (bottom panels),  when surface gravity is set as a fixed parameter (left panels) and when surface gravity is set as a free parameter (right panels), color-coded based on five groups with different  metallicity ranges, using the \textbf{normal method}. The number of stars in each metallicity group, N, is \textbf{panel ``a'':}  N$_\textrm{{Black}}$=105,  N$_\textrm{{Yellow}}$=2849, N$_\textrm{{Blue}}$=593, N$_\textrm{{Red}}$=85, and N$_\textrm{{Green}}$=113, \textbf{panel ``b'':}  N$_\textrm{{Black}}$=784,  N$_\textrm{{Yellow}}$=2278, N$_\textrm{{Blue}}$=477, N$_\textrm{{Red}}$=107, and N$_\textrm{{Green}}$=99,    \textbf{panel ``c'':}  N$_\textrm{{Black}}$=82,  N$_\textrm{{Yellow}}$=1943, N$_\textrm{{Blue}}$=461, N$_\textrm{{Red}}$=83, and N$_\textrm{{Green}}$=110,     \textbf{panel ``d'':}  N$_\textrm{{Black}}$=502,  N$_\textrm{{Yellow}}$=1599, N$_\textrm{{Blue}}$=378, N$_\textrm{{Red}}$=103, and N$_\textrm{{Green}}$=97.}} 
   \end{figure*}

\begin{figure*}\centering
\subfloat
       [Fixed surface gravity]{\includegraphics[ height=5.7cm, width=9.1cm]{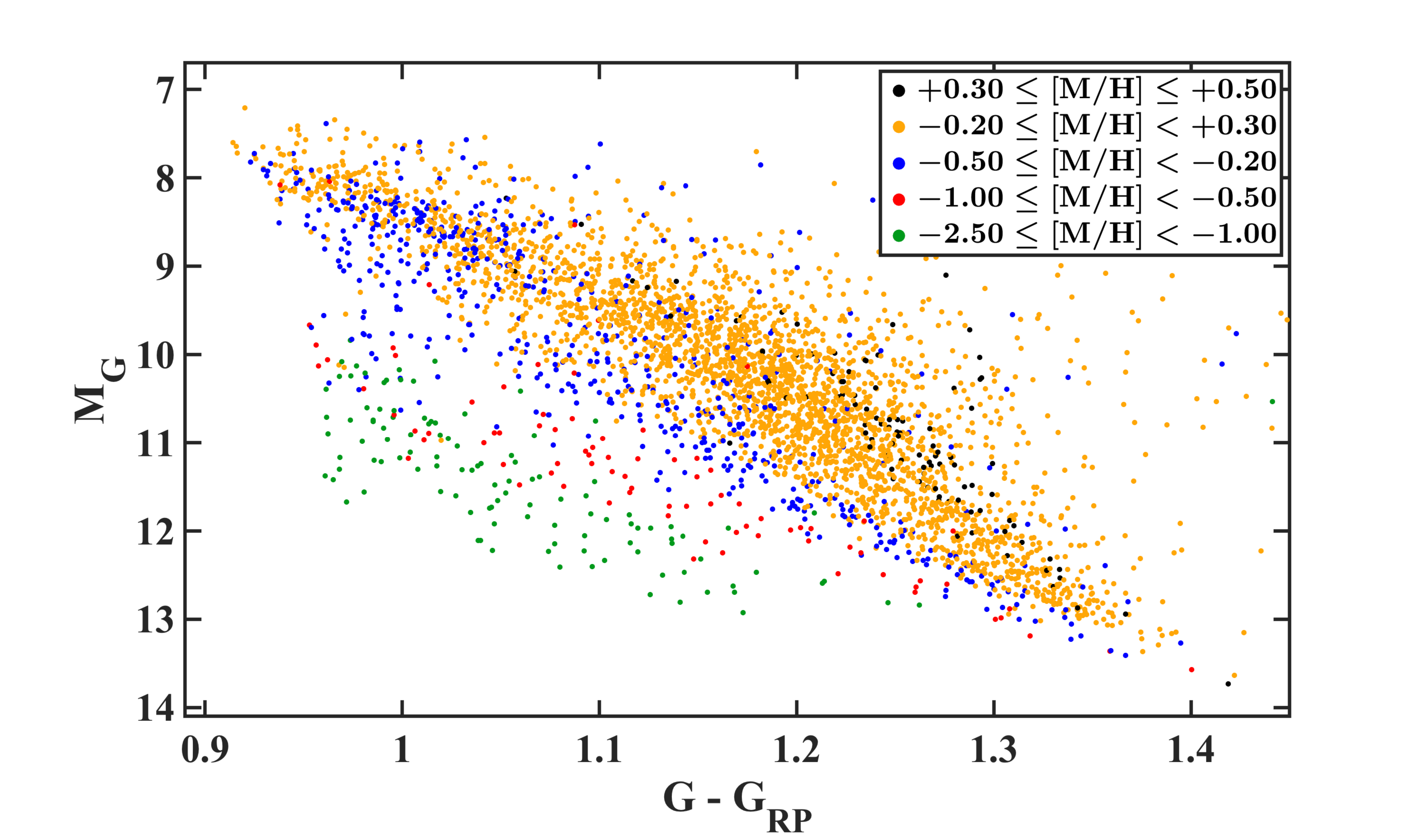}}
 \hspace{-0.35cm} 
 \subfloat      
        [Variable surface gravity]{\includegraphics[ height=5.7cm, width=9.1cm]{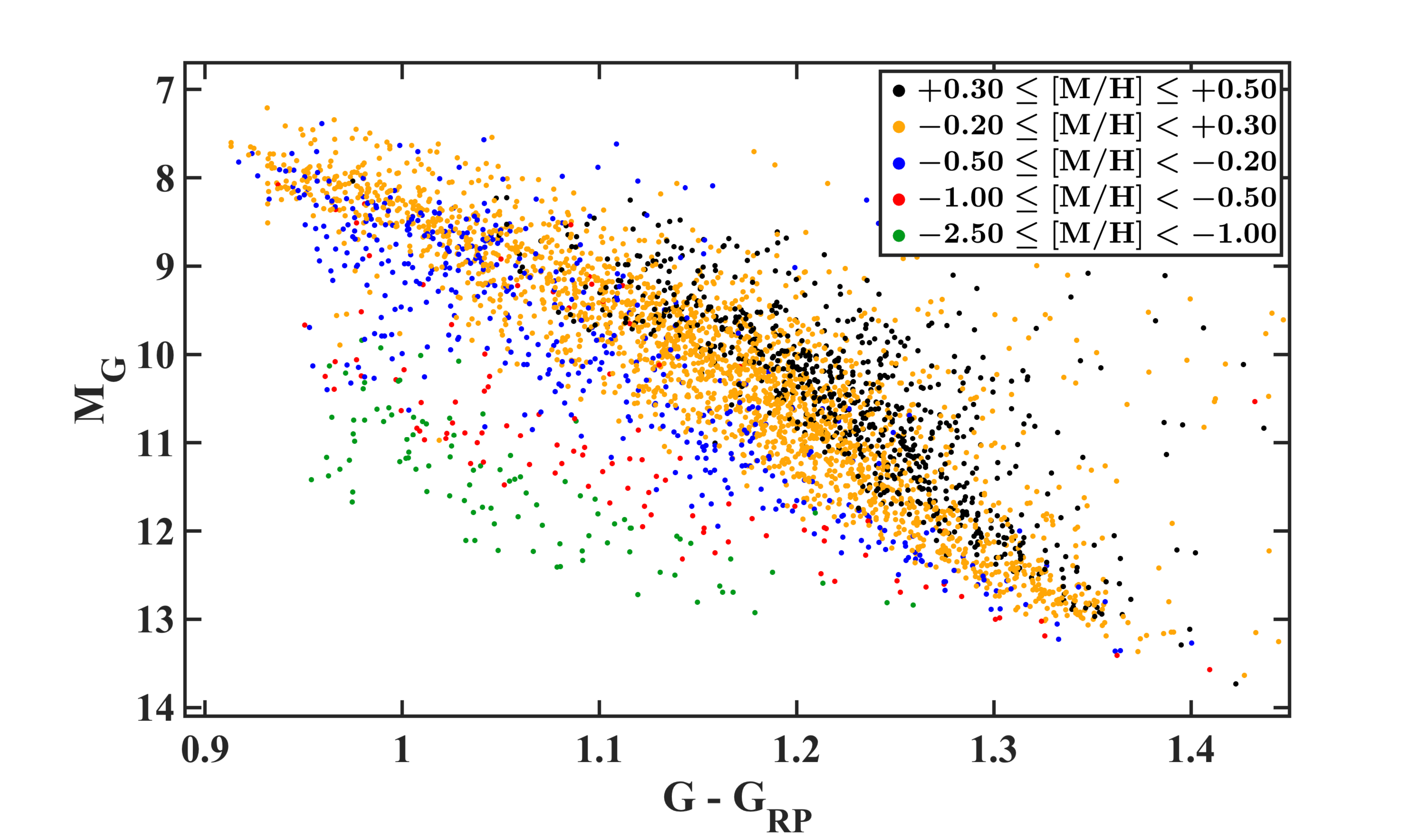}}
  \vspace{-0.35cm}

 \subfloat
         [Fixed surface gravity]{\includegraphics[ height=5.7cm, width=9.1cm]{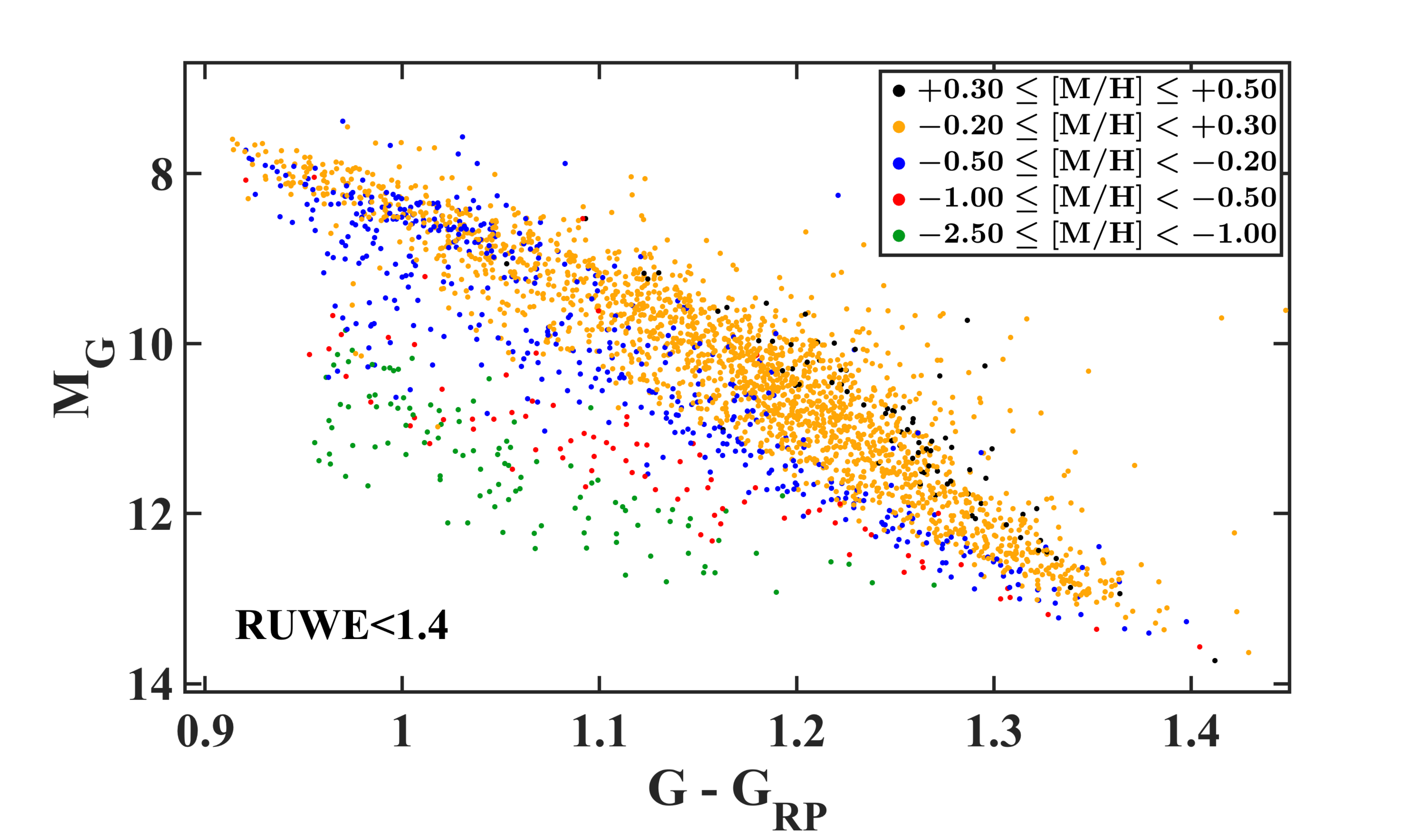}}   
 \hspace{-0.35cm} 
 \subfloat 
         [Variable surface gravity]{\includegraphics[height=5.7cm, width=9.1cm]{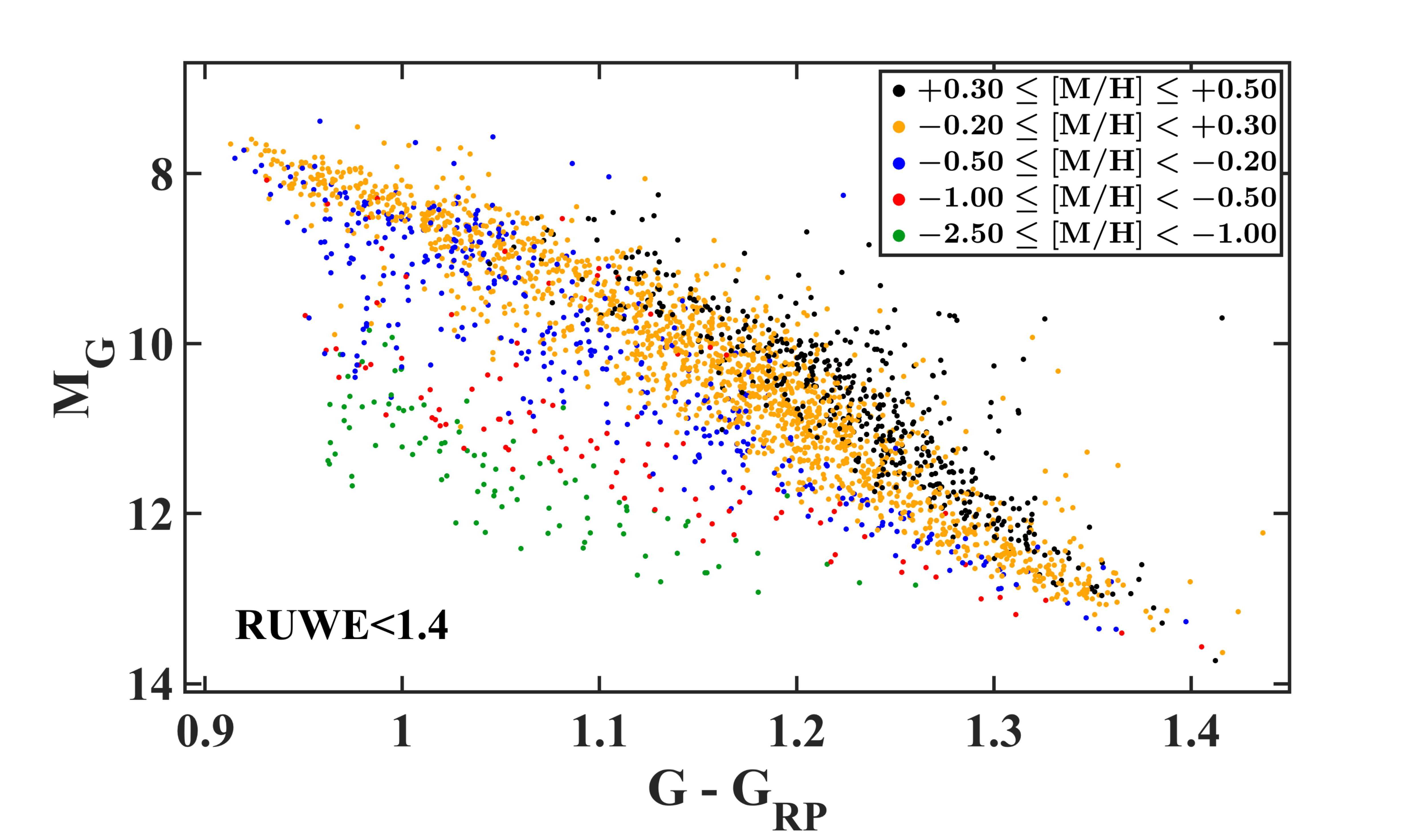}}  
\caption
        {\footnotesize{HR diagram,  \textit{M}$_\textrm{\footnotesize{G}}$ versus \textit{G} - \textit{G}$_\textrm{\footnotesize{RP}}$, of the 3745 stars (top panels)  and a subset of 2679 stars with RUWE<1.4 (bottom panels),  when surface gravity is set as a fixed parameter (left panels) and when surface gravity is set as a free parameter (right panels), color-coded based on five groups with different  metallicity ranges, using the \textbf{normal method}. The number of stars in each metallicity group is the same as that shown in Figure 23 associated with the respective panels.}} 
   \end{figure*}

\begin{figure*}\centering
\subfloat
       [Fixed surface gravity]{\includegraphics[ height=5.5cm, width=8.4cm]{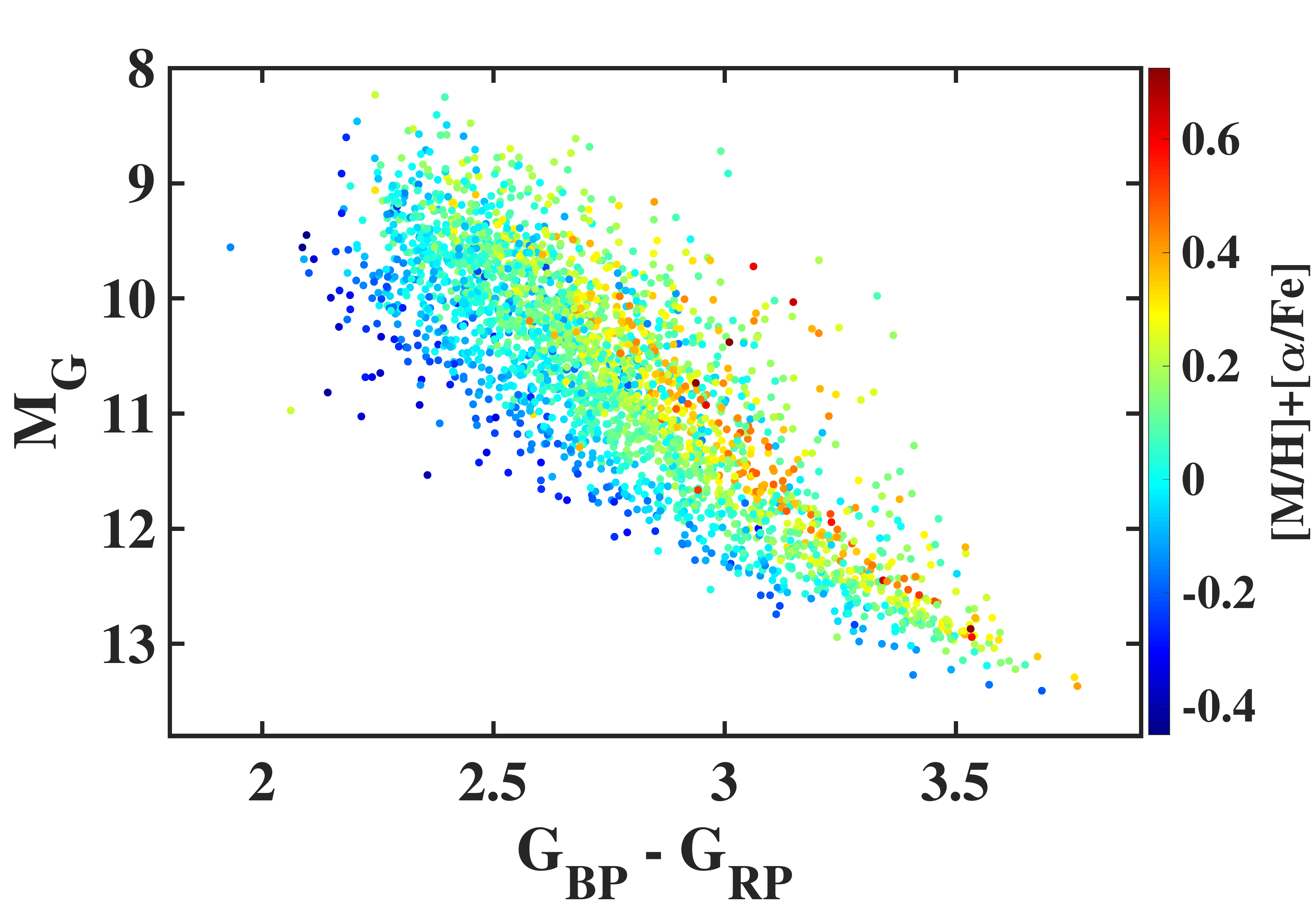}}
 \hspace{0cm} 
 \subfloat        
       [Variable surface gravity]{\includegraphics[height=5.5cm, width=8.4cm]{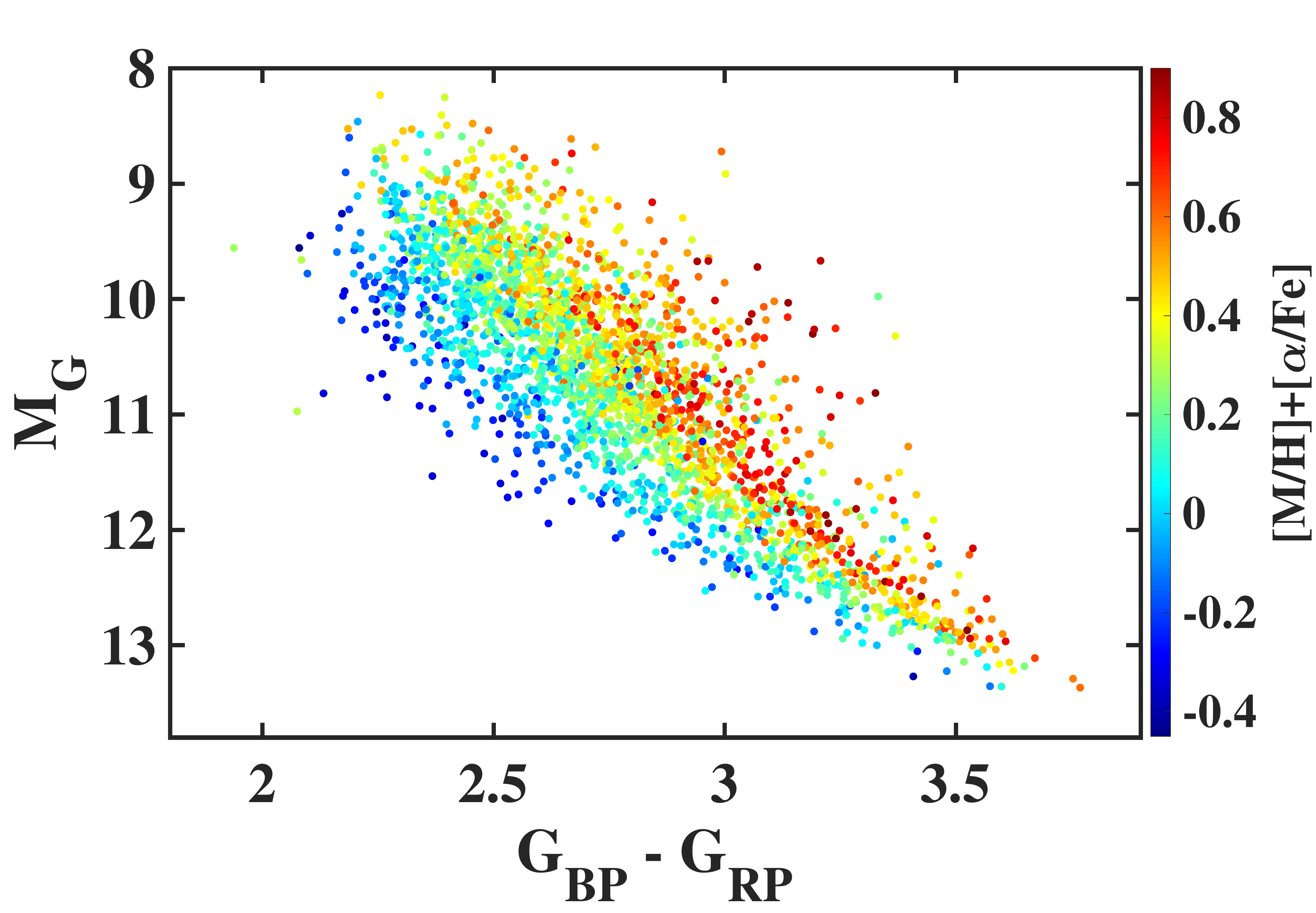}}
 \vspace{-0.25cm}

\subfloat
       [Fixed surface gravity]{\includegraphics[height=5.5cm, width=8.4cm]{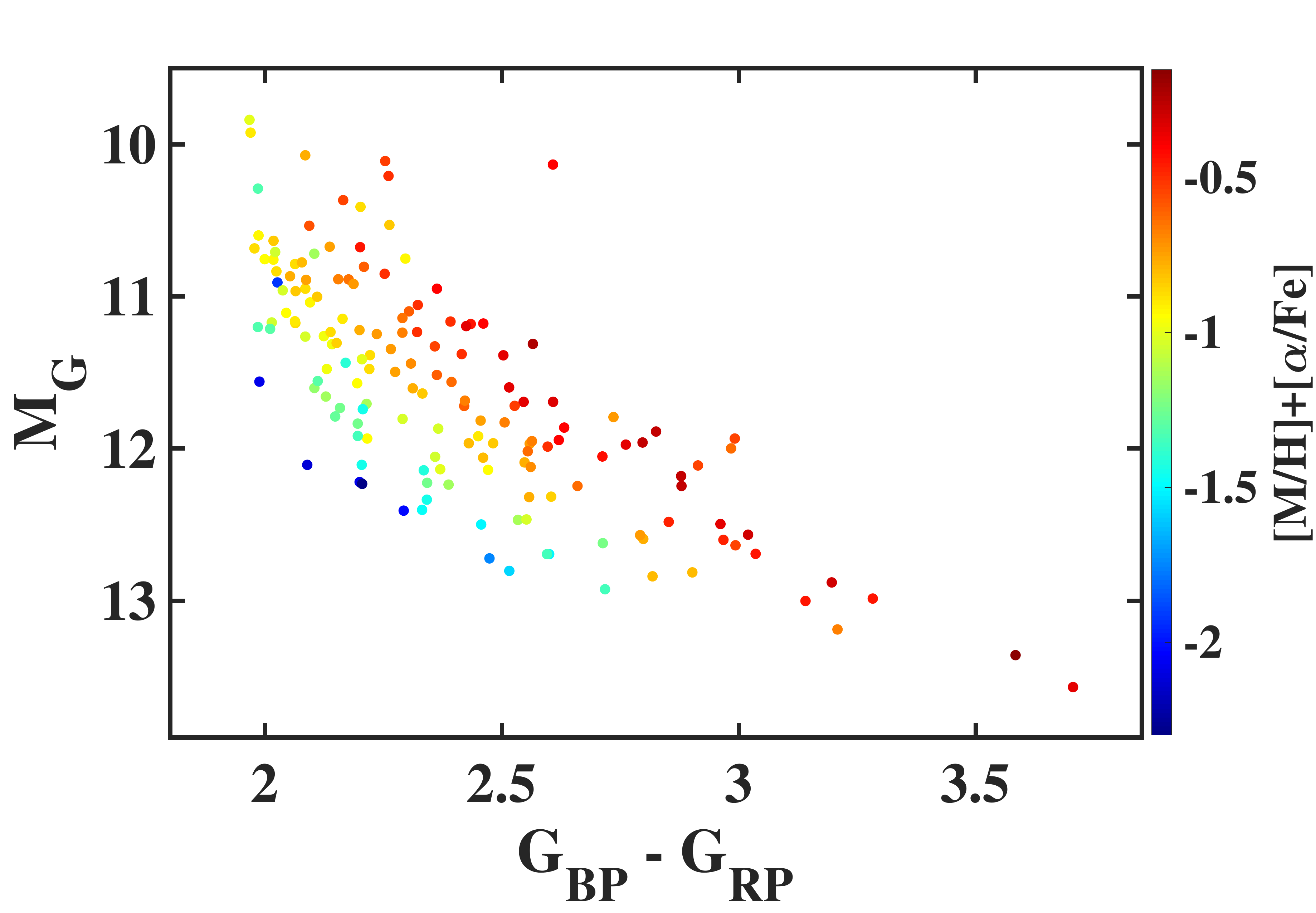}}
 \hspace{0cm}         
 \subfloat
       [Variable surface gravity]{\includegraphics[height=5.5cm, width=8.4cm]{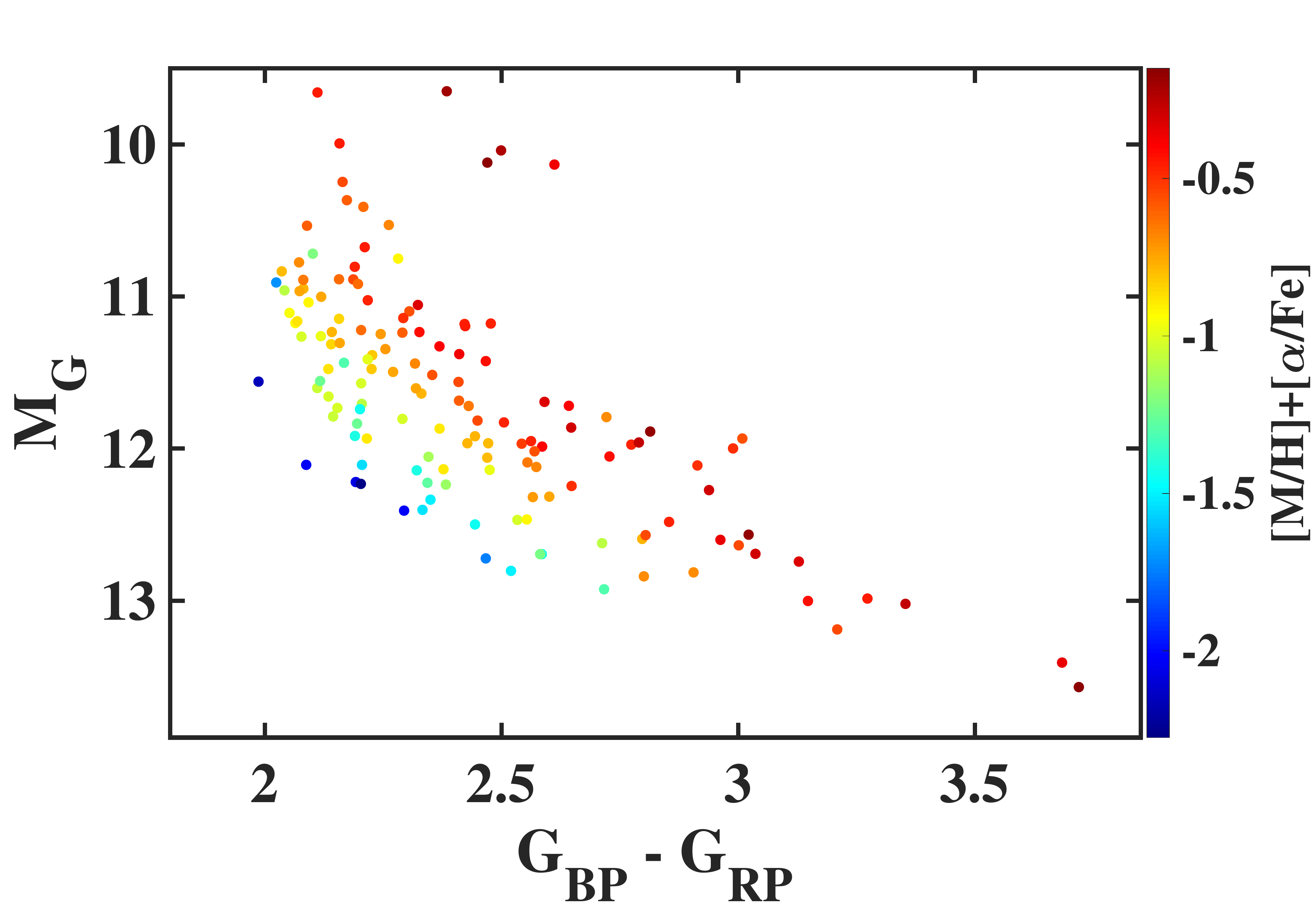}}         
 \caption
        {\footnotesize{HR diagram for  the stars with T$_\textrm{\footnotesize{eff}}$$\leq$3550 K, divided into two groups with [M/H]$\geq$$-$0.5 dex (top panels) and  [M/H]<$-$0.5 dex  (bottom panels), when the parameters are determined with fixed surface gravity  (a: 2714 stars and c: 162 stars) and with variable surface gravity (b: 2701 stars and d: 143 stars), color-mapped with associated values of the combined chemical parameter [M/H]+[$\alpha$/Fe], using the \textbf{normal method}. The values of RUWE are not taken into account in this figure.}}
\end{figure*}

\begin{figure*}\centering
\subfloat 
     []{\includegraphics[ height=4.3cm, width=5.65cm]{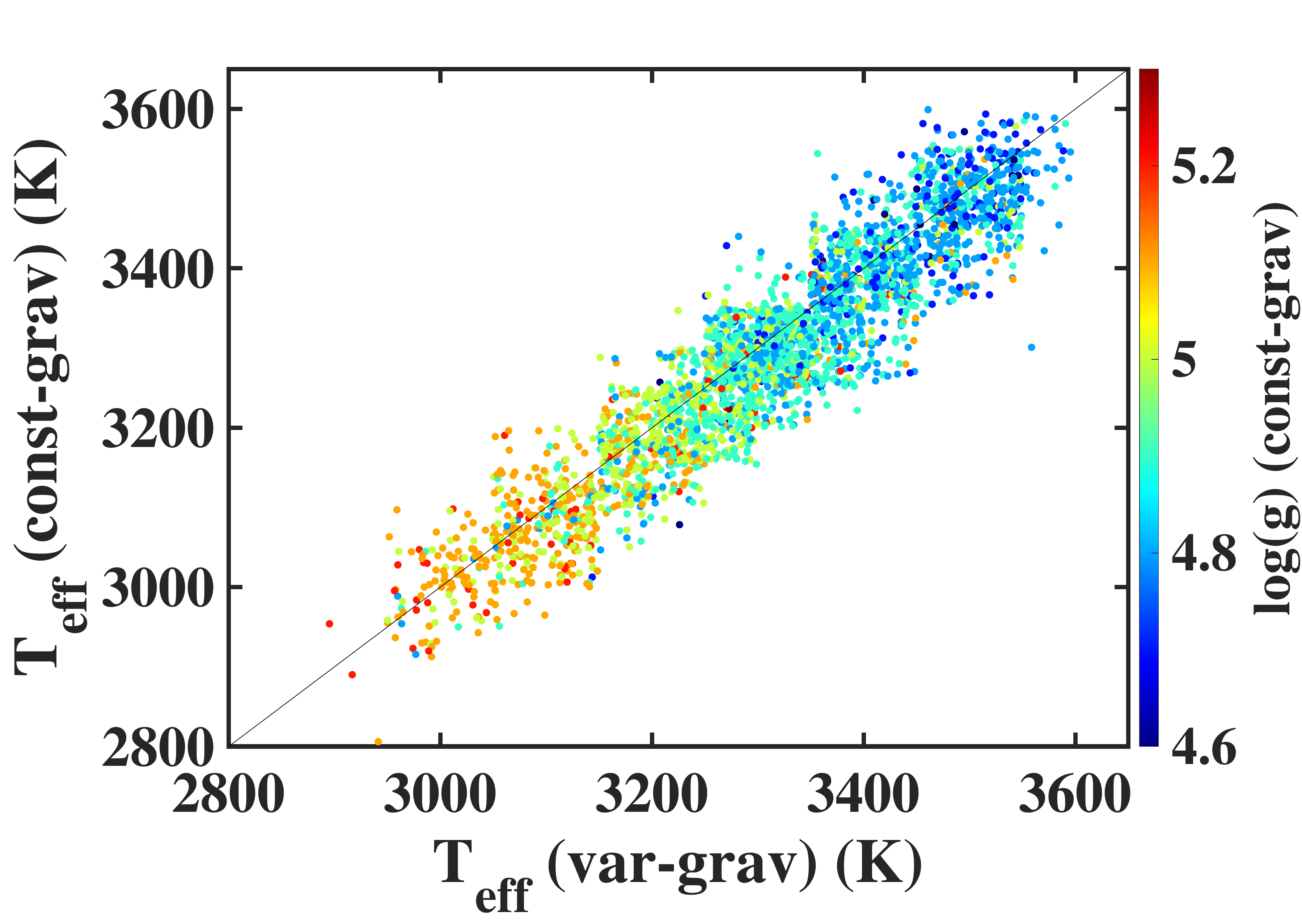}}               
\hspace{0.38cm}
\subfloat 
      []{\includegraphics[ height=4.3cm, width=5.65cm]{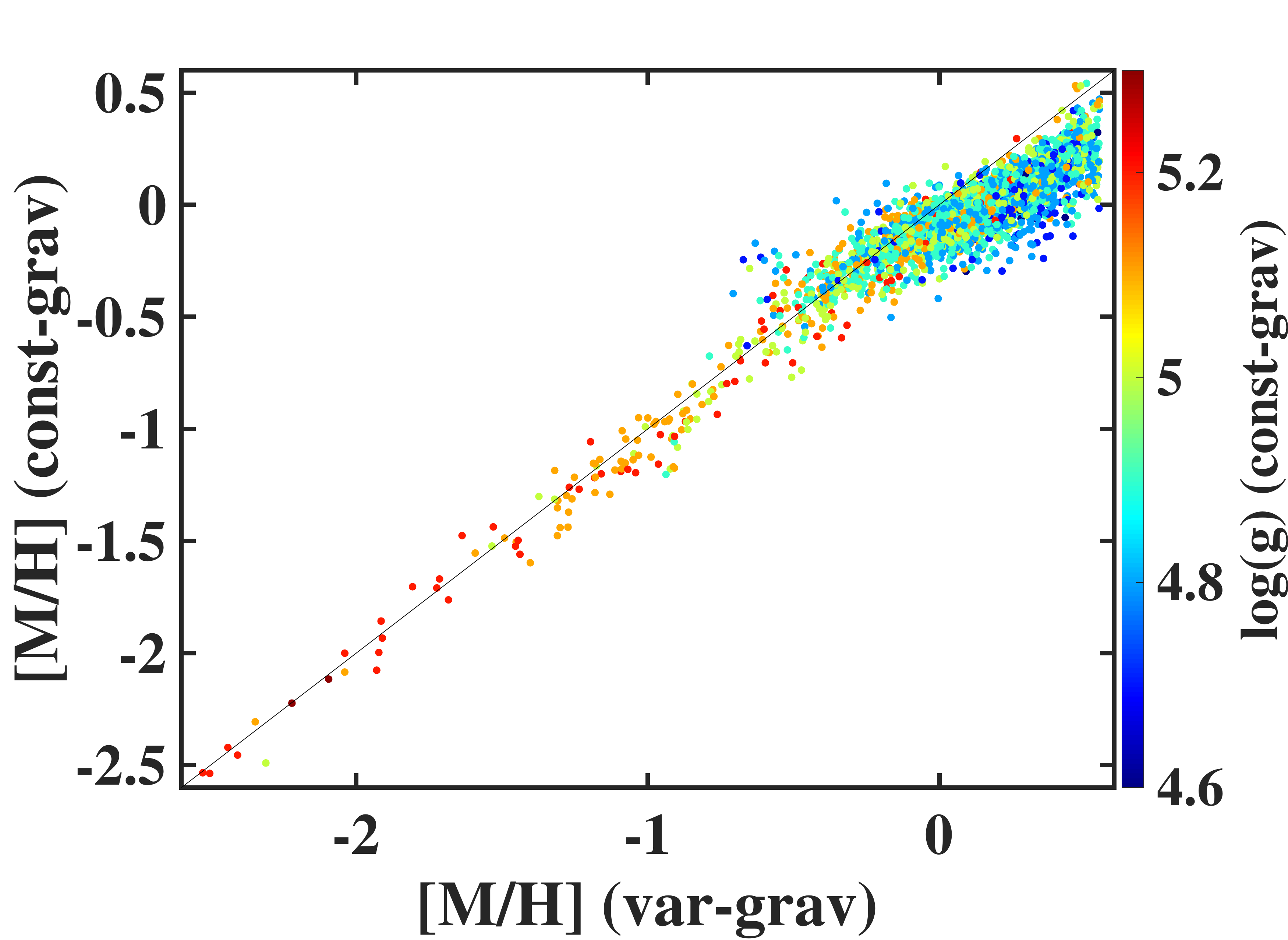}}              
\hspace{0.38cm} 
 \subfloat      
         []{\includegraphics[ height=4.3cm, width=5.65cm]{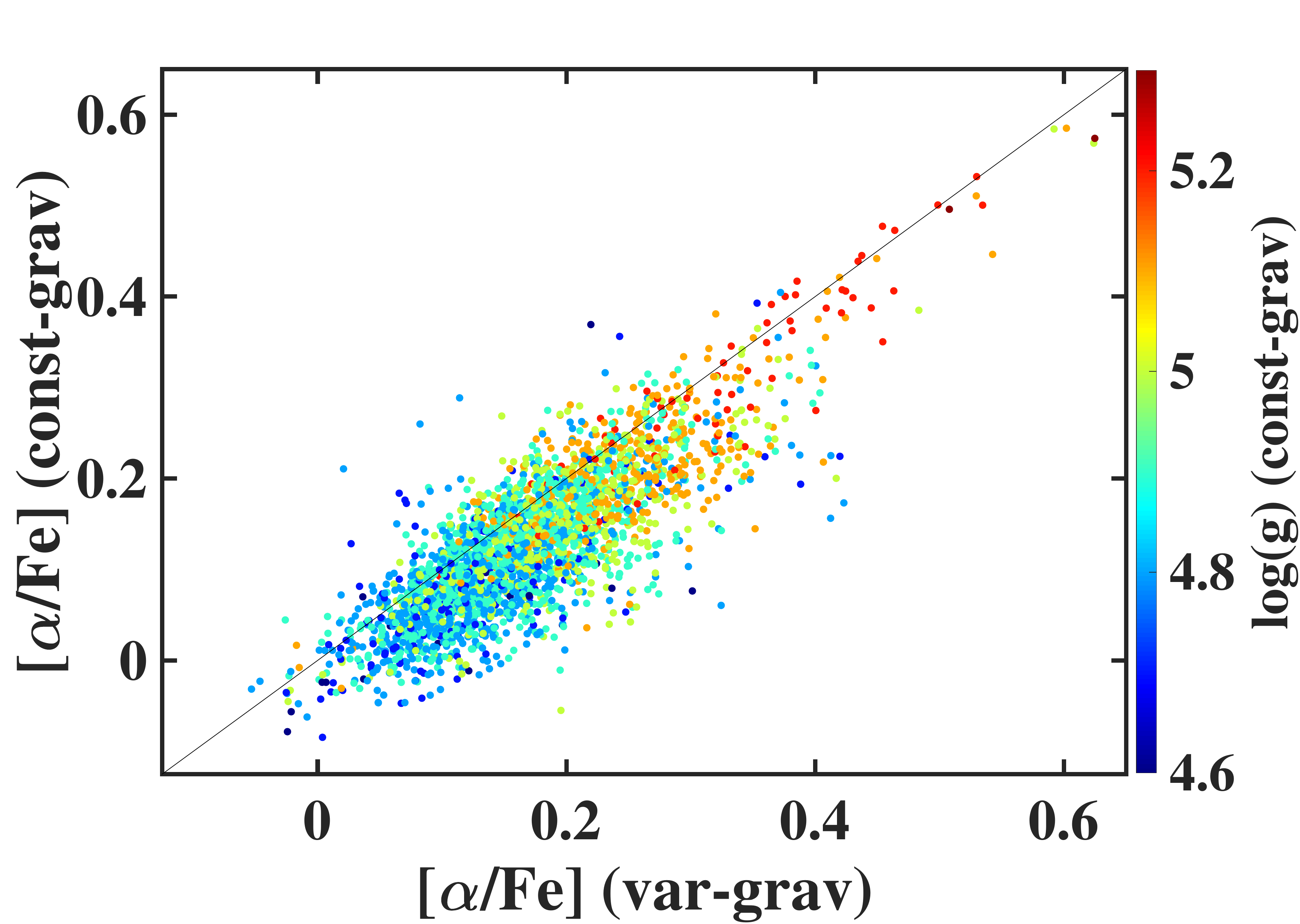}}
 \vspace{-0.25cm}

 \subfloat 
      []{\includegraphics[ height=4.3cm, width=5.65cm]{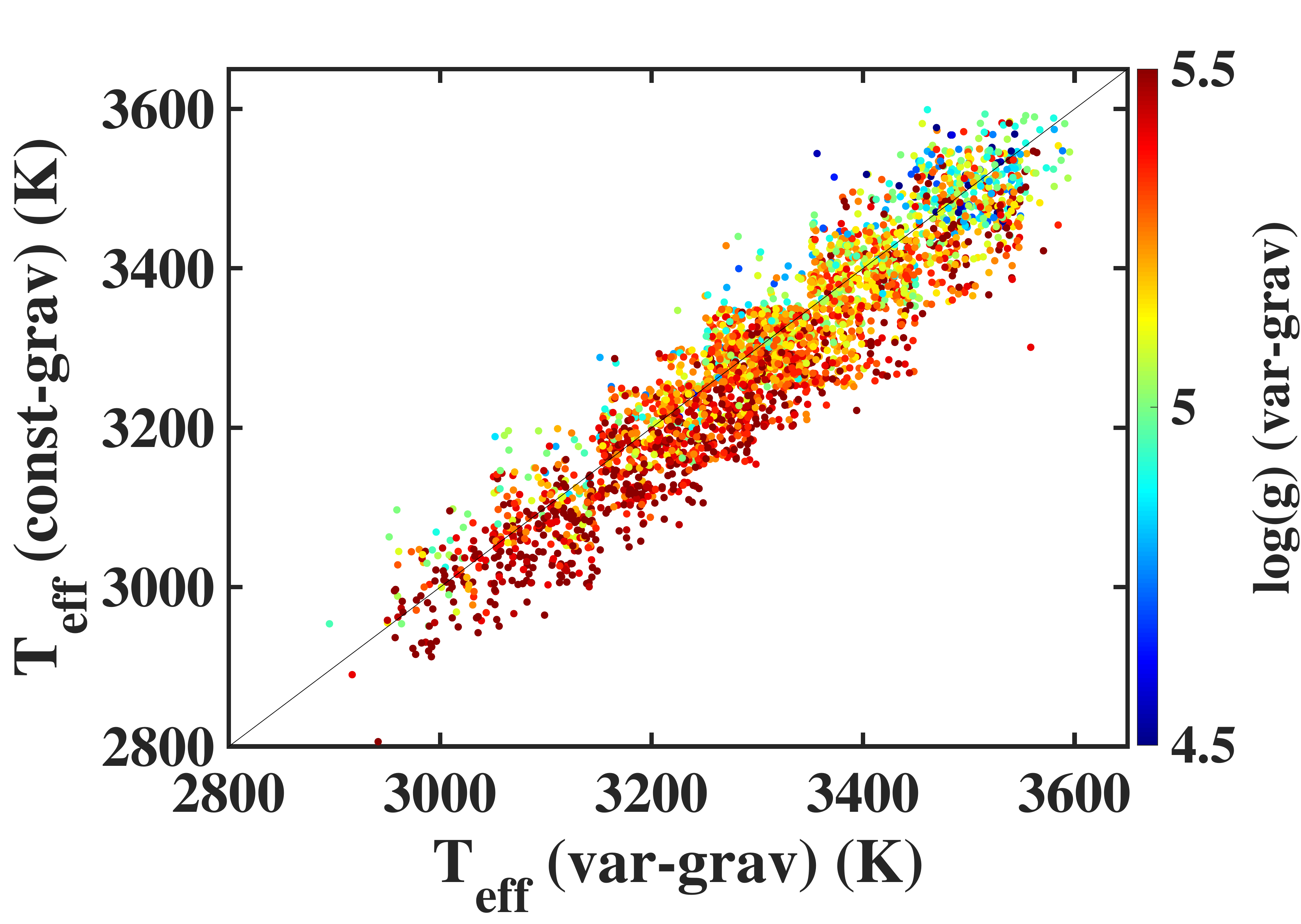}}               
\hspace{0.38cm}
\subfloat 
      []{\includegraphics[ height=4.3cm, width=5.65cm]{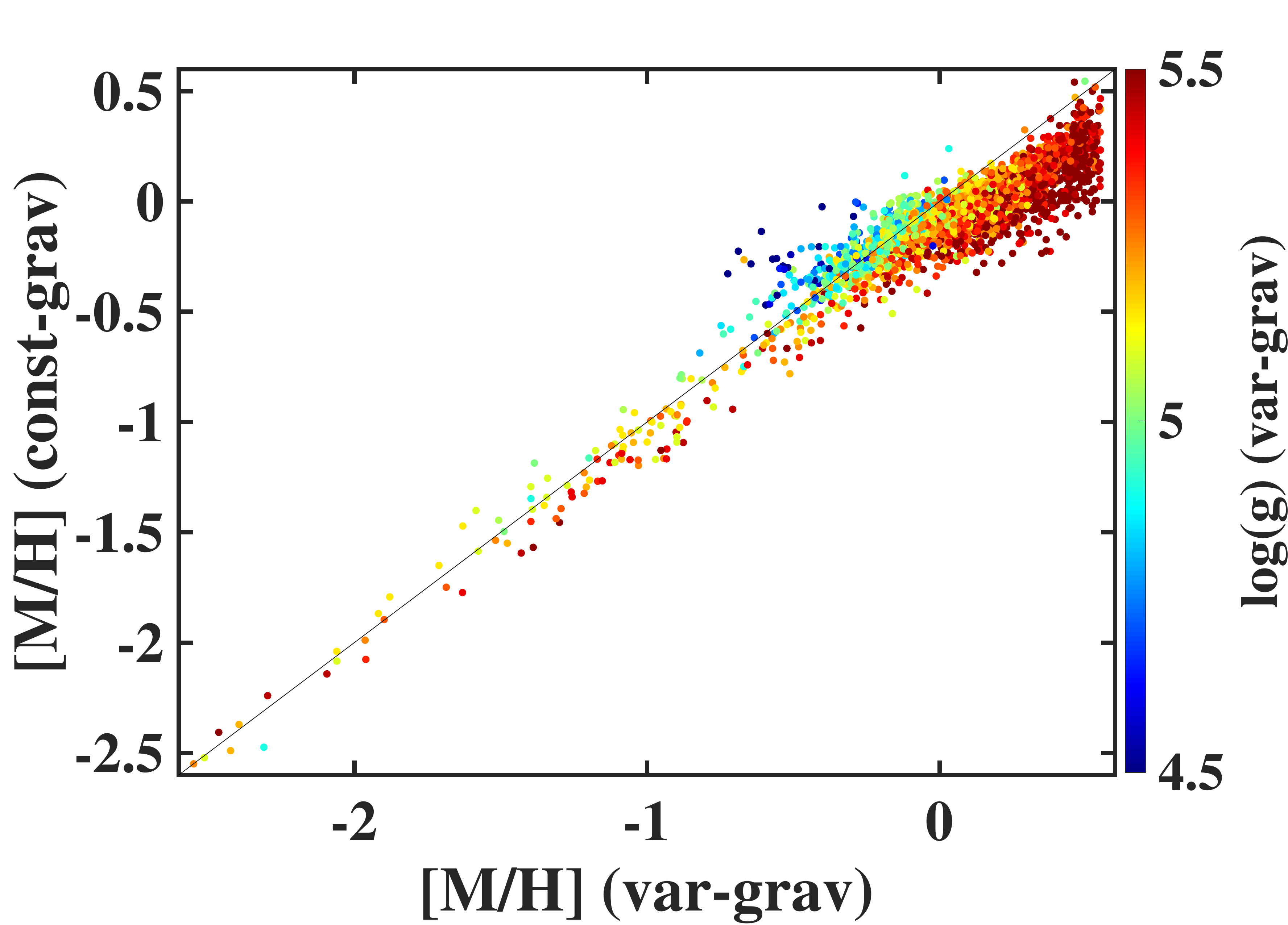}}               
\hspace{0.38cm} 
 \subfloat      
         []{\includegraphics[ height=4.3cm, width=5.65cm]{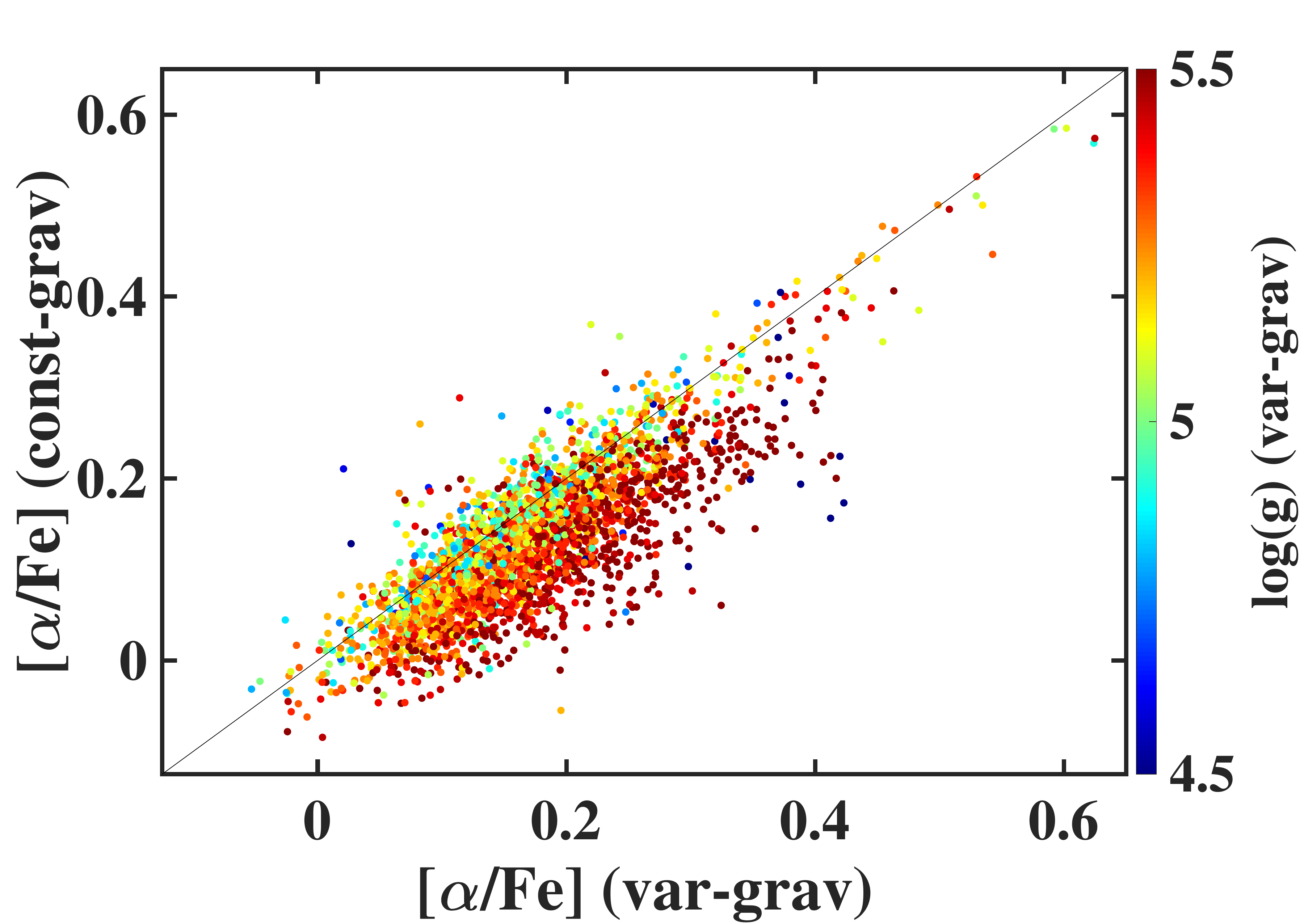}}      
\vspace{-0.25cm}

 \subfloat 
      []{\includegraphics[ height=4cm, width=5.4cm]{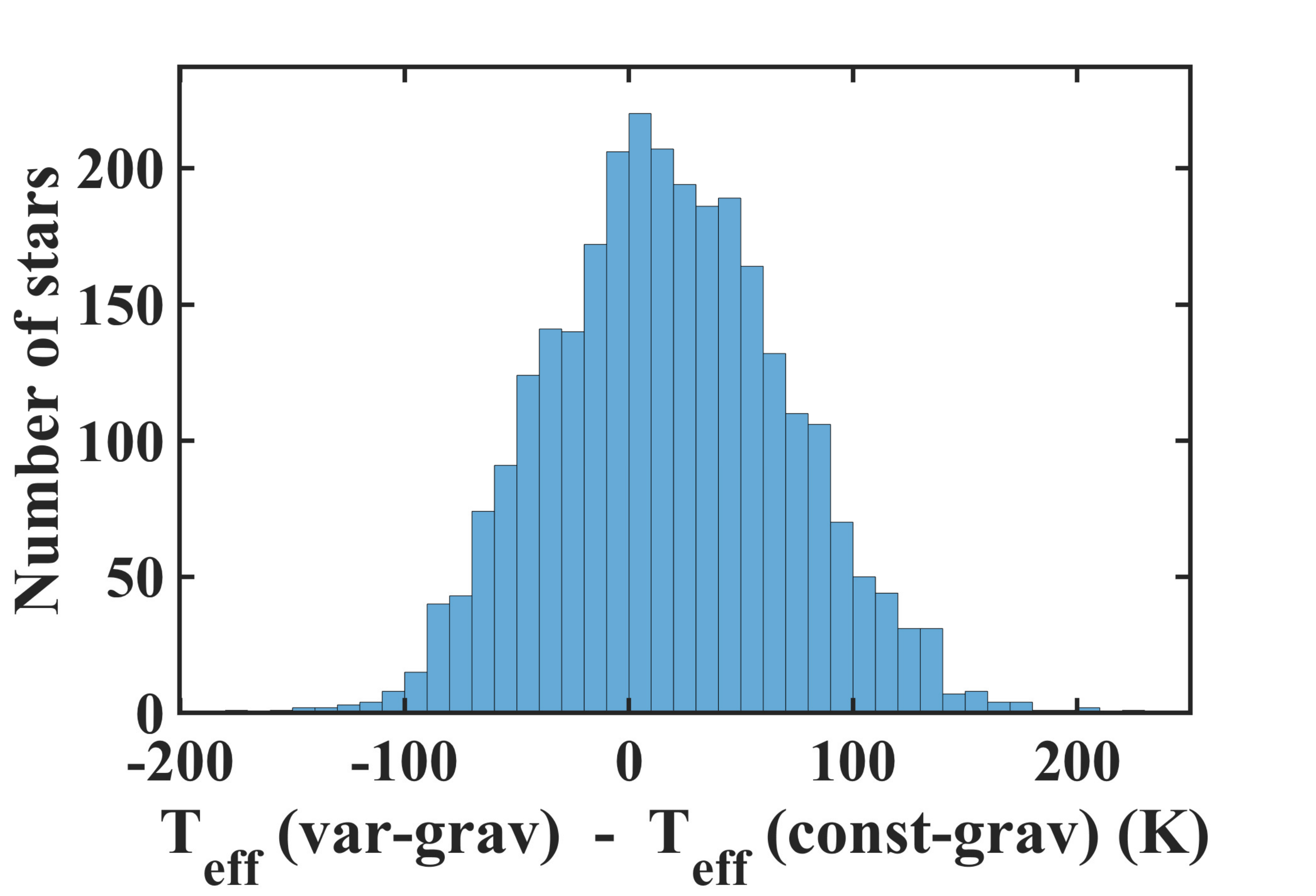}}               
\hspace{0.75cm}
\subfloat 
      []{\includegraphics[ height=4cm, width=5.4cm]{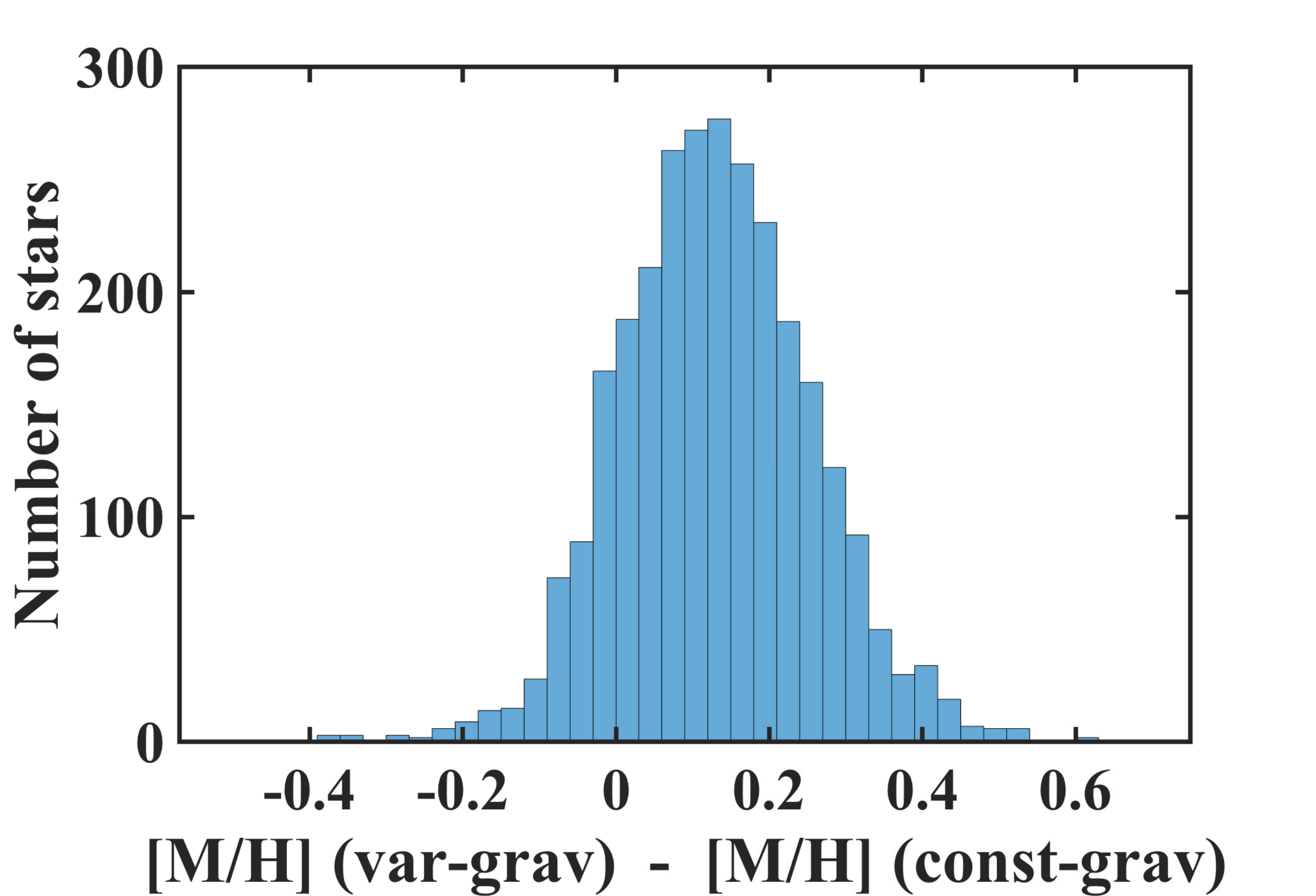}}               
\hspace{0.75cm} 
 \subfloat      
         []{\includegraphics[ height=4cm, width=5.4cm]{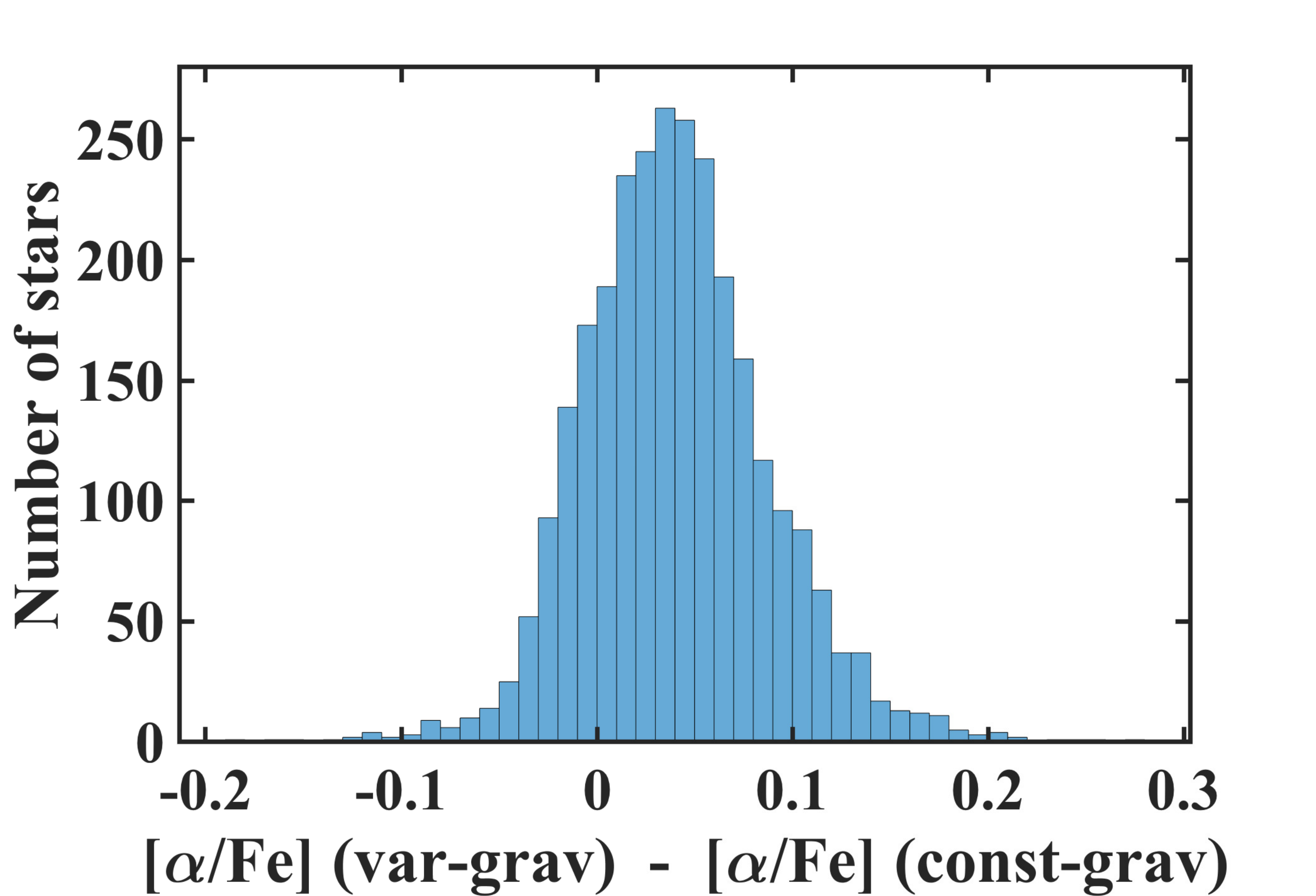}}        
 \caption        
        {\footnotesize{Comparison between the inferred model-fit parameters, T$_\textrm{\footnotesize{eff}}$, [M/H], and  [$\alpha$/Fe]  when surface gravity is a fixed parameter  and their corresponding values when surface gravity is a free parameter, for the 2829 stars with T$_\textrm{\footnotesize{eff}}$ $\leq$3550 K in both gravity cases, color-mapped based on the surface gravity values from the former approach  (top panels) and from the latter approach (middle panels), using the \textbf{normal method}. The bottom panels show the histograms of the differences between corresponding values inferred from the two gravity-modeling approaches.}}
\end{figure*}

\begin{figure}\centering
\subfloat
        [Fixed surface gravity]{\includegraphics[ height=4.3cm, width=7.5cm]{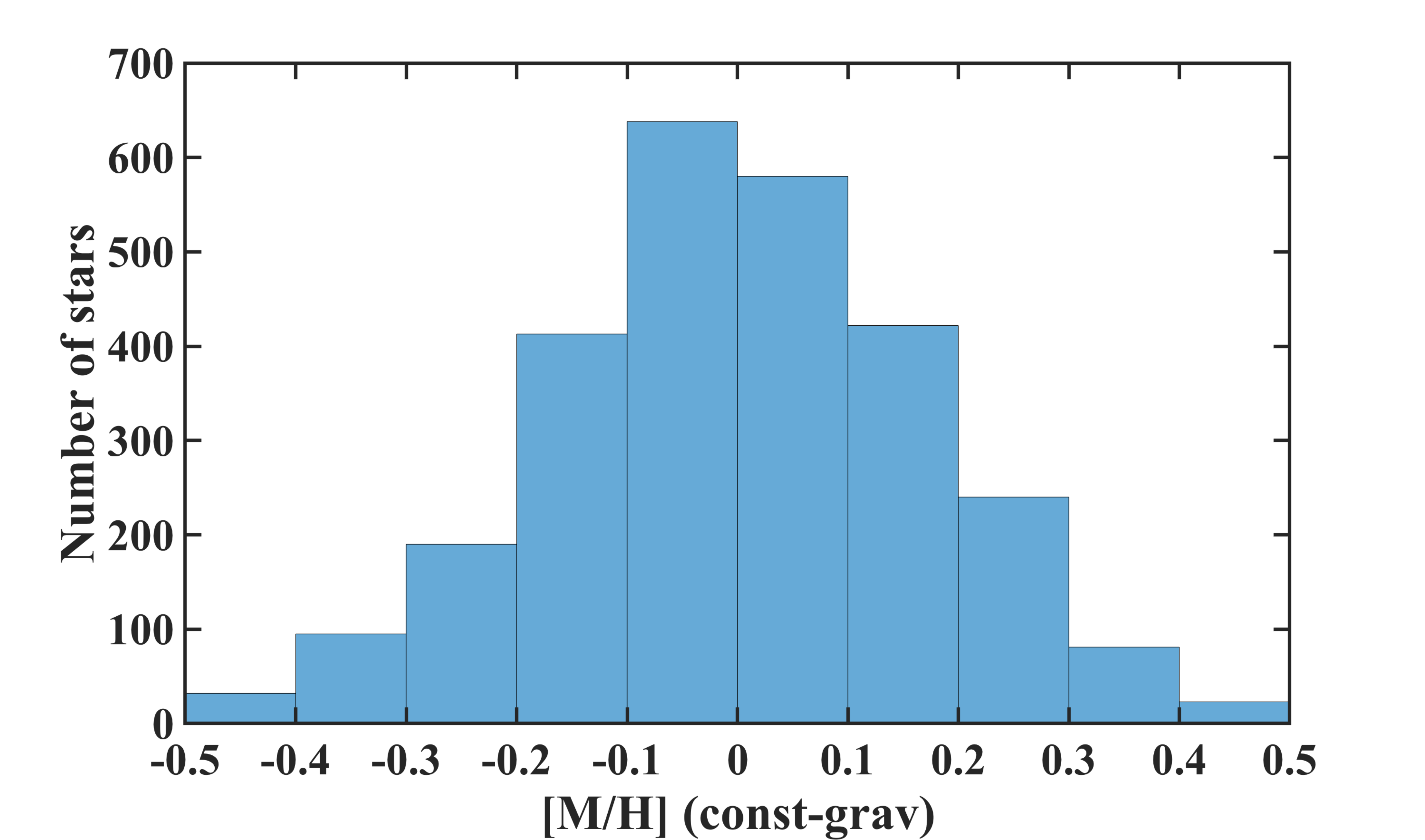}}
\vspace{-0.25cm}

\subfloat       
        [Variable surface gravity]{\includegraphics[ height=4.3cm, width=7.5cm]{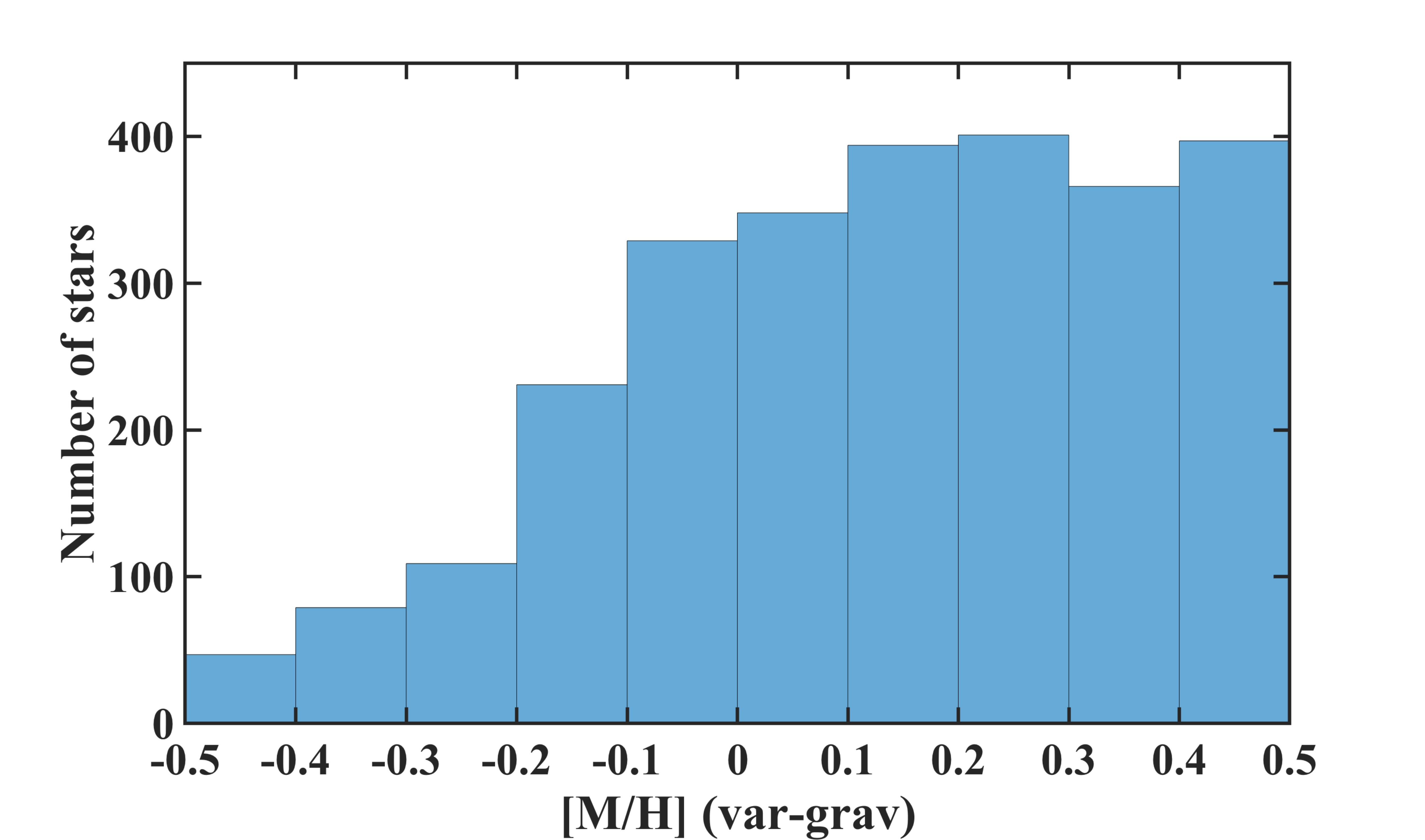}}
 \caption
        {\footnotesize{Histogram of estimated metallicities for the stars with  T$_\textrm{\footnotesize{eff}}$$\leq$3550 K and [M/H]$\geq$$-$0.5 dex from the approach when surface gravity is constant  (2714 stars, top panel) and when surface gravity is varying (2701 stars, bottom panel), using the \textbf{normal method}.}}
\end{figure}

\begin{figure*}\centering
\subfloat
        [Fixed surface gravity]{\includegraphics[ height=3.7cm, width=8.8cm]{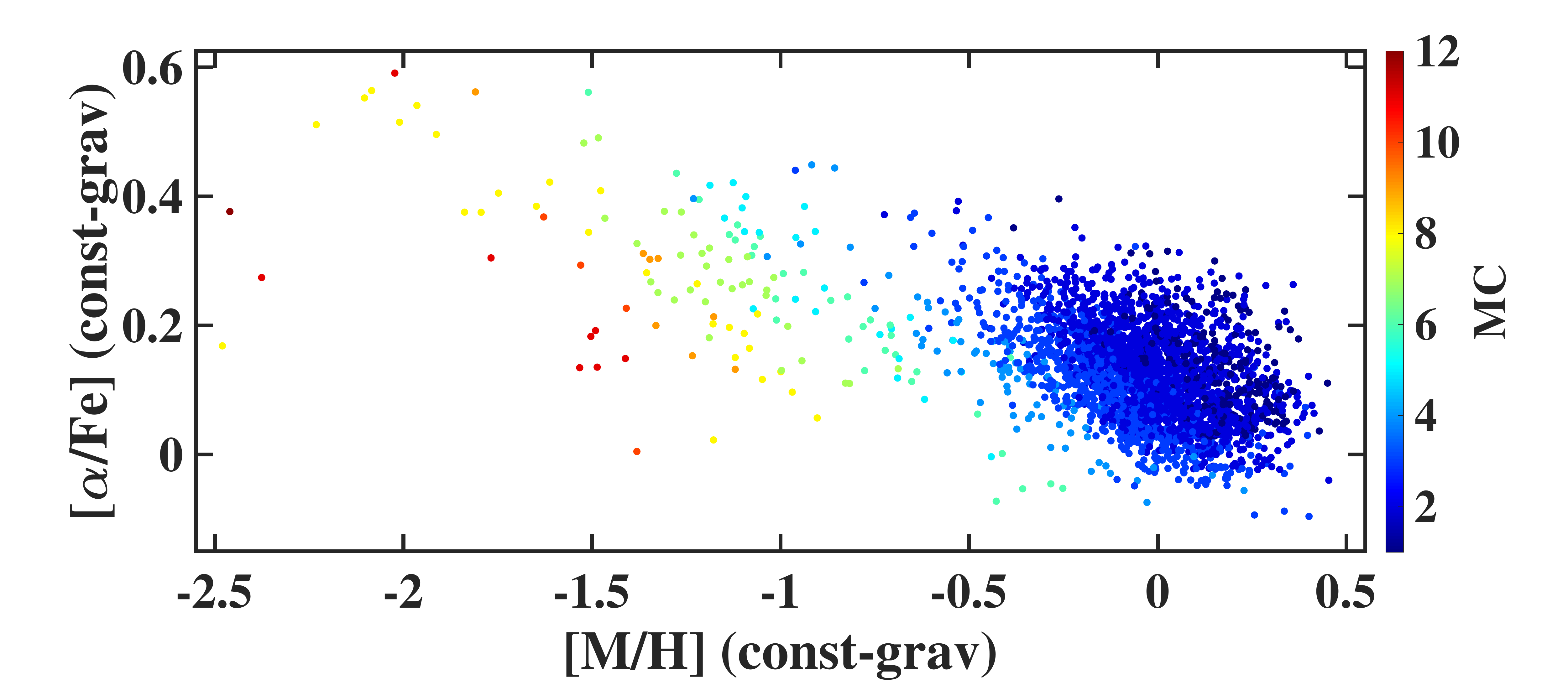}}
 \hspace{0cm} 
 \subfloat 
         [Variable surface gravity]{\includegraphics[ height=3.7cm, width=8.8cm]{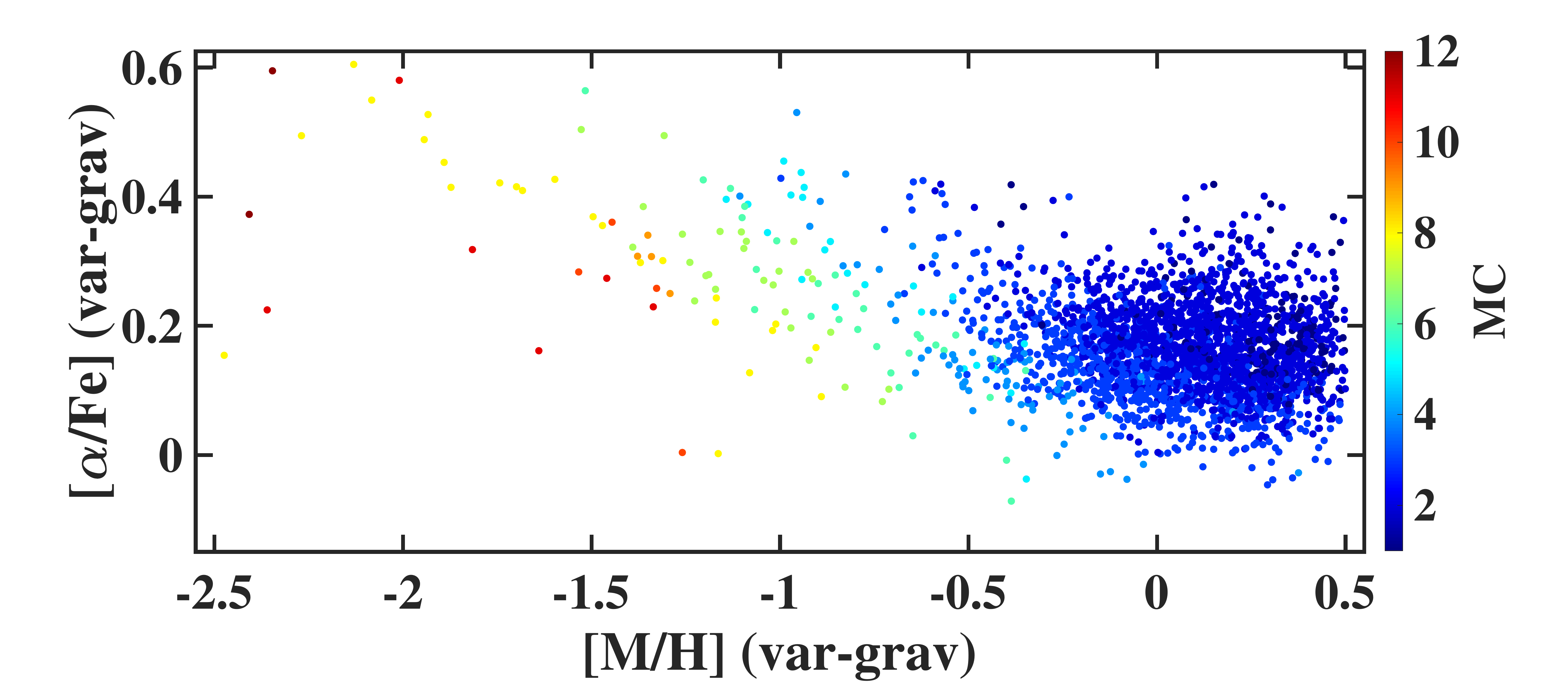}}         
\vspace{-0.35cm}

\subfloat
        [Fixed surface gravity]{\includegraphics[ height=3.7cm, width=8.8cm]{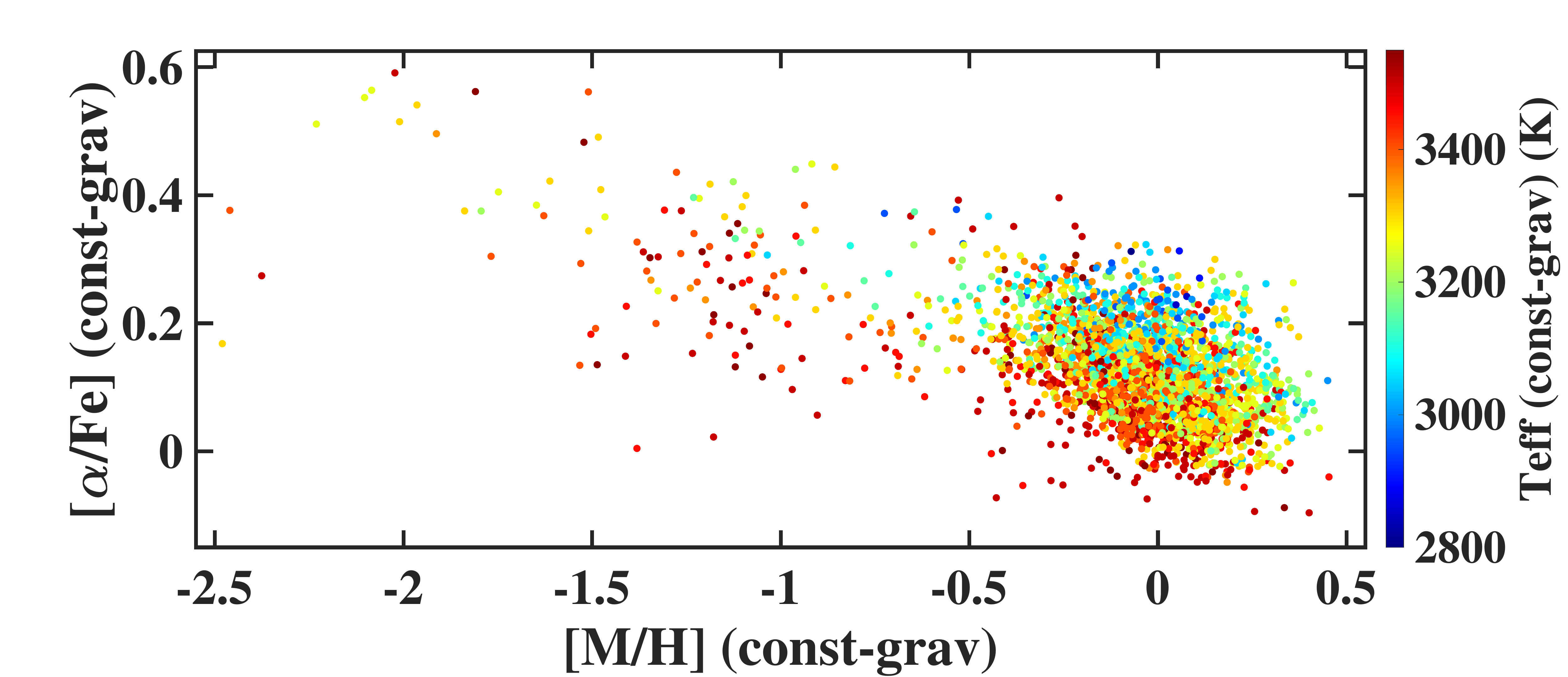}}
 \hspace{0cm} 
 \subfloat 
         [Variable surface gravity]{\includegraphics[ height=3.7cm, width=8.8cm]{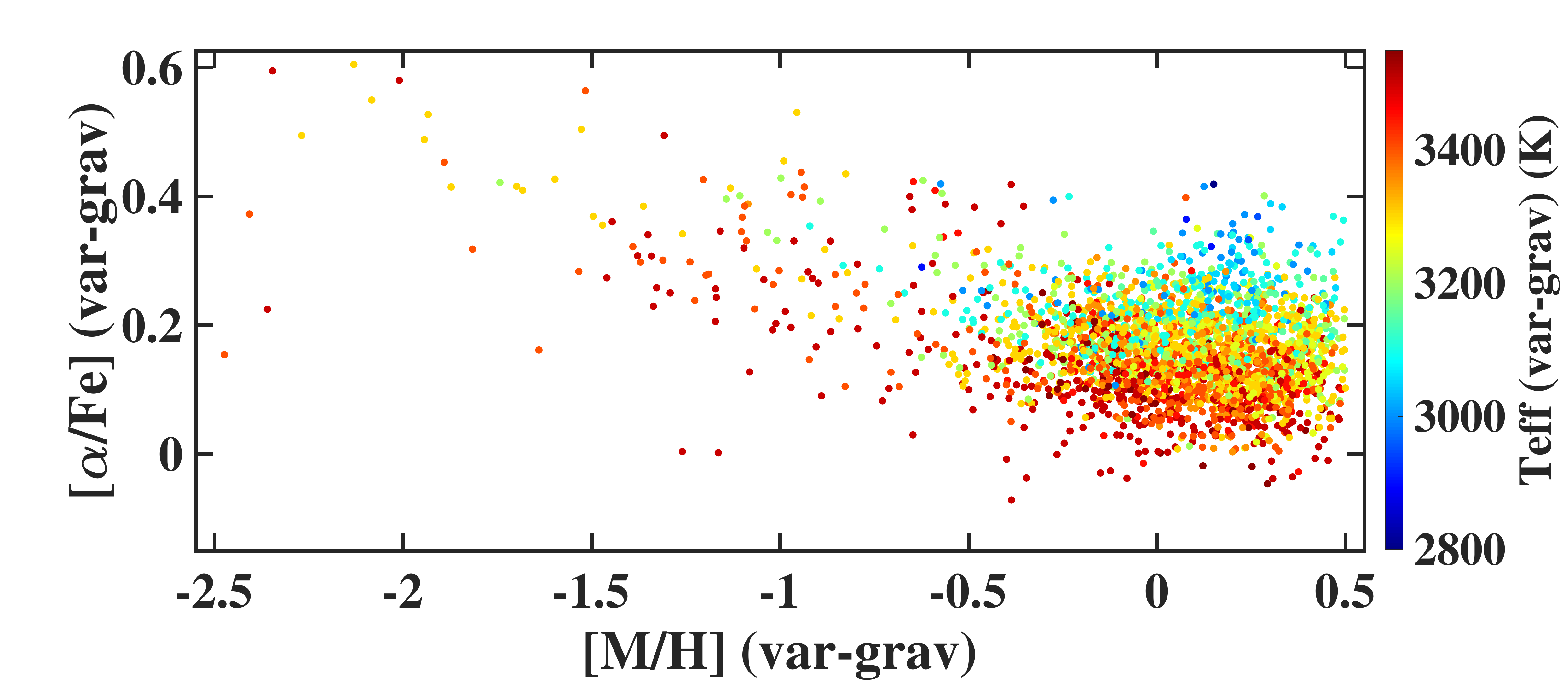}}         
\vspace{-0.35cm}

\subfloat
         [Fixed surface gravity]{\includegraphics[ height=3.7cm, width=8.8cm]{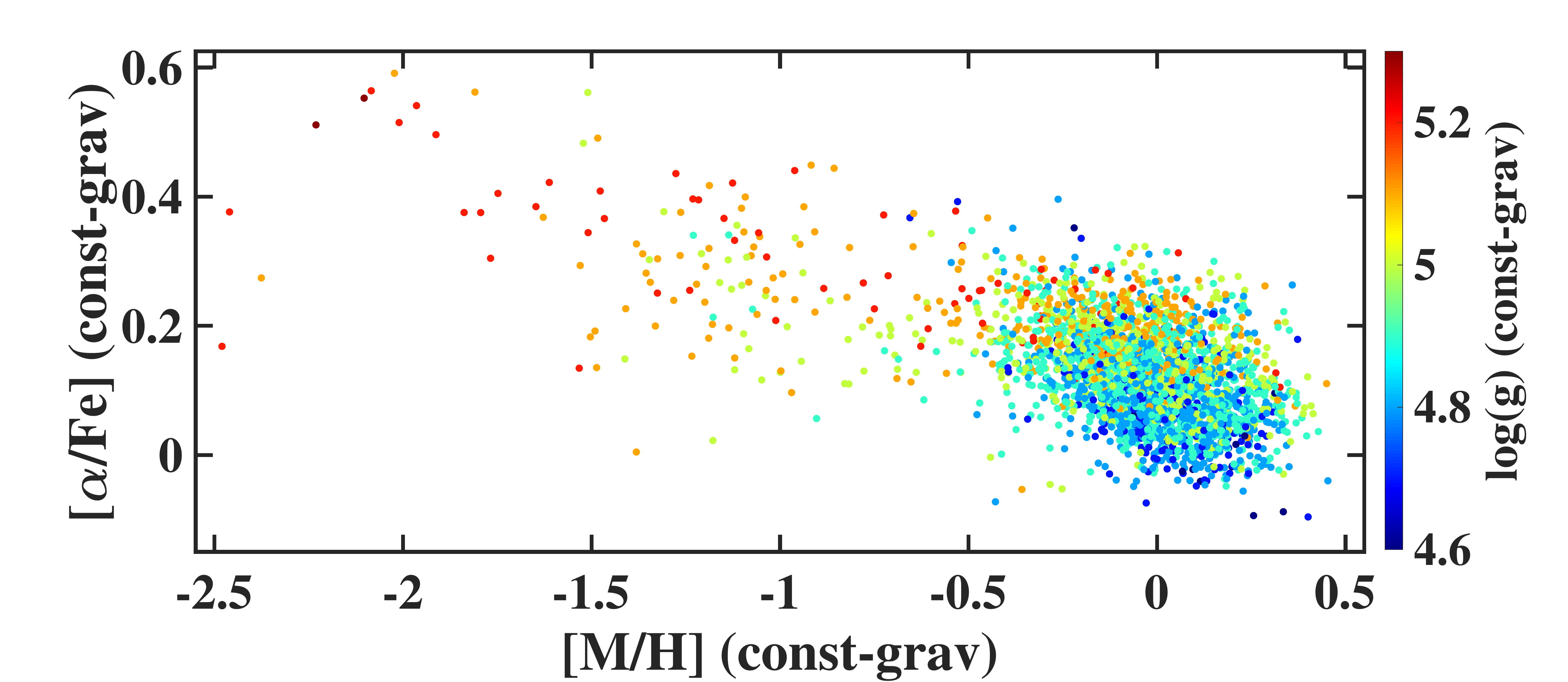}}
 \hspace{0cm}  
 \subfloat       
          [Variable surface gravity]{\includegraphics[ height=3.7cm, width=8.8cm]{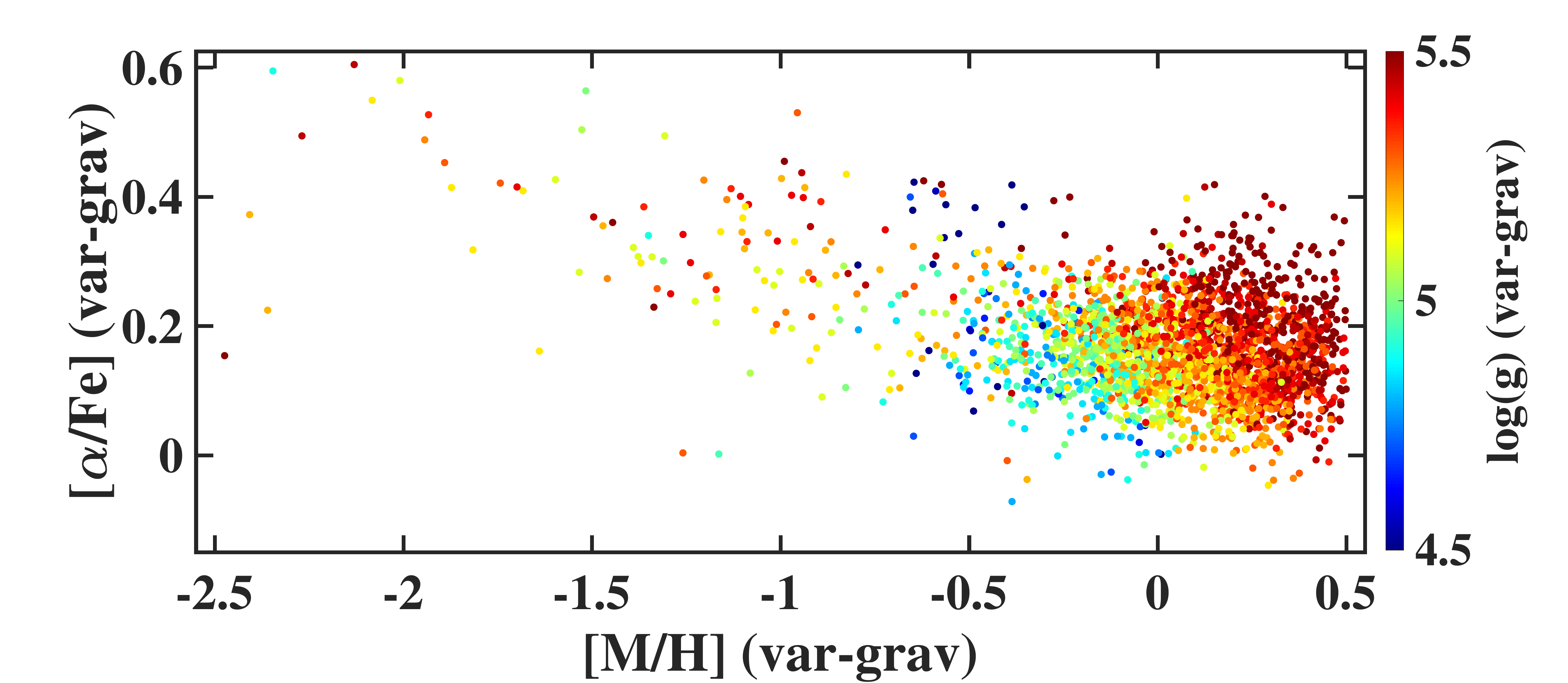}}
 \caption
        {\footnotesize{[$\alpha$/Fe] vs. [M/H] diagram for the stars with T$_\textrm{\footnotesize{eff}}$$\leq$3550 K and $-$2.5<[M/H]<$+$0.5 dex when surface gravity is a fixed parameter (left panels, 2868 stars) and when surface gravity is a free parameter (right panels, 2668 stars), color mapped based on metallicity class (top panels), T$_\textrm{\footnotesize{eff}}$ (middle panels), and log \emph{g} (bottom panels), using the \textbf{normal method}.}}
\end{figure*}

\begin{figure*}\centering
\subfloat
        [Fixed surface gravity]{\includegraphics[ height=3.7cm, width=8.8cm]{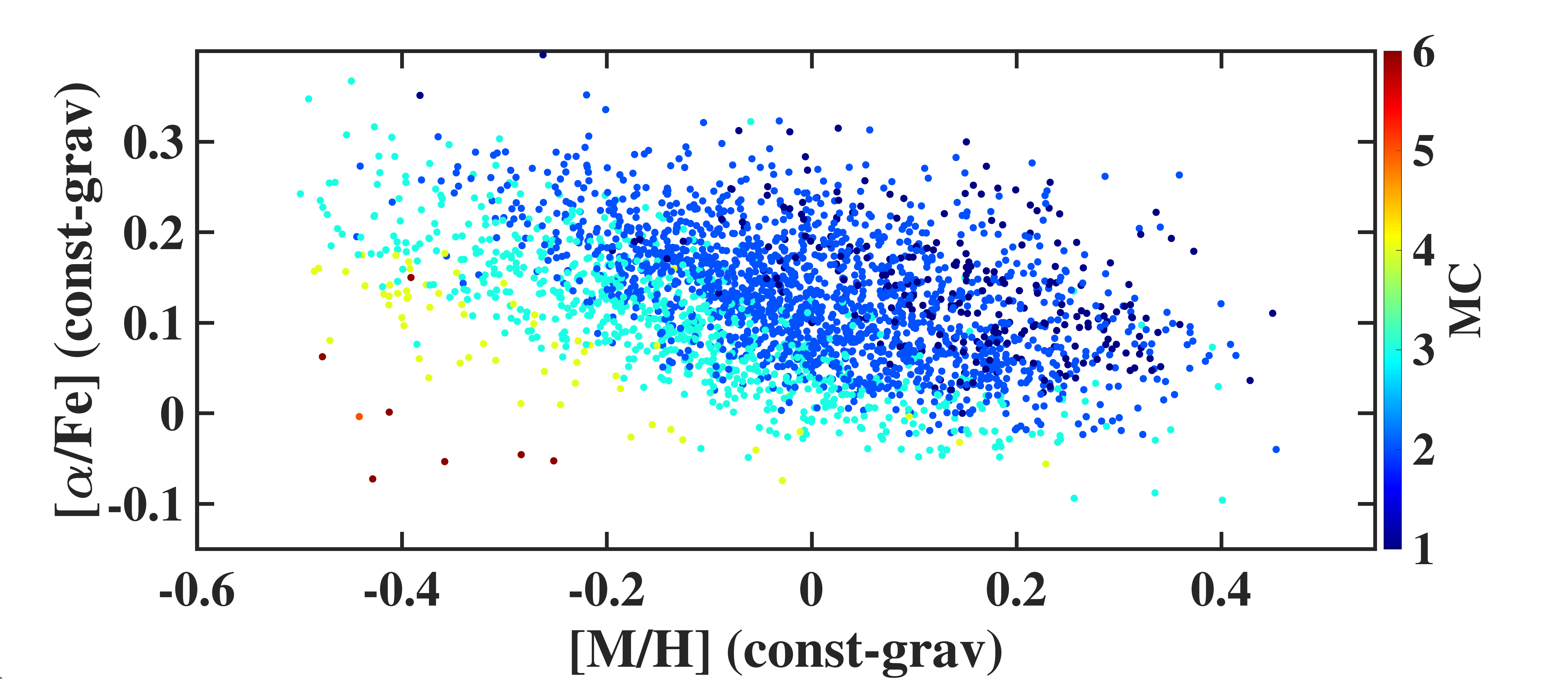}}
 \hspace{0cm} 
 \subfloat 
         [Variable surface gravity]{\includegraphics[ height=3.7cm, width=8.8cm]{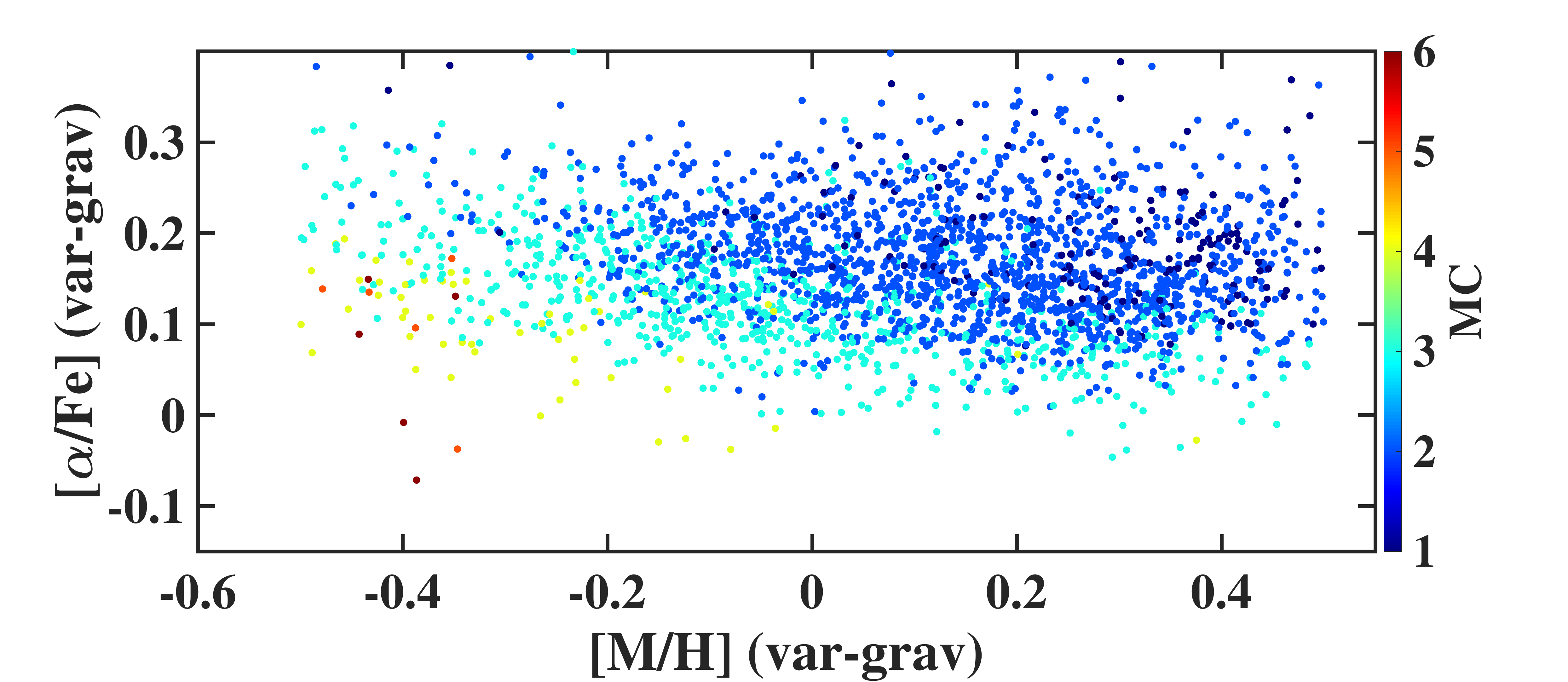}}         
\vspace{-0.35cm}

\subfloat
        [Fixed surface gravity]{\includegraphics[ height=3.7cm, width=8.8cm]{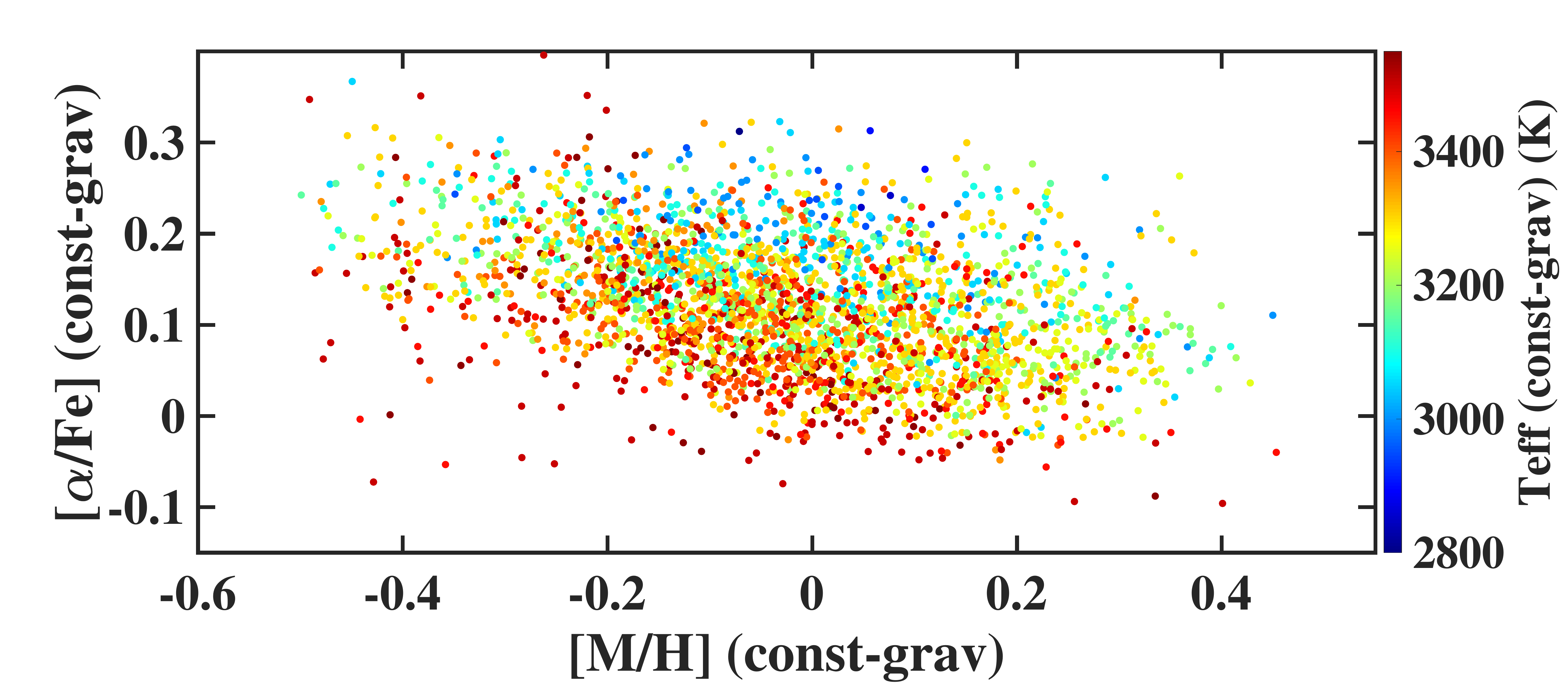}}
 \hspace{0cm} 
 \subfloat 
         [Variable surface gravity]{\includegraphics[ height=3.7cm, width=8.8cm]{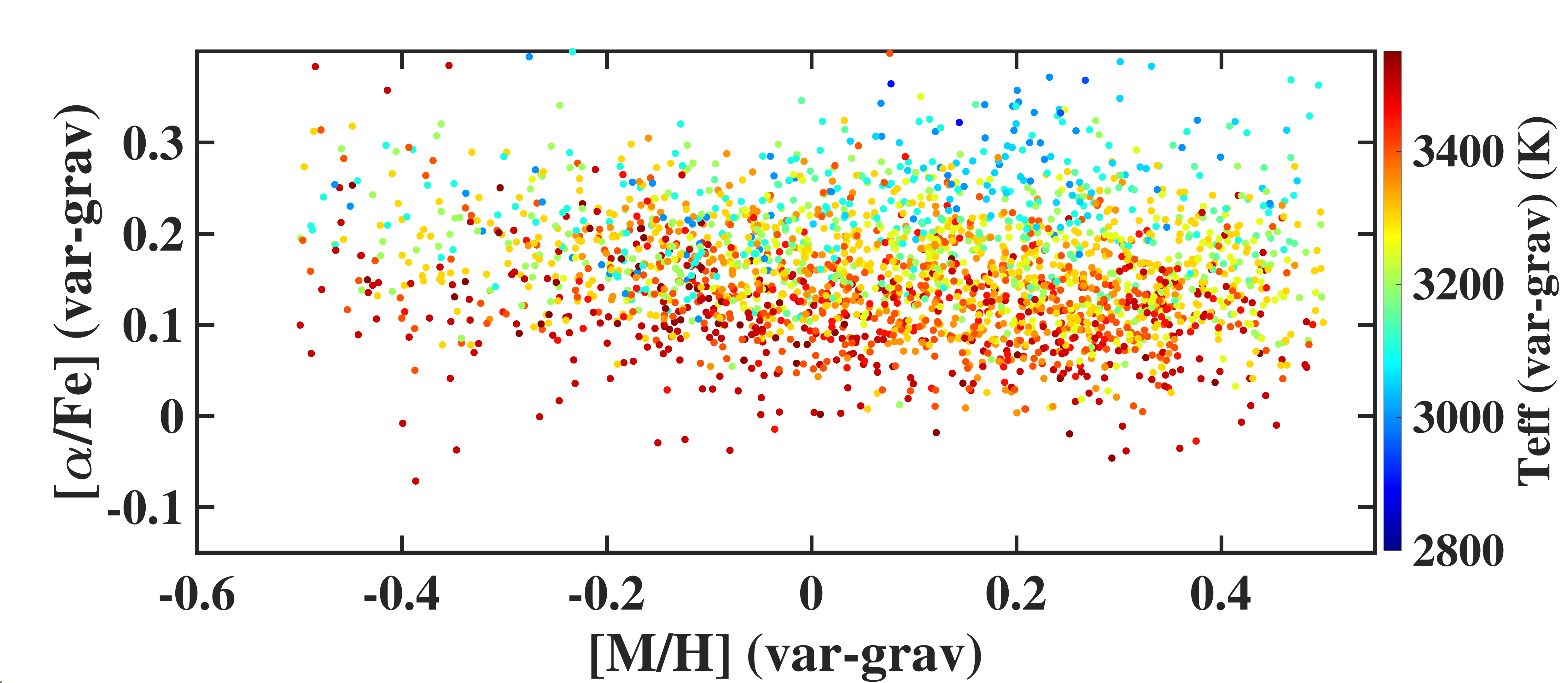}}         
\vspace{-0.35cm}

\subfloat
         [Fixed surface gravity]{\includegraphics[ height=3.7cm, width=8.8cm]{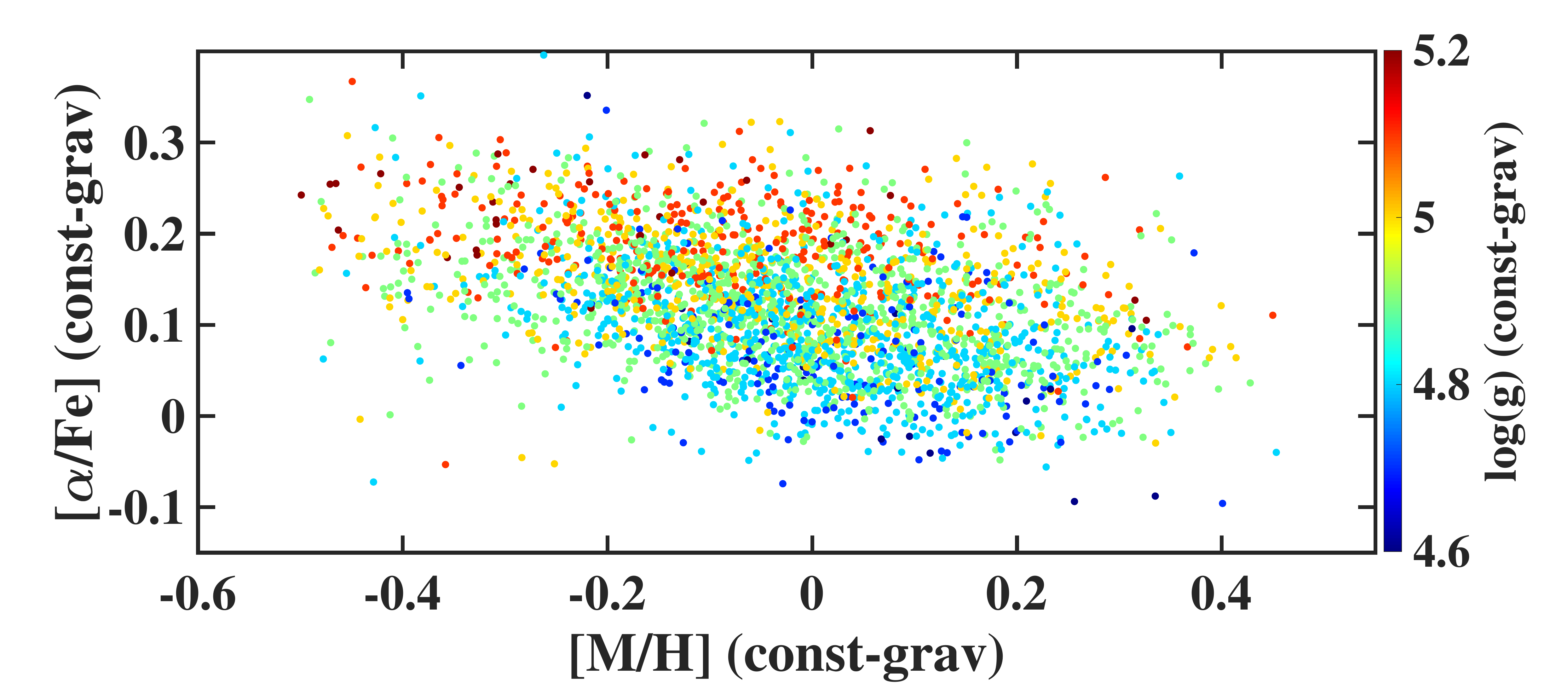}}
 \hspace{0cm}  
 \subfloat       
          [Variable surface gravity]{\includegraphics[ height=3.7cm, width=8.8cm]{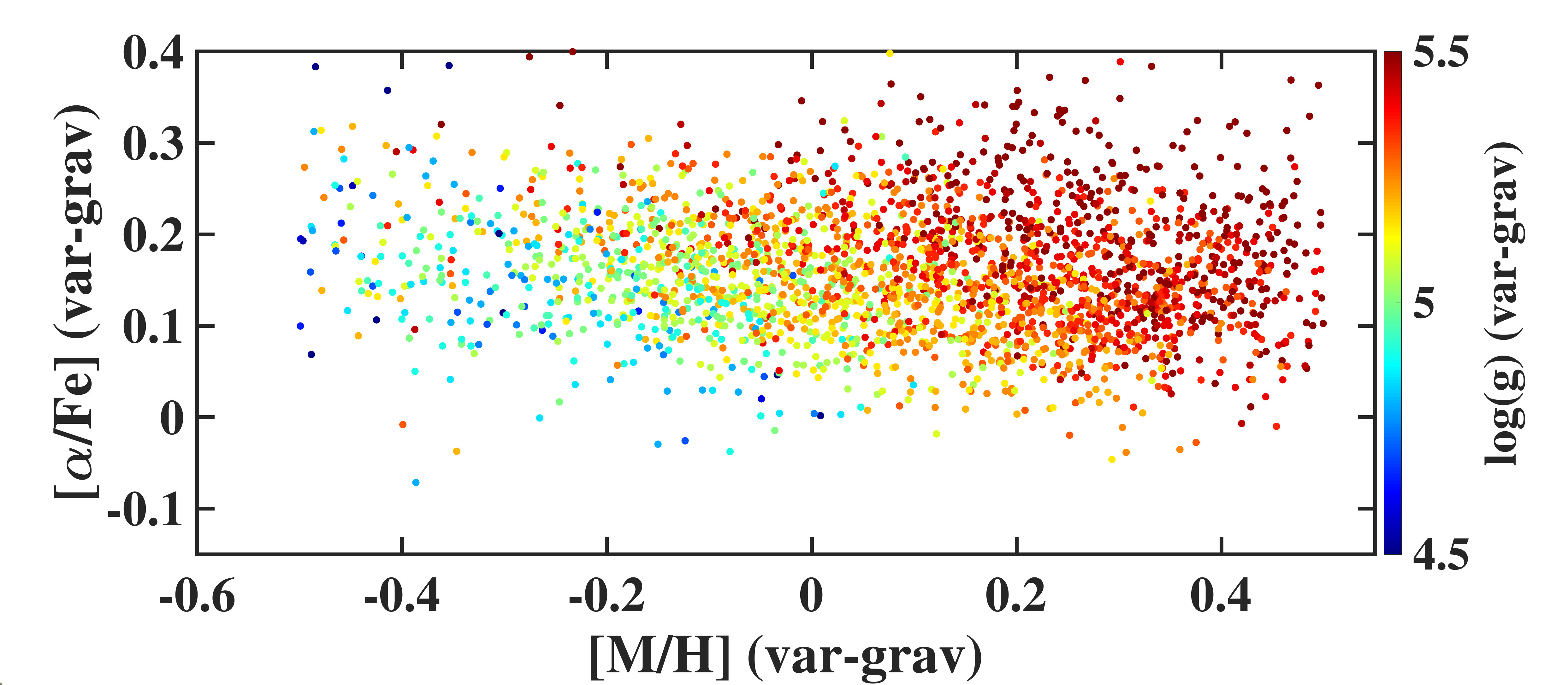}}
 \caption
        {\footnotesize{[$\alpha$/Fe] vs. [M/H] diagram for the stars with T$_\textrm{\footnotesize{eff}}$$\leq$3550 K and $-$0.5$\leq$[M/H]<$+$0.5 dex when surface gravity is a fixed parameter (left panels, 2698 stars) and when surface gravity is a free parameter (right panels, 2511 stars), color mapped based on metallicity class (top panels), T$_\textrm{\footnotesize{eff}}$ (middle panels), and log \emph{g} (bottom panels), using the \textbf{normal method}.}}
\end{figure*}

\begin{figure}\centering
        {\includegraphics[ height=5cm, width=7.5cm]{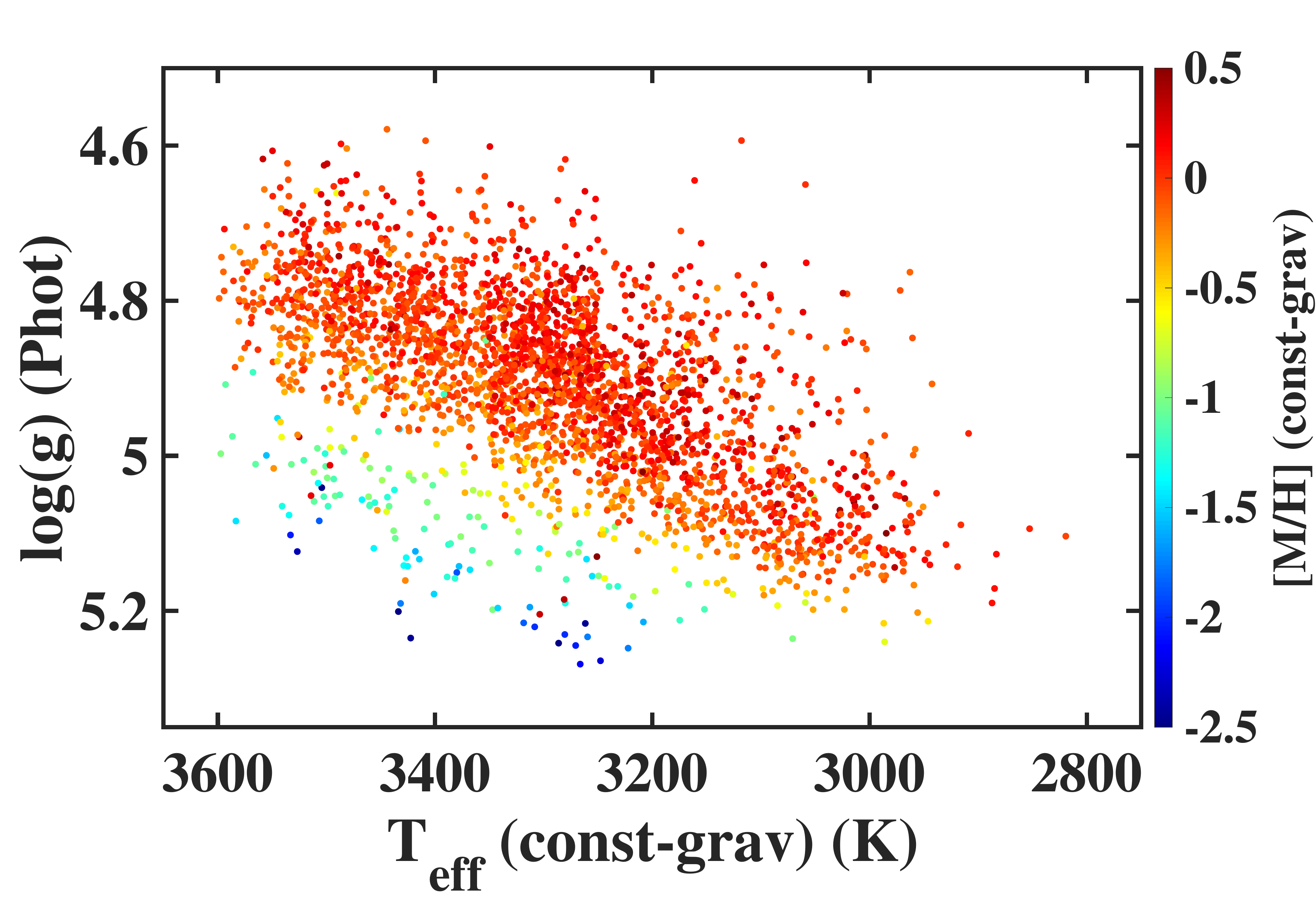}}
\vspace{-0.35cm}

\subfloat   
        {\includegraphics[ height=5cm, width=7.5cm]{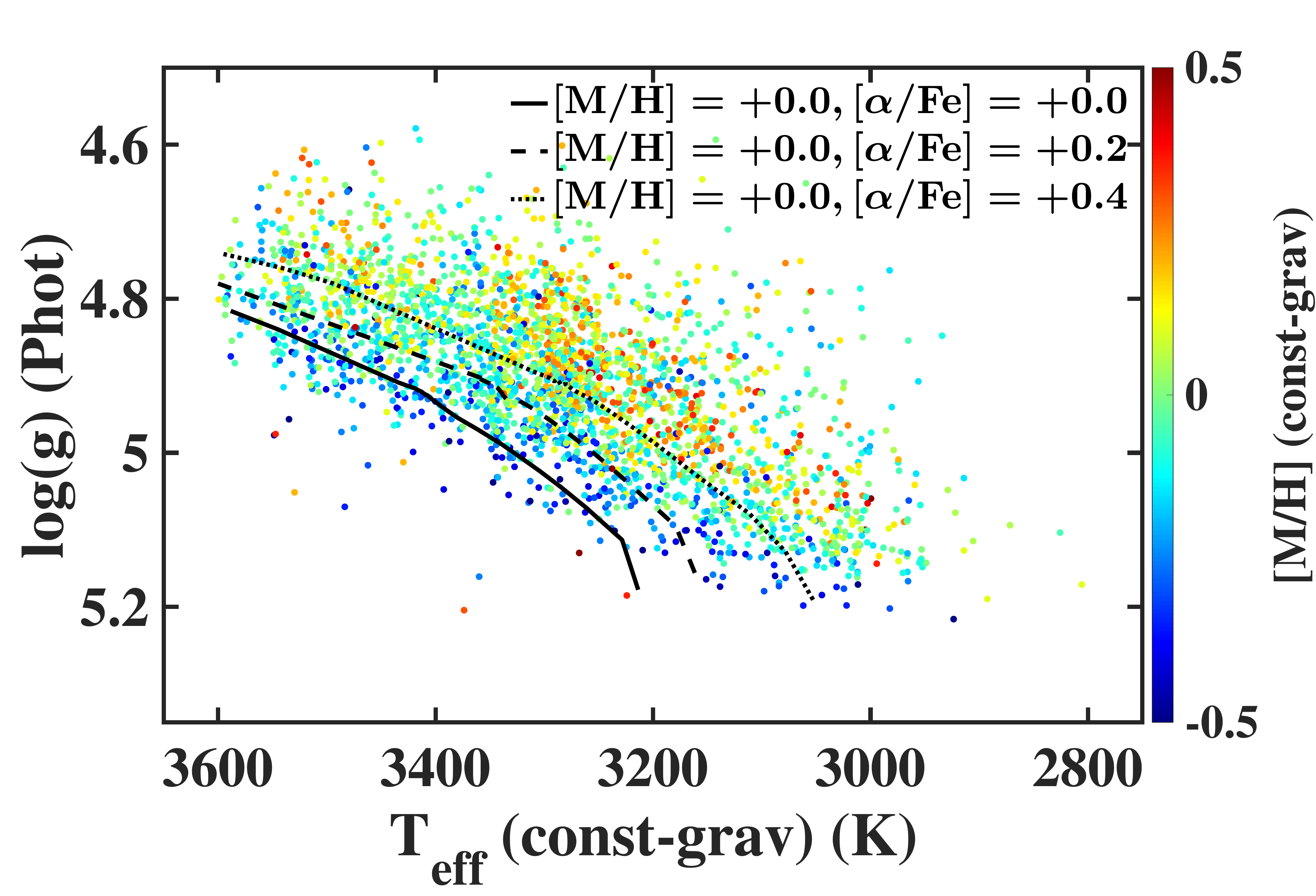}}   
 \caption
        {\footnotesize{Top panel: photometric surface gravity versus inferred model-fit effective temperature for 2877 stars  with T$_\textrm{\footnotesize{eff}}$$\leq$3550 K, color-mapped based on the metallicity of  stars. All best-fit values are derived using the \textbf{normal method} when surface gravity is fixed and equal to the photometric values. Bottom panel: the same plot as shown in the top panel but for stars with [M/H]$\geq$$-$0.5 dex along with solar-metallicity isochrones related to  three different values of  [$\alpha$/Fe]=+0.0, +0.2, and +0.4 dex (solid, dashed and dotted black lines, respectively) from the Dartmouth stellar evolutionary models.}}
\end{figure}

\begin{figure}\centering
\subfloat
        [Fixed surface gravity]{\includegraphics[ height=5cm, width=7.5cm]{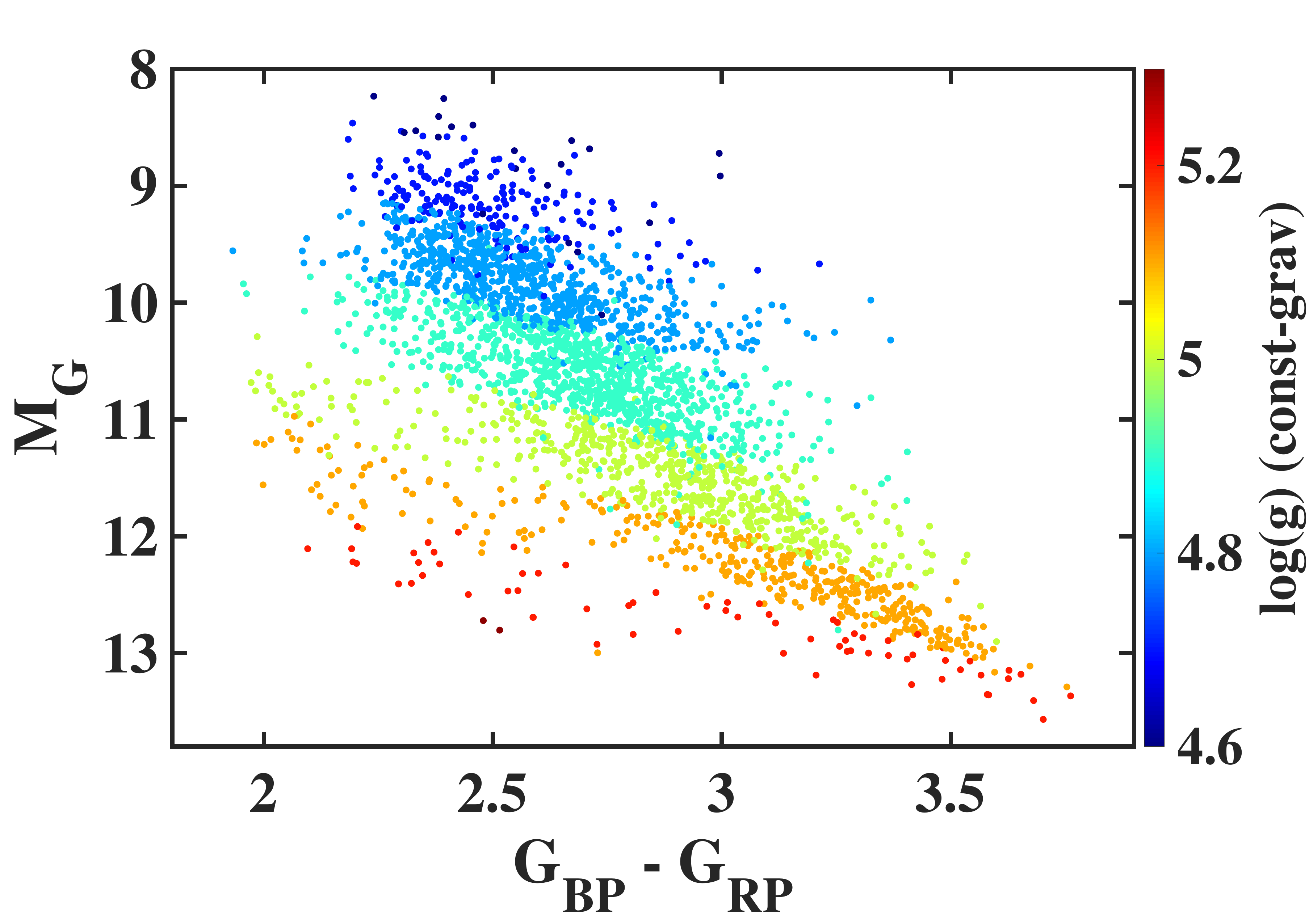}}
\vspace{-0.25cm}

\subfloat       
        [Variable surface gravity]{\includegraphics[ height=5cm, width=7.5cm]{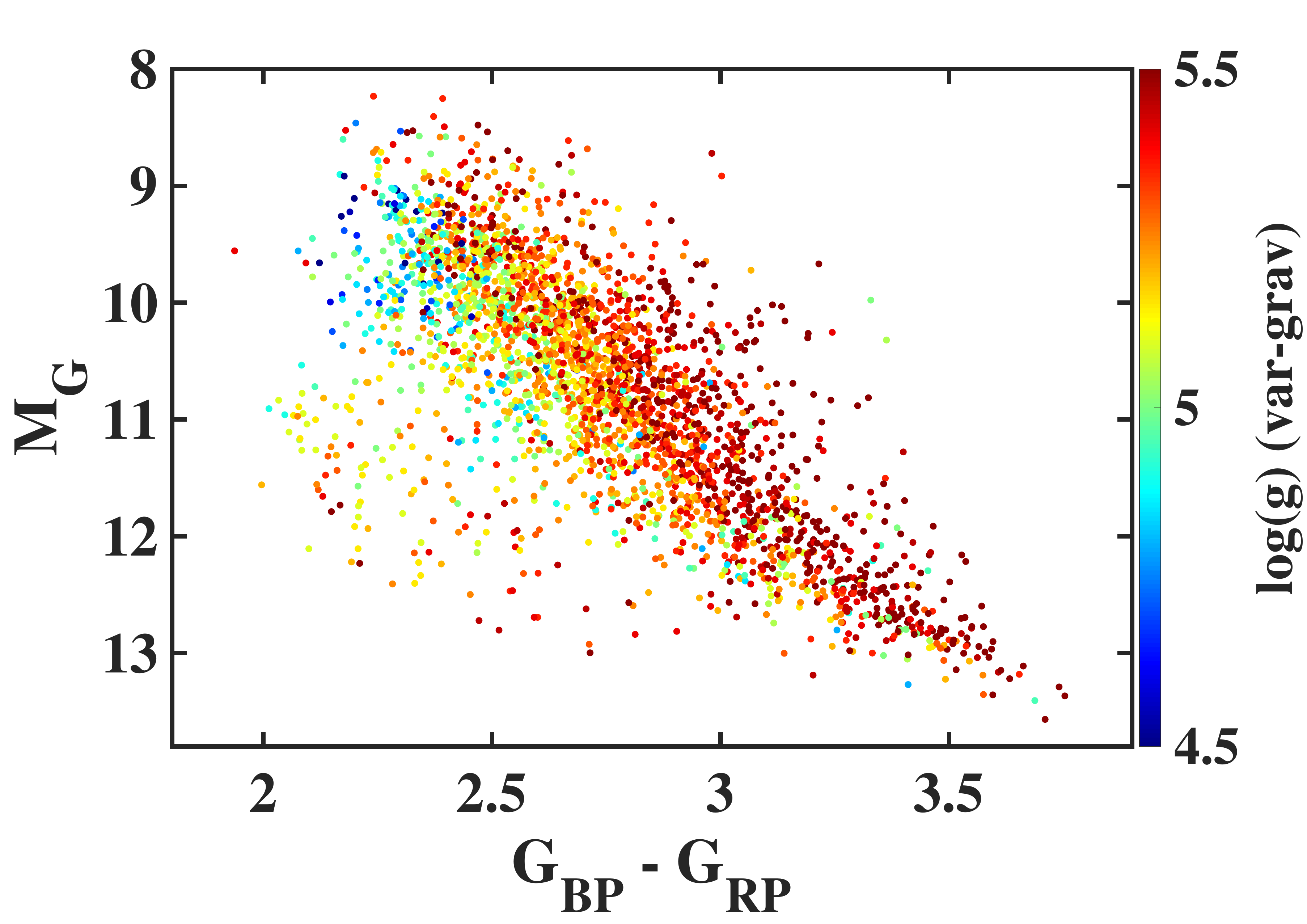}}
 \caption
        {\footnotesize{HR diagram for the stars with  T$_\textrm{\footnotesize{eff}}$$\leq$3550 K from the approach when surface gravity is  constant  (2877 stars, top panel) and when surface gravity is varying (2851 stars, bottom panel), color-mapped based on surface gravity values, using the \textbf{normal method}.}}
\end{figure}

Figures 24 and 25 show the same HR diagram panels as in Figures 22 and 23, respectively, but color-coded according to five groups  of stars with different  metallicity ranges, as noted in the legends. Results from the constant-gravity and variable-gravity methods mostly differ in the range of best-fit values, with the variable-gravity yielding a wider range of metallicities, notably at the metal-rich end. On the other hand, both methods show a clear stratification in the color-absolute magnitude diagram. This signifies that our pipeline can separate different metallicity populations with high precision, whether surface gravity is a fixed or a free parameter in the fitting process.  While the two gravity approaches do not  make a considerable  difference  in the resulting model-fit effective temperatures, the derived metallicities, particularly in the near-solar metallicity regime,  are  significantly dependent on which approach is used. As shown in Sections 3 and 5, surface gravity  is more correlated with  metallicity  than effective temperature, and as a result, the best-fit values of metallicity are  remarkably affected by the variation of   surface gravity (Sections 6.3 and 6.4).

Both Figures 24 and 25 also indicate that, in either gravity approach, only very few  observed  spectra are matched with high-temperature, high-metallicity models (T$_\textrm{\footnotesize{eff}}$>3550 K and [M/H]>+0.1), and our high-temperature stars (whose best-fit models resemble the atmosphere of  late K dwarfs with a weaker level of molecular opacity, rather than M dwarfs) are mostly concentrated in the metallicity range $-$0.5$\leq$[M/H]$\leq$+0.1 dex. This systematic concentration  is found to be  independent of the selected initial values, the $\chi$$^\textrm{\footnotesize{2}}$ formalism (e.g., whether the observed flux errors are included in the $\chi$$^\textrm{\footnotesize{2}}$ expression or not), the number of iterative processes,  and the extent of the examination grid. Therefore, the lack of high-temperature, high-metallicity stars in our sample  is likely due to the inconsistency between models and observed spectra, rather than issues in the model-fit pipeline.

Figure 26 presents the same HR diagrams  as  shown in the top panels of Figure 24, but for stars with T$_\textrm{\footnotesize{eff}}$$\leq$3550 K (that are believed to  have more accurate parameters),  for two metallicity ranges, i.e., [M/H]$\geq$$-$0.5 dex (top panels) and [M/H]<$-$0.5 dex  (bottom panels), and for two gravity approaches, i.e.,  when surface gravity is a fixed parameter (left panels) and when surface gravity is a free parameter (right panels). All these plots are color-mapped based on the combined parameter [M/H]+[$\alpha$/Fe]. As can be seen, the distribution of  [M/H]+[$\alpha$/Fe] values is well stratified in both near-solar metallicity and low-metallicity regimes, which again indicates a high precision in   the best-fit values of  chemical-parameters, no matter which surface gravity approach is used.

\subsection{Constant versus Variable Gravity}
In this section we examine how the variation of surface gravity  in the model-fitting affects the values of the other parameters. As explained  in Section 6.2, the inferred parameter values of high-temperature stars (T$_\textrm{\footnotesize{eff}}$>3550 K), whose best-fit models are closer to spectra of late-K dwarfs than they are to M dwarfs (i.e., they have relatively  weak molecular bands), are most likely to suffer from large systematic uncertainties as a result. Because of this, we exclude those stars from our present analysis.

 Figure 27 compares the best-fit values of T$_\textrm{\footnotesize{eff}}$, [M/H], and [$\alpha$/Fe]  obtained from the two gravity-modeling  approaches, as specified by ``const-grav'' and ``var-grav'', for the 2829 stars with T$_\textrm{\footnotesize{eff}}$ $\leq$ 3550. To ease the interpretation, the best-fit values for each parameter are randomized within the corresponding step size of the model grid (otherwise they would fall along the evenly spaced grid points). The  color map in the top panels is based on log \emph{g} (const-grav) and in the middle panels is based on log \emph{g} (var-grav).  The bottom panels show the histograms of the differences between the values of the corresponding parameters obtained from the two gravity-modeling approaches.  Although there is a significant change  in the best-fit values of  log \emph{g} when being treated as a free parameter  relative to those constant values (i.e., photometric surface gravities), there appears to be only a small (downward) shift in the distribution of temperatures with respect to the line 1:1, as shown in the left panels ``a'' or ``d''.  The histogram plotted in the bottom left panel ``g'' is also slightly off relative to the zero value towards higher values by $\sim$10-20 K. This is the reason why there is no significant difference in the variation of T$_\textrm{\footnotesize{eff}}$ in the HR diagram from the two gravity-modeling approaches (Figures 22 and 23), which is consistent with the weak degeneracy effect between effective temperature and the other parameters (Section 3.4).

 Clearly, the variation of surface gravity has a  more important effect on the best-fit values of metallicity,  as illustrated in the middle panels of Figure 27.  As shown in Section 3.4, the simultaneous variation of log \emph{g} and [M/H] can make a substantial degeneracy effect in the inferred model-fit  values, particularly for near-solar metallicity spectra. Accordingly,  once surface gravity becomes a free parameter,  the  surface gravity and metallicity  estimates  of a significant number of stars with near-solar metallicities shift to higher values, while  there is no noticeable change   in these two parameters for metal-poor stars.  This can clearly be seen from the middle panels (``b'' and ``e'') that show a relatively large fraction of stars which  are shifted  below  the 1:1 line (black line). The histogram in the panel ``h'' also shows an overall shift of the distribution towards higher values by $\sim$0.12-0.13 dex relative to the values obtained from the fixed-gravity model-fit. During the model-fit process, the values of log \emph{g} and [M/H] competitively increase in each iteration, and for some stars, may even drift to the upper limits of the  model grid  (i.e., log \emph{g}=5.5 dex and/or  [M/H]=+0.5 dex), which are unlikely to represent realistic values of these parameters for the stars in our sample. There is also an overall  increase in [$\alpha$/Fe], mainly  associated with near-solar metallicity stars,  which is also owing to the degeneracy effect between log \emph{g} and [$\alpha$/Fe] (Section 3.4).  This can be observed in  the  right panels  ``c'' or ``f'', that show how the overall distribution of stars  are systematically off below the 1:1 line,  mostly for [$\alpha$/Fe]$\lesssim$+0.4 dex. This effect  yields  a shift in the overall histogram distribution shown in  the panel ``i'' towards higher values by $\sim$0.03-0.04 dex relative to values obtained with the fixed-gravity method.

 The end effect is quite dramatic: in Figure 28, we show the metallicity distribution of stars with [M/H]$\geq$$-$0.5 dex  and T$_\textrm{\footnotesize{eff}}$$\leq$3550 K when surface gravity is kept constant  (2714 stars, top panel) and when surface gravity is allowed to vary (2701 stars, bottom panel). There is a clear shift of metallicities towards higher values when  surface gravity becomes a free parameter in the model fit to synthetic spectra. The constant gravity approach results in  just six stars with metallicities at  the upper limit of the model grid, i.e., [M/H]=+0.5 dex, and  98 stars  with +0.3$\leq$[M/H]<+0.5 dex. On the other hand,  the varying gravity approach yields 181 stars having metallicities at  [M/H]=+0.5 dex and 582 stars with +0.3$\leq$[M/H]<+0.5 dex. The large fraction of stars with  high metallicity values  ([M/H]$\gtrsim$+0.3 dex) in our volume-limited  sample derived from the latter method  is not consistent with the metallicity distribution of more massive (FGK) disk stars  in the solar neighborhood. On the other hand, the metallicity distribution obtained from the constant gravity method  is similar to those of previous studies for nearby dwarf stars (e.g., Rocha-Pinto \& Maciel 1998; Haywood 2001; Nordstr\"om et al. 2004, hereafter N04; Du et al. 2004; Hejazi et al. 2015; An 2019). This implies that the high-metallicity fits  from the varying gravity approach  most likely suffer from large systematic errors and  do not seem to represent the realistic measurements of metal content.

In Figure 29, we examine the distribution of our stars in the [$\alpha$/Fe]  versus [M/H] diagram, color-mapped by metallicity class (top panels), effective temperature (middle panels), and surface gravity (bottom panels), for the two  gravity-modeling approaches, i.e., when log \emph{g} is kept fixed (2868 stars, left panels) and when gravity is free to change (2668 stars, right panels).  This time, we also exclude all stars with best-fits values of [M/H]=+0.5 dex,  because these stars have best-fit values in one parameter that are at the edge of the model grid and we assume that all best-fit parameters (notably gravity) may be biased one way or another. The values of [M/H] and [$\alpha$/Fe] for the remaining stars are randomized within the associated step sizes of the model grid, again for ease of interpretation. In order to better present the detailed distribution of the near-solar metallicity (disk) stars,  Figure 30 shows the same panels as those in Figure 29, but zoomed-in over the metallicity range $-$0.5$\leq$[M/H]<$+$0.5 dex.

There is a global trend in the distribution of stars for both gravity-modeling methods: the higher-metallicity stars, on average, have lower values of $\alpha$-element enhancement, while  lower-metallicity stars, on average, tend to have higher values of $\alpha$-element enhancement, which is consistent with current models of  the Milky Way's enrichment history  (e.g., Croswell 1995; Pagel 1997; Chiappini et al. 1999, 2001; Chiappini 2001). The time scales of chemical enrichment, particularly by type II supernovae (SNe II) and type Ia supernovae (SNe Ia),  are critical to establish the evolutionary models for elemental abundance ratios such as [$\alpha$/Fe];  $\alpha$-elements are mostly produced by SNe II over time scales of 3 to 30 Myr, while iron is primarily  generated by SNe Ia  over larger time-scales ranging from 30 Myr to the Hubble time (Pipino \& Matteucci 2009). During  the chemical  enrichment  of  the interstellar medium (ISM)   by core-collapse  SNe II,  heavy elements are  highly  rich in  $\alpha$-elements. However, after the ``time delay''  ($\sim$ a few billion years, e.g., Heringer et al. 2019), when white dwarfs in binary systems   grow up to  the Chandrasekhar mass by accretion from their companions, which leads to SNe Ia explosions,  the ejected materials start to enrich the ISM with iron, and accordingly  [$\alpha$/Fe]  begins to drop.  This figure  also shows how  stars get redistributed when log \emph{g} becomes a free parameter, shifting a considerable number of  stars towards  higher values of [M/H], log \emph{g} and/or higher values of  [$\alpha$/Fe].

There is a very clear correlation between the location of a star in this diagram and the star's metallicity class (MC) estimated from spectral classification, as seen from the color-coding on the top panels. The trend is nearly  independent of the gravity-modeling approach used in the synthetic fitting. Regions with similar values of MC are tilted from higher-[M/H] and lower-[$\alpha$/Fe] to lower-[M/H] and higher-[$\alpha$/Fe] values, which indicates that the metallicity class depends on  both metallicity and $\alpha$-element enhancement. The spectral metallicity classification based on the TiO and CaH molecular bands (Paper I) is thus determined by both chemical parameters, and is not merely a measure  of ``metallicity'' in the general sense (i.e., of [M/H]). We find that the distribution  of stars with nearly the same values of log \emph{g}, for both gravity-modeling approaches, are also tilted in nearly the same direction as the distribution of stars with similar MC values, which shows a tight correlation between [M/H] and [$\alpha$/Fe] (Sections 3 and 5).

More perplexing is the presence of a systematic trend between T$_\textrm{\footnotesize{eff}}$ and [$\alpha$/Fe], primarily for stars with [M/H]$\geq$$-$0.5 dex, that occurs while using both gravity-modeling  approaches, as seen from the middle panels in Figures 29 and 30. The higher-temperature stars tend to  have lower values of [$\alpha$/Fe] and lower-temperature stars have higher values of  [$\alpha$/Fe]. This correlation cannot have any physical basis because all stars are drawn from the same local population, and it is not possible that their chemical abundances be somehow correlated with their effective temperature. This resulting trend is independent of the initial values, $\chi$$^\textrm{\footnotesize{2}}$ expression (i.e., whether weighted with flux errors or not), and the extent of the searching grid in the model-fit pipeline. The range of values over which there is a potential degeneracy in the synthetic spectra while T$_\textrm{\footnotesize{eff}}$  and [$\alpha$/Fe] are varied (see Section 3) are not sufficiently extensive to cause a noticeable systematic trend in the diagram. Moreover, even if these ranges were large enough, the resulting trend would be in the opposite direction, i.e., higher-temperature stars would shift toward higher values of [$\alpha$/Fe], and inversely,  lower-temperature stars would shift towards lower values of  [$\alpha$/Fe]. Since  this trend is independent on the gravity-modeling approach, the correlation between surface gravity and   either T$_\textrm{\footnotesize{eff}}$ or [$\alpha$/Fe] does not have an important  role in the correlation between these two parameters. As a result, we believe that this trend is more likely due to some issues in the model atmospheres themselves (Section 3.4).

The bottom panels in Figures 29 and 30 show the same distribution as the upper panels, but now color-mapped according to the best-fit surface gravity of stars, for both gravity-modeling approaches. There is a clear trend between surface gravity and $\alpha$-element enhancement using the constant-gravity approach  (panel ``e'') and between surface gravity and metallicity using the variable-gravity approach (panel ``f''), in the  near-solar metallicity regime. The former trend can be attributed to the correlation between surface gravity, that are equal to the photometric values,  and effective temperature, as shown in Figure 31.   The surface gravity values seem to follow quite the same trend in [$\alpha$/Fe] as effective temperatures do (panel ``c''), which   indicates the tight correlation between   log \emph{g} and T$_\textrm{\footnotesize{eff}}$;  stars with   lower values of  log \emph{g} have higher values of T$_\textrm{\footnotesize{eff}}$  (as predicted by theoretical isochrones, Figures 11 and 31), and accordingly,  lower values of [$\alpha$/Fe].  However, the latter trend is  most likely due to systematic errors caused by the strong degeneracy  between the two parameters log \emph{g} and [M/H]. When gravity is allowed to vary during model-fitting, a significant number of stars  shift towards high values of [M/H]  and  log \emph{g}. In addition,  some metal-rich  stars move toward the region with higher values of [$\alpha$/Fe], which is mostly due to  the degeneracy between log \emph{g} and [$\alpha$/Fe]. This shift is clearly inconsistent with previous analyses of FGK-dwarf distributions  in the [$\alpha$/Fe]  versus [M/H] diagram (e.g., Lee et al. 2011b; Adibekyan et al. 2012, 2013; Recio-Blanco et al. 2014); metal-rich dwarfs, that belong to the Galactic disk and are among the young stellar population, are expected to have low values of  [$\alpha$/Fe], as compared to the old halo stars that  are believed to be rich in $\alpha$ elements. 

We find that our model-fit values from the pipeline in which surface gravity is kept fixed  are generally  in  good agreement with those of previous analyses. More importantly, the  distribution of  the chemical parameter estimates  inferred from the constant-gravity approach in the abundance diagram of [$\alpha$/Fe]  versus [M/H]  is  comparable with that from  the latest (third) release of the Galactic Archaeology with HERMES (GALAH) survey  (Buder et al.  2021, Figure 5). The GALAH DR3 provides accurate atmospheric parameters and individual chemical abundances of 30 elements (11 of which are based on NLTE computations) from high-resolution optical spectroscopy for  mostly nearby  $\sim$500000 stars\footnote{The sample includes 65{\%} dwarfs, 34{\%} giants, and 1{\%} as unclassified objects.} in the Galactic disk. This indicates that our best-fit  estimates  are sufficiently precise to show trends compatible with those from other studies using much more extensive samples.

 As shown in the top panel of Figure 31, the parameter values derived from the constant-gravity approach can show a clear contrast in log \emph{g} between metal-rich and metal-poor stars.  The metal-poor stars have larger values of log \emph{g}, as compared to the metal-rich stars with the same T$_\textrm{\footnotesize{eff}}$. For a given effective temperature, low-metallicity  M subdwarfs have been shown to be as much as five times smaller than their solar-metallicity counterparts (Kesseli et al. 2019). Consequently,  the surface gravity of these metal-poor dwarfs is expected to be larger than that of metal-rich stars, for a specific  temperature. However, we find neither  a trend in log \emph{g} between metal-poor and metal-rich stars nor a relationship between log \emph{g} and T$_\textrm{\footnotesize{eff}}$ using parameter values obtained from the variable-gravity approach, as these values have large systematic uncertainties. The bottom panel of Figure 31 shows the same plot as presented in the top panel, but for near-solar metallicity stars ([M/H]$\geq$$-$0.5 dex) with a more detailed metallicity stratification. Overplotted are the solar-metallicity isochrones associated with a stellar age of 10 Gyr  and three different values of [$\alpha$/Fe]=+0.0, +0.2, and +0.4 dex from the Dartmouth stellar evolutionary models. The isochrones with higher values of [$\alpha$/Fe], i.e., +0.2 and +0.4  seem to be in better agreement with the model-fit metallicities, as compared to the isochrone with [$\alpha$/Fe]=+0.0 dex\footnote{Since the number of metal-poor stars in our sample is not statistically large, the comparison between the isochrones  and the distribution of stars with specific metallicity values in low-metallicity regime is not appropriate.}. However, since the photometric surface gravities, model-fit parameter values, and isochrones may not be completely accurate, no clear conclusion can be made for the location of the isochrones in the diagram. Nevertheless, the values of T$_\textrm{\footnotesize{eff}}$  show a decreasing trend with increasing log \emph{g}, which  is consistent with that of the  theoretical  isochrones, regardless of their location relative to the stars' distribution.

Figure 32 compares HR diagrams of the stars in our subset, color-mapped according to the values of surface gravities derived from the constant-gravity approach (or photometric values, 2877 stars, top panel) and from the variable-gravity approach (2851 stars, bottom panel). The variation of photometric surface gravities in the color-absolute magnitude diagram is consistent with the relationship between these values and effective temperatures, as predicted by theoretical evolutionary models (Figures 11 and 31). Furthermore,  metal-poor M subdwarfs tend to have higher values of  surface gravity, which is expected from their smaller size, as compared to their metal-rich counterparts. However, the values of log \emph{g} obtained from the variable-gravity method, that are affected by large systematic uncertainties, show no acceptable trend in  log \emph{g}.

\section{Results: Reduced-Correlation Method}
To see whether a different fitting method may change the results, we apply the ``reduced-correlation method'' to the 3745 M dwarfs/subdwarfs  again for two gravity-modeling approaches, i.e., when surface gravity is kept constant and when surface gravity is allowed to vary. We analyze  the distribution of the  best-fit estimates using  some key diagrams and investigate   the influence of surface gravity on the values of the  other parameters, similar to our examination of resulting values from  the ``normal method'' (Section 6). We also compare the  results  from the two methods and discuss how the reducing of the residual correlation between wavelength datapoints can affect the derived parameter values.

\subsection{Effective Temperature Distribution}

Figure 33 depicts the same comparison as shown in Figure 19, but using the inferred temperature estimates from the ``reduced-correlation method''.  The scatter  around the 1:1 lines (black) shown in each panel is very similar to the one presented in the corresponding panel in Figure 19. This indicates that the ``reduced-correlation method'' does not make  a major difference  in the best-fit temperature values relative to those obtained from the ``normal method''.

\subsection{Gaia HR Diagram}

The top  and the bottom panels of Figure 34 are the same as the top panels of Figures 22 and 23, respectively, but color-mapped using the temperature estimates from the ``reduced-correlation method''. Similarly, the top  and the bottom panels of Figure 35 are the same as the top panels of Figures 24 and 25, respectively, but color-coded based on  the metallicity estimates from the ``reduced-correlation method''. The RUWE values are not taken into consideration in Figures 34 and 35. Despite some slight differences, the variation of temperatures and metallicities in these color-absolute magnitude diagrams are much the same as what can be seen from the corresponding plots in the top panels of  Figures 22-25. The stratification in T$_\textrm{\footnotesize{eff}}$ and [M/H] is still prominent, which again suggests that our model-fit method is relatively reliable. However, there  seems to be again  the same heavy concentration of high-temperature  stars (T$_\textrm{\footnotesize{eff}}$$\gtrsim$3600 K) in the narrow metallicity range $-$0.5$\lesssim$[M/H]$\lesssim$+0.1 dex, which is likely  owing to issues in the models of M dwarf atmospheres.

We do not include  any further plots that are  equivalent to those of Figures 26-32 when using the  best-fit values derived from the ``reduced-correlation method''  because such plots are all remarkably similar to the corresponding ones using the  best-fit values inferred from the ``normal method'', and  consequently,  would provide no more information about the  distributions of resulting parameter values.

\subsection{Normal versus Reduced-Correlation Method}

Figure 36 compares the best-fit estimates of T$_\textrm{\footnotesize{eff}}$, [M/H], and [$\alpha$/Fe] obtained from the normal and ``reduced-correlation method''  for stars with T$_\textrm{\footnotesize{eff}}$$\leq$3550 K using the constant gravity approach (top panels).  All parameter values are randomized within the corresponding bin size of our model grid. These plots are color-mapped based on the constant (photometric) surface  gravity values.  The bottom panels show the histograms of the differences between the values of respective parameters derived from the two methods. Regardless of very slight offsets, the scatter of stars around the 1:1 lines is almost symmetric. The histogram panels indicate nearly normal distributions with an average around zero value.  We calculate the standard deviation of these differences: $\sim$44 K  in T$_\textrm{\footnotesize{eff}}$,  0.08 dex in [M/H], and 0.04 dex in [$\alpha$/Fe].

Figure 37 shows the same comparison as presented in Figure 36, but using the variable gravity approach. The plots are color-mapped based on the surface gravity values inferred from the ``normal method''. We find a little larger shift from the 1:1 line for effective temperature (panel ``a'') and metallicity (panel ``b''), as compared to the equivalent panels in Figure 36. On the other hand, there is no noticeable offset relative to the 1:1 line for $\alpha$-element enhancement (panel ``c''). These are also clear from the histograms in the bottom panels; the distributions show a $\sim$20-30 K offset from the zero value in the panel ``d'' and a $\sim$0.02-0.03 dex  offset from the zero value in the panel ``e'' towards higher values, while there is no  shift in the panel ``f''. We determine the standard deviation of these differences around the average  values: $\sim$46 K  in T$_\textrm{\footnotesize{eff}}$, 0.1 dex in [M/H], and and 0.04 dex in [$\alpha$/Fe].

In general, there is a deviation within  $\sim$45 K in T$_\textrm{\footnotesize{eff}}$, within $\sim$0.1 dex in [M/H], and within $\sim$0.04 dex in [$\alpha$/Fe] between the values obtained from the normal and ``reduced-correlation method''. As a result, the ``reduced-correlation method'', which decreases the correlation of the residuals between spectral datapoints due to instrumental limitations, does not make a significant difference in the model-fit parameter values and their resulting distributions, which is likely due to the low spectral resolution of our spectra. The uncertainties caused by these correlations are thus unlikely  to be of importance in the fitting of synthetic spectra to our observed spectra (or in general, to low-resolution M dwarf spectra). This suggests that systematic errors in the model-fit method are unlikely to be due to the fitting technique itself, but more likely due to issues in the models and/or degeneracies in the synthetic spectra discussed in previous sections.

\begin{figure}\centering
\subfloat
        [Fixed surface gravity]{\includegraphics[  height=4.8cm, width=6.8cm]{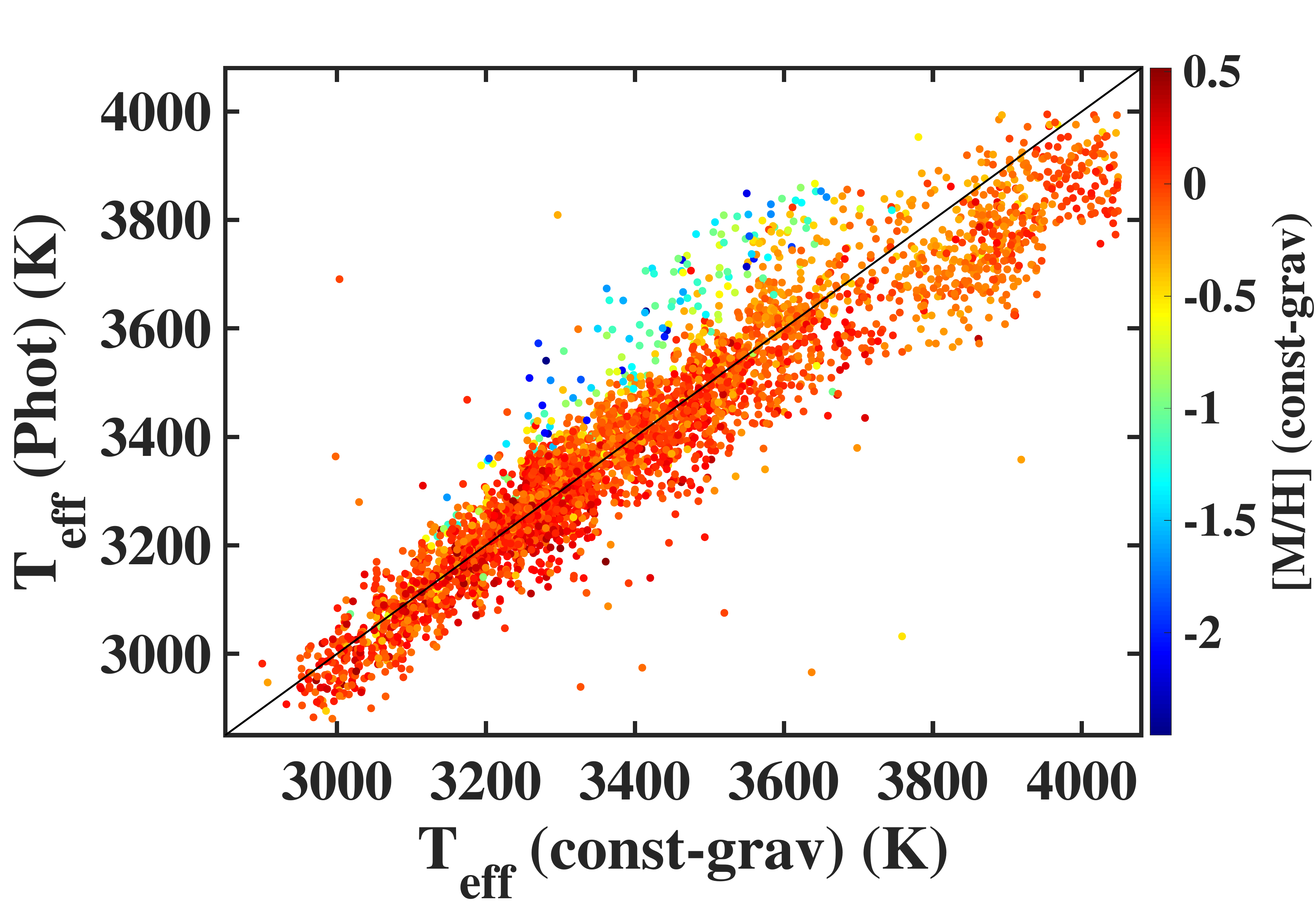}}
\vspace{-0.25cm}

\subfloat       
        [Variable surface gravity]{\includegraphics[ height=4.8cm, width=6.8cm]{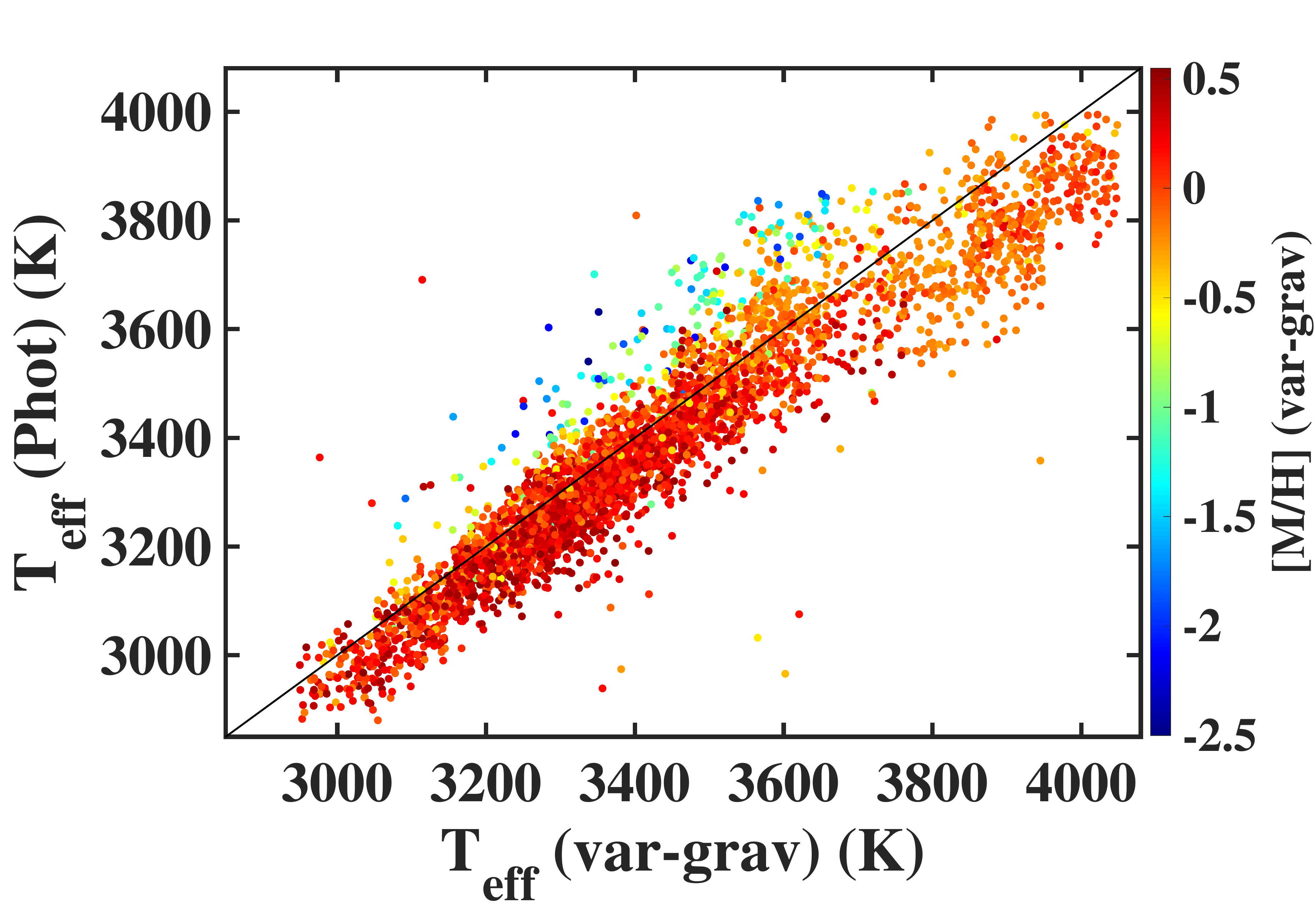}}
 \caption
        {\footnotesize{Comparison between the photometric temperatures and the inferred model-fit values for the 3745 stars, when surface gravity is considered as a fixed parameter  (top panel) and when surface gravity is considered as a free parameter (bottom panel), using the \textbf{reduced-correlation method}.}}
\end{figure}

\begin{figure*}\centering
\subfloat
       [Fixed surface gravity]{\includegraphics[ height=5.1cm, width=8.2cm]{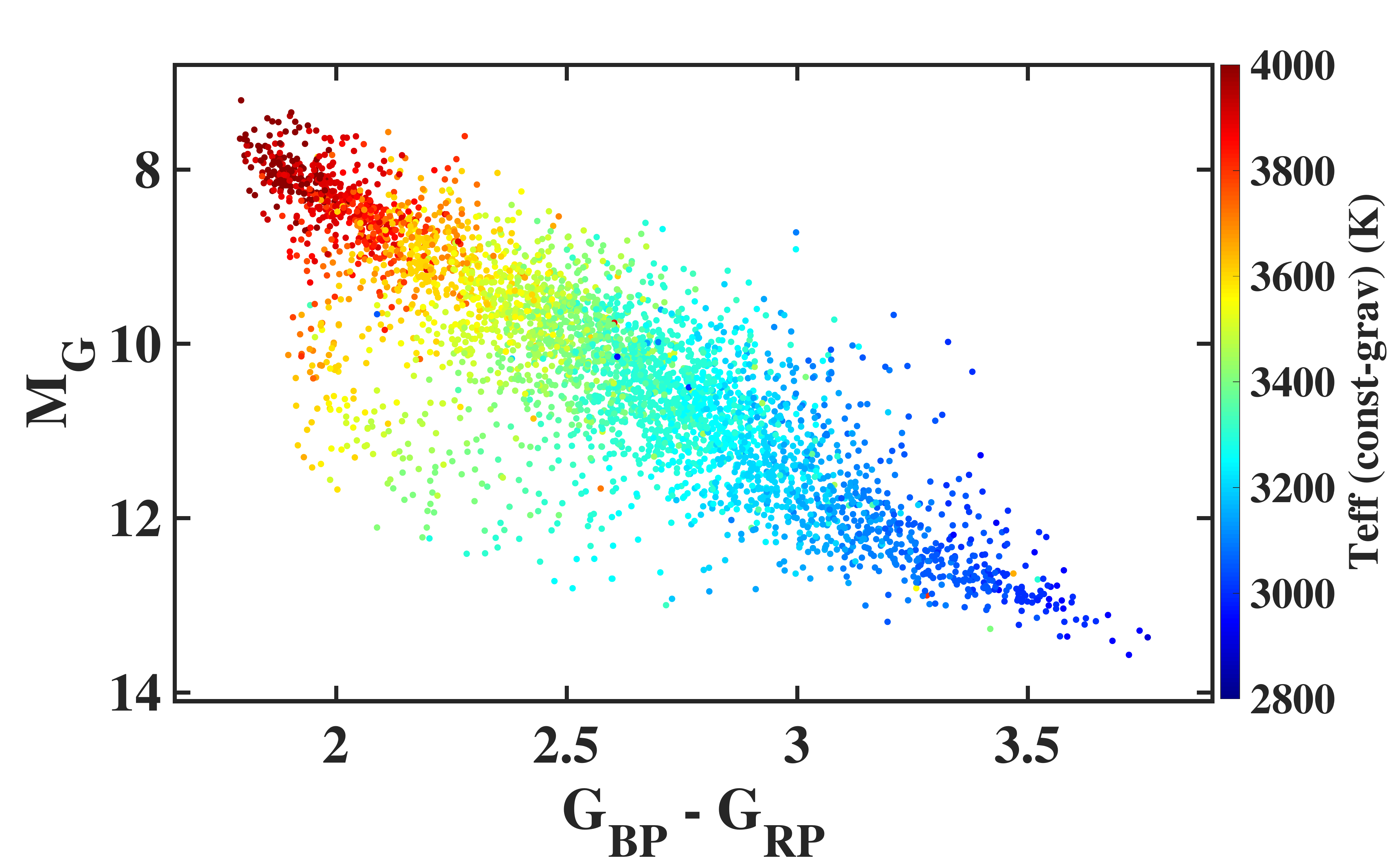}}
 \hspace{0.3cm} 
 \subfloat      
        [Variable surface gravity]{\includegraphics[ height=5.1cm, width=8.2cm]{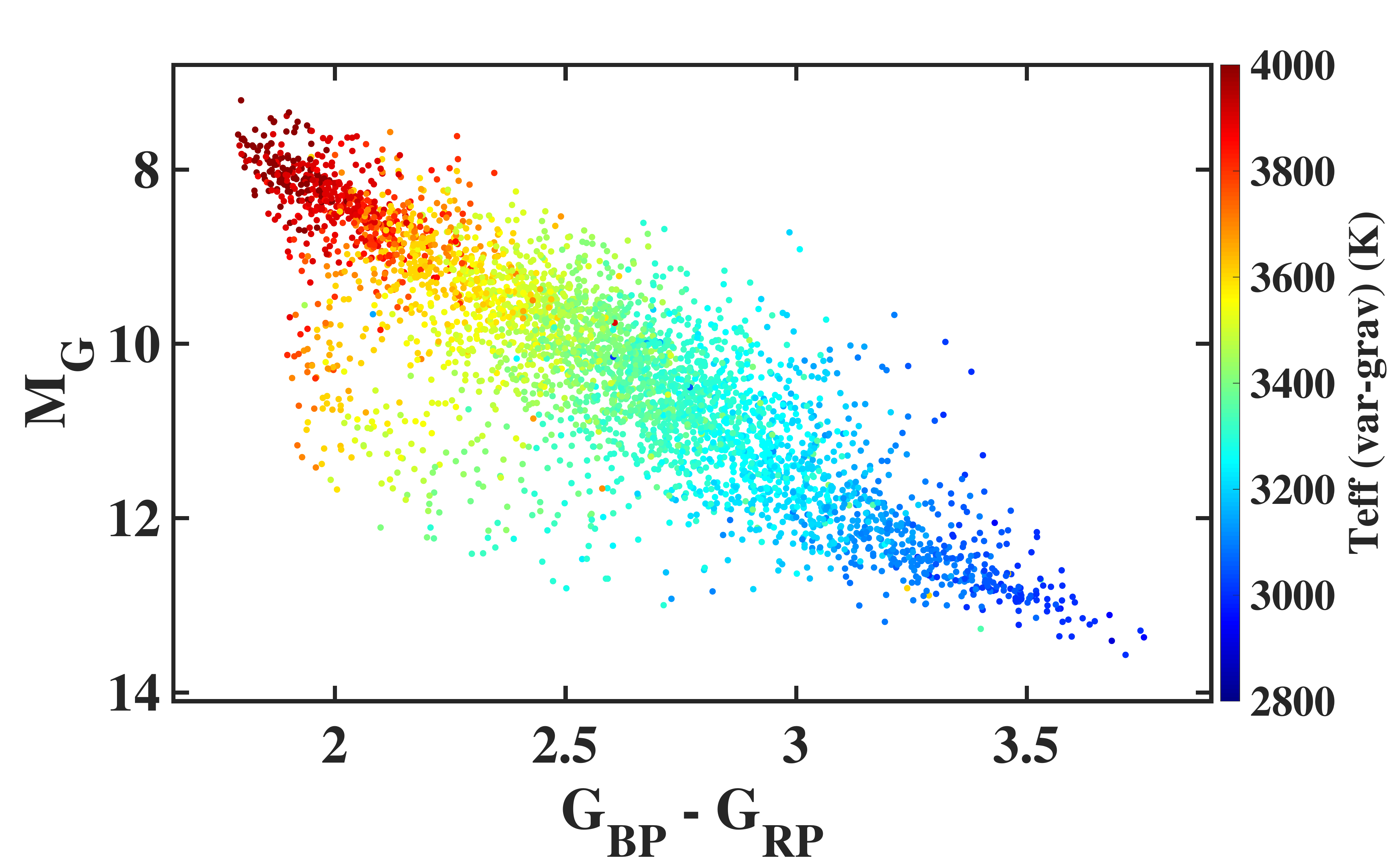}}
  \vspace{-0.4cm}

 \subfloat
         [Fixed surface gravity]{\includegraphics[ height=5.1cm, width=8.2cm]{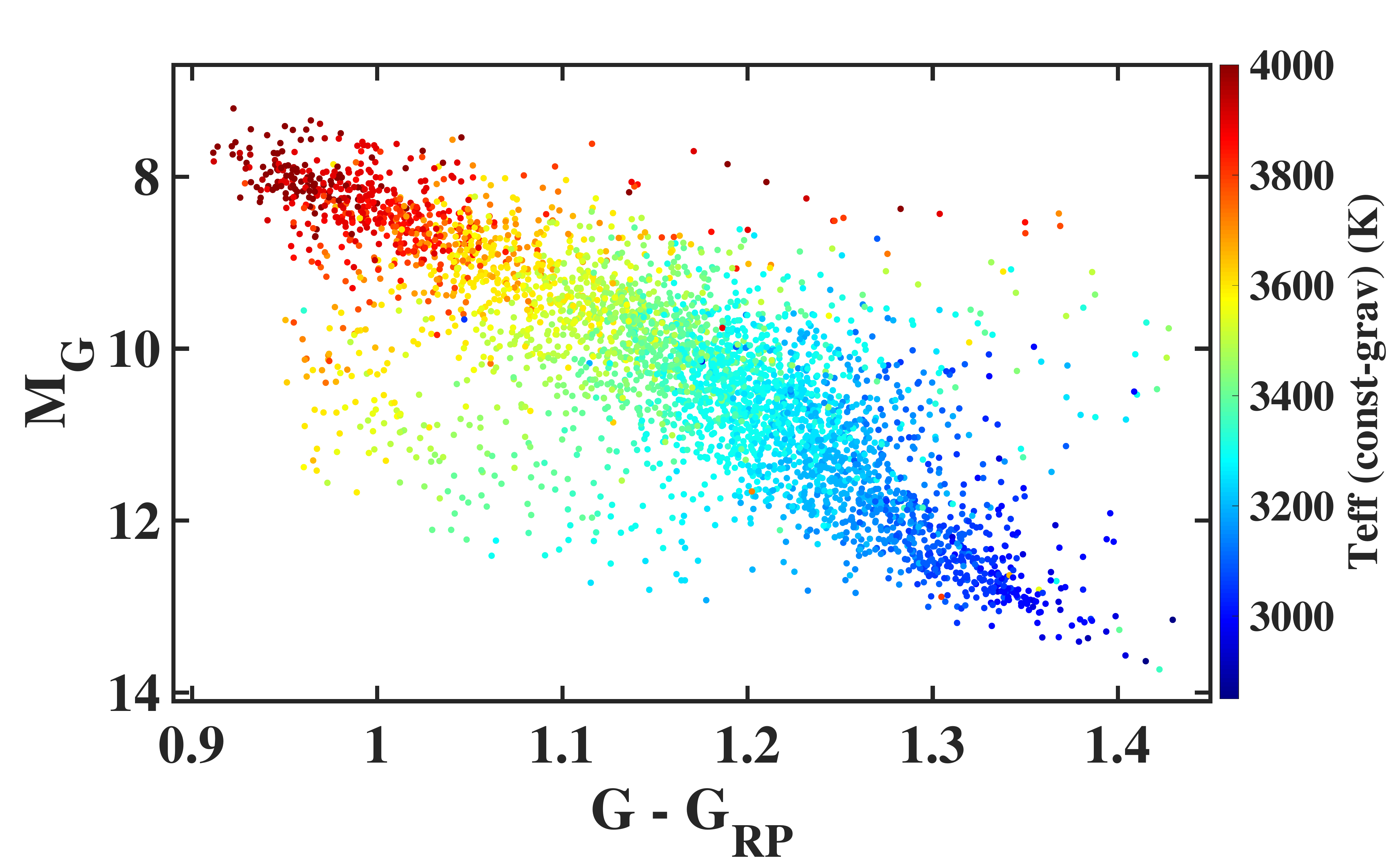}}   
 \hspace{0.3cm} 
 \subfloat 
         [Variable surface gravity]{\includegraphics[height=5.1cm, width=8.2cm]{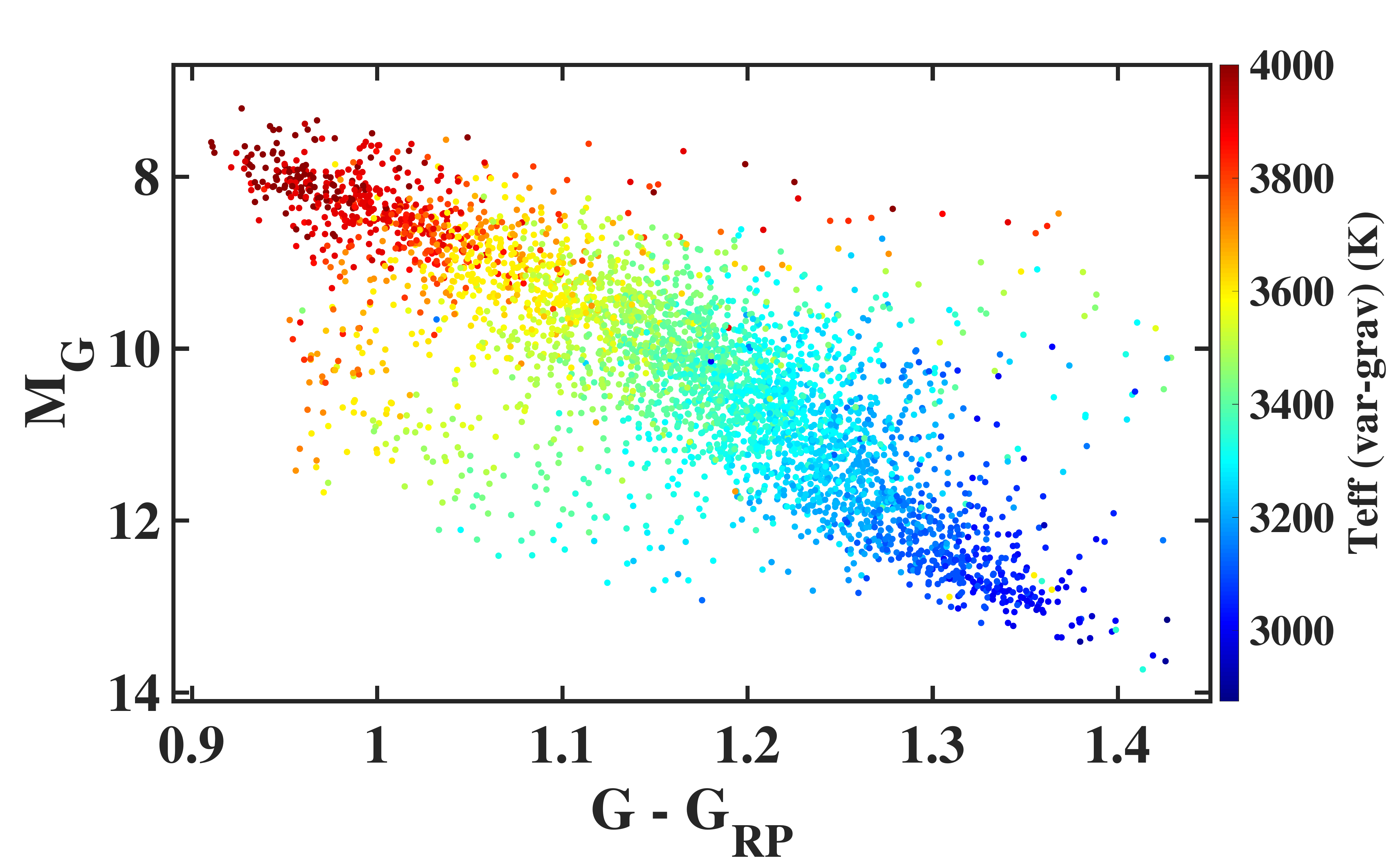}}  
\caption
 {\footnotesize{HR diagram,  \textit{M}$_\textrm{\footnotesize{G}}$ versus \textit{G}$_\textrm{\footnotesize{BP}}$ - \textit{G}$_\textrm{\footnotesize{RP}}$ (top panels) and  \textit{M}$_\textrm{\footnotesize{G}}$ versus \textit{G} - \textit{G}$_\textrm{\footnotesize{RP}}$ (bottom panels), of the 3745 stars,  when surface gravity is set a fixed parameter (left panels) and when surface gravity is set a free parameter (right panels), color-mapped based on  effective temperatures, using the \textbf{reduced-correlation method}. The values of RUWE are not taken into account in this figure.}} 
   \end{figure*}

\begin{figure*}\centering
\subfloat
       [Fixed surface gravity]{\includegraphics[  height=5.7cm, width=9.1cm]{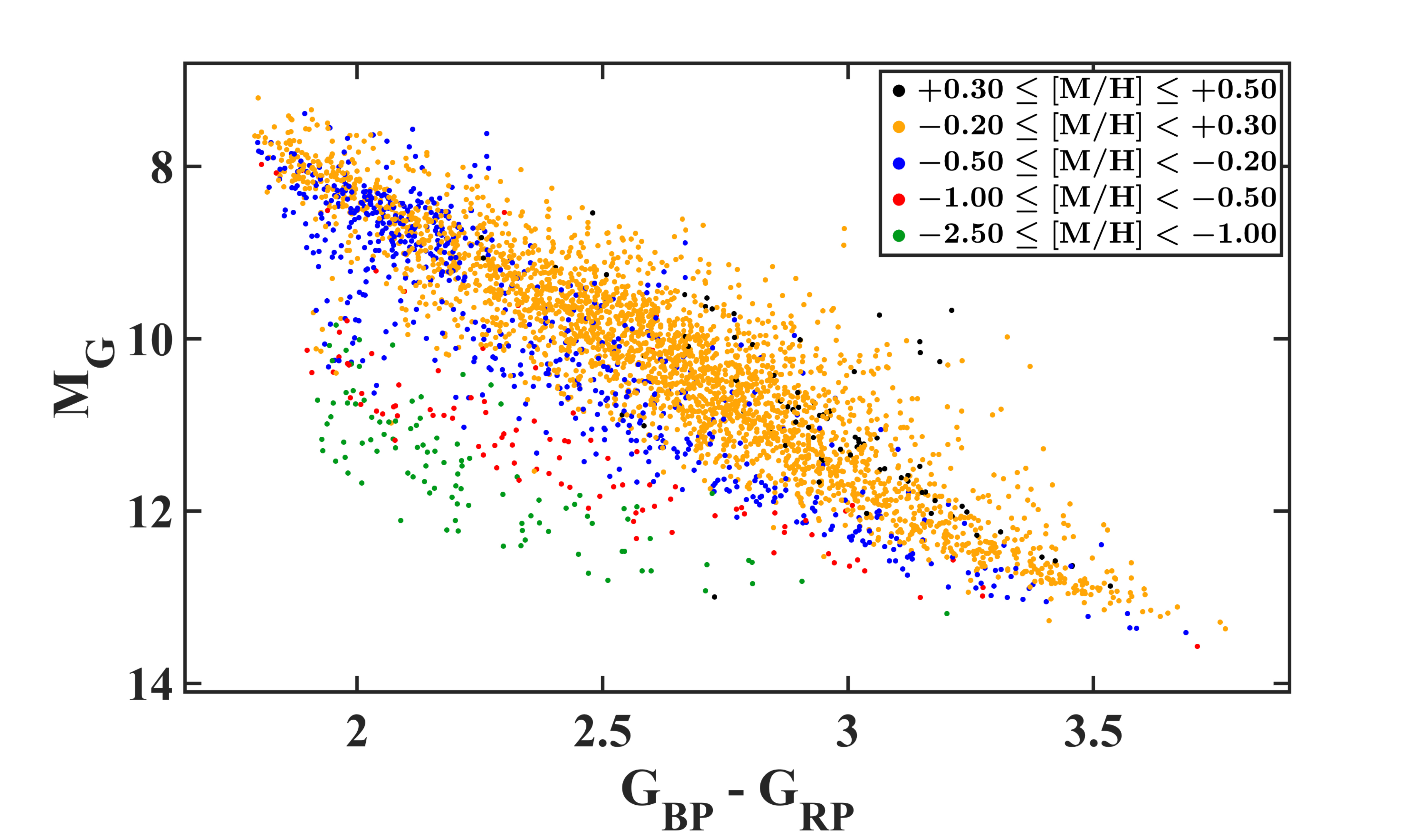}}
 \hspace{-0.35cm} 
 \subfloat      
        [Variable surface gravity]{\includegraphics[  height=5.7cm, width=9.1cm]{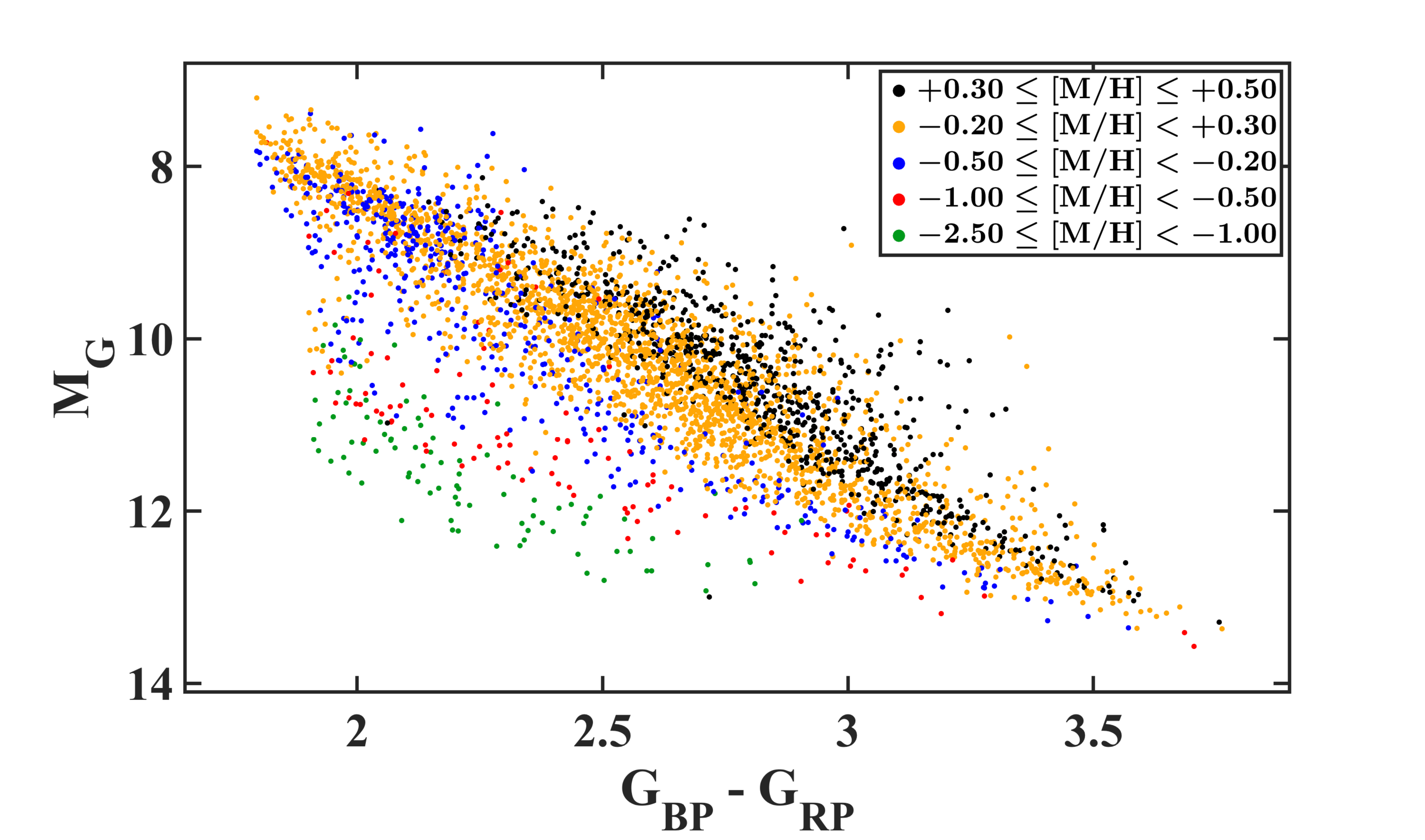}}
  \vspace{-0.35cm}

 \subfloat
         [Fixed surface gravity]{\includegraphics[  height=5.7cm, width=9.1cm]{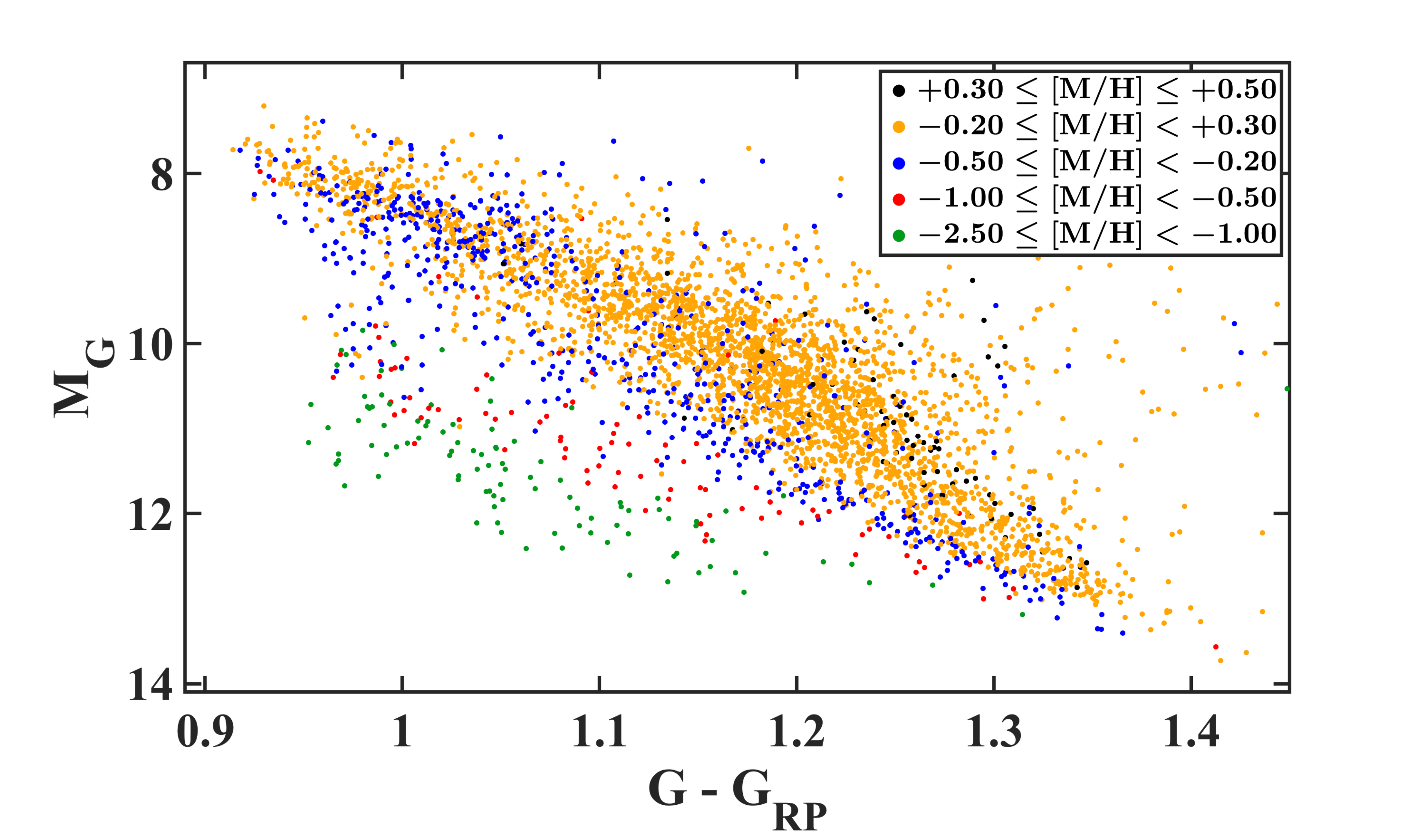}}   
 \hspace{-0.35cm} 
 \subfloat 
         [Variable surface gravity]{\includegraphics[ height=5.7cm, width=9.1cm]{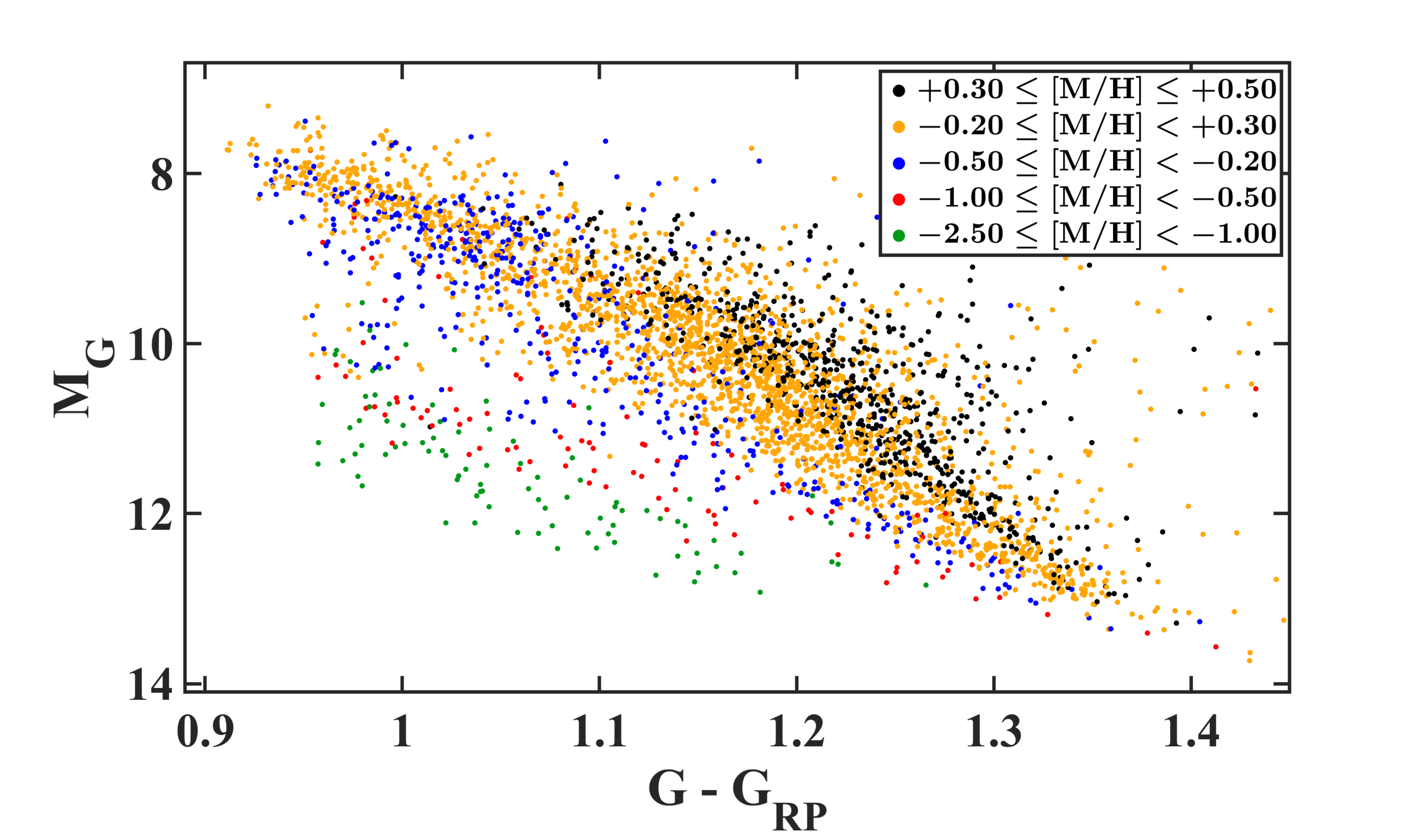}}  
\caption
        {\footnotesize{HR diagram,  \textit{M}$_\textrm{\footnotesize{G}}$ versus \textit{G}$_\textrm{\footnotesize{BP}}$ - \textit{G}$_\textrm{\footnotesize{RP}}$ (top panels) and  \textit{M}$_\textrm{\footnotesize{G}}$ versus \textit{G} - \textit{G}$_\textrm{\footnotesize{RP}}$ (bottom panels), of the 3745 stars,  when surface gravity is set a fixed parameter (left panels) and when surface gravity is set a free parameter (right panels), color-coded based on five groups with different  metallicity ranges, using the \textbf{reduced-correlation method}. The number of stars in each metallicity group, N, is \textbf{panels ``a'' and ``c'':}  N$_\textrm{{Black}}$=78,  N$_\textrm{{Yellow}}$=2822, N$_\textrm{{Blue}}$=644, N$_\textrm{{Red}}$=98, and N$_\textrm{{Green}}$=103, \textbf{panel ``b'' and ``d'':}  N$_\textrm{{Black}}$=757,  N$_\textrm{{Yellow}}$=2294, N$_\textrm{{Blue}}$=494, N$_\textrm{{Red}}$=108, and N$_\textrm{{Green}}$=92. The values of RUWE are not taken into account in this figure.} }
   \end{figure*}

\begin{figure*}\centering
\subfloat 
      []{\includegraphics[ height=4.1cm, width=5.65cm]{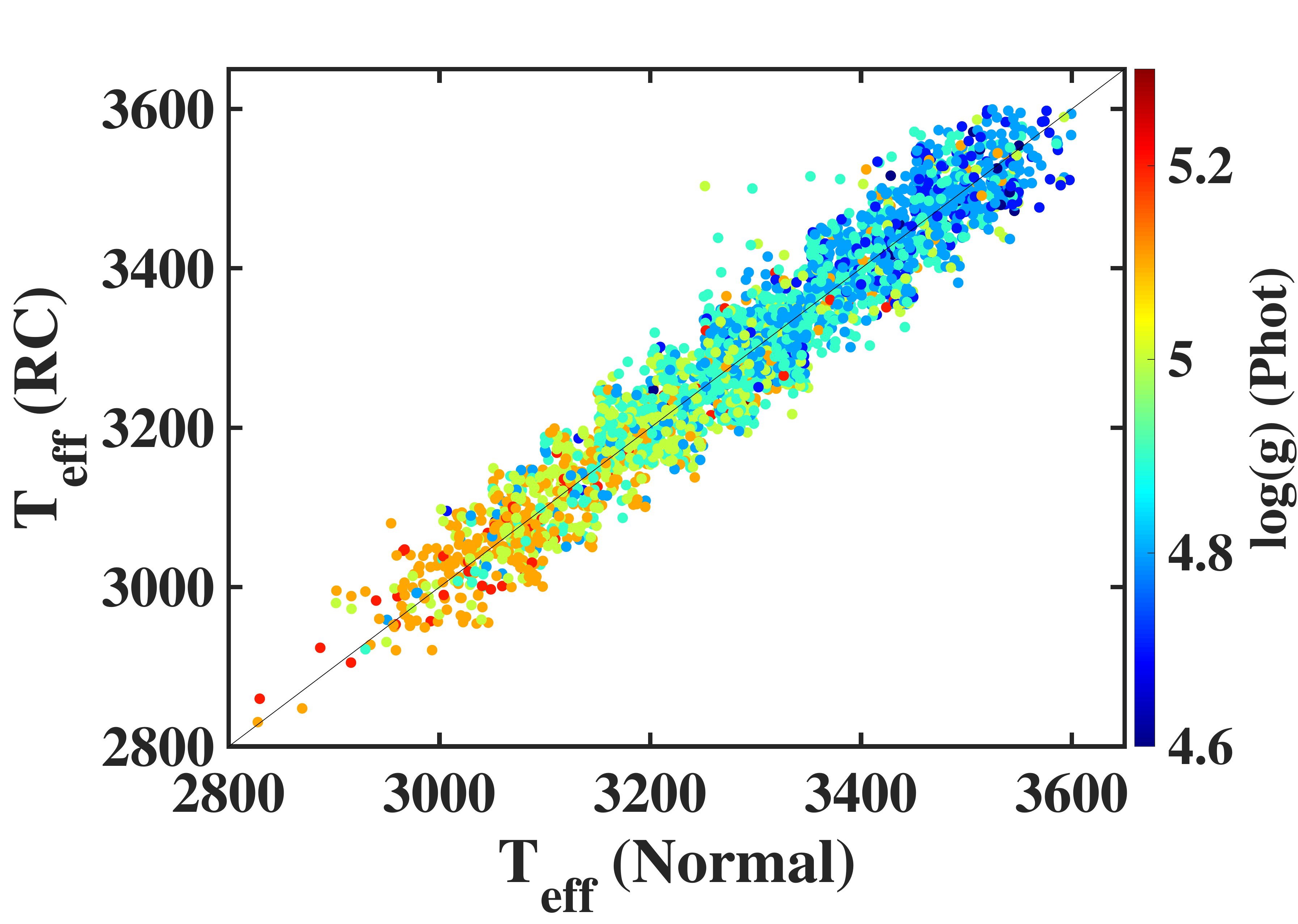}}               
\hspace{0.38cm}
\subfloat 
      []{\includegraphics[ height=4.1cm, width=5.65cm]{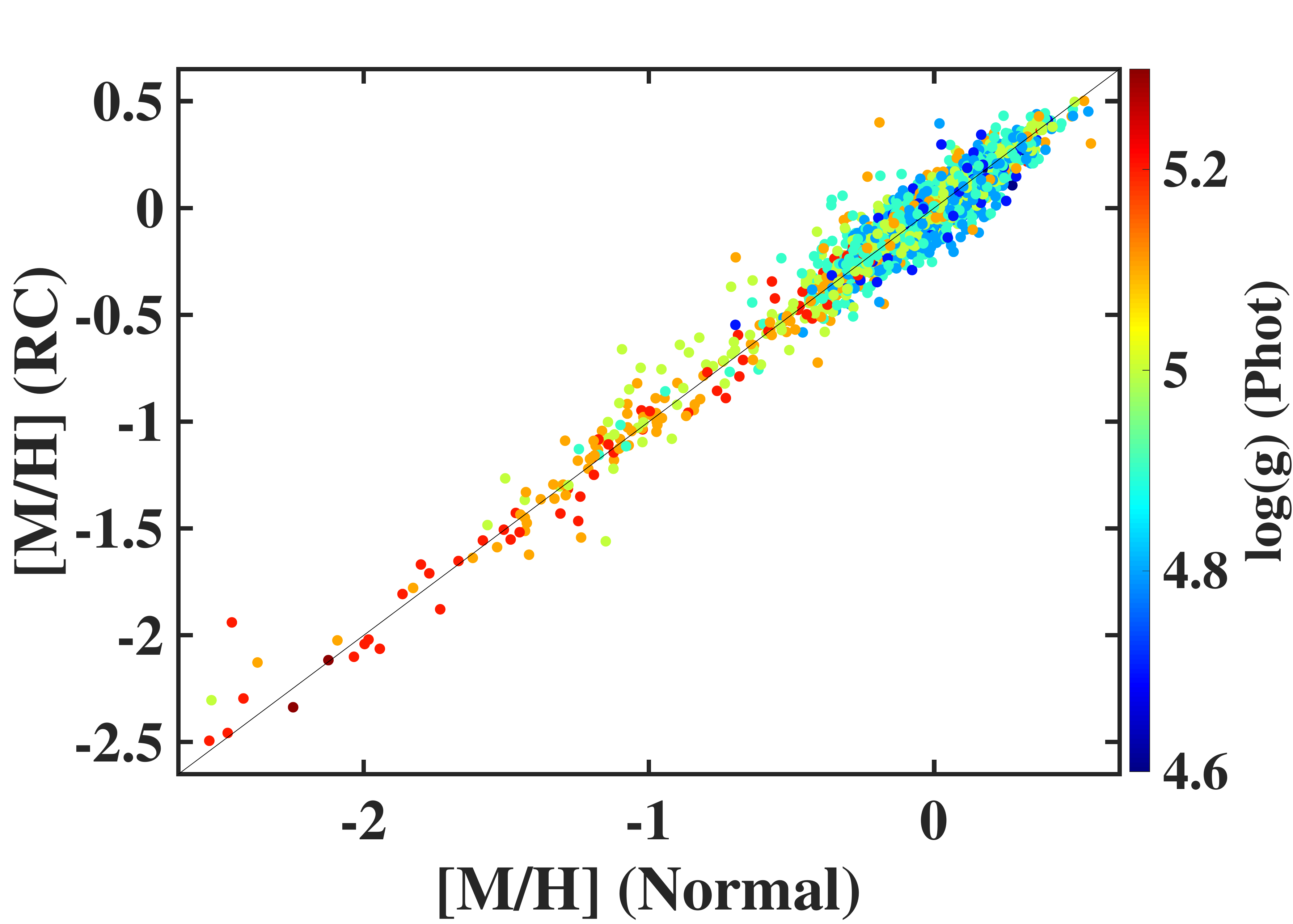}}               
\hspace{0.38cm} 
 \subfloat      
         []{\includegraphics[ height=4.1cm, width=5.65cm]{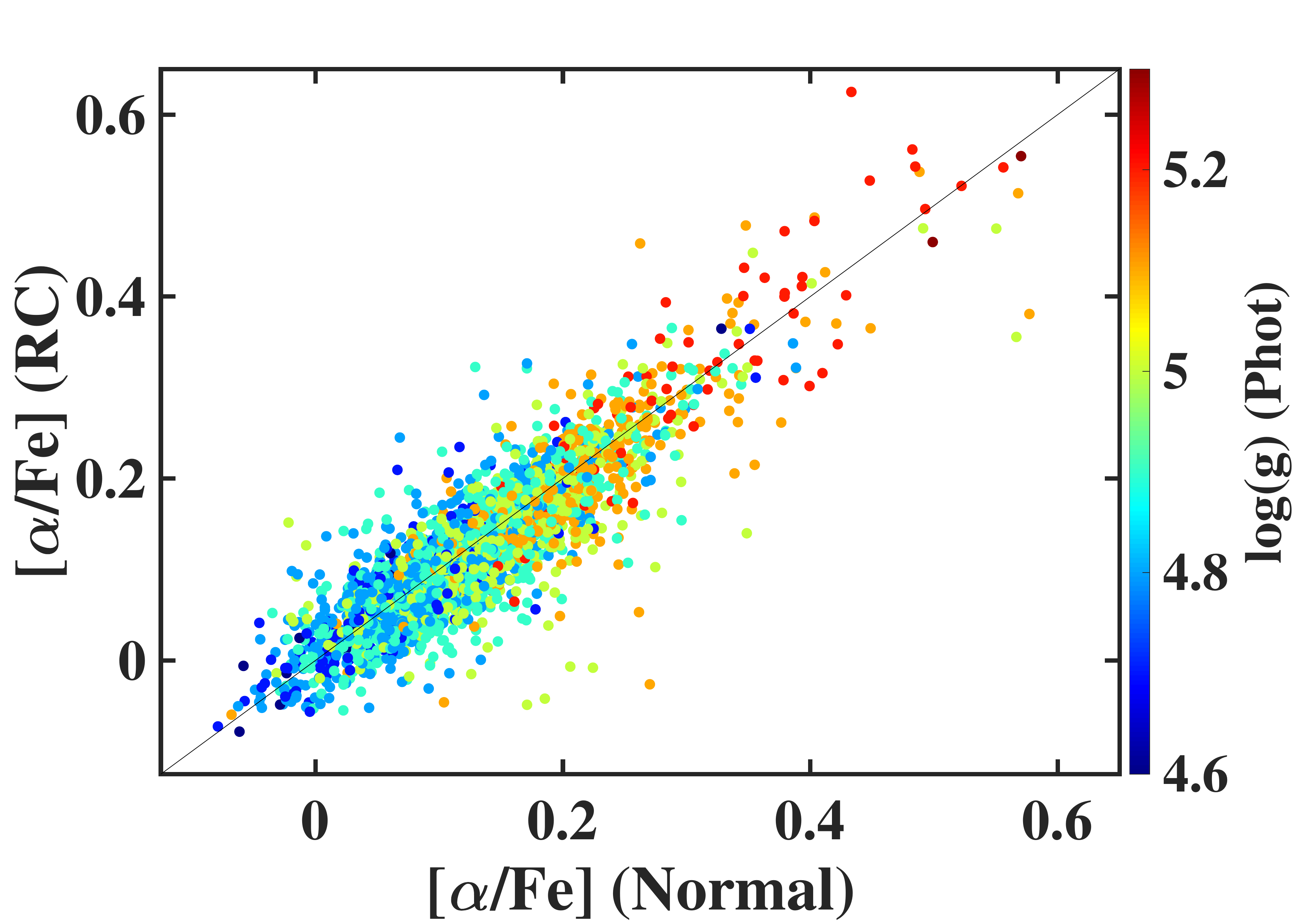}}
 \vspace{-0.25cm}

 \subfloat 
      []{\includegraphics[ height=4cm, width=5.4cm]{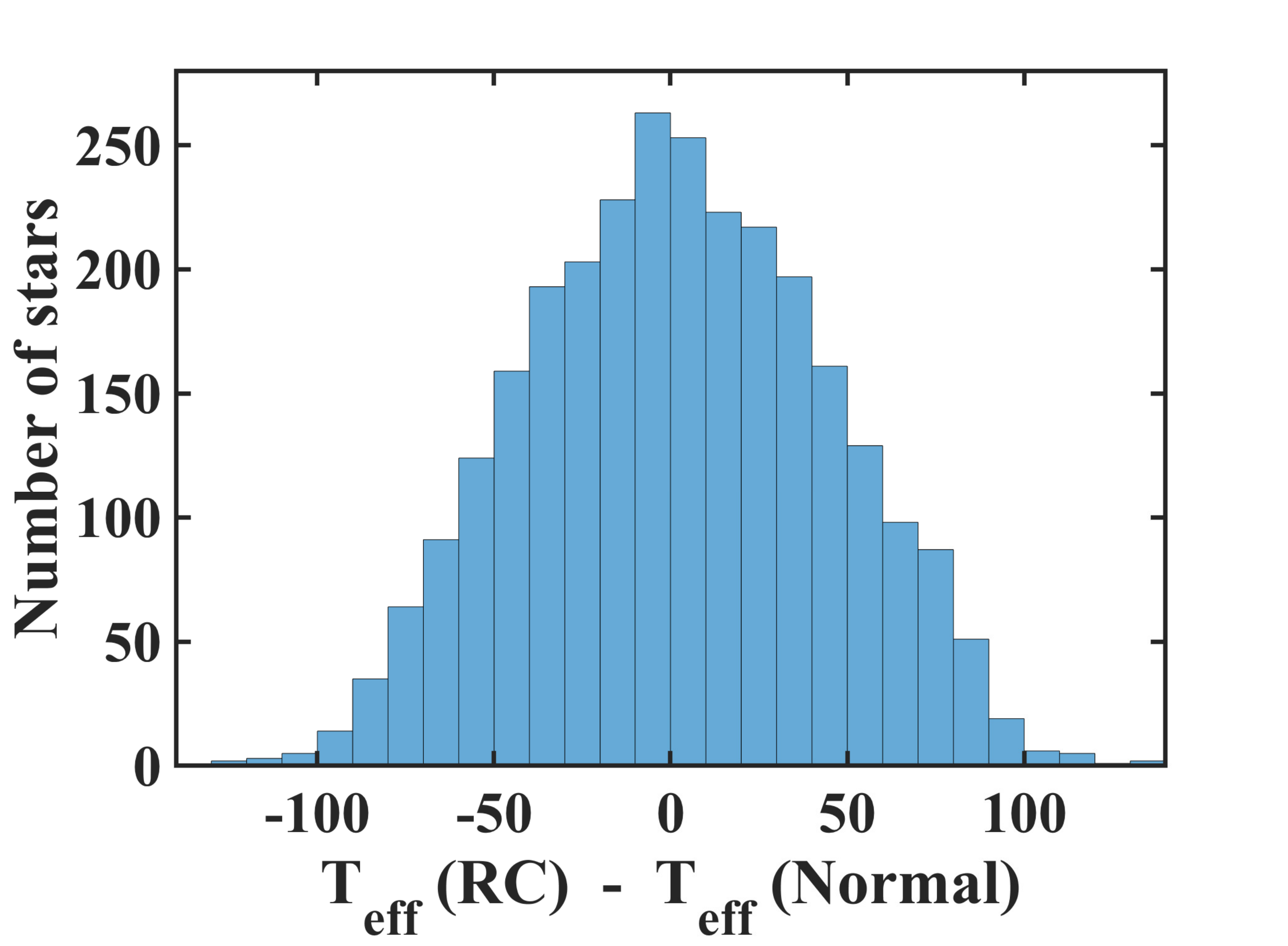}}               
\hspace{0.38cm}
\subfloat 
      []{\includegraphics[ height=4cm, width=5.4cm]{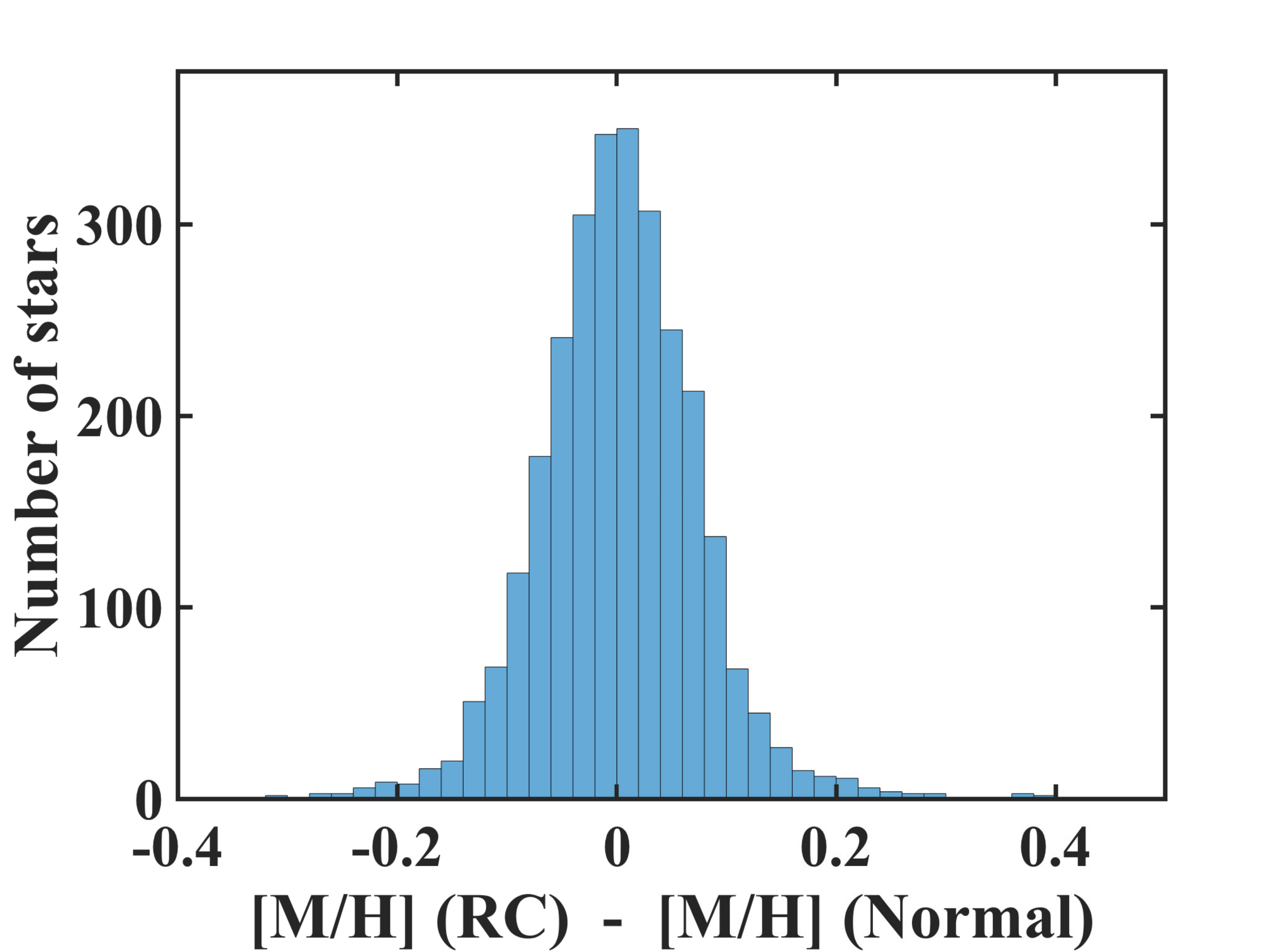}}               
\hspace{0.38cm} 
 \subfloat      
         []{\includegraphics[ height=4cm, width=5.4cm]{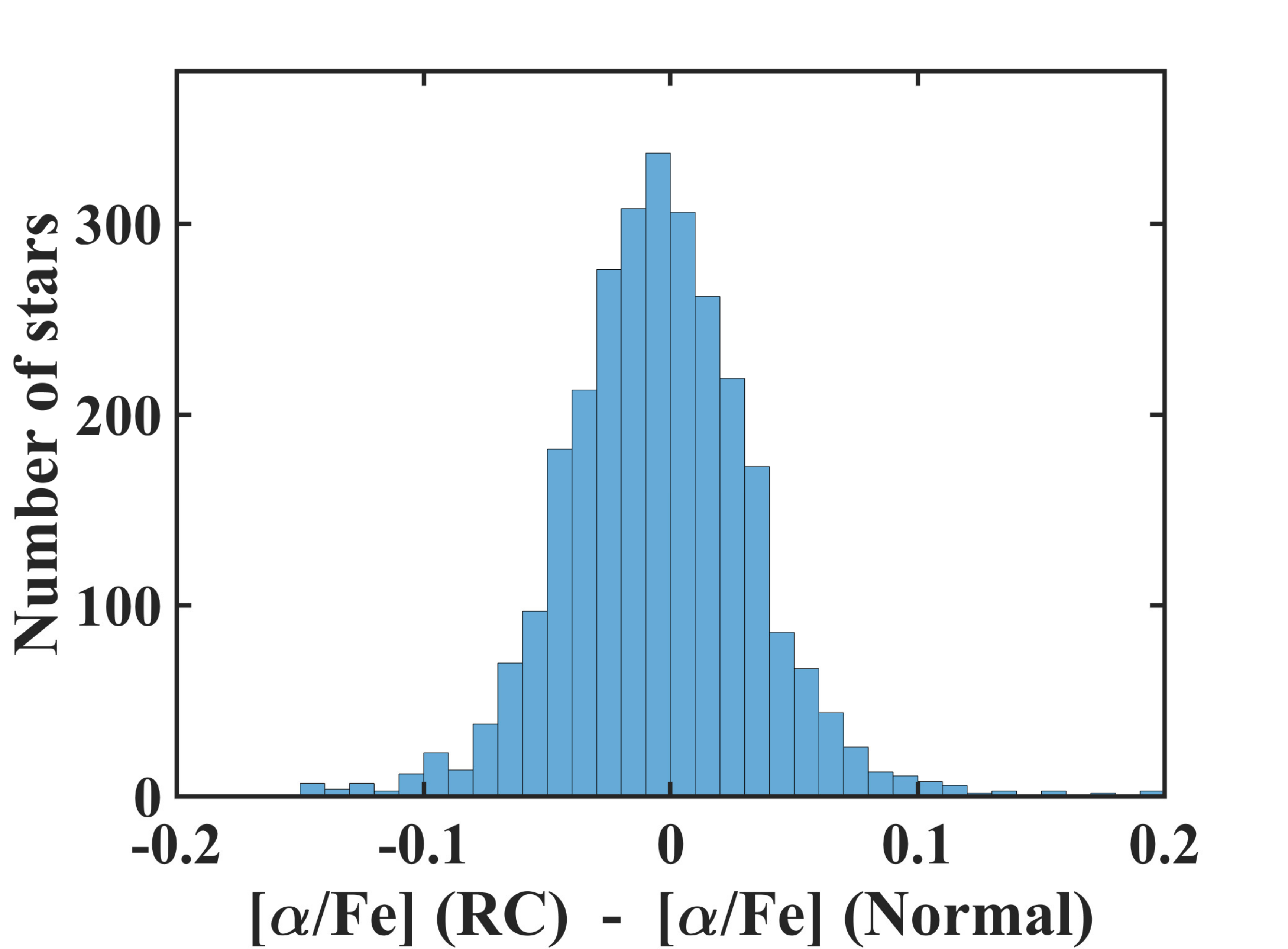}}        
 \caption
        {\footnotesize{Comparison between inferred model-fit parameters, T$_\textrm{\footnotesize{eff}}$, [M/H], and  [$\alpha$/Fe] using the \textbf{reduced-correlation method} and corresponding values using the \textbf{normal method} for 2836 stars with T$_\textrm{\footnotesize{eff}}$$\leq$3550 K (top panels). Surface gravity is treated as a constant parameter in both methods. The bottom panels show the histograms of the differences between  corresponding values inferred from the two methods.}}
\end{figure*}

\begin{figure*}\centering
\subfloat 
      []{\includegraphics[ height=4.1cm, width=5.65cm]{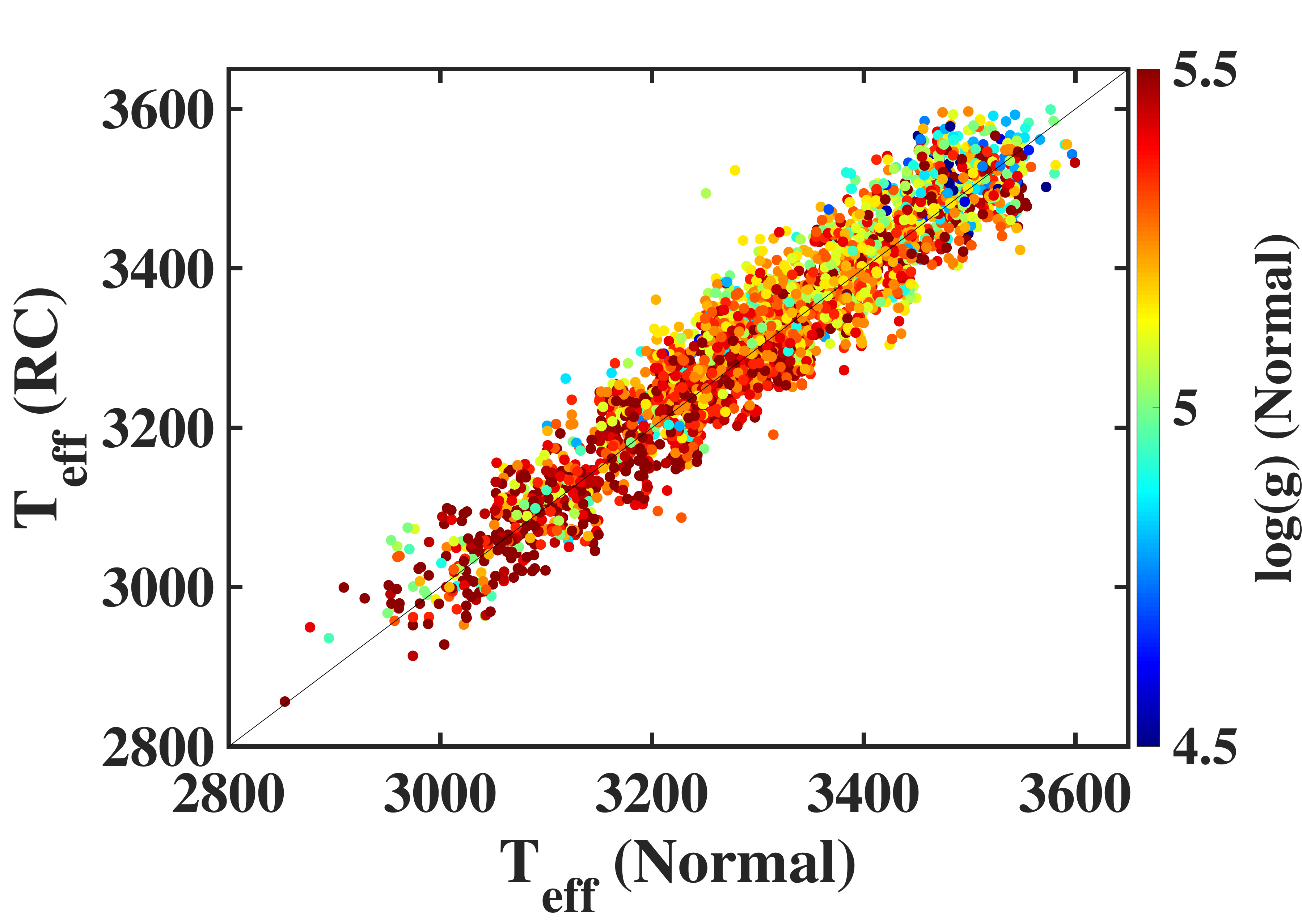}}               
\hspace{0.38cm}
\subfloat 
      []{\includegraphics[ height=4.1cm, width=5.65cm]{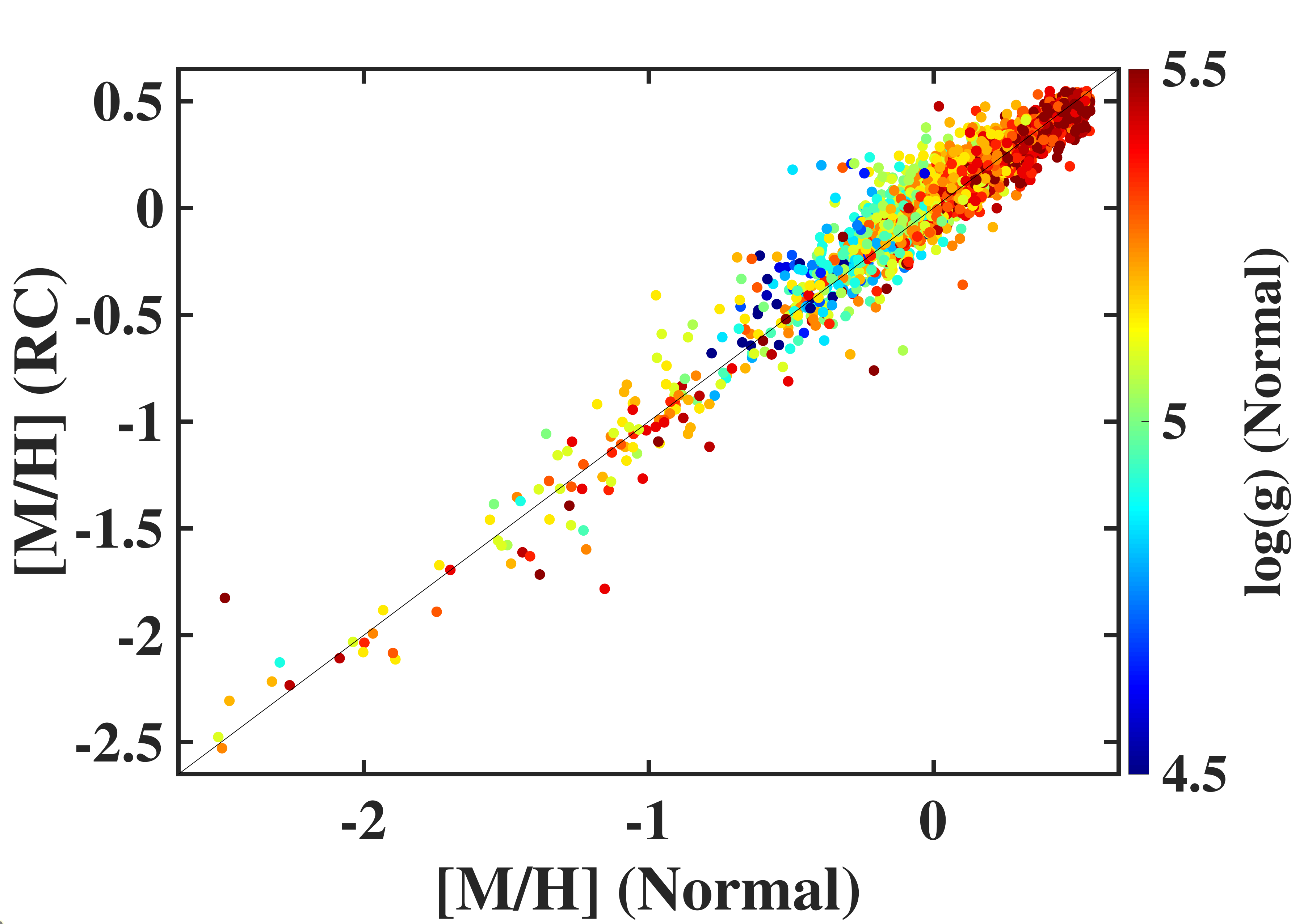}}               
\hspace{0.38cm} 
 \subfloat      
         []{\includegraphics[ height=4.1cm, width=5.65cm]{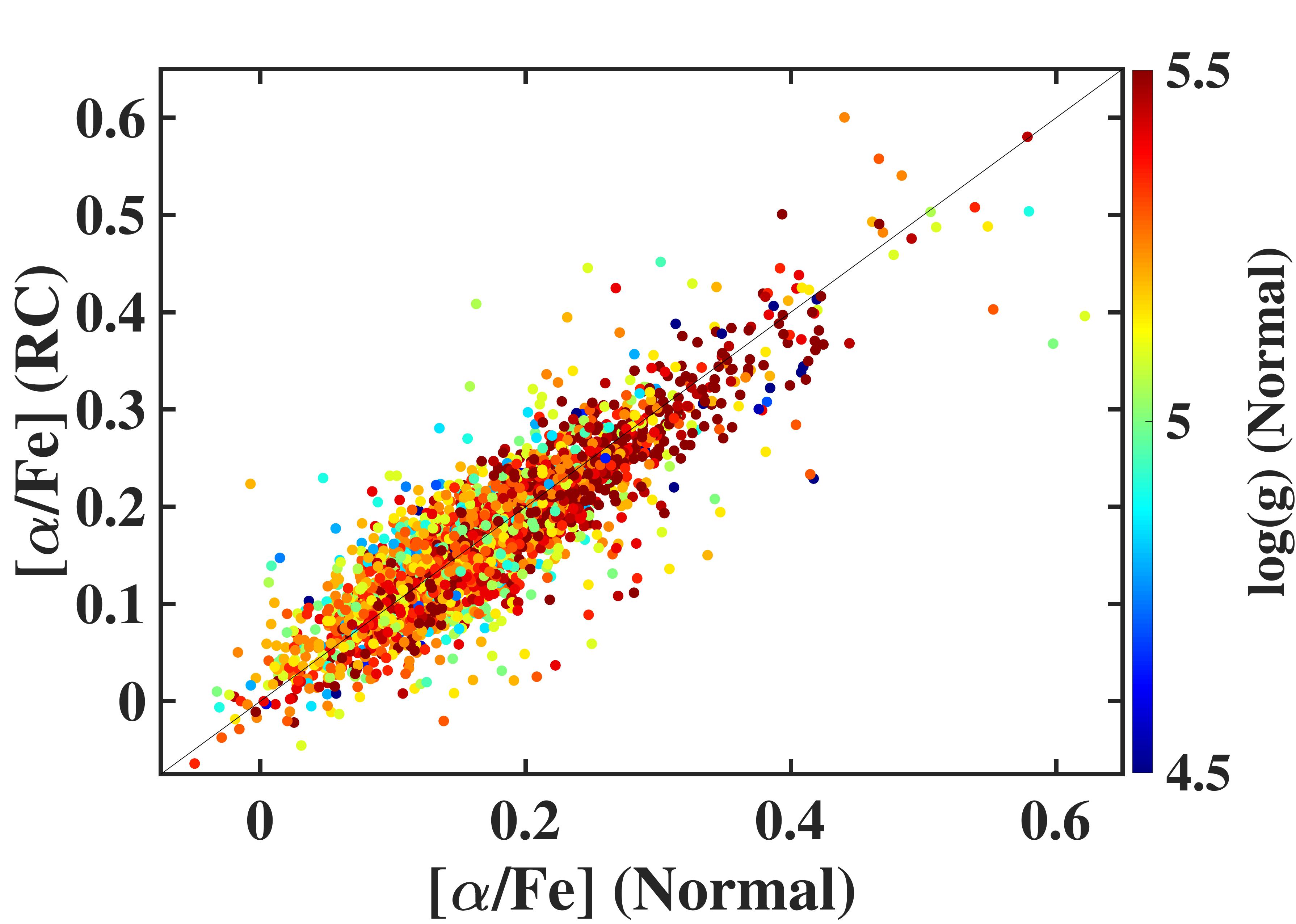}}
 \vspace{-0.25cm}

 \subfloat 
      []{\includegraphics[height=4cm, width=5.4cm]{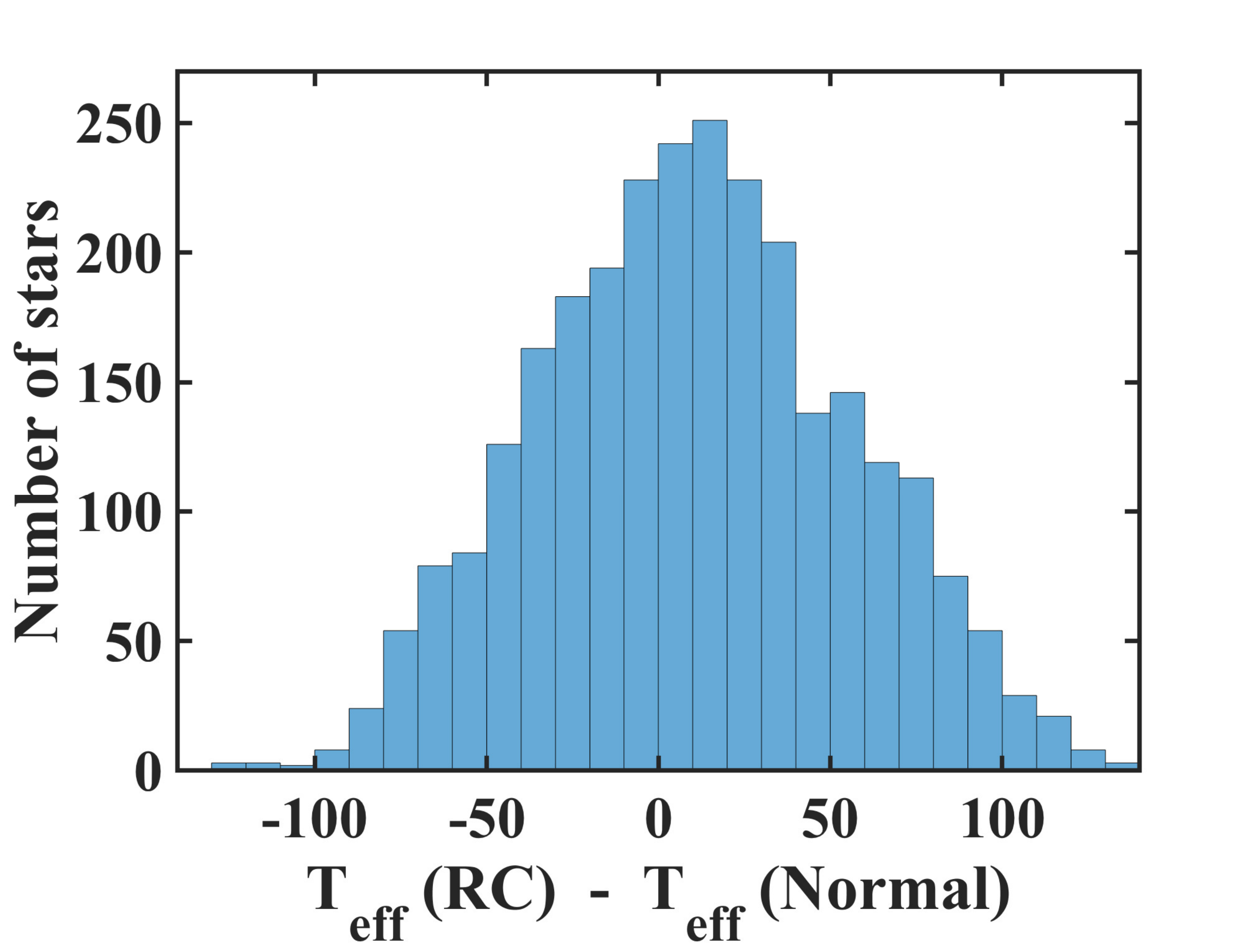}}               
\hspace{0.38cm}
\subfloat 
      []{\includegraphics[ height=4cm, width=5.4cm]{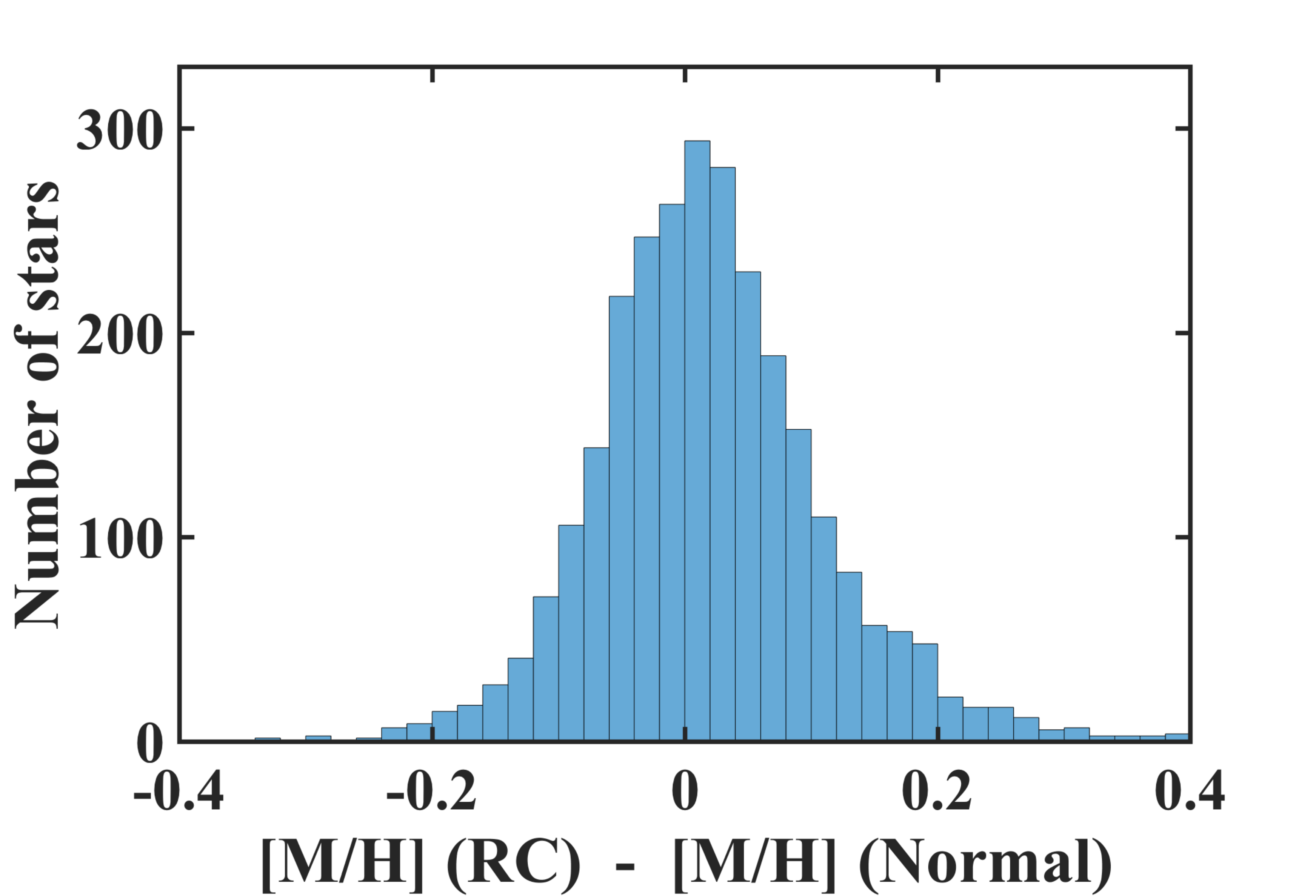}}               
\hspace{0.38cm} 
 \subfloat      
         []{\includegraphics[ height=4cm, width=5.4cm]{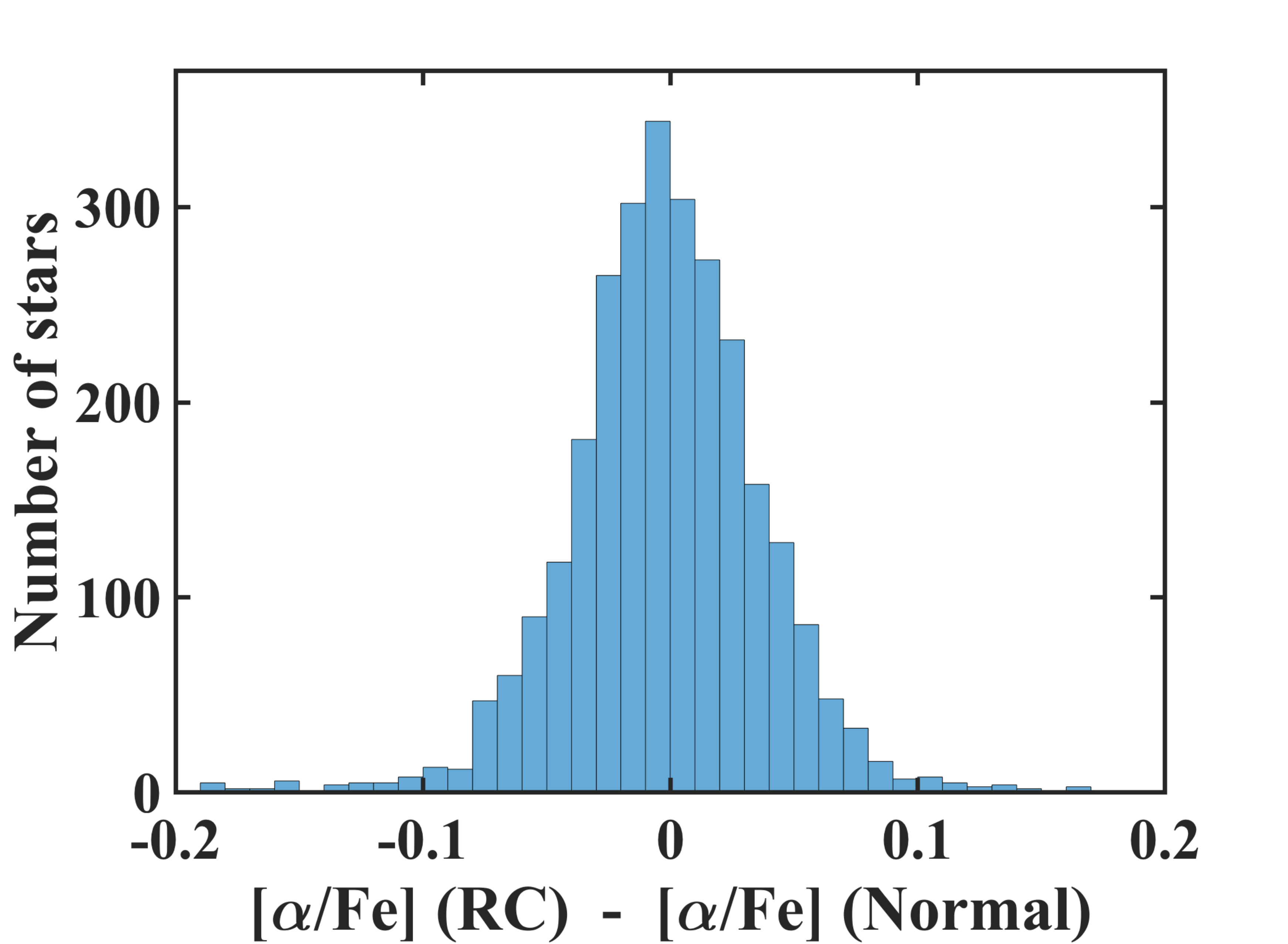}}        
 \caption
        {\footnotesize{Comparison between inferred model-fit chemical  parameters, T$_\textrm{\footnotesize{eff}}$, [M/H], and  [$\alpha$/Fe] using the \textbf{reduced-correlation method} and corresponding values using the \textbf{normal method} for 2788 stars with T$_\textrm{\footnotesize{eff}}$$\leq$3550 K (top panels). Surface gravity is treated as a free parameter in both methods. The bottom panels show the histograms of the differences between corresponding values inferred from the two methods.}}
\end{figure*}

\begin{figure*}\centering
\subfloat 
      [Fixed surface gravity]{\includegraphics[height=4.4cm, width=5.5cm]{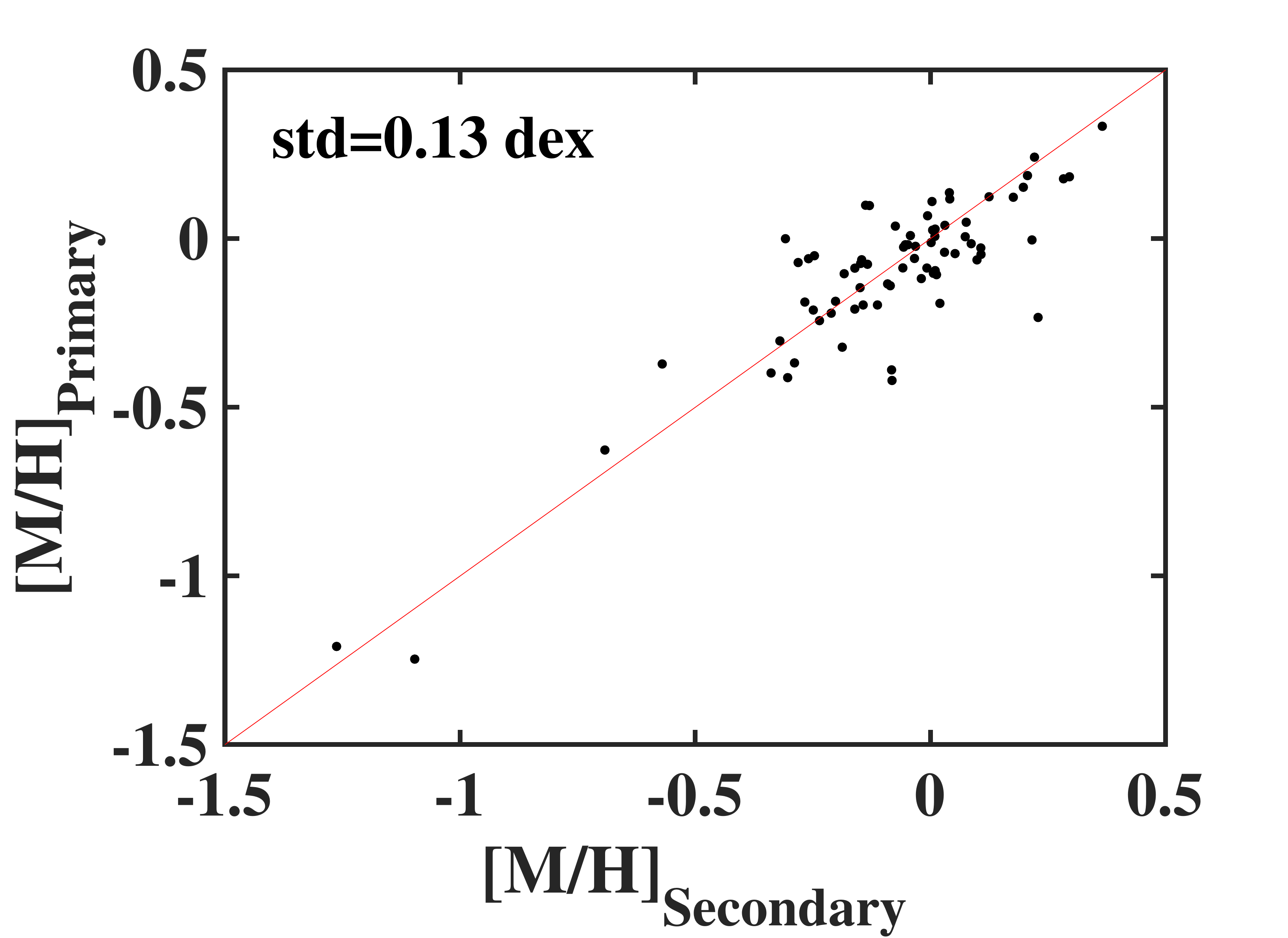}}               
\hspace{0.25cm}
\subfloat 
      [Fixed surface gravity]{\includegraphics[ height=4.4cm, width=5.5cm]{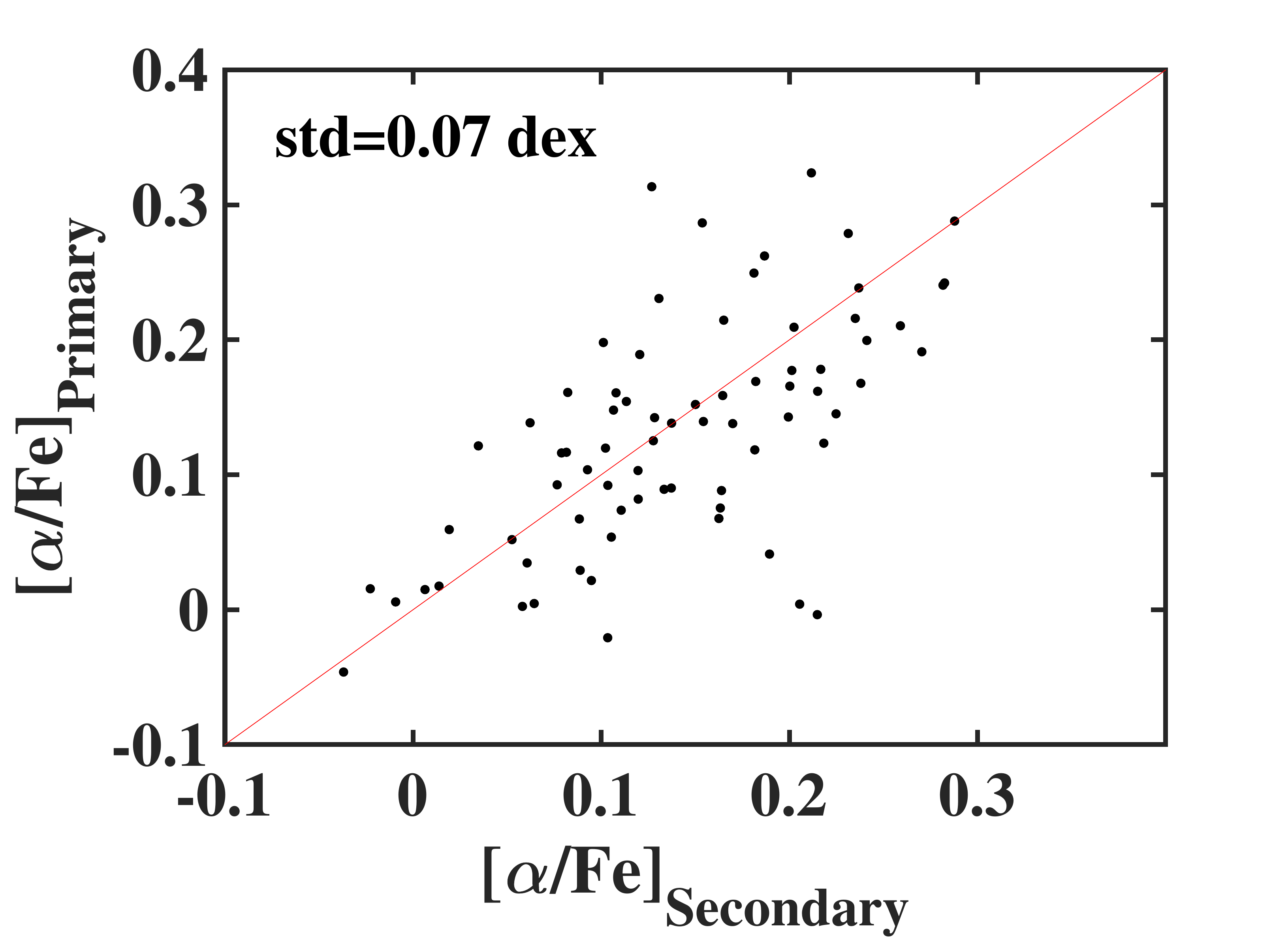}}               
\hspace{0.25cm} 
 \subfloat      
         [Fixed surface gravity]{\includegraphics[ height=4.4cm, width=5.5cm]{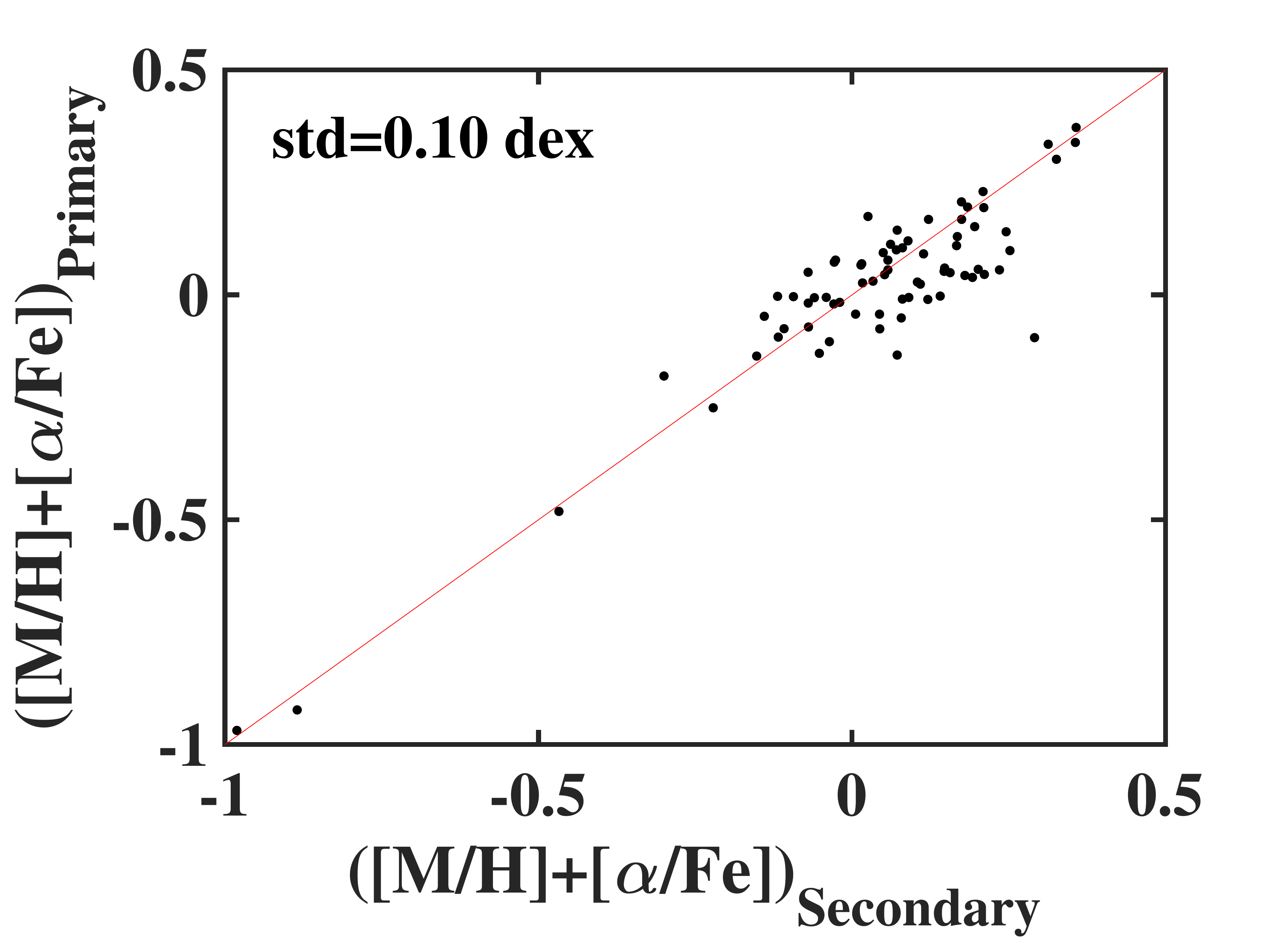}}
 \vspace{-0.25cm}

 \subfloat 
      [Variable surface gravity]{\includegraphics[ height=4.4cm, width=5.5cm]{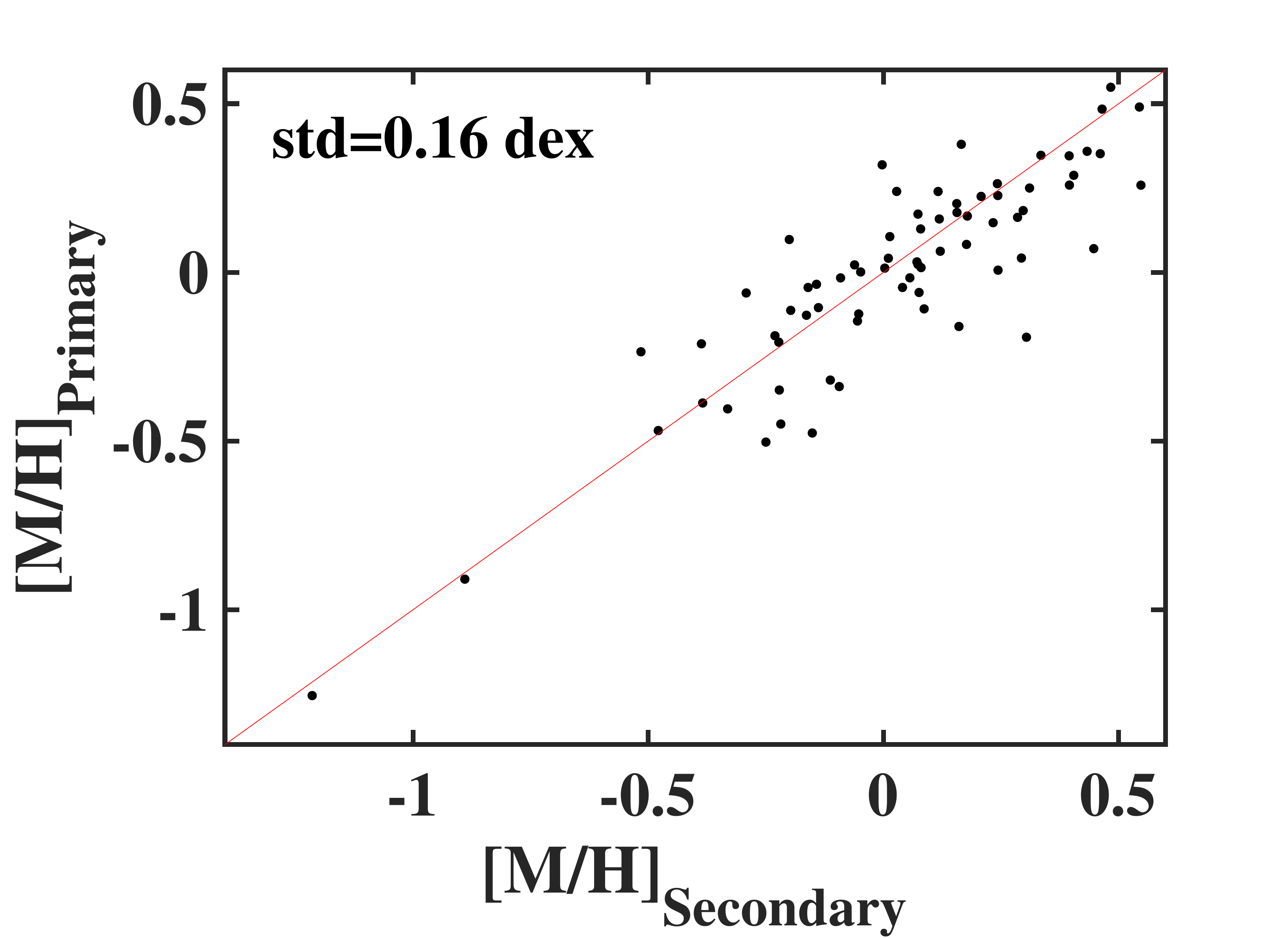}}               
\hspace{0.38cm}
\subfloat 
      [Variable surface gravity]{\includegraphics[ height=4.4cm, width=5.5cm]{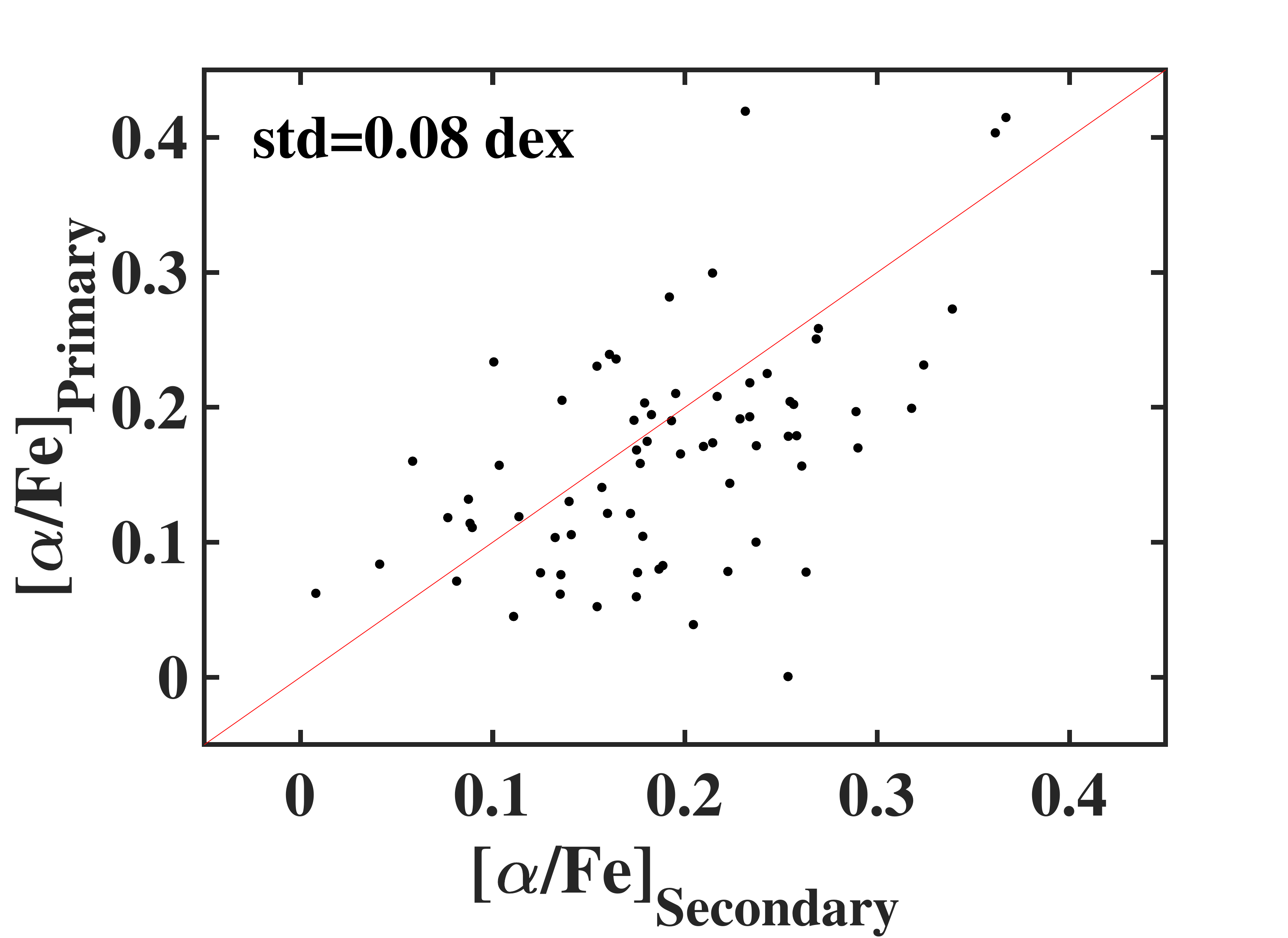}}               
\hspace{0.38cm} 
 \subfloat      
         [Variable surface gravity]{\includegraphics[height=4.4cm, width=5.5cm]{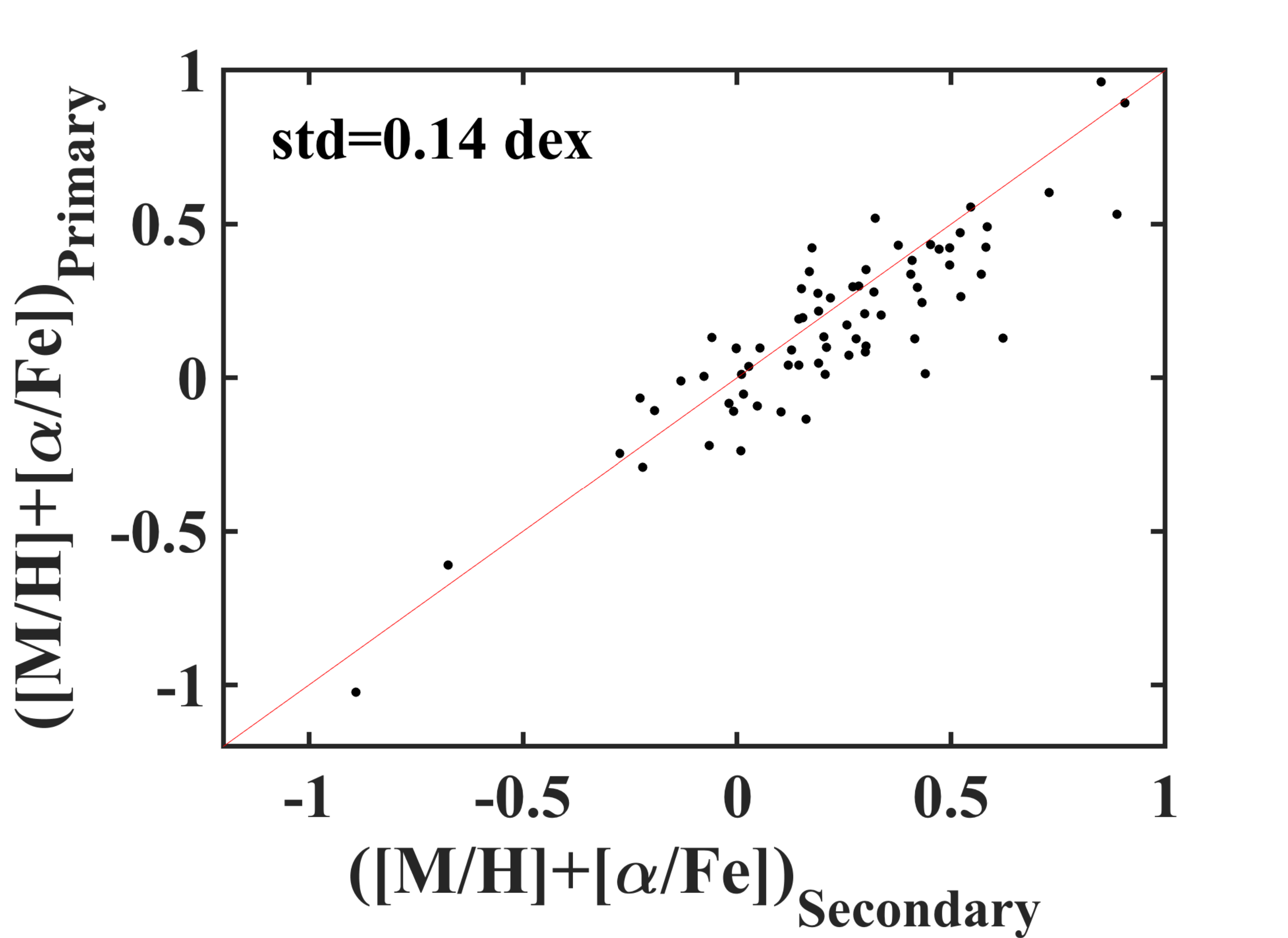}}   
 \caption
        {\footnotesize{Comparison between the inferred model-fit values of  [M/H], [$\alpha$/Fe], and the combined parameter [M/H]+[$\alpha$/Fe] of the primaries and  those of their respective companions, for binary systems whose both components have T$_\textrm{\footnotesize{eff}}$$\leq$3550 K, when surface gravity is a constant parameter (top panels, 74 binaries) and when surface gravity is a free parameter (bottom panels, 71 binaries) in the model fitting process, using the \textbf{normal method}.}}
\end{figure*}

\begin{figure*}\centering
\subfloat 
      [Fixed surface gravity]{\includegraphics[height=4.4cm, width=5.5cm]{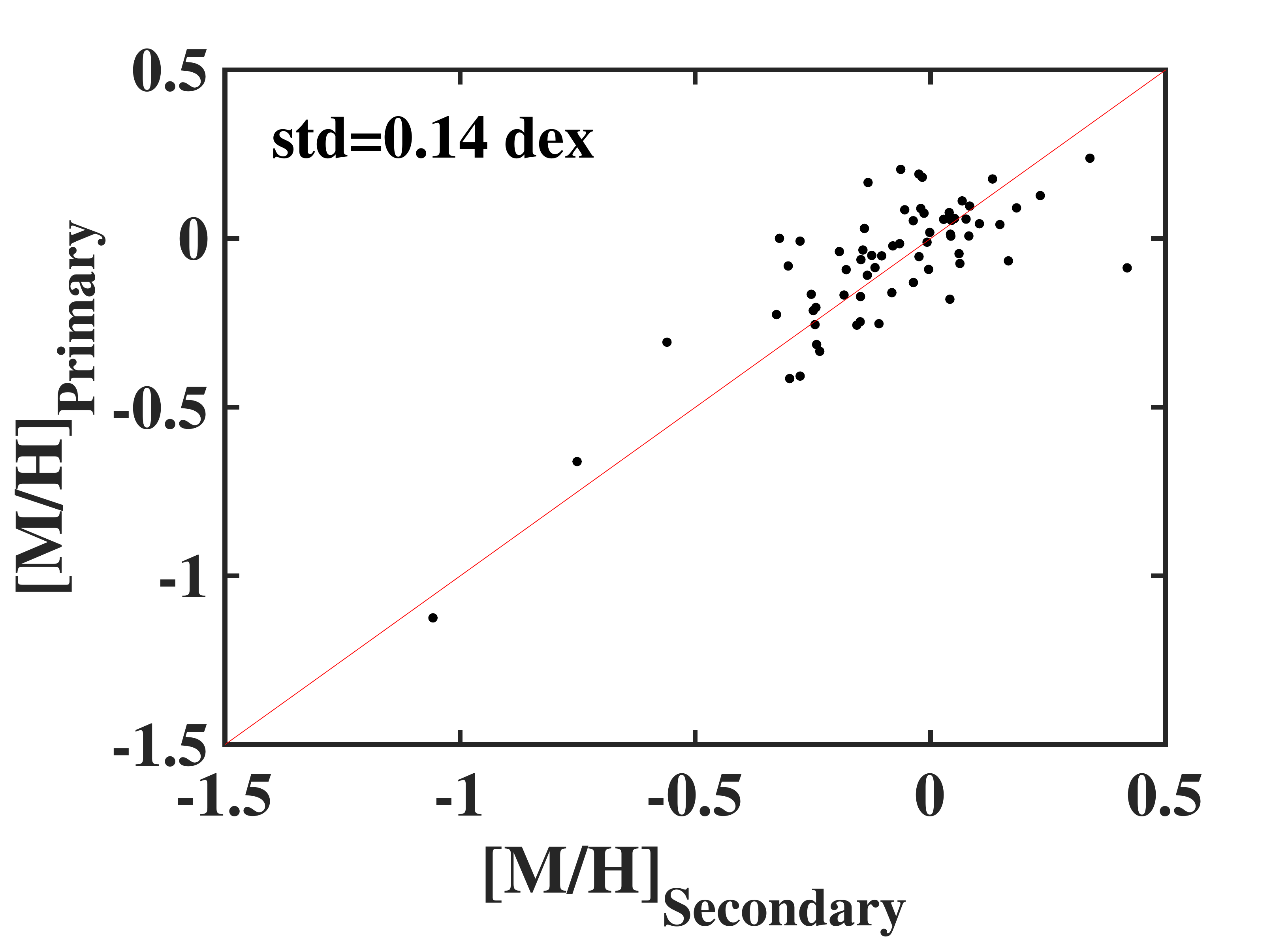}}               
\hspace{0.25cm}
\subfloat 
      [Fixed surface gravity]{\includegraphics[ height=4.4cm, width=5.5cm]{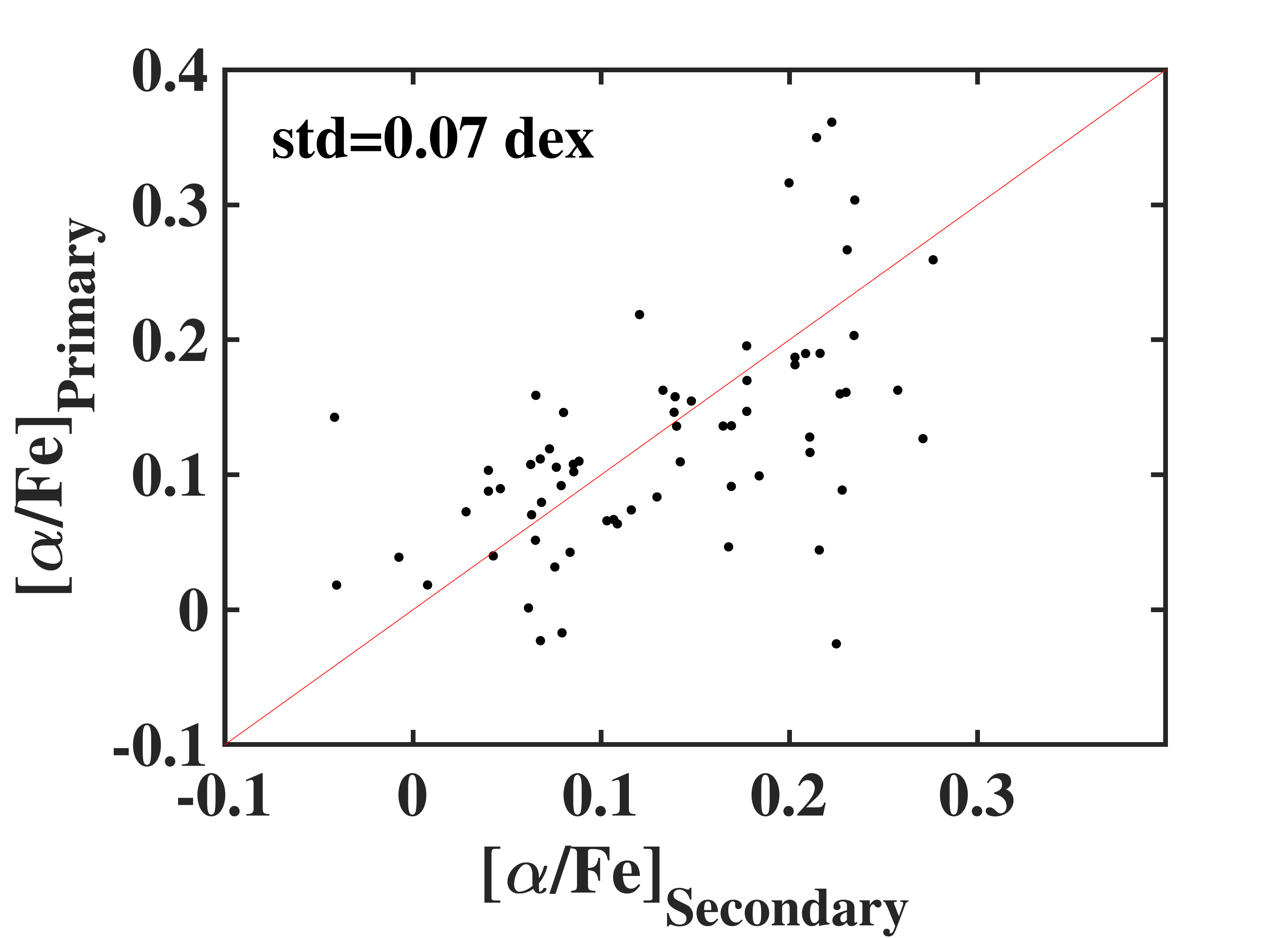}}               
\hspace{0.25cm} 
 \subfloat      
         [Fixed surface gravity]{\includegraphics[ height=4.4cm, width=5.5cm]{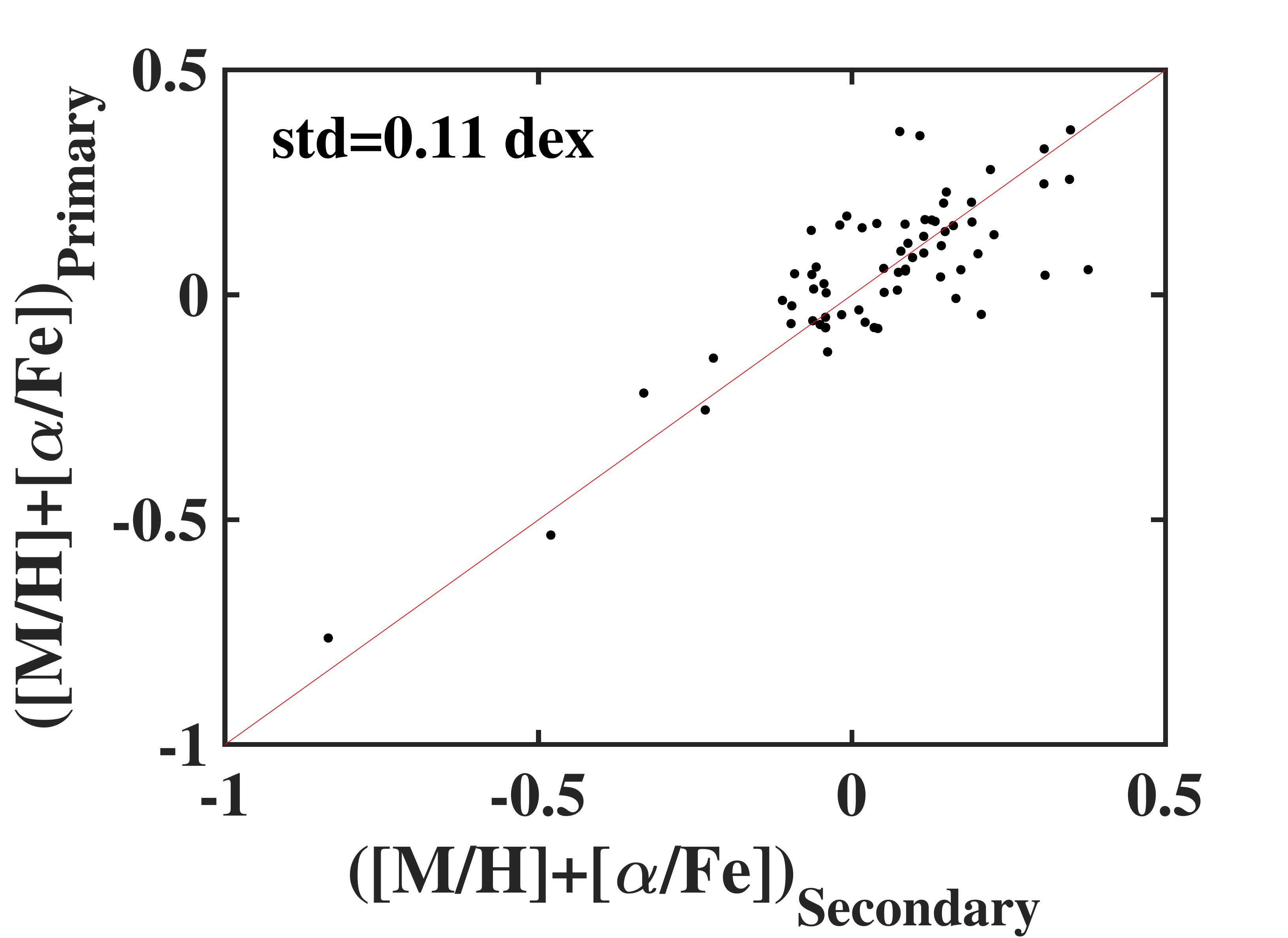}}
 \vspace{-0.25cm}

 \subfloat 
      [Variable surface gravity]{\includegraphics[ height=4.4cm, width=5.5cm]{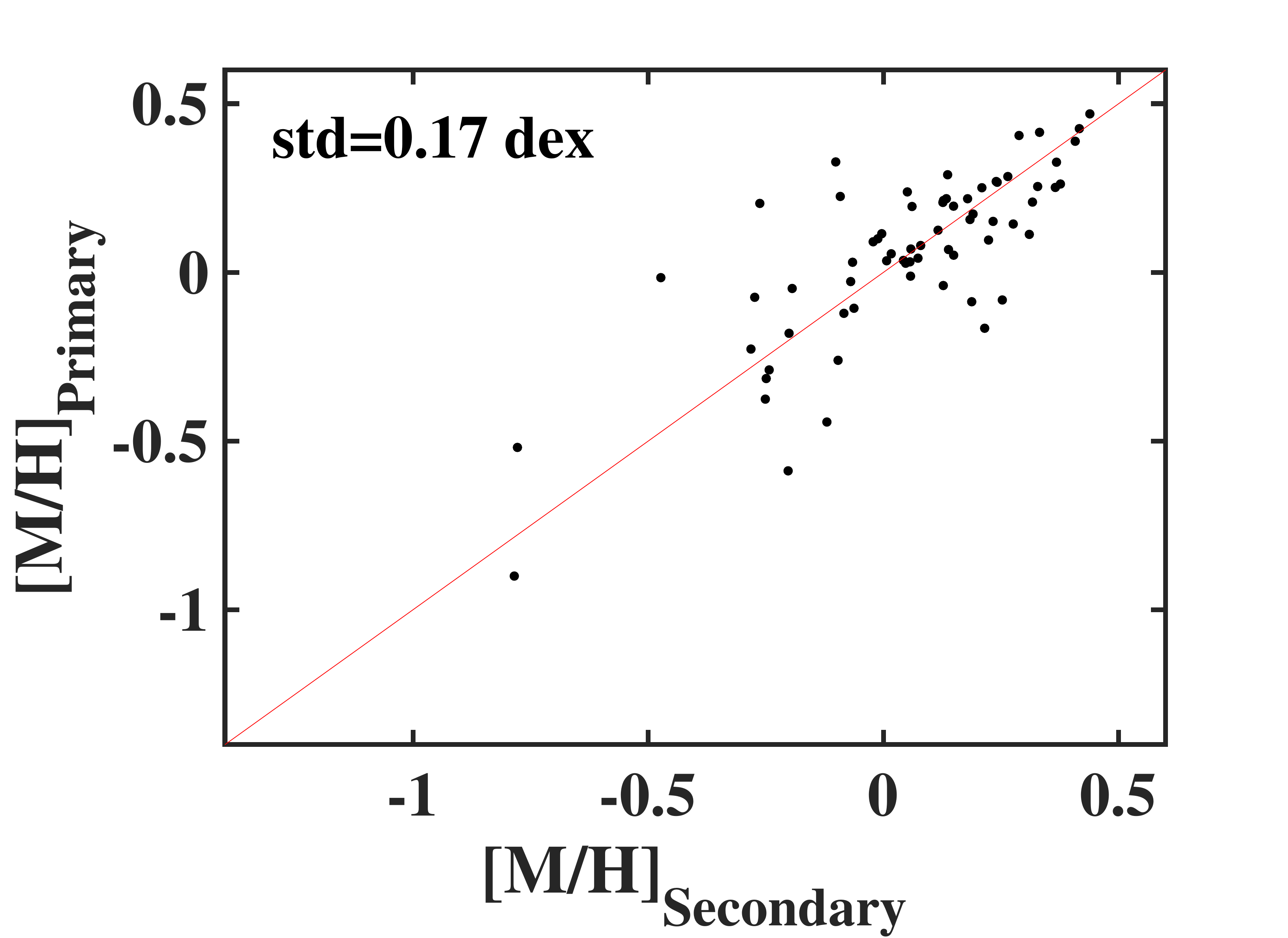}}               
\hspace{0.38cm}
\subfloat 
      [Variable surface gravity]{\includegraphics[ height=4.4cm, width=5.5cm]{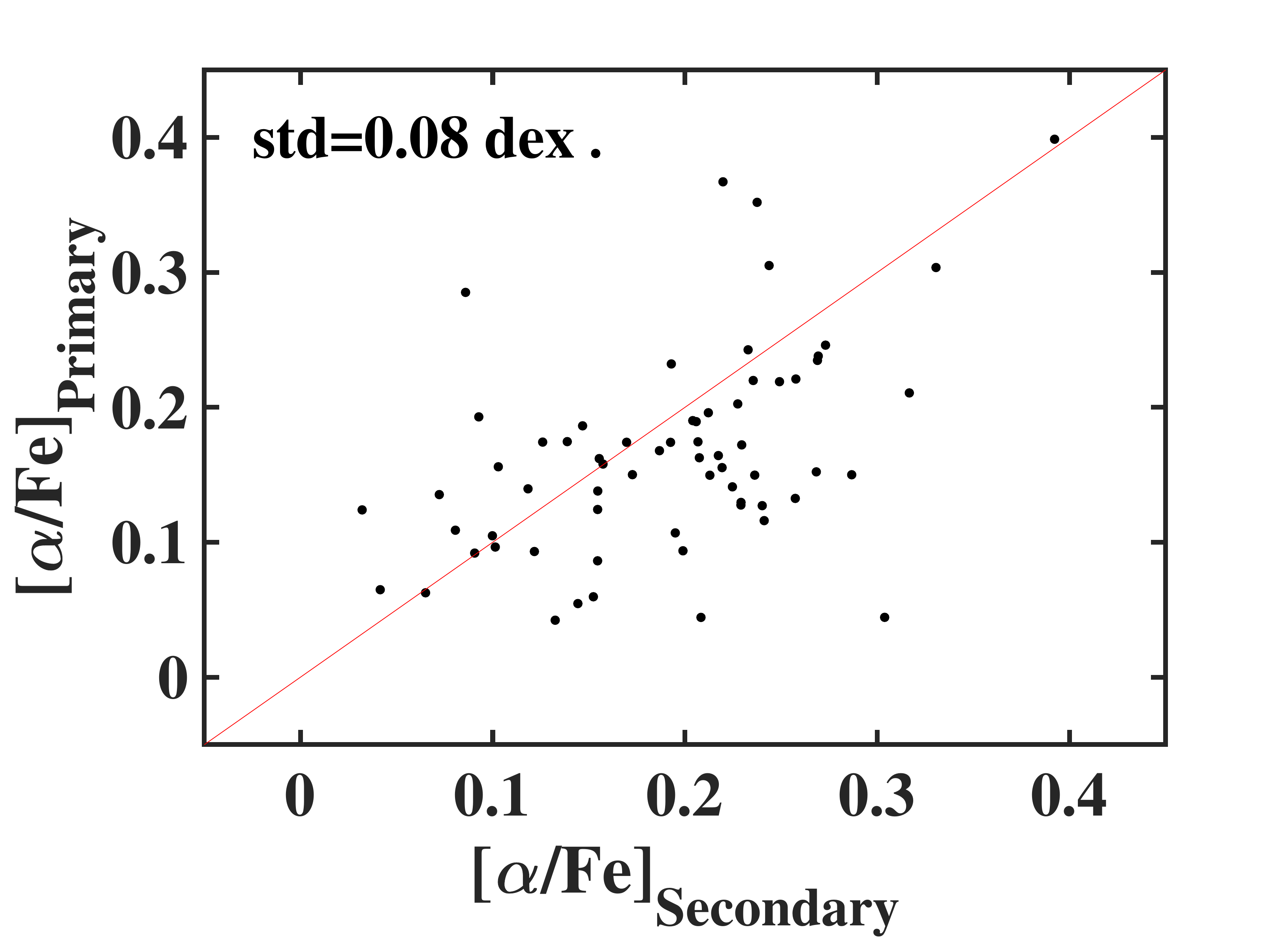}}               
\hspace{0.38cm} 
 \subfloat      
         [Variable surface gravity]{\includegraphics[height=4.4cm, width=5.5cm]{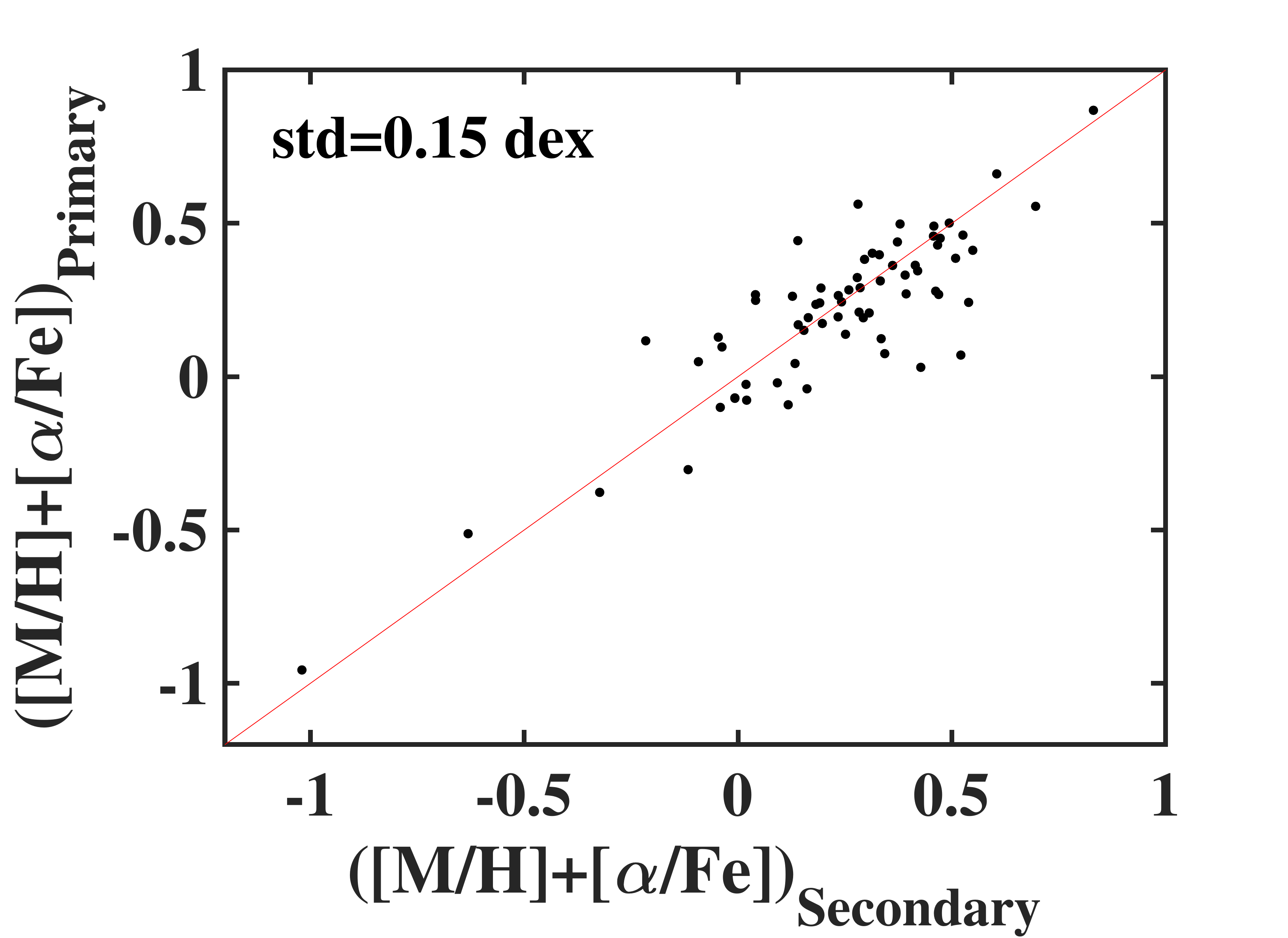}}   
 \caption
        {\footnotesize{Comparison between the inferred model-fit values of  [M/H], [$\alpha$/Fe], and the combined parameter [M/H]+[$\alpha$/Fe] of the primaries and those of their respective companions, for binary systems whose both components have T$_\textrm{\footnotesize{eff}}$$\leq$3550 K, when surface gravity is a constant parameter (top panels, 68 binaries) and when surface gravity is a free parameter (bottom panels, 68 binaries) in the model fitting process, using the \textbf{reduced-correlation method}.}}
\end{figure*}

\begin{figure*}\centering
 \subfloat
       []{\includegraphics[height=4.3cm, width=5.85cm]{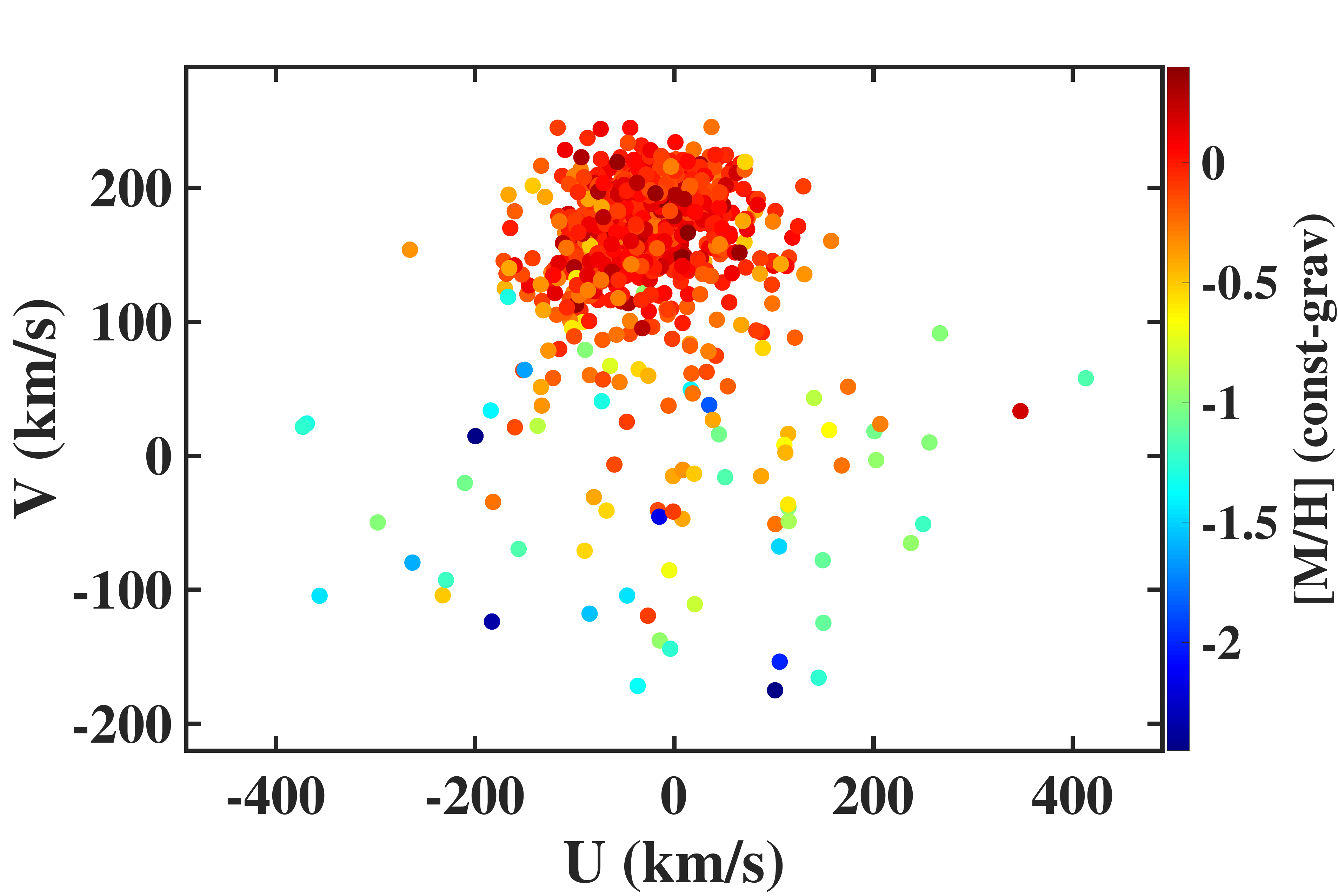}}
 \hspace{0.13cm} 
 \subfloat      
        []{\includegraphics[height=4.3cm, width=5.85cm]{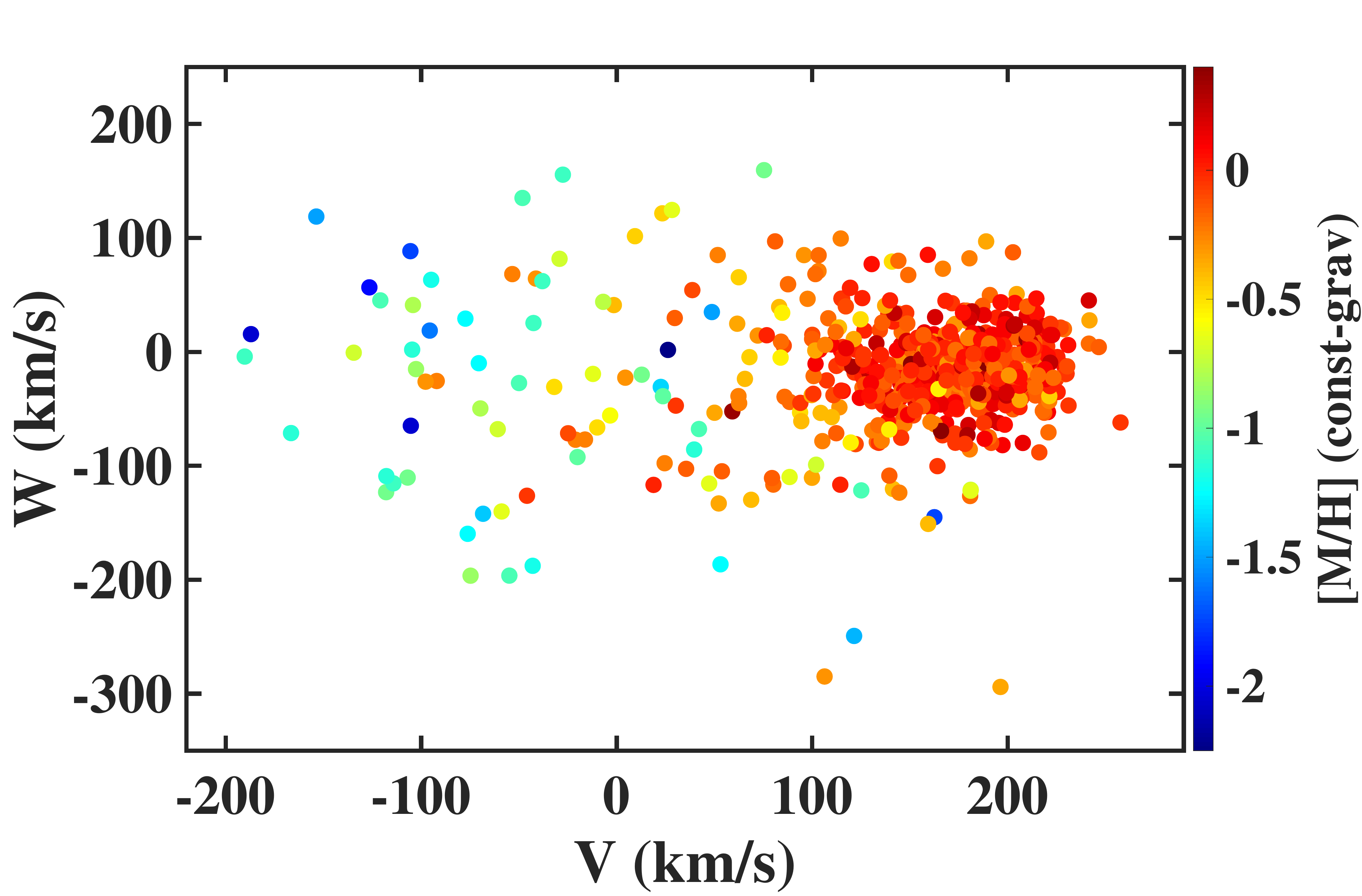}}
\hspace{0.13cm} 
 \subfloat      
        []{\includegraphics[height=4.3cm, width=5.85cm]{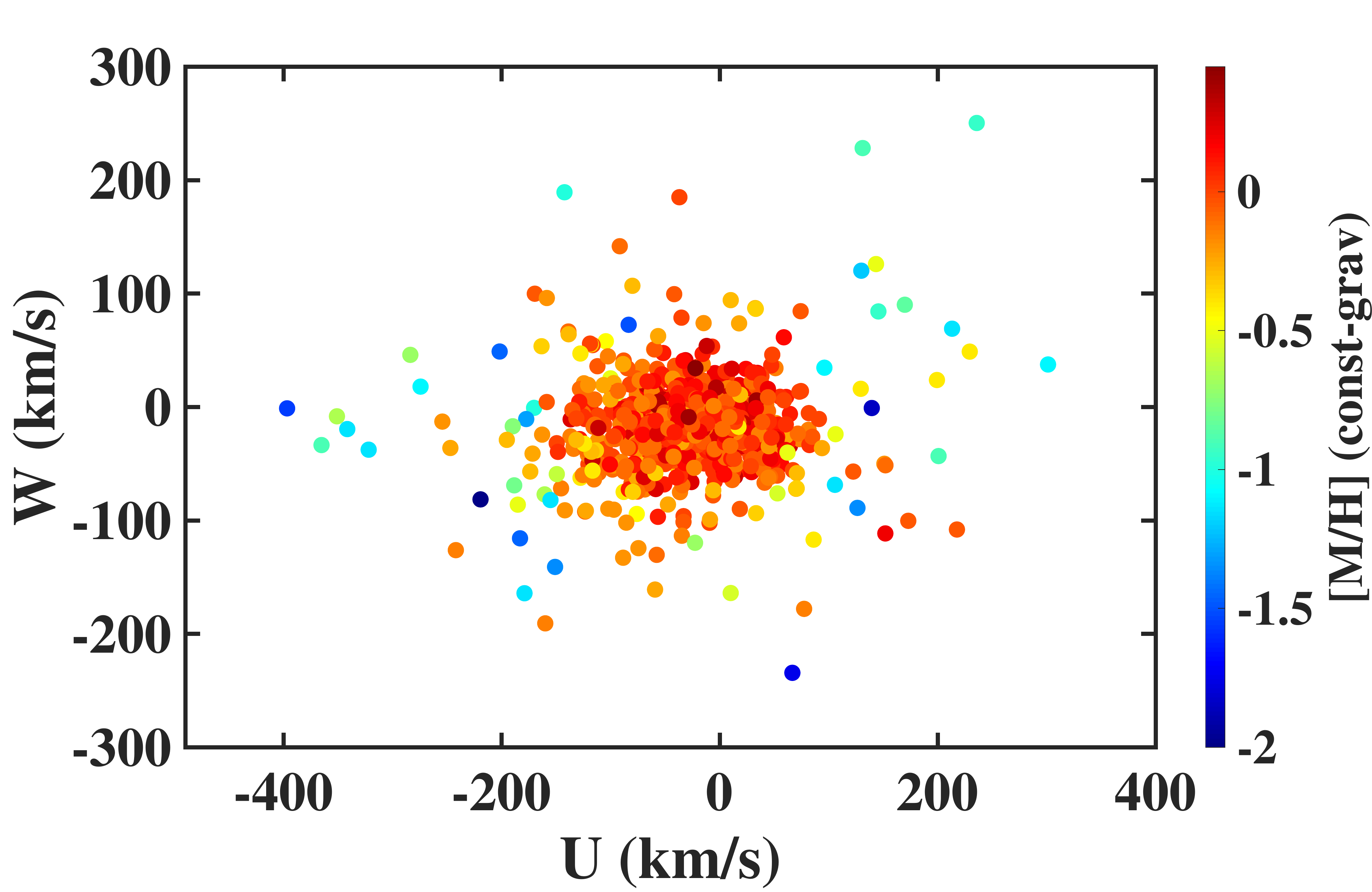}}            
  \caption
        {\footnotesize{Kinematic distributions of the stars in three different directions of the sky:  the Galactic poles (Group UV, left panels),  the Galactic center and anti-center (Group VW, middle panels), and  the solar apex and anti-apex (Group UW, right panels).   These panels show the projected motion in the Galactic UV, VW, and UW planes for a subset of stars having  T$_\textrm{\footnotesize{eff}}$$\leq$3550 K and  $-$2.5<[M/H]<+0.5 dex, that are color-mapped according to their metallicities,  inferred  from the  \textbf{normal method} when surface gravity is regarded as a fixed parameter: 1013 stars (Group UV), 803 stars (Group VW), and 1051 stars (Group UW). }}
 \end{figure*}

\begin{figure}\centering
\subfloat
        []{\includegraphics[ height=4cm, width=9cm]{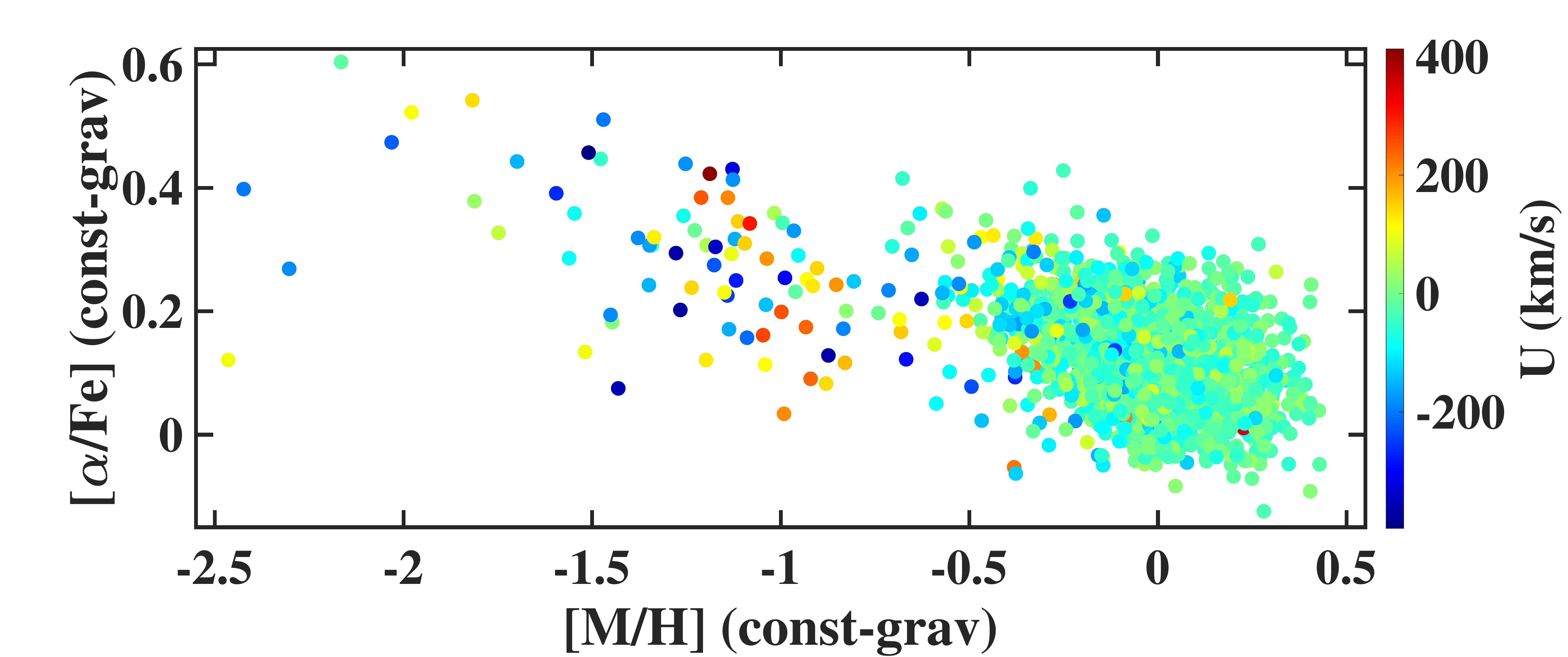}}
 \vspace{-0.35cm}

\subfloat
        []{\includegraphics[height=4cm, width=9cm]{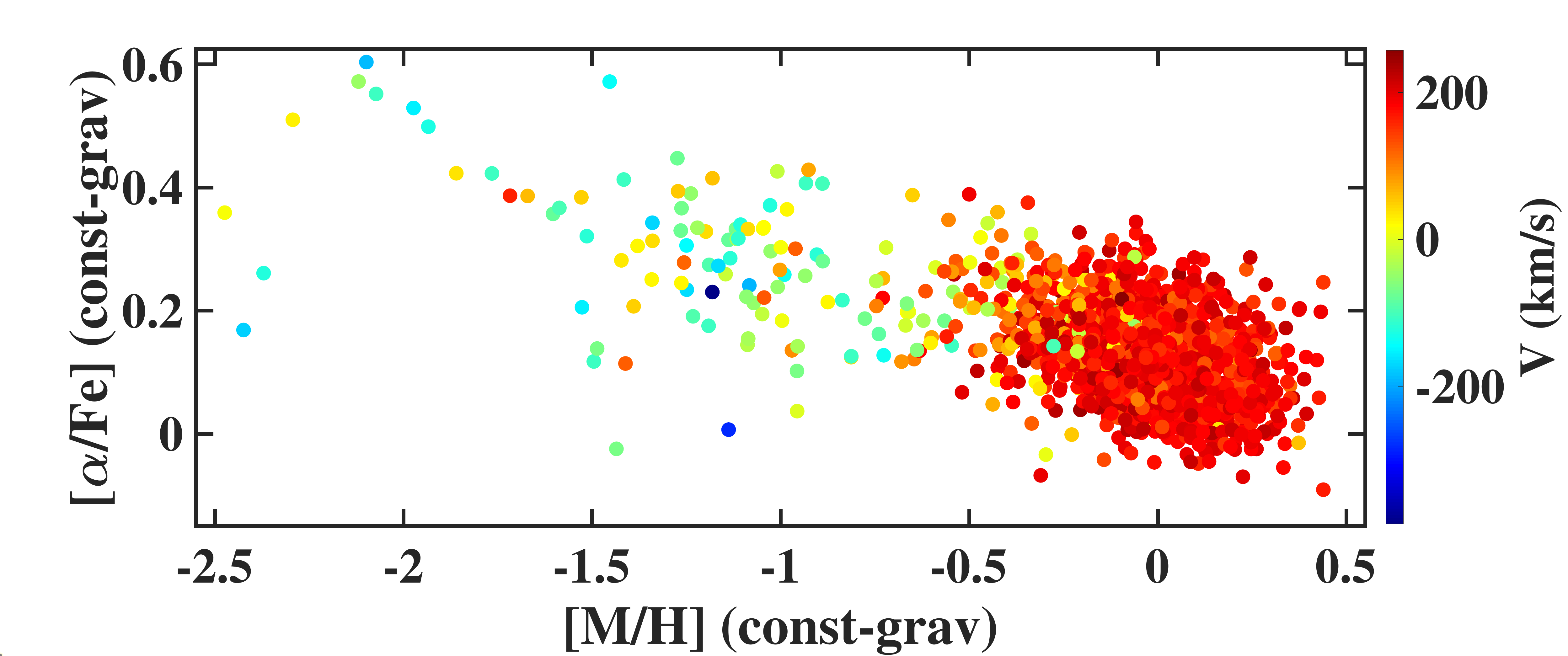}}
 \vspace{-0.35cm}

\subfloat
         []{\includegraphics[ height=4cm, width=9cm]{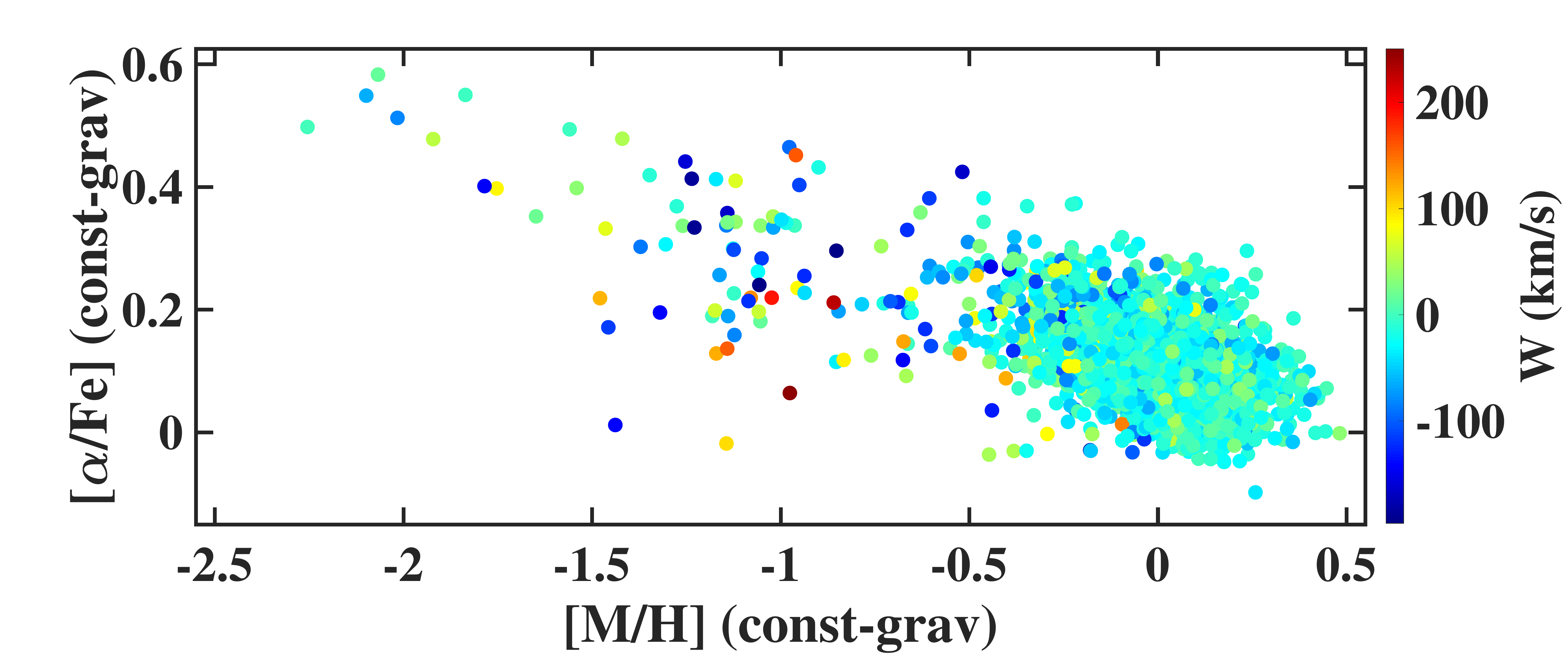}}
  \caption 
  {\footnotesize{[$\alpha$/Fe] vs. [M/H] diagram for the stars with T$_\textrm{\footnotesize{eff}}$$\leq$3550 K and $-$2.5<[M/H]<+0.5 dex, in Group UV+UW (top panels), Group UV+VW (middle panels), and Group UW+VW (bottom panels). The parameter values are inferred from  the \textbf{normal method} using the constant-gravity approach: 2065 stars in Group UV+UW, 1817 stars in Group UV+VW, and 1849 stars in  Group UW+VW. The stars are color-mapped based on U, V, and W velocity components for  Groups UV+UW, UV+VW, and UW+VW, respectively.}}
\end{figure}

\begin{figure*}\centering
 \subfloat
       []{\includegraphics[height=3.2cm, width=5.8cm]{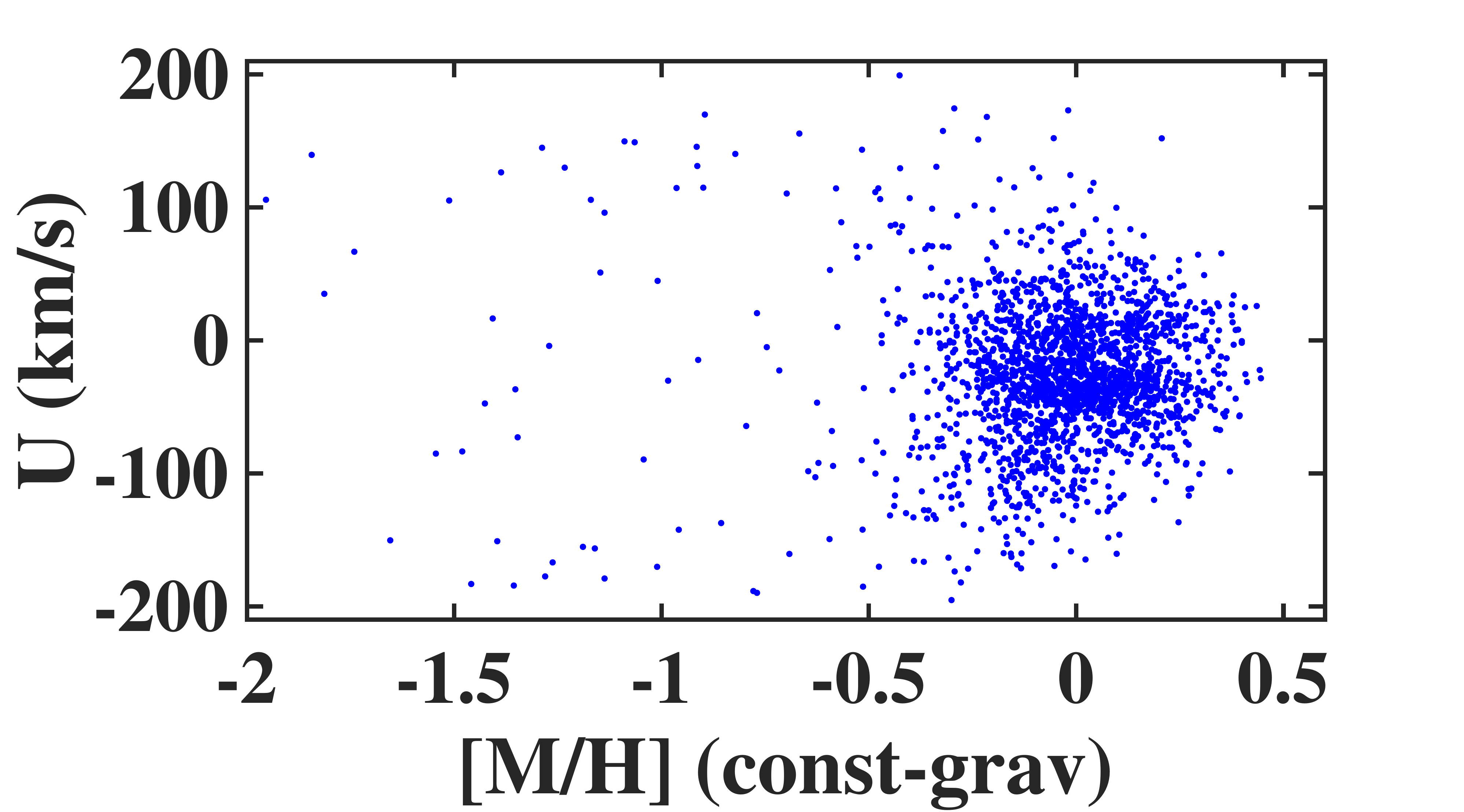}}
 \hspace{0.1cm} 
 \subfloat      
        []{\includegraphics[height=3.2cm, width=5.8cm]{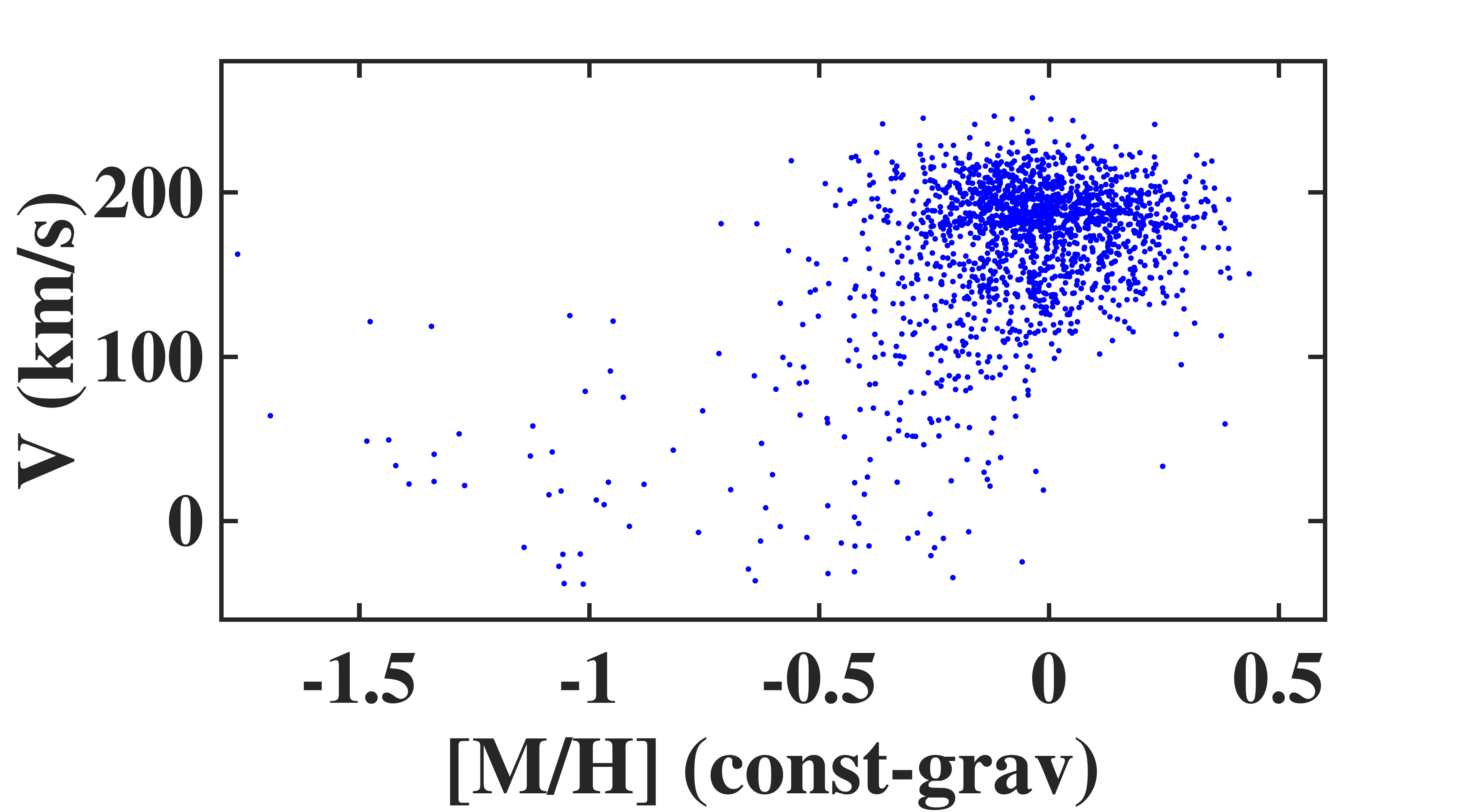}}
\hspace{0.1cm} 
 \subfloat      
        []{\includegraphics[height=3.2cm, width=5.8cm]{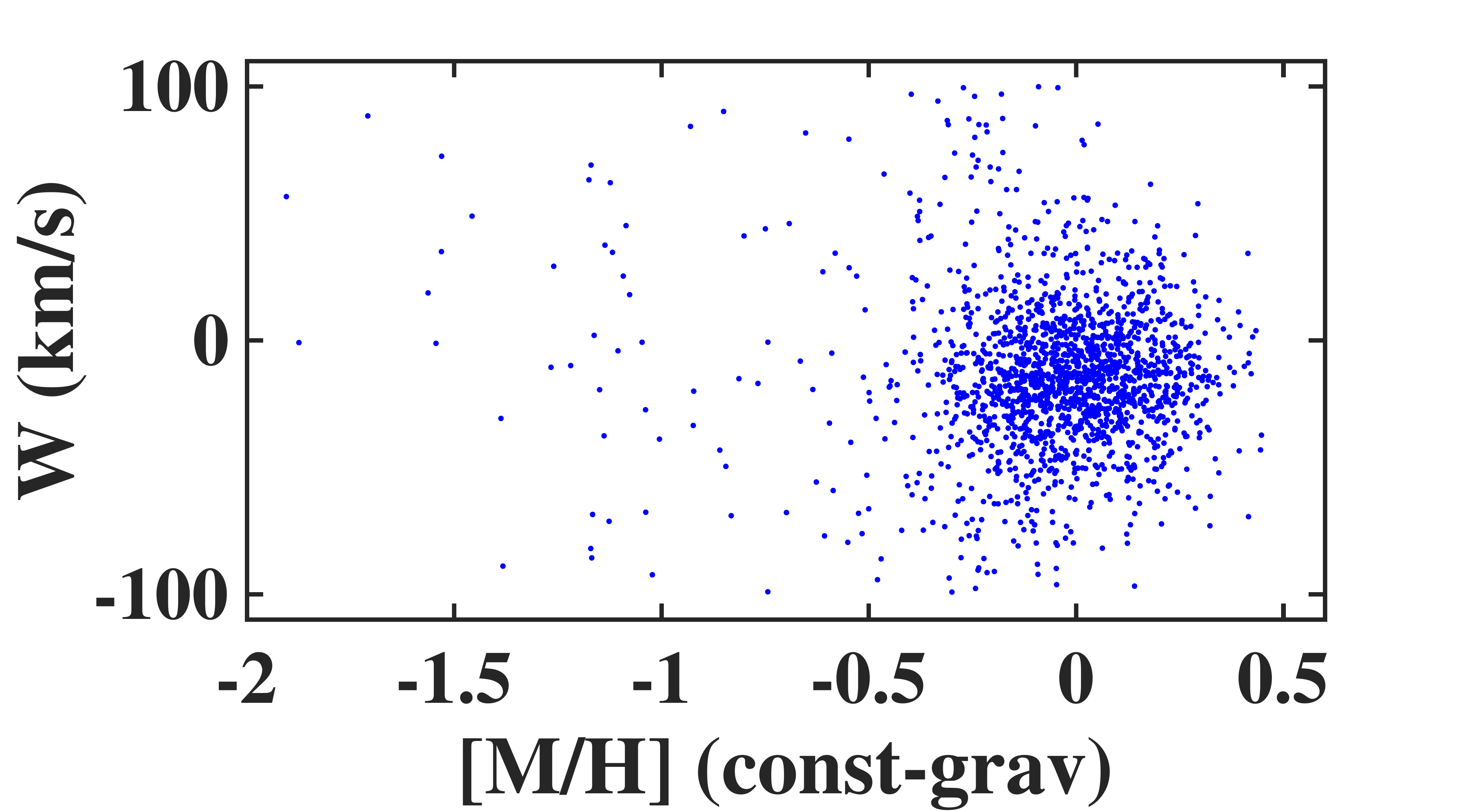}}            
 \vspace{0.1cm}

\subfloat
       []{\includegraphics[height=3.2cm, width=5.8cm]{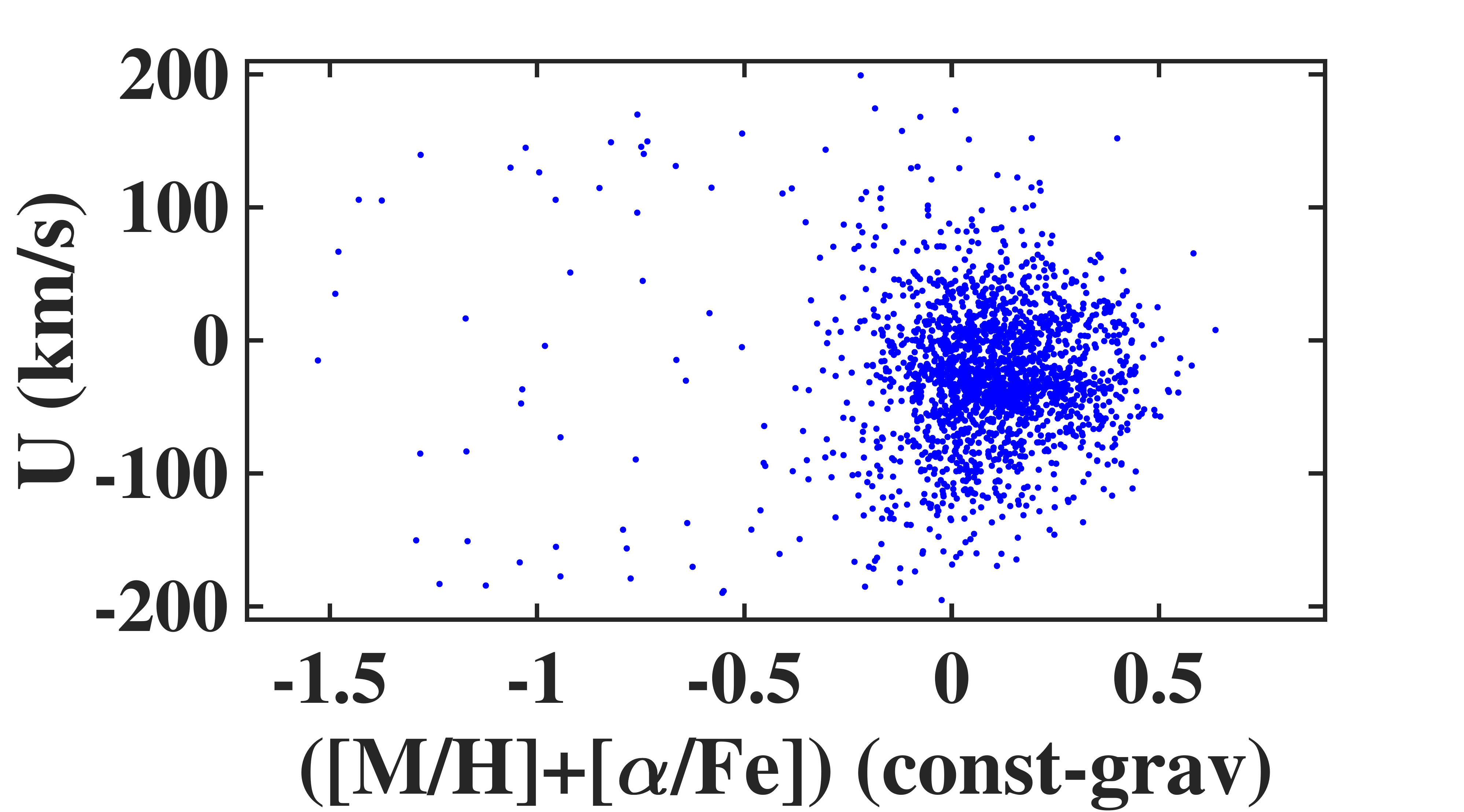}}
 \hspace{0.1cm} 
 \subfloat      
        []{\includegraphics[height=3.2cm, width=5.8cm]{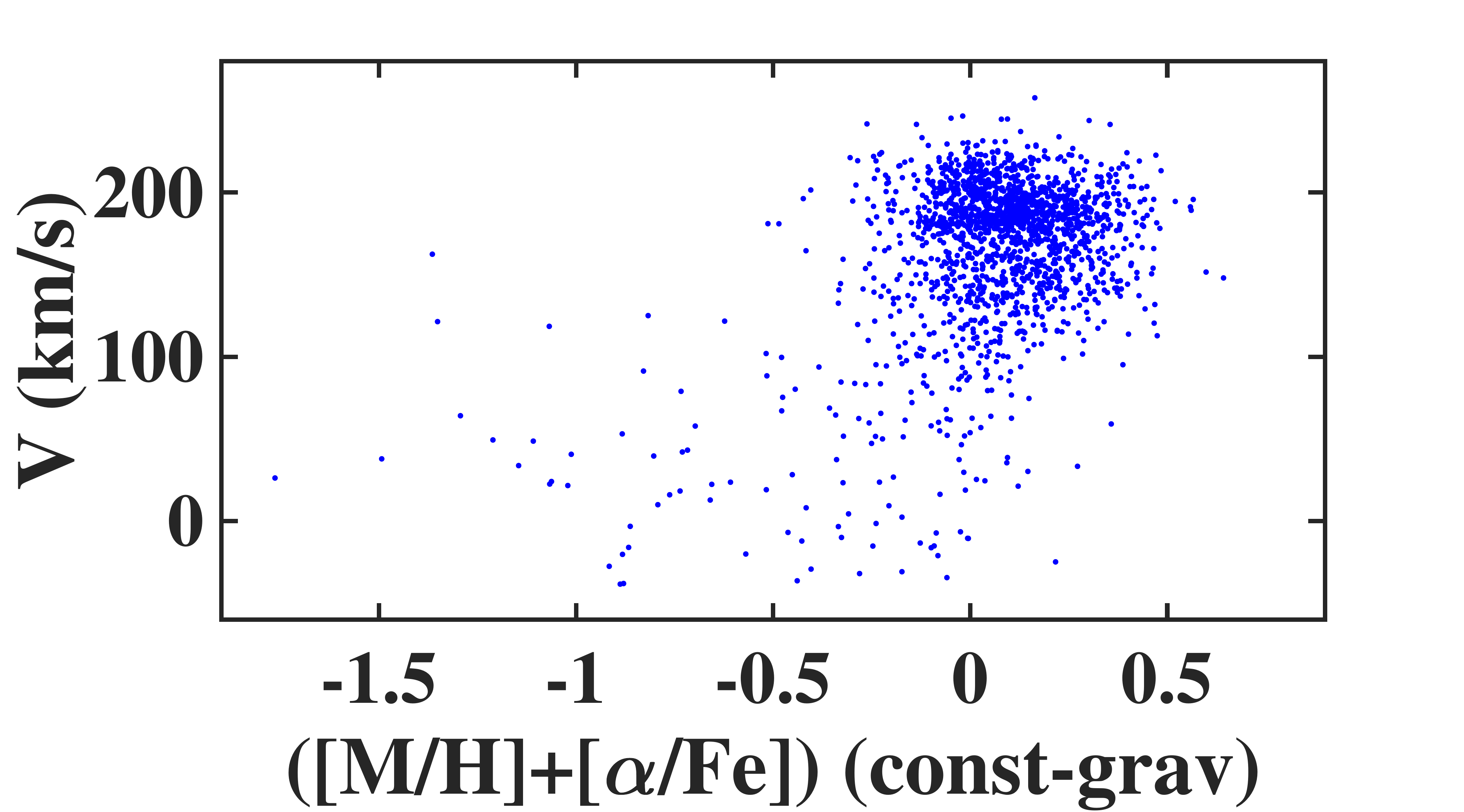}}
\hspace{0.1cm} 
 \subfloat      
        []{\includegraphics[height=3.2cm, width=5.8cm]{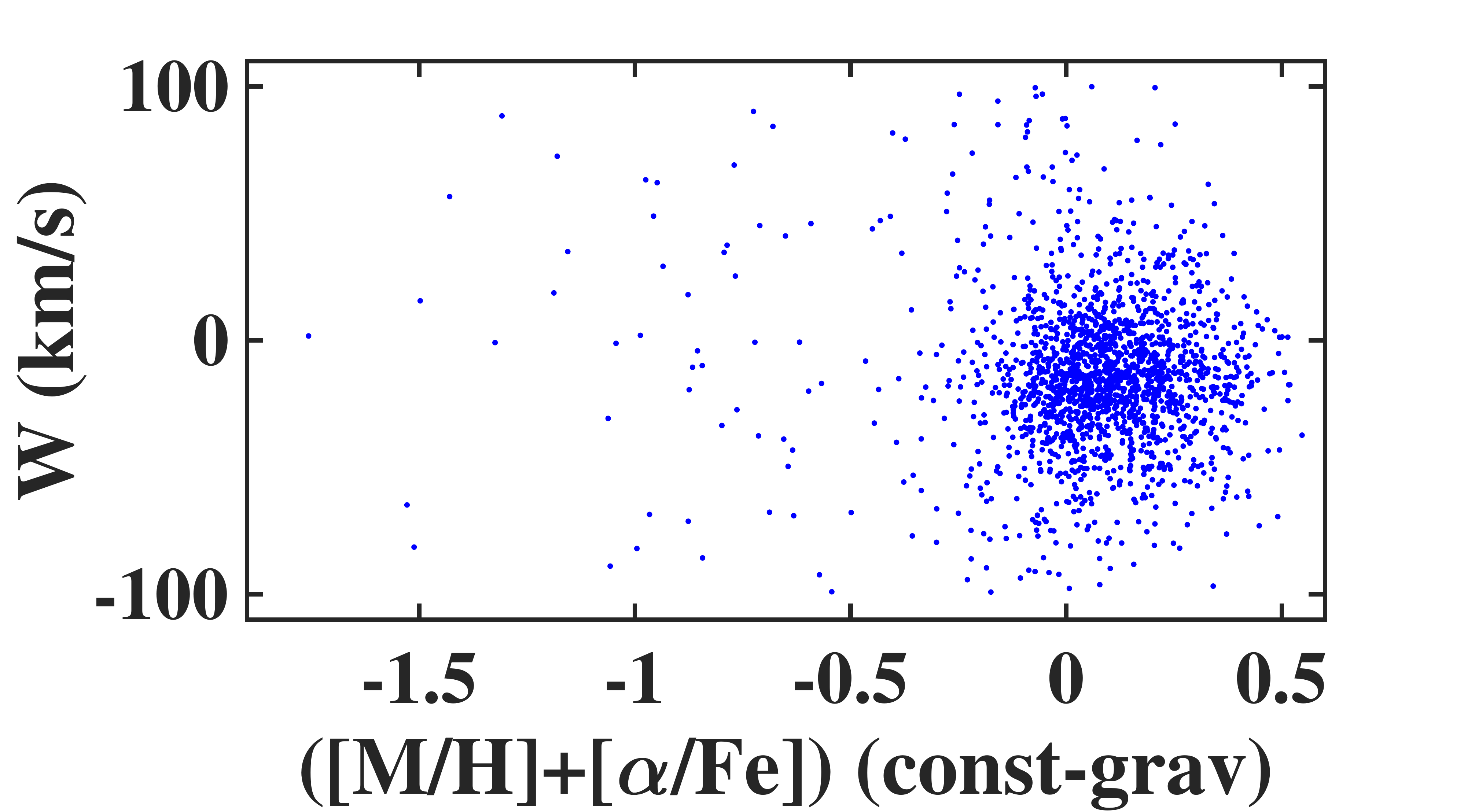}}            
 \caption
        {\footnotesize{Velocity component  U (left panels),  V (middle panels), and W (right panels) versus [M/H] (top panels) and [M/H]+[$\alpha$/Fe] (bottom panels) of the stars  with T$_\textrm{\footnotesize{eff}}$$\leq$3550 K and $-$2.5<[M/H]<+0.5 dex in the respective combined Group UV+UW,  UV+VW, and UW+VW, as  those shown in Figure 41, but with narrower ranges of velocities mainly associated with the disk stars:  $-$200$\leq$U$\leq$200 km/s (2028 stars in Group UV+UW), $-$40$\leq$V$\leq$260 km/s (1747 stars in Group UV+VW) , and $-$100$\leq$W$\leq$100 km/s (1780 stars in Group UW+VW). The parameter values are inferred from the  \textbf{normal method} in which surface gravity is  a fixed parameter.}}
 \end{figure*}

\begin{figure}\centering
        {\includegraphics[ height=5.5cm, width=8.2cm]{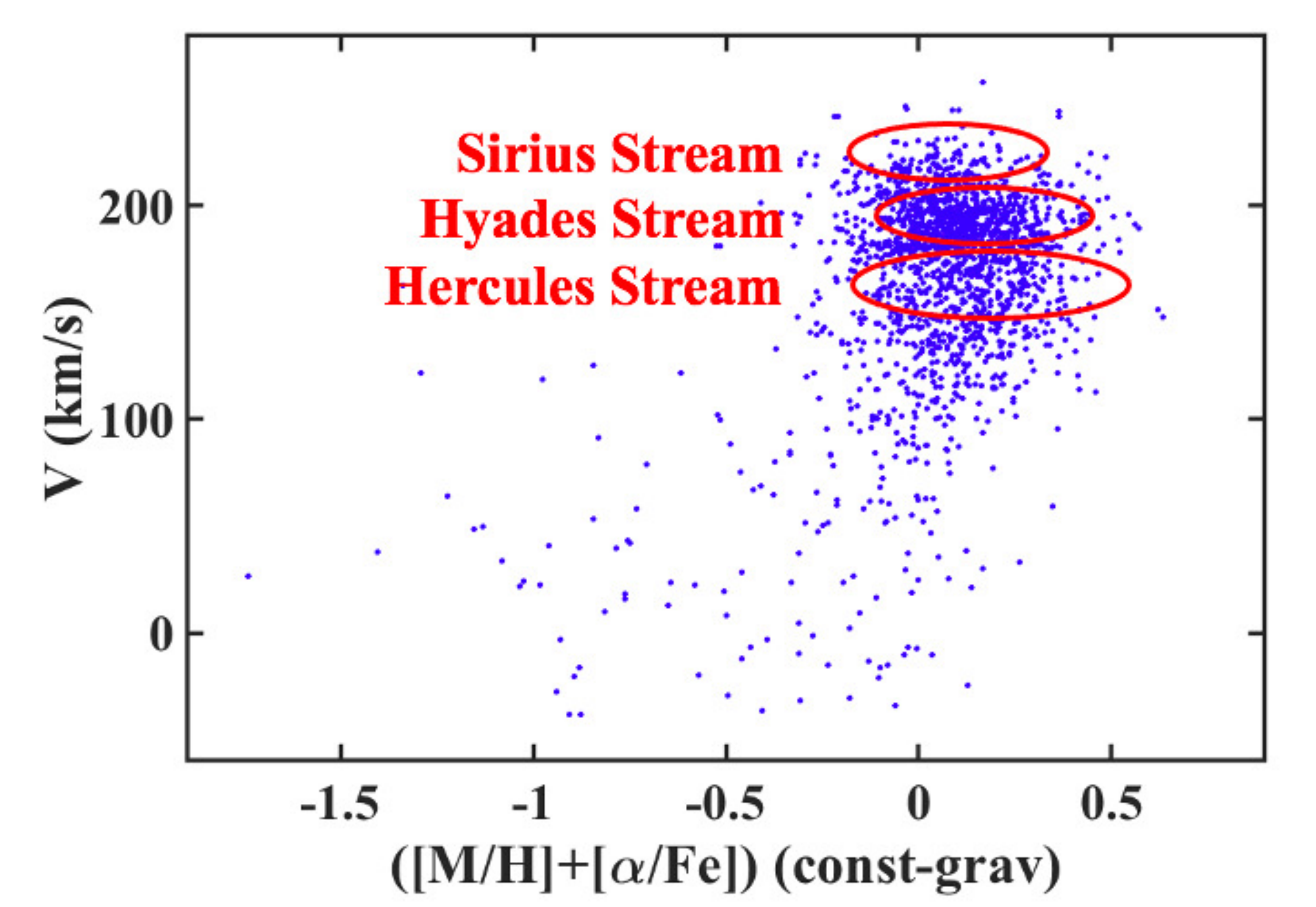}}
 \caption
        {\footnotesize{Velocity component V versus [M/H]+[$\alpha$/Fe] of the stars  with T$_\textrm{\footnotesize{eff}}$$\leq$3550 K and $-$2.5<[M/H]<+0.5 dex in the combined Group UV+VW, as shown in the bottom middle panel of Figure 42.  Three stellar streams are marked in red.}} 
\end{figure}

\section{Precision Analysis from Common Proper-Motion Pairs}

Wide binaries are believed to be formed in different processes (see e.g., Hawkins et al. 2019 for more details). Pairs with smaller  separations, i.e., from  a hundred to a few thousand AU, are more likely to form during the turbulent  fragmentation of a molecular core (e.g. Offner et al. 2010; Lee et al. 2017). Wider pairs with separations between 0.01 to 1 pc  are thought to be born  by the dynamical scattering of  one component in  triple systems (Reipurth \& Mikkola 2012), the dissipation of stellar clusters (e.g. Kouwenhoven et al. 2010; Moeckel \& Clarke 2011), or the pairing of  adjacent pre-stellar  cores  that move slowly enough relative to each other (e.g. Tokovinin 2017).  In most cases, it can be assumed that the two components in wide binary systems are formed from the same molecular cloud and have nearly the same chemical composition, and this is generally borne out by most observations of co-moving pairs. Occasional observed discrepancies in the elemental abundances of the components in some co-moving pairs have been attributed to  the general dynamical coldness of the thin disk or resonances induced by the Galactic tidal field (e.g., Simpson et al. 2019). It has been argued that these pairs of stars are ``coincidentally'' on similar orbits because of such dynamical interactions. Detailed investigations of chemical abundance distributions of stellar groups or pairs  are of great importance to understand the circumstances under which the members exhibit significant abundance differences and to find out whether these stars truly originate from a single parent cloud or have coincidentally converged to a similar orbit.

We perform a proper-motion pair searching over our sample  of the 3745  M dwarfs/subdwarfs using their Gaia coordinates, proper motions, and parallaxes \footnote{Our routine first separates any two stars with a small difference in their  equatorial coordinates, i.e., right ascension and declination. These pairs are then checked based on their parallaxes and proper motion components along right ascension and declination to find common proper-motion pairs. Our inspection confirms that  these pairs are real binary systems}. We identify 130 common proper-motion pairs, most of which   have metallicity estimates in the near-solar metallicity range ($-$0.5$\leq$[M/H]$\leq$+0.5 dex), but with a few systems having metal-poor metal-poor components. Figures 38 and 39 compare the [M/H], [$\alpha$/Fe] and [M/H]+[$\alpha$/Fe] of the primaries with those  of their respective companions, using the normal and ``reduced-correlation method'', respectively. For each method, we also examined the parameter values using the two gravity-modeling approaches, i.e., when surface gravity is regarded as a fixed parameter (top panels) and when surface gravity is allowed to vary (bottom panels). In all cases, we exclude any pair with at least one component having T$_\textrm{\footnotesize{eff}}$$>$3550K, whose best-fit estimates may suffer from larger systematic uncertainties.  The standard deviation of  the differences in parameter values associated with the primaries and their companions is shown in each panel.

As can be seen from these figures,   the model-fit values inferred from the constant-gravity approach  generally show a  better agreement for the two components in each binary system, as compared to those obtained from the variable-gravity approach. In addition, the ``reduced-correlation method'' does not necessarily improve the precision of the resulting best-fit estimates;  there is only a slight difference between the standard deviation of the differences in parameter values from the normal and ``reduced-correlation method'' for a given gravity-modeling approach. This again indicates that the residual correlations between wavelength datapoints have no considerable effect on the derived parameters  and  systematic issues in modeling stellar atmospheres can  cause  significant  uncertainties in the synthetic fitting using low-resolution spectra. Nevertheless, our measured  parameters are  still sufficiently precise  to present  a reasonable level of chemical homogeneity between the components of the binary systems. Considering its high efficiency,  our pipeline  can therefore be used for large samples of common proper-motion pairs to examine the chemical parameters of their components.

The primary and companion of the  74 pairs along with their inferred classification and chemical parameters [M/H] and [$\alpha$/Fe] from the ``normal method'' using the constant-gravity approach are listed in Table 3.

\section{Kinematic Analysis}

We calculate  the Galactic velocities U (towards the Galactic center), V (in the direction of Galactic rotation), and  W (towards the north of the Galactic plane) of our stars and then plot the  two-dimension projections on  the UV, VW and UW planes. Although we do derive a radial velocity correction for our stars to find their respective  best-fit models, these velocities are unreliable because the long-slit spectra were collected primarily for spectral classification purposes, and no radial velocity calibration standards were collected, which makes the absolute wavelength calibration of the stars relatively uncertain. In  addition, the radial velocities that are inferred from synthetic fitting are dependent on the atomic and molecular line list used in model atmospheres. We find  some offsets  between the radial velocities inferred from the model fitting using the PHOENIX and MARCS (Gustafsson et al. 2008) model atmospheres that use different sources of line lists. Although such velocity corrections are precise enough to determine best-fit model spectra, they are not reliable to calculate the Galactic velocity components U, V, and W. Moreover, none of  the M dwarf/subdwarfs in our subset have radial velocity from the Gaia catalog (they are too faint for this). 

For this reason, we compute these components of motion  to a first approximation assuming radial velocity $v_{r}$=0. The computed  U, V, and W velocities  are shifted to a Galactocentric reference frame assuming that the Sun has a component of motion [11.1 km/s, 12.24 km/s, 7.25 km/s] relative to the local standard of rest (Sch\"onrich et al. 2010), and that the local standard of rest has a component of motion [U, V, W]=[0, +220km/s, 0] relative to the Galactic center. To construct the UV, VW and UW kinematic plots, we use only stars in specific sectors of the sky to minimize the effect of the missing radial velocities\footnote{This can be perceived from the formulation of U, V, and W:  \newline U= $v_{r}$cos(\textit{l})cos(\textit{b})$-$$v_{l}$sin(\textit{l})$-$$v_{b}$cos(\textit{l})sin(\textit{b}), \newline  V= $v_{r}$sin(\textit{l})cos(\textit{b})+$v_{l}$cos(\textit{l})$-$$v_{b}$sin(\textit{l})sin(\textit{b}), \newline W= $v_{r}$sin(\textit{b})+$v_{b}$cos(\textit{b}), \newline where $v_{r}$ is the radial velocity, \textit{l} is the Galactic longitude, \textit{b} is the Galactic latitude, and $v_{l}$ and $v_{b}$ are the tangential velocity components along \textit{l} and \textit{b}, respectively. The position of stars on the three sectors of the sky,  with respect to \textit{l} and \textit{b}, makes the contribution of the first term that includes $v_{r}$ significantly small, for each of  the UV, VW, and UW cases.}. To this end,  we divide the 3745 stars into three groups  according to  their positions  on three different  sectors of the sky, as described below:

\begin{itemize}
\item 1311 stars with positions on the sky closest to the Galactic poles are only used for the UV plots (hereafter,  Group UV).
\item 1048 stars with positions closest to the Galactic center and anti-center are only used for the VW plots (hereafter, Group VW).
\item 1386 stars with positions closest to the solar apex and anti-apex are used only for the UW plots (hereafter, Group UW).
\end{itemize}

Figure 40 presents the projected motions in the two-dimensional UV, VW, and UW  diagrams for subsets of the respective Group UV (left panels), VW (middle panels), and UW (right panels), including all stars with  T$_\textrm{\footnotesize{eff}}$$\leq$3550 K and   $-$2.5<[M/H]<+0.5 dex,  color-mapped based on their [M/H] values  inferred from the  ``normal method'' when surface gravity is a constant parameter.  In all panels, the  metal-poor  stars are well separated both chemically and kinematically  from the   metal-rich  stars,
which is consistent with the metal-poor M subdwarfs ([M/H]<$-$0.5) generally being members of the local Galactic halo population, while the metal-rich M dwarfs ([M/H]$\geq$$-$0.5) are members of the local Galactic disk population.  As expected, the metal-poor stars, that belong to older (halo) stellar populations, show larger dispersions in the Galactic  velocities U, V, and W. On the other hand, the metal-rich stars, that are younger, have smaller dispersions in their velocity components. There is no noticeable  difference in the plots using the model-fit values from the ``reduced-correlation method'',  as compared to those in Figure 40, and we do not include them in this paper.

An alternative way to show the contrast between the halo and disk stars in chemodynamical space is to color-map stars in stars in the [$\alpha$/Fe]  versus [M/H] diagram according to their U, V, and W velocity components. For this purpose, we combine  Groups UV and UW (hereafter, Group UV+UW), Groups UV and VW (hereafter, Group UV+VW), and Groups UW and VW (hereafter, Group UW+VW), and then examine the distribution of the combined groups for stars with  T$_\textrm{\footnotesize{eff}}$$\leq$3550 K and   $-$2.5<[M/H]<+0.5 dex in the [$\alpha$/Fe]-[M/H] plane, as shown in Figure 41. The parameter model-fit estimates are those derived from the pipeline where surface gravity is kept fixed and randomized within the corresponding step size in the model grid. The distributions are color-mapped according to U, V, and W velocity components for  Groups UV+UW, UV+VW, and UW+VW, respectively\footnote{In order to make a better color map with a clearer contrast between the two populations, i.e., metal-poor and metal-rich stars, for the W velocity components, we exclude 5 stars with W<$-$200 km/s and W>350 km/s.}.  

The top panel shows  a systematic difference in U velocity  between the  metal-rich and   the metal-poor stars;  the higher-metallicity stars tend to have lower velocity components towards the Galactic center, which is  consistent with the motion of the Galactic disk  stars. On the other hand,   the lower-metallicity stars possess more randomly directed velocities with higher values of U component, which is compatible with the motion of the Galactic halo stars. The middle panel also confirms that the metal-rich stars have velocities that are characteristic  of  the disk  with  higher rotational  velocities, whereas the  metal-poor stars  show  random motions with lower V components. As seen in the bottom panel, the clear contrast in the W components  between the metal-rich and the metal-poor stars is again consistent with the difference between the   motion of  the  disk and halo populations. In  other words,  the metal-rich stars typically have lower velocity components perpendicular to the Galactic plane, which is an indication of the stellar motions associated with  the Galactic disk, while the metal-poor stars show higher vertical velocities, which is compatible with the more random orbits of the halo stars. The variation of Galactic velocity components in the chemical parameter space suggests a transition between the disk and halo stars around [M/H]$\approxeq$$-$0.5 dex.

The color maps in both Figures 40 and 41 emphasize the distinction between metal-poor halo stars and metal-rich disk stars, but the details of certain correlations between metallicity and velocity for the disk stars are mostly missed in these diagrams. For this reason, we plot the velocity components U, V, and W versus [M/H] and [M/H]+[$\alpha$/Fe] for the above-defined combined  Groups UV+UW, UV+VW, and UW+VW, respectively, but with narrower ranges of velocities for each group:  $-$200$\leq$U$\leq$200 km/s (Group UV+UW), $-$40$\leq$V$\leq$260 km/s (Group UV+VW) , and $-$100$\leq$W$\leq$100 km/s (Group UW+VW),  using the best-fit chemical parameter estimates inferred from the constant-gravity approach, as shown in Figure 42. The chemical parameter values are again randomized within the respective step size of the model grid. The plots related to the ``reduced-correlation method'' do not convey independent information and are not contained in this paper.  Apart from the offset  in the V values by $\sim$220 km/s due to our converted velocities to  the Galactic rest frame, the distributions are quite similar to the metallicity-velocity plots of   N04 (Figure 32 in that paper) for their subset of ~14000 nearby F and G dwarfs, and the distribution of our M dwarfs/subdwarfs  shows the same chemodynamical substructure. 

We identify three stellar streams in the bottom middle panel (``e'') of Figure 42, as  zoomed in Figure 43. There is a group of stars at the top of the distribution (V$\sim$220 km/s) which has an average value of  [M/H]+[$\alpha$/Fe]$\sim$0.0 dex and appears slightly shifted to the left on the plot relative to the overall distribution  of the disk stars.  These stars are  associated with the ``Sirius Stream''  that belongs to  the outer disk and are expected to be  slightly more metal-poor than the average disk stars. This group can be compared to the bulk of the disk stars (V$\sim$180-190 km/s) that have  an average value of  [M/H]+[$\alpha$/Fe]$\sim$0.1-0.2 dex and are marginally  more metal-rich than the stars in the Sirius Stream. Two other streams, i.e., Hyades and Hercules streams, are also marked in Figure 43. Additional substructure may originate from moving groups, or as mentioned in N04, from inhomogeneous Galactic potentials, e.g., by stochastic spiral waves (e.g., De Simone et al. 2004) or the bar (e.g., Fux 2001). 

In general, the range of metallicities for the disk stars ($\sim$0.5-0.6 dex) is remarkably similar to that of N04, which reinforces that our model-fit parameter values are highly precise. Furthermore, the average metallicity of our disk stars  has only an offset $\sim$0.1-0.2 dex relative to the average value of N04 (while the chemical parameter values from the variable-gravity approach  show a larger difference in average metallicity $\sim$0.3-0.35 dex with respect to that of N04), which indicates that our parameter estimates using the constant-gravity approach are accurate at the level of  $\sim$0.1-0.2 dex. Although the photometric surface gravities are relatively approximate and perhaps not entirely reliable (notably for unresolved binary stars), these values still yield reasonably accurate  estimates of T$_\textrm{\footnotesize{eff}}$, [M/H] and [M/H]+[$\alpha$/Fe]. If these estimates had very large uncertainties, we would not find subtle features in abundance-velocity diagrams and we would not reach  such a consistency with other previous studies.


\begin{deluxetable*}{lcccccc} [h!]

 \tablenum{1}
\tablecaption{Spectroscopic Catalog of the 3745 stars: Astrometry and Kinematics}
\tablewidth{0pt}
\tablehead{
\colhead{\footnotesize{Object}}   &  \colhead{\footnotesize{R.A. (Deg)}}  & \colhead{\footnotesize{Decl (Deg)}} &
 \colhead{$\mu_\textrm{\scriptsize{R.A.}}$ \footnotesize{($\arcsec$$\textrm{yr}^{-1}$}) } & \colhead{$\mu_\textrm{\scriptsize{Decl}}$ \footnotesize{($\arcsec$$\textrm{yr}^{-1}$}) }& \colhead{prlx \footnotesize{($\arcsec$)}}  & \colhead{\footnotesize{Gaia Flag}}}
\startdata
\footnotesize{PM J00012+0659} &  \footnotesize{0.31397} &  \footnotesize{6.99282} &  \footnotesize{$-$0.4363} &  \footnotesize{$-$0.0834} &  \footnotesize{0.04278} &  \footnotesize{DR3}\\
\footnotesize{PM J00031+0616} &  \footnotesize{0.78061} &  \footnotesize{6.27530} &  \footnotesize{ { }0.2389} &  \footnotesize{$-$0.5128} &  \footnotesize{0.03430} &  \footnotesize{DR3}\\
\footnotesize{PM J00051+4547} &  \footnotesize{1.30092} & \footnotesize{45.78589} &  \footnotesize{ { }0.8707} &  \footnotesize{$-$0.1514} &  \footnotesize{0.08693} &  \footnotesize{DR3}\\
\footnotesize{PM J00077+6022} &  \footnotesize{1.93035} & \footnotesize{60.38166} &  \footnotesize{ { }0.3216} &  \footnotesize{$-$0.0031} &  \footnotesize{0.06500} &  \footnotesize{DR3}\\
\footnotesize{PM J00078+6736} &  \footnotesize{1.96105} &  \footnotesize{67.60664} &  \footnotesize{$-$0.0559} &  \footnotesize{$-$0.0911} &  \footnotesize{0.03965} &  \footnotesize{DR3}\\
\footnotesize{PM J00081+4757} &  \footnotesize{2.02595} &  \footnotesize{47.95068} &  \footnotesize{$-$0.1241} &  \footnotesize{ { }0.0044} &  \footnotesize{0.05362} & \footnotesize{DR3}\\
\footnotesize{PM J00088+2050} &  \footnotesize{2.22444} &  \footnotesize{20.83929} &  \footnotesize{$-$0.0486} &  \footnotesize{$-$0.2602} &  \footnotesize{0.05526} &  \footnotesize{DR2}\\
\footnotesize{PM J00094+6549} &  \footnotesize{2.35085} &  \footnotesize{65.81710} &  \footnotesize{$-$0.0827} &  \footnotesize{$-$0.0453} &  \footnotesize{0.00977} &  \footnotesize{DR3}\\
\footnotesize{PM J00095+6548} &  \footnotesize{2.38440} &  \footnotesize{65.80898} &  \footnotesize{$-$0.0788} &  \footnotesize{$-$0.0450} &  \footnotesize{0.00963} &  \footnotesize{DR3}\\
\footnotesize{PM J00101+1327} &  \footnotesize{2.54048} &  \footnotesize{13.45336} &  \footnotesize{$-$0.2956} &  \footnotesize{$-$0.4169} &  \footnotesize{0.02281} &  \footnotesize{DR3}\\
\footnotesize{PM J00110+0420} &  \footnotesize{2.75393} &  \footnotesize{4.33799} &  \footnotesize{ { }0.1485} &  \footnotesize{$-$0.5234} &  \footnotesize{0.01229} &  \footnotesize{DR3}\\
\footnotesize{PM J00119+3303} &  \footnotesize{2.98221} &  \footnotesize{33.05296} &  \footnotesize{$-$0.5566} &  \footnotesize{$-$0.3965} &  \footnotesize{0.04775} &  \footnotesize{DR3}\\
\footnotesize{PM J00120+0206} &  \footnotesize{3.01249} &  \footnotesize{2.10687} &  \footnotesize{ { }0.3022} &  \footnotesize{ { }0.0658} &  \footnotesize{0.04209} &  \footnotesize{DR3}\\
\footnotesize{PM J00122+3028} &  \footnotesize{3.05619} &  \footnotesize{30.47884} &  \footnotesize{ { }0.0584} &  \footnotesize{ $-$0.0231} &  \footnotesize{ 0.01166} &  \footnotesize{DR2}\\
\footnotesize{PM J00123+6142} &  \footnotesize{3.07600} &  \footnotesize{61.71170} &  \footnotesize{$-$0.0764} &  \footnotesize{ { }0.0379} &  \footnotesize{0.02986} &  \footnotesize{DR3}\\
\footnotesize{PM J00131+7023} &  \footnotesize{3.29868} &  \footnotesize{70.39858} &  \footnotesize{ { }0.0432} &  \footnotesize{ { }0.1392} & \footnotesize{0.03153} &  \footnotesize{DR3}\\
\footnotesize{PM J00133+3908} &  \footnotesize{3.33305} & \footnotesize{39.14572} &  \footnotesize{$-$0.4503} &  \footnotesize{$-$0.1882} &  \footnotesize{0.02713} &  \footnotesize{DR3}\\
\footnotesize{PM J00137+8039} &  \footnotesize{3.43557} &  \footnotesize{80.66449} &  \footnotesize{ { }0.2508} &  \footnotesize{ { }0.1944} &  \footnotesize{0.05203} &  \footnotesize{DR2}\\
\footnotesize{PM J00138+3537} &  \footnotesize{3.45507} &  \footnotesize{35.61685} &  \footnotesize{$-$0.1917} &  \footnotesize{$-$0.4274} &  \footnotesize{0.03687} &  \footnotesize{DR2}\\
\footnotesize{PM J00144+3609N} &  \footnotesize{3.60644} &  \footnotesize{36.15699} &  \footnotesize{ { }0.1998} &  \footnotesize{$-$0.0905} &  \footnotesize{0.00607} &  \footnotesize{DR3}\\
\enddata
\tablecomments{This table is available in its entirety in a machine-readable  form in the online journal. A portion is shown here for guidance regarding its form and content.}
\end{deluxetable*}

\begin{deluxetable*}{lcccccccc} [h!]
 \tablenum{2}
\tablecaption{Spectroscopic Catalog of the 3745 stars: Classification and Photometry}
\tablewidth{0pt}
\tablehead{
\colhead{\footnotesize{Object}} & \colhead{\footnotesize{Spectral Type }}  & \colhead{\footnotesize{Metallicity Class}}     
 & \colhead{\footnotesize{\textit{G}}}     & \colhead{\footnotesize{\textit{G}$_\textrm{\footnotesize{BP}}$}}  & \colhead{\footnotesize{\textit{G}$_\textrm{\footnotesize{RP}}$}}   & \colhead{\footnotesize{\textit{J}}} & \colhead{\footnotesize{\textit{H}}} & \colhead{\footnotesize{\textit{K}}}}
\startdata
\footnotesize{PM J00012+0659} & \footnotesize{dM6.0} & \footnotesize{2} & \footnotesize{14.67} & \footnotesize{16.77} & \footnotesize{13.33} & \footnotesize{11.29} & \footnotesize{10.74} & \footnotesize{10.42}\\
\footnotesize{PM J00031+0616} & \footnotesize{dM4.5} & \footnotesize{2} & \footnotesize{14.00} & \footnotesize{15.67} & \footnotesize{12.76} & \footnotesize{11.04} & \footnotesize{10.53} & \footnotesize{10.30}\\
\footnotesize{PM J00051+4547} & \footnotesize{dM1.5} & \footnotesize{2} & \footnotesize{9.09} & \footnotesize{10.18} & \footnotesize{8.05} & \footnotesize{6.70} & \footnotesize{6.10} & \footnotesize{5.85}\\
\footnotesize{PM J00077+6022} & \footnotesize{dM5.0} & \footnotesize{3} & \footnotesize{12.40} & \footnotesize{13.64} & \footnotesize{10.70} & \footnotesize{8.91} & \footnotesize{8.33} & \footnotesize{8.05}\\
\footnotesize{PM J00078+6736} & \footnotesize{dM2.0} & \footnotesize{2} & \footnotesize{11.03} & \footnotesize{12.12} & \footnotesize{9.81} & \footnotesize{8.36} & \footnotesize{7.73} & \footnotesize{7.51}\\
\footnotesize{PM J00081+4757} & \footnotesize{dM4.0} & \footnotesize{2} & \footnotesize{11.42} & \footnotesize{12.99} & \footnotesize{10.20} & \footnotesize{8.52} & \footnotesize{8.00} & \footnotesize{7.68}\\
\footnotesize{PM J00088+2050} & \footnotesize{dM5.0} & \footnotesize{2} & \footnotesize{11.99} & \footnotesize{13.77} & \footnotesize{10.72} & \footnotesize{8.87} & \footnotesize{8.26} & \footnotesize{8.01}\\
\footnotesize{PM J00094+6549} & \footnotesize{dM0.5} & \footnotesize{3} & \footnotesize{12.66} & \footnotesize{13.56} & \footnotesize{11.71} & \footnotesize{10.50} & \footnotesize{9.85} & \footnotesize{9.67}\\
\footnotesize{PM J00095+6548} & \footnotesize{dM3.5} & \footnotesize{2} & \footnotesize{14.79} & \footnotesize{16.13} & \footnotesize{13.65} & \footnotesize{12.14} & \footnotesize{11.56} & \footnotesize{11.33}\\
\footnotesize{PM J00101+1327} & \footnotesize{dM4.5} & \footnotesize{3} & \footnotesize{14.92} & \footnotesize{16.44} & \footnotesize{13.72} & \footnotesize{12.12} & \footnotesize{11.59} & \footnotesize{11.34}\\
\footnotesize{PM J00110+0420} & \footnotesize{esdM6.0} & \footnotesize{8} & \footnotesize{16.89} & \footnotesize{18.14} & \footnotesize{15.79} & \footnotesize{14.34} & \footnotesize{13.81} & \footnotesize{13.76}\\
\footnotesize{PM J00119+3303} & \footnotesize{dM4.0} & \footnotesize{2} & \footnotesize{11.83} & \footnotesize{13.27} & \footnotesize{10.65} & \footnotesize{9.07} & \footnotesize{8.40} & \footnotesize{8.16}\\
\footnotesize{PM J00120+0206} & \footnotesize{dM4.5} & \footnotesize{2} & \footnotesize{13.69} & \footnotesize{15.45} & \footnotesize{12.43} & \footnotesize{10.59} & \footnotesize{10.01} & \footnotesize{9.72}\\
\footnotesize{PM J00122+3028} & \footnotesize{dM6.5} & \footnotesize{2} & \footnotesize{13.39} & \footnotesize{15.10} & \footnotesize{12.11} & \footnotesize{10.24} & \footnotesize{9.68} & \footnotesize{9.41}\\
\footnotesize{PM J00123+6142} & \footnotesize{dM4.5} & \footnotesize{2} & \footnotesize{13.17} & \footnotesize{14.70} & \footnotesize{11.93} & \footnotesize{9.97} & \footnotesize{9.30} & \footnotesize{8.95}\\
\footnotesize{PM J00131+7023} & \footnotesize{dM1.5} & \footnotesize{3} & \footnotesize{10.44} & \footnotesize{11.40} & \footnotesize{9.47} & \footnotesize{8.26} & \footnotesize{7.59} & \footnotesize{7.39}\\
\footnotesize{PM J00133+3908} & \footnotesize{dM3.0} & \footnotesize{3} & \footnotesize{12.66} & \footnotesize{13.86} & \footnotesize{11.58} & \footnotesize{10.18} & \footnotesize{9.61} & \footnotesize{9.36}\\
\footnotesize{PM J00137+8039} & \footnotesize{dM6.5} & \footnotesize{3} & \footnotesize{14.42} & \footnotesize{15.65} & \footnotesize{12.93} & \footnotesize{10.94} & \footnotesize{10.37} & \footnotesize{10.06}\\
\footnotesize{PM J00138+3537} & \footnotesize{dM6.5} & \footnotesize{2} & \footnotesize{13.69} & \footnotesize{15.36} & \footnotesize{12.45} & \footnotesize{10.68} & \footnotesize{10.12} & \footnotesize{9.86}\\
\footnotesize{PM J00144+3609N} & \footnotesize{dM3.0} & \footnotesize{3} & \footnotesize{15.93} & \footnotesize{17.14} & \footnotesize{14.84} & \footnotesize{13.45} & \footnotesize{12.93} & \footnotesize{12.71}\\
\enddata
\tablecomments{This table is available in its entirety in a machine-readable  form in the online journal. A portion is shown here for guidance regarding its form and content.}
\end{deluxetable*}

\setcounter{table}{2}
\begin{table*} [ht]  
\caption {Classification and Chemical Parameters of the 74 Common Proper-Motion Pairs}
\begin{center} 
\Rotatebox{90}{
\begin{tabular}{llccc|llccc}
 \hline
\multicolumn{5}{c  |}{\footnotesize{Companion}}  &  \multicolumn{5}{c}{\footnotesize{Primary}}\\
\hline
\footnotesize{Name} & \footnotesize{Sp-Type} & \footnotesize{MC} & \footnotesize{[M/H]} & \footnotesize{[$\alpha$/Fe]} & \footnotesize{Name} & \footnotesize{Sp-Type} & \footnotesize{MC} & \footnotesize{[M/H]} & \footnotesize{[$\alpha$/Fe]} \\
\hline
\footnotesize{PM J00144+3609S} &\footnotesize{dM4.0} &\footnotesize{3} &\footnotesize{$-$0.25} &\footnotesize{+0.175} &\footnotesize{PM J00144+3609N} &\footnotesize{dM3.0} &\footnotesize{3} &\footnotesize{$-$0.10} &\footnotesize{+0.075}\\
\footnotesize{PM J00162+1951W} &\footnotesize{dM4.5} &\footnotesize{2} &\footnotesize{+0.15} &\footnotesize{+0.200} &\footnotesize{PM J00162+1951E} &\footnotesize{dM4.0} &\footnotesize{1} &\footnotesize{+0.10} &\footnotesize{+0.250}\\
\footnotesize{PM J00277+0936N} &\footnotesize{dM4.0} &\footnotesize{2} &\footnotesize{+0.00} &\footnotesize{+0.125} &\footnotesize{PM J00277+0936S} &\footnotesize{dM4.0} &\footnotesize{2} &\footnotesize{+0.15} &\footnotesize{+0.100}\\
\footnotesize{PM J00358+5241N} &\footnotesize{dM4.5} &\footnotesize{2} &\footnotesize{$-$0.10} &\footnotesize{+0.175} &\footnotesize{PM J00358+5241S} &\footnotesize{dM3.0} &\footnotesize{2} &\footnotesize{$-$0.10} &\footnotesize{+0.100}\\
\footnotesize{PM J00512+5840} &\footnotesize{dM4.5} &\footnotesize{2} &\footnotesize{$-$0.05} &\footnotesize{+0.150} &\footnotesize{PM J00512+5839} &\footnotesize{dM3.5} &\footnotesize{2} &\footnotesize{$-$0.05} &\footnotesize{+0.125}\\
\footnotesize{PM J01024+4101} &\footnotesize{dM4.5} &\footnotesize{2} &\footnotesize{$-$0.15} &\footnotesize{+0.100} &\footnotesize{PM J01024+4102} &\footnotesize{dM3.0} &\footnotesize{3} &\footnotesize{+0.05} &\footnotesize{$-$0.025}\\
\footnotesize{PM J01544+5741N} &\footnotesize{dM4.0} &\footnotesize{2} &\footnotesize{+0.05} &\footnotesize{+0.075} &\footnotesize{PM J01544+5741S} &\footnotesize{dM4.0} &\footnotesize{2} &\footnotesize{+0.00} &\footnotesize{+0.050}\\
\footnotesize{PM J02325+5313} &\footnotesize{dM5.0} &\footnotesize{2} &\footnotesize{$-$0.20} &\footnotesize{+0.225} &\footnotesize{PM J02324+5313} &\footnotesize{dM3.0} &\footnotesize{2} &\footnotesize{$-$0.25} &\footnotesize{+0.175}\\
\footnotesize{PM J02340+0950W} &\footnotesize{dM6.5} &\footnotesize{3} &\footnotesize{$-$0.10} &\footnotesize{+0.150} &\footnotesize{PM J02340+0950E} &\footnotesize{dM3.0} &\footnotesize{3} &\footnotesize{$-$0.40} &\footnotesize{+0.300}\\
\footnotesize{PM J02498+3945E} &\footnotesize{dM3.5} &\footnotesize{2} &\footnotesize{$-$0.15} &\footnotesize{+0.125} &\footnotesize{PM J02498+3945W} &\footnotesize{dM3.5} &\footnotesize{2} &\footnotesize{$-$0.10} &\footnotesize{+0.150}\\
\footnotesize{PM J02515+5922E} &\footnotesize{dM4.0} &\footnotesize{2} &\footnotesize{$-$0.15} &\footnotesize{+0.175} &\footnotesize{PM J02515+5922W} &\footnotesize{dM4.0} &\footnotesize{2} &\footnotesize{$-$0.10} &\footnotesize{+0.200}\\
\footnotesize{PM J03055+1149S} &\footnotesize{dM3.5} &\footnotesize{3} &\footnotesize{+0.00 }&\footnotesize{+0.050} &\footnotesize{PM J03055+1149N} &\footnotesize{dM3.0} &\footnotesize{3} &\footnotesize{+0.10} &\footnotesize{+0.000}\\
\footnotesize{PM J03101$-$0046N} &\footnotesize{dM3.5} &\footnotesize{3} &\footnotesize{$-$0.20} &\footnotesize{+0.225} &\footnotesize{PM J03101$-$0046S} &\footnotesize{dM3.5} &\footnotesize{2} &\footnotesize{$-$0.35} &\footnotesize{+0.300}\\
\footnotesize{PM J03265+5908W} &\footnotesize{dM4.5} &\footnotesize{2} &\footnotesize{$-$0.10} &\footnotesize{+0.100} &\footnotesize{PM J03265+5908E} &\footnotesize{dM4.0} &\footnotesize{3} &\footnotesize{$-$0.05} &\footnotesize{+0.075}\\
\footnotesize{PM J03375+1751S} &\footnotesize{dM4.0} &\footnotesize{2} &\footnotesize{+0.10} &\footnotesize{+0.025} &\footnotesize{PM J03375+1751N} &\footnotesize{dM3.0} &\footnotesize{3} &\footnotesize{+0.00} &\footnotesize{+0.000}\\
\footnotesize{PM J03576+5715E} &\footnotesize{sdM5.0} &\footnotesize{5} &\footnotesize{$-$0.70} &\footnotesize{+0.200} &\footnotesize{PM J03576+5715W} &\footnotesize{sdM4.0} &\footnotesize{6} &\footnotesize{$-$0.65} &\footnotesize{+0.150}\\
\footnotesize{PM J04059+7116W} &\footnotesize{dM6.0} &\footnotesize{2} &\footnotesize{$-$0.05} &\footnotesize{+0.225} &\footnotesize{PM J04059+7116E} &\footnotesize{dM4.5} &\footnotesize{2} &\footnotesize{+0.05} &\footnotesize{+0.175}\\
\footnotesize{PM J04248+0106E} &\footnotesize{dM5.0} &\footnotesize{2} &\footnotesize{+0.05} &\footnotesize{+0.150} &\footnotesize{PM J04248+0106W} &\footnotesize{dM3.0} &\footnotesize{2} &\footnotesize{$-$0.15} &\footnotesize{+0.150}\\
\footnotesize{PM J04278+2630S} &\footnotesize{dM4.5} &\footnotesize{2} &\footnotesize{$-$0.05} &\footnotesize{+0.200} &\footnotesize{PM J04278+2630N} &\footnotesize{dM3.5} &\footnotesize{2} &\footnotesize{$-$0.05} &\footnotesize{+0.125}\\
\footnotesize{PM J05342+1019S} &\footnotesize{dM4.0} &\footnotesize{2} &\footnotesize{+0.05} &\footnotesize{+0.150} &\footnotesize{PM J05342+1019N} &\footnotesize{dM3.5 }&\footnotesize{2} &\footnotesize{$-$0.05} &\footnotesize{+0.125}\\
\footnotesize{PM J05422$-$2219} &\footnotesize{dM6.5} &\footnotesize{3} &\footnotesize{$-$0.30} &\footnotesize{+0.250} &\footnotesize{PM J05422$-$2220} &\footnotesize{dM4.0} &\footnotesize{3} &\footnotesize{$-$0.35} &\footnotesize{+0.225}\\
\footnotesize{PM J06007+6808} &\footnotesize{dM4.0} &\footnotesize{2} &\footnotesize{$-$0.05} &\footnotesize{+0.150} &\footnotesize{PM J06008+6809} &\footnotesize{dM4.0} &\footnotesize{2} &\footnotesize{$-$0.10} &\footnotesize{+0.150}\\
\footnotesize{PM J06381+2219E} &\footnotesize{dM4.0} &\footnotesize{2} &\footnotesize{$-$0.20} &\footnotesize{+0.225} &\footnotesize{PM J06381+2219W} &\footnotesize{dM3.0} &\footnotesize{2} &\footnotesize{$-$0.05} &\footnotesize{+0.100}\\
\footnotesize{PM J07390+7734W} &\footnotesize{dM5.5} &\footnotesize{2} &\footnotesize{+0.05} &\footnotesize{+0.175} &\footnotesize{PM J07390+7734E} &\footnotesize{dM4.5} &\footnotesize{2} &\footnotesize{+0.00} &\footnotesize{+0.150}\\
\footnotesize{PM J07430+5109E} &\footnotesize{dM3.5} &\footnotesize{3} &\footnotesize{+0.05} &\footnotesize{0.075} &\footnotesize{PM J07430+5109W} &\footnotesize{dM3.0} &\footnotesize{2} &\footnotesize{+0.00} &\footnotesize{+0.050}\\
\footnotesize{PM J08398+0856E} &\footnotesize{dM4.5} &\footnotesize{3} &\footnotesize{$-$0.15} &\footnotesize{+0.075} &\footnotesize{PM J08398+0856W} &\footnotesize{sdM3.0} &\footnotesize{4} &\footnotesize{$-$0.05} &\footnotesize{+0.025}\\
\footnotesize{PM J09070+7226} &\footnotesize{dM5.5} &\footnotesize{2} &\footnotesize{+0.00} &\footnotesize{+0.200} &\footnotesize{PM J09069+7226} &\footnotesize{dM4.5} &\footnotesize{2} &\footnotesize{+0.00} &\footnotesize{+0.175}\\
\footnotesize{PM J09112+0127N} &\footnotesize{dM4.5} &\footnotesize{2} &\footnotesize{+0.25} &\footnotesize{+0.100} &\footnotesize{PM J09112+0127S} &\footnotesize{dM4.0} &\footnotesize{1} &\footnotesize{+0.15} &\footnotesize{+0.125}\\
\footnotesize{PM J09593+4350E} &\footnotesize{dM5.0} &\footnotesize{3} &\footnotesize{$-$0.20} &\footnotesize{+0.125} &\footnotesize{PM J09593+4350W} &\footnotesize{dM4.5} &\footnotesize{3} &\footnotesize{$-$0.10} &\footnotesize{+0.075}\\
\footnotesize{PM J10182$-$2028W} &\footnotesize{dM6.5} &\footnotesize{2} &\footnotesize{+0.25} &\footnotesize{+0.250} &\footnotesize{PM J10182$-$2028E} &\footnotesize{dM5.5} &\footnotesize{2} &\footnotesize{+0.20} &\footnotesize{+0.225}\\
\footnotesize{PM J10260+5027N} &\footnotesize{dM4.0} &\footnotesize{2} &\footnotesize{+0.00} &\footnotesize{+0.100} &\footnotesize{PM J10260+5027S} &\footnotesize{dM4.0} &\footnotesize{2} &\footnotesize{+0.00} &\footnotesize{+0.125}\\
\footnotesize{PM J10344+4618} &\footnotesize{dM5.5} &\footnotesize{2} &\footnotesize{+0.15} &\footnotesize{+0.125} &\footnotesize{PM J10345+4618} &\footnotesize{dM3.5} &\footnotesize{2} &\footnotesize{+0.00} &\footnotesize{+0.100}\\
\footnotesize{PM J10449+3224} &\footnotesize{dM5.0} &\footnotesize{2} &\footnotesize{+0.05} &\footnotesize{+0.125} &\footnotesize{PM J10448+3224} &\footnotesize{dM3.5} &\footnotesize{2} &\footnotesize{$-$0.10} &\footnotesize{+0.100}\\
\footnotesize{PM J11285+5643N} &\footnotesize{dM4.5} &\footnotesize{2} &\footnotesize{$-$0.05} &\footnotesize{+0.125} &\footnotesize{PM J11285+5643S} &\footnotesize{dM4.5} &\footnotesize{2} &\footnotesize{+0.00} &\footnotesize{+0.125}\\
\footnotesize{PM J11488+1800E} &\footnotesize{dM4.0} &\footnotesize{2} &\footnotesize{+0.05} &\footnotesize{+0.100} &\footnotesize{PM J11488+1800W} &\footnotesize{dM4.0} &\footnotesize{2} &\footnotesize{+0.00} &\footnotesize{+0.100}\\
\footnotesize{PM J12091+4735} &\footnotesize{dM5.0} &\footnotesize{2} &\footnotesize{$-$0.05} &\footnotesize{+0.150} &\footnotesize{PM J12091+4736} &\footnotesize{dM5.0} &\footnotesize{2} &\footnotesize{$-$0.05} &\footnotesize{+0.125}\\
\footnotesize{PM J14282+6349} &\footnotesize{dM4.0} &\footnotesize{3} &\footnotesize{$-$0.15} &\footnotesize{+0.075} &\footnotesize{PM J14283+6349} &\footnotesize{dM4.0} &\footnotesize{3} &\footnotesize{$-$0.20} &\footnotesize{+0.100}\\
\hline
\end{tabular}  
}
\end{center}
\end{table*}

\setcounter{table}{2}
\begin{table*} [ht]  
\caption {Continued}
\begin{center} 
\Rotatebox{90}{
\begin{tabular}{llccc|llccc}
 \hline
\multicolumn{5}{c  |}{\footnotesize{Companion}}  &  \multicolumn{5}{c }{\footnotesize{Primary}}\\
\hline
\footnotesize{Name} & \footnotesize{Sp-Type} & \footnotesize{MC} & \footnotesize{[M/H]} & \footnotesize{[$\alpha$/Fe]} & \footnotesize{Name} & \footnotesize{Sp-Type} & \footnotesize{MC} & \footnotesize{[M/H]} & \footnotesize{[$\alpha$/Fe]} \\
\hline
\footnotesize{PM J14372+7536} &\footnotesize{sdM3.0} &\footnotesize{4} &\footnotesize{$-$0.35} &\footnotesize{+0.075} &\footnotesize{PM J14371+7536} &\footnotesize{dM3.0} &\footnotesize{3} &\footnotesize{$-$0.40} &\footnotesize{+0.175}\\
\footnotesize{PM J14518+5147W} &\footnotesize{dM5.0} &\footnotesize{3} &\footnotesize{$-$0.55} &\footnotesize{+0.275} &\footnotesize{PM J14518+5147E} &\footnotesize{dM3.0} &\footnotesize{3} &\footnotesize{$-$0.40} &\footnotesize{+0.175}\\
\footnotesize{PM J14567+1245} &\footnotesize{dM4.0} &\footnotesize{3} &\footnotesize{+0.20} &\footnotesize{$-$0.025} &\footnotesize{PM J14568+1245} &\footnotesize{dM4.0} &\footnotesize{3} &\footnotesize{+0.20} &\footnotesize{$-$0.025}\\
\footnotesize{PM J15091+1513E} &\footnotesize{dM6.5} &\footnotesize{2} &\footnotesize{+0.00} &\footnotesize{+0.200} &\footnotesize{PM J15091+1513W} &\footnotesize{dM3.0} &\footnotesize{2} &\footnotesize{+0.15} &\footnotesize{+0.025}\\
\footnotesize{PM J15308+5608N} &\footnotesize{esdM6.0} &\footnotesize{7} &\footnotesize{$-$1.30} &\footnotesize{+0.275} &\footnotesize{PM J15308+5608S} &\footnotesize{esdM6.0} &\footnotesize{7} &\footnotesize{$-$1.25} &\footnotesize{+0.250}\\
\footnotesize{PM J15353+1743} &\footnotesize{dM5.0} &\footnotesize{2} &\footnotesize{$-$0.30} &\footnotesize{+0.225} &\footnotesize{PM J15353+1742} &\footnotesize{dM3.5} &\footnotesize{3} &\footnotesize{$-$0.15} &\footnotesize{+0.150}\\
\footnotesize{PM J15378+0017E} &\footnotesize{sdM5.0} &\footnotesize{5} &\footnotesize{$-$1.05} &\footnotesize{+0.225} &\footnotesize{PM J15378+0017W} &\footnotesize{esdM4} &\footnotesize{7} &\footnotesize{$-$1.20} &\footnotesize{+0.325}\\
\footnotesize{PM J17023+1647N} &\footnotesize{dM3.5} &\footnotesize{3} &\footnotesize{$-$0.30} &\footnotesize{+0.200} &\footnotesize{PM J17023+1647S} &\footnotesize{dM3.5} &\footnotesize{3} &\footnotesize{$-$0.10} &\footnotesize{+0.050}\\
\footnotesize{PM J17184$-$2246} &\footnotesize{dM5.0} &\footnotesize{2} &\footnotesize{+0.00} &\footnotesize{+0.125} &\footnotesize{PM J17184$-$2245} &\footnotesize{dM4.0} &\footnotesize{3 }&\footnotesize{$-$0.15} &\footnotesize{+0.200}\\
\footnotesize{PM J17193$-$2949N} &\footnotesize{dM5.0} &\footnotesize{2} &\footnotesize{$-$0.05} &\footnotesize{+0.175} &\footnotesize{PM J17193$-$2949S} &\footnotesize{dM4.5} &\footnotesize{2} &\footnotesize{+0.00} &\footnotesize{+0.150}\\
\footnotesize{PM J17277+5200N} &\footnotesize{dM5.0} &\footnotesize{2} &\footnotesize{+0.05} &\footnotesize{+0.100} &\footnotesize{PM J17277+5200S} &\footnotesize{dM5.0} &\footnotesize{2} &\footnotesize{+0.05} &\footnotesize{+0.100}\\
\footnotesize{PM J18180+3846W} &\footnotesize{dM4.5} &\footnotesize{2} &\footnotesize{$-$0.25} &\footnotesize{+0.275} &\footnotesize{PM J18180+3846E} &\footnotesize{dM3.5} &\footnotesize{2} &\footnotesize{$-$0.25} &\footnotesize{+0.225}\\
\footnotesize{PM J18216+3840S} &\footnotesize{dM5.0} &\footnotesize{2} &\footnotesize{+0.05} &\footnotesize{+0.100} &\footnotesize{PM J18216+3840N} &\footnotesize{dM4.0} &\footnotesize{3} &\footnotesize{$-$0.05} &\footnotesize{+0.075}\\
\footnotesize{PM J18427+5937S} &\footnotesize{dM4.0} &\footnotesize{2} &\footnotesize{$-$0.20} &\footnotesize{+0.275} &\footnotesize{PM J18427+5937N} &\footnotesize{dM3.5} &\footnotesize{2} &\footnotesize{$-$0.25} &\footnotesize{+0.300}\\
\footnotesize{PM J18447+3238} &\footnotesize{dM5.5} &\footnotesize{3} &\footnotesize{+0.20} &\footnotesize{+0.000} &\footnotesize{PM J18447+3237} &\footnotesize{dM3.0} &\footnotesize{2} &\footnotesize{+0.00} &\footnotesize{+0.050}\\
\footnotesize{PM J19096+1401E} &\footnotesize{dM3.5} &\footnotesize{2} &\footnotesize{+0.00} &\footnotesize{+0.075} &\footnotesize{PM J19096+1401W} &\footnotesize{dM3.0} &\footnotesize{2} &\footnotesize{+0.05} &\footnotesize{+0.025}\\
\footnotesize{PM J19113+5224E} &\footnotesize{dM3.5} &\footnotesize{2} &\footnotesize{+0.00} &\footnotesize{+0.100} &\footnotesize{PM J19113+5224W} &\footnotesize{dM3.5} &\footnotesize{1} &\footnotesize{+0.00} &\footnotesize{+0.125}\\
\footnotesize{PM J19312+3607} &\footnotesize{dM5.0} &\footnotesize{2} &\footnotesize{+0.15} &\footnotesize{+0.200} &\footnotesize{PM J19311+3608} &\footnotesize{dM4.5} &\footnotesize{2} &\footnotesize{+0.10} &\footnotesize{+0.200}\\
\footnotesize{PM J19317+4506E} &\footnotesize{dM4.0} &\footnotesize{2} &\footnotesize{+0.15} &\footnotesize{+0.000} &\footnotesize{PM J19317+4506W} &\footnotesize{dM4.0} &\footnotesize{2} &\footnotesize{+0.15} &\footnotesize{+0.025}\\
\footnotesize{PM J19388+3512E} &\footnotesize{dM5.0} &\footnotesize{3} &\footnotesize{$-$0.35} &\footnotesize{+0.200} &\footnotesize{PM J19388+3512W} &\footnotesize{dM4.0} &\footnotesize{3} &\footnotesize{+0.00} &\footnotesize{+0.000}\\
\footnotesize{PM J19433+5433E} &\footnotesize{dM4.0} &\footnotesize{2} &\footnotesize{+0.35} &\footnotesize{+0.000} &\footnotesize{PM J19433+5433W} &\footnotesize{dM4.0} &\footnotesize{3} &\footnotesize{+0.30} &\footnotesize{+0.000}\\
\footnotesize{PM J20231+1844E} &\footnotesize{dM4.0} &\footnotesize{3} &\footnotesize{$-$0.30} &\footnotesize{+0.200} &\footnotesize{PM J20231+1844W} &\footnotesize{dM3.5} &\footnotesize{2} &\footnotesize{$-$0.35} &\footnotesize{+0.275}\\
\footnotesize{PM J21013+3314} &\footnotesize{dM4.0} &\footnotesize{3} &\footnotesize{$-$0.15} &\footnotesize{+0.100} &\footnotesize{PM J21012+3314} &\footnotesize{dM3.5} &\footnotesize{3} &\footnotesize{$-$0.05} &\footnotesize{+0.050}\\
\footnotesize{PM J21080+3116S} &\footnotesize{dM5.5} &\footnotesize{2} &\footnotesize{$-$0.20} &\footnotesize{+0.200} &\footnotesize{PM J21080+3116N} &\footnotesize{dM3.0} &\footnotesize{2} &\footnotesize{$-$0.15} &\footnotesize{+0.150}\\
\footnotesize{PM J21160+2951E} &\footnotesize{dM4.0} &\footnotesize{2} &\footnotesize{$-$0.10} &\footnotesize{+0.050} &\footnotesize{PM J21160+2951W} &\footnotesize{dM4.0} &\footnotesize{2} &\footnotesize{$-$0.15} &\footnotesize{+0.100}\\
\footnotesize{PM J22173$-$0848S} &\footnotesize{dM5.0} &\footnotesize{2} &\footnotesize{$-$0.20} &\footnotesize{+0.200} &\footnotesize{PM J22173-0848N} &\footnotesize{dM5.0} &\footnotesize{3} &\footnotesize{$-$0.20} &\footnotesize{+0.175}\\
\footnotesize{PM J22176+6010W} &\footnotesize{dM4.0} &\footnotesize{2} &\footnotesize{+0.00} &\footnotesize{+0.125} &\footnotesize{PM J22176+6010E} &\footnotesize{dM3.5} &\footnotesize{2} &\footnotesize{+0.00} &\footnotesize{+0.100}\\
\footnotesize{PM J22297+2226N} &\footnotesize{dM6.5} &\footnotesize{2} &\footnotesize{+0.20} &\footnotesize{+0.050} &\footnotesize{PM J22297+2226S} &\footnotesize{dM3.0} &\footnotesize{3} &\footnotesize{$-$0.20} &\footnotesize{+0.125}\\
\footnotesize{PM J22343+3129} &\footnotesize{dM4.0} &\footnotesize{2} &\footnotesize{$-$0.10} &\footnotesize{+0.175} &\footnotesize{PM J22342+3128} &\footnotesize{dM3.5} &\footnotesize{2} &\footnotesize{+0.05} &\footnotesize{+0.075}\\
\footnotesize{PM J22412$-$1513E} &\footnotesize{dM3.5} &\footnotesize{1} &\footnotesize{$-$0.05} &\footnotesize{+0.150} &\footnotesize{PM J22412$-$1513W} &\footnotesize{dM3.5} &\footnotesize{1} &\footnotesize{$-$0.05} &\footnotesize{+0.225}\\
\footnotesize{PM J23257+1735} &\footnotesize{dM6.5} &\footnotesize{2} &\footnotesize{$-$0.35} &\footnotesize{+0.225} &\footnotesize{PM J23258+1735} &\footnotesize{dM4.5} &\footnotesize{3} &\footnotesize{$-$0.30} &\footnotesize{+0.200}\\
\footnotesize{PM J23293+4127} &\footnotesize{dM4.5} &\footnotesize{2} &\footnotesize{$-$0.05} &\footnotesize{+0.100} &\footnotesize{PM J23293+4128} &\footnotesize{dM4.0} &\footnotesize{2} &\footnotesize{+0.05} &\footnotesize{+0.050}\\
\footnotesize{PM J23310+0842N} &\footnotesize{dM4.0} &\footnotesize{2} &\footnotesize{$-$0.10} &\footnotesize{+0.100} &\footnotesize{PM J23310+0842S} &\footnotesize{dM3.0} &\footnotesize{2} &\footnotesize{$-$0.15} &\footnotesize{+0.150}\\
\footnotesize{PM J23578+7836} &\footnotesize{dM3.5} &\footnotesize{3} &\footnotesize{$-$0.10} &\footnotesize{+0.000} &\footnotesize{PM J23580+7836} &\footnotesize{dM3.0} &\footnotesize{3} &\footnotesize{$-$0.10} &\footnotesize{+0.025}\\
\footnotesize{PM J02340+0950W} &\footnotesize{dM6.5} &\footnotesize{3 }&\footnotesize{$-$0.10} &\footnotesize{+0.150} &\footnotesize{PM J02340+0950E} &\footnotesize{dM3.0} &\footnotesize{3} &\footnotesize{$-$0.40} &\footnotesize{+0.300}\\
\footnotesize{PM J10182$-$2028W} &\footnotesize{dM6.5} &\footnotesize{2} &\footnotesize{+0.25} &\footnotesize{+0.250} &\footnotesize{PM J10182$-$2028E} &\footnotesize{dM5.5} &\footnotesize{2} &\footnotesize{+0.20} &\footnotesize{+0.225}\\
\footnotesize{PM J17184$-$2246} &\footnotesize{dM5.0} &\footnotesize{2 }&\footnotesize{+0.00} &\footnotesize{+0.125} &\footnotesize{PM J17184$-$2245} &\footnotesize{dM4.0} &\footnotesize{3} &\footnotesize{$-$0.15} &\footnotesize{+0.200}\\
\hline
\end{tabular}  
}
\end{center}
\end{table*}

\setcounter{table}{3}
\begin{table*} [ht]  
\caption {Spectroscopic Catalog of the 3745 stars: Model-Fit Stellar Parameters\\
\scriptsize{This table is available in its entirety in a machine-readable  form in the online journal. A portion is shown here for guidance regarding its form and content.}}
\begin{center} 
\Rotatebox{90}{
\begin{tabular}{lccccccccccccc}
\hline
\hline
\scriptsize{Object} & \scriptsize{($\textrm{T}_\textrm{eff}$)$_\textrm{N}$ (K)} & \scriptsize{$\sigma$($\textrm{T}_\textrm{eff}$)$_\textrm{N}$ (K)} &   \scriptsize{[M/H]$_\textrm{N}$}  &   \scriptsize{$\sigma$[M/H]$_\textrm{N}$} & \scriptsize{[$\alpha$/Fe]$_\textrm{N}$} & \scriptsize{$\sigma$[$\alpha$/Fe]$_\textrm{N}$} &  \scriptsize{($\textrm{T}_\textrm{eff}$)$_\textrm{RC}$ (K)}  &  \scriptsize{$\sigma$($\textrm{T}_\textrm{eff}$)$_\textrm{RC}$ (K)} &   \scriptsize{[M/H]$_\textrm{RC}$} & \scriptsize{$\sigma$[M/H]$_\textrm{RC}$} & \scriptsize{[$\alpha$/Fe]$_\textrm{RC}$} & \scriptsize{$\sigma$[$\alpha$/Fe]$_\textrm{RC}$} & \scriptsize{log(g)$_\textrm{Phot}$}\\
\hline
\scriptsize{PM J00012+0659} & \scriptsize{2950.0} & \scriptsize{25.1} & \scriptsize{$-$0.15} & \scriptsize{0.18} & \scriptsize{+0.200} & \scriptsize{0.102} & \scriptsize{2950.0} & \scriptsize{{ }0.0} & \scriptsize{$-$0.18} & \scriptsize{0.12} & \scriptsize{+0.215} & \scriptsize{0.049} & \scriptsize{5.12}\\
\scriptsize{PM J00031+0616} & \scriptsize{3200.0} & \scriptsize{33.5} & \scriptsize{+0.00} & \scriptsize{0.18} & \scriptsize{+0.150} & \scriptsize{0.097} & \scriptsize{3191.7} & \scriptsize{20.4} & \scriptsize{$-$0.04} & \scriptsize{0.14} & \scriptsize{+0.150} & \scriptsize{0.089} & \scriptsize{5.03}\\
\scriptsize{PM J00051+4547} & \scriptsize{3850.0} & \scriptsize{132.9} & \scriptsize{$-$0.10} & \scriptsize{0.18} & \scriptsize{+0.075} & \scriptsize{0.085} & \scriptsize{3816.7} & \scriptsize{66.1} & \scriptsize{$-$0.12} & \scriptsize{0.08} & \scriptsize{+0.100} & \scriptsize{0.055} & \scriptsize{4.72}\\
\scriptsize{PM J00077+6022} & \scriptsize{3150.0} & \scriptsize{37.2} & \scriptsize{$-$0.30} & \scriptsize{0.18} & \scriptsize{+0.125} & \scriptsize{0.097} & \scriptsize{3150.0} & \scriptsize{{ }0.0} & \scriptsize{$-$0.30} & \scriptsize{0.07} & \scriptsize{+0.113} & \scriptsize{0.018} & \scriptsize{4.90}\\
\scriptsize{PM J00078+6736} & \scriptsize{3700.0} & \scriptsize{127.5} & \scriptsize{$-$0.05} & \scriptsize{0.16} & \scriptsize{+0.125} & \scriptsize{0.093} & \scriptsize{3700.0} & \scriptsize{ { }0.0} & \scriptsize{$-$0.08} & \scriptsize{0.04} & \scriptsize{+0.138} & \scriptsize{0.038} & \scriptsize{4.69}\\
\scriptsize{PM J00081+4757} & \scriptsize{3300.0} & \scriptsize{87.2} & \scriptsize{$-$0.25} & \scriptsize{0.19} & \scriptsize{+0.200} & \scriptsize{0.100} & \scriptsize{3300.0} & \scriptsize{{ }0.0} & \scriptsize{$-$0.27} & \scriptsize{0.06} & \scriptsize{+0.214} & \scriptsize{0.028} & \scriptsize{4.82}\\
\scriptsize{PM J00088+2050} & \scriptsize{3100.0} & \scriptsize{24.8} & \scriptsize{$-$0.10} & \scriptsize{0.18} & \scriptsize{+0.125} & \scriptsize{0.093} & \scriptsize{3100.0} & \scriptsize{{ }0.0} & \scriptsize{$-$0.09 }& \scriptsize{0.05} & \scriptsize{+0.119} & \scriptsize{0.039} & \scriptsize{4.87}\\
\scriptsize{PM J00094+6549} & \scriptsize{4000.0} & \scriptsize{81.1} & \scriptsize{+0.00} & \scriptsize{0.20} & \scriptsize{+0.075} & \scriptsize{0.081} & \scriptsize{4000.0} & \scriptsize{{ }0.0} & \scriptsize{$-$0.12} & \scriptsize{0.13} & \scriptsize{+0.167} & \scriptsize{0.052} & \scriptsize{4.58}\\
\scriptsize{PM J00095+6548} & \scriptsize{3400.0} & \scriptsize{86.1} & \scriptsize{$-$0.10} & \scriptsize{0.17} & \scriptsize{+0.150} & \scriptsize{0.093} & \scriptsize{3433.3} & \scriptsize{57.7} & \scriptsize{$-$0.07} & \scriptsize{0.10} & \scriptsize{+0.150} & \scriptsize{0.000} & \scriptsize{4.81}\\
\scriptsize{PM J00101+1327} & \scriptsize{3250.0} & \scriptsize{32.9} & \scriptsize{$-$0.40} & \scriptsize{0.17} & \scriptsize{+0.200} & \scriptsize{0.101} & \scriptsize{3250.0} & \scriptsize{{ }0.0} & \scriptsize{$-$0.42} & \scriptsize{0.08} & \scriptsize{+0.222} & \scriptsize{0.042} & \scriptsize{5.06}\\
\scriptsize{PM J00110+0420} & \scriptsize{3300.0} & \scriptsize{37.6} & \scriptsize{$-$1.85} & \scriptsize{0.25} & \scriptsize{+0.400} & \scriptsize{0.119} & \scriptsize{3307.1} & \scriptsize{18.9} & \scriptsize{$-$1.85} & \scriptsize{0.19} & \scriptsize{+0.396} & \scriptsize{0.051} & \scriptsize{5.22}\\
\scriptsize{PM J00119+3303} & \scriptsize{3300.0} & \scriptsize{75.3} & \scriptsize{+0.05} & \scriptsize{0.16} & \scriptsize{+0.050} & \scriptsize{0.078} & \scriptsize{3300.0} & \scriptsize{{ }0.0} & \scriptsize{+0.04} & \scriptsize{0.07} & \scriptsize{+0.050} & \scriptsize{0.038}& \scriptsize{4.85}\\
\scriptsize{PM J00120+0206} & \scriptsize{3200.0} & \scriptsize{38.5} & \scriptsize{+0.20} & \scriptsize{0.17} & \scriptsize{+0.050} & \scriptsize{0.081} & \scriptsize{3188.9} & \scriptsize{22.1} & \scriptsize{+0.17} & \scriptsize{0.07} & \scriptsize{+0.056} & \scriptsize{0.037} & \scriptsize{5.01}\\
\scriptsize{PM J00122+3028} & \scriptsize{3100.0} & \scriptsize{68.1} & \scriptsize{+0.00} & \scriptsize{0.21} & \scriptsize{+0.100} & \scriptsize{0.095} & \scriptsize{3100.0} & \scriptsize{{ }0.0} & \scriptsize{+0.00} & \scriptsize{0.00} & \scriptsize{+0.100} & \scriptsize{0.000} & \scriptsize{4.59}\\
\scriptsize{PM J00123+6142} & \scriptsize{3300.0} & \scriptsize{92.9} & \scriptsize{+0.00} & \scriptsize{0.17} & \scriptsize{+0.225} & \scriptsize{0.094} & \scriptsize{3316.7} & \scriptsize{35.4} & \scriptsize{$-$0.01} & \scriptsize{0.07} & \scriptsize{+0.236} & \scriptsize{0.052} & \scriptsize{4.82}\\
\scriptsize{PM J00131+7023} & \scriptsize{3900.0} & \scriptsize{93.1} & \scriptsize{+0.25} & \scriptsize{0.21} & \scriptsize{$-$0.025} & \scriptsize{0.069} & \scriptsize{3900.0} & \scriptsize{{ }0.0} & \scriptsize{+0.25} & \scriptsize{0.04} & \scriptsize{$-$0.025} & \scriptsize{0.000} & \scriptsize{4.61}\\
\scriptsize{PM J00133+3908} & \scriptsize{3500.0} & \scriptsize{104.5} & \scriptsize{$-$0.05} & \scriptsize{0.20} & \scriptsize{+0.000} & \scriptsize{0.075} & \scriptsize{3500} & \scriptsize{{ }0.0} & \scriptsize{$-$0.14} & \scriptsize{0.18} & \scriptsize{+0.025} & \scriptsize{0.100} & \scriptsize{4.85}\\
\scriptsize{PM J00137+8039} & \scriptsize{3300.0} & \scriptsize{99.3} & \scriptsize{+0.50} & \scriptsize{0.09} & \scriptsize{+0.125} & \scriptsize{0.083} & \scriptsize{3327.8} & \scriptsize{83.3} & \scriptsize{+0.47} & \scriptsize{0.07} & \scriptsize{+0.142} & \scriptsize{0.076} & \scriptsize{5.13}\\
\scriptsize{PM J00138+3537} & \scriptsize{3200.0} & \scriptsize{37.7} & \scriptsize{+0.00} & \scriptsize{0.18} & \scriptsize{+0.100} & \scriptsize{0.093} & \scriptsize{3200.0} & \scriptsize{{ }0.0} & \scriptsize{+0.08} & \scriptsize{0.04} & \scriptsize{+0.063} & \scriptsize{0.018} & \scriptsize{5.00}\\
\scriptsize{PM J00144+3609N} & \scriptsize{3500.0} & \scriptsize{95.5} & \scriptsize{$-$0.10} & \scriptsize{0.17} & \scriptsize{+0.075} & \scriptsize{0.087} & \scriptsize{3533.3} & \scriptsize{35.4} & \scriptsize{$-$0.11} & \scriptsize{0.15} & \scriptsize{+0.108} & \scriptsize{0.082} & \scriptsize{4.86}\\
\hline
\end{tabular}  
}
\end{center}
\end{table*}

\section{Summary and Conclusion}
We present a synthetic  fitting pipeline that can be applied to low-resolution spectra of M dwarfs/subdwarfs to infer their physical parameters. Since such spectra are available in large numbers, their derived chemical parameter values  can provide excellent clues on the Galactic chemodynamical  evolution. Moreover,  the atmosphere of these low-mass stars has remained in a near-pristine chemical state since they formed and their abundance properties  can be therefore used as reliable  probes of   the Galaxy's early chemical enrichment. However, we describe some complications  in our low-resolution model-fit pipeline that have to be taken into account when interpreting the results, as summarized below:

\begin{enumerate}

\item As can be seen from Figures 22-25,  the stars generally resist to be matched with high-temperature, high-metallicity models (T$_\textrm{\footnotesize{eff}}$>3550 K and [M/H]>+0.1 dex), which is most likely due to issues in high-temperature synthetic models. Our high-temperature stars are systematically fitted with metallicity values within a specific  range (i.e., $-$0.5$\leq$[M/H]$\leq$+0.1), and are thus excluded from the present analysis.  

\item  As presented in Figures 24, 25, 27, 28, and the bottom panels of Figures 29 and 30, compared to the results using the constant (photometric) surface gravities, when log \emph{g} becomes a free parameter, due to the strong degeneracy between this parameter and [M/H], a large number of stars with near-solar metallicities move towards higher values (or even higher ends) of log \emph{g} and [M/H]. This shows that the variable-gravity approach  may cause  large systematic uncertainties in the inferred parameter values.

\item As shown in Figure 27 and the bottom panels of Figures 29 and 30, in comparison with the results using the constant (photometric) surface gravities, when log \emph{g} is allowed to vary, a significant number of stars with near-solar metallicities shift towards higher values of log \emph{g} and  [$\alpha$/Fe], which is due to the  degeneracy effect caused by these two parameters. This again indicates the unreliability of the derived parameter values from the variable-gravity approach.

\item  For either gravity-modeling approach, there is a systematic trend between T$_\textrm{\footnotesize{eff}}$ and [$\alpha$/Fe]. As it can clearly be seen from the middle panels of Figures 29 and 30, the higher-temperature stars tend to be fitted with the models having  lower values of [$\alpha$/Fe] (that are less affected by molecular bands) and vice versa, which is likely owing to deficiencies in the model spectra. 

\item  The tight correlation between the resulting values of  T$_\textrm{\footnotesize{eff}}$  and log \emph{g} inferred from the constant-gravity approach  appears as a relation between log \emph{g} and [$\alpha$/Fe], which  can be perceived from the panels ``c'' and ``e'' in Figures 29 and 30. When  log \emph{g} is kept fixed,  the derived values of [$\alpha$/Fe] follow nearly the same trend with log \emph{g} (panel ``e'') as with T$_\textrm{\footnotesize{eff}}$ (panel ``c''). Since log \emph{g}  is constant, the degeneracy between log \emph{g} and [$\alpha$/Fe] plays no role in the relation between these two parameters.

\item For either gravity-modeling approach, as shown in the bottom panels of Figures 29 and 30,  the tilted distributions of stars with nearly the same values of log \emph{g} is a reflection of the tight anti-correlation between [M/H] and [$\alpha$/Fe].

\end{enumerate}

It should be noted that all these complications and subtleties  are independent of the selected set of initial values, the formulation of  $\chi$$^\textrm{\footnotesize{2}}$  (i.e., whether including the observational flux error or not), and the extent of the grids over which the minimization routines are performed.  Limitations  in the model atmospheres to correctly address the effect of molecular bands  caused by T$_\textrm{\footnotesize{eff}}$ and the two chemical parameters, i.e., [M/H] and  [$\alpha$/Fe], have raised some systematic trends in the distributions of our resulting parameter values. Although the exclusion  of high-temperature stars (T$_\textrm{\footnotesize{eff}}$ >3550 K) reduces  these systematics to a great extent, there is still a trend between T$_\textrm{\footnotesize{eff}}$ and [$\alpha$/Fe], even using the constant-gravity approach,  which is unlikely  due to  sample selection effects or problems in the model-fit pipeline. Despite this systematic relation, we find the best-fit values of  T$_\textrm{\footnotesize{eff}}$, [M/H], and [M/H]+[$\alpha$/Fe] highly precise. This precision is confirmed by

\begin{enumerate}

\item The clear stratification of T$_\textrm{\footnotesize{eff}}$ in the HR diagram (Figures 22 and 23),

\item The clear stratification of [M/H] and [M/H]+[$\alpha$/Fe] in the HR diagram (Figures 24-26),

\item The similarity between the distribution of our stars (left panels in Figures 29 and 30)  and that of  other studies (e.g., GALAH DR3,  Buder et al. 2021, Figure 5) in the abundance diagram of [$\alpha$/Fe] versus [M/H]. Regardless of  some systematic offsets in parameter values between the two studies, which is due to the difference in the spectral resolutions, methods and model atmospheres, both distributions show nearly the same trend between [M/H] and  [$\alpha$/Fe].

\item Chemical homogeneity between the components of a set of binary systems (top panels in Figure 38), 

\item Revealing some substructure like stellar “streams” in abundance-velocity diagrams (Figure 43),

\item The notable similarity between the distributions of our stars (top panels in Figure 42) and those of other studies (e.g., N04, Figure 32) in  metallicity-velocity diagrams.

\end{enumerate}

While the photometric surface gravities may not be accurate, these values still yield reasonable results when being kept fixed  in the synthetic fitting pipeline. It should be recalled that the relation between these gravity values and the temperatures derived from the constant-gravity approach is predicted by theoretical isochrones. In addition, the evident contrast in surface gravity between metal-poor M subdwarfs and their metal-rich counterparts is consistent with the relative size of these stars (Section 6.3). To achieve more accurate stellar parameters from low-resolution spectroscopy, more reliable values of surface gravity from future extensive studies will thus be required.

Table 4 lists the best-fit  values of T$_\textrm{\footnotesize{eff}}$, [M/H], and [$\alpha$/Fe] inferred from both the normal (with a subscript ``N'') and the reduced-correlation (with a subscript ``RC'') methods using the constant-gravity approach for the 3745 stars. $\sigma$X$_\textrm{\footnotesize{N}}$ where X is T$_\textrm{\footnotesize{eff}}$, [M/H], or [$\alpha$/Fe] stands for the corresponding parameter  error  that is calculated based on the method described in Section 5. $\sigma$X$_\textrm{\footnotesize{RC}}$ denotes the standard deviation of the ten measurements from the mean value for each parameter T$_\textrm{\footnotesize{eff}}$, [M/H], and [$\alpha$/Fe] (Sections 4 and 5). The photometric surface gravities are also listed in this table.  It is important to note that all effective temperatures are included in the table, however, caution should be taken when using high temperature stars (T$_\textrm{\footnotesize{eff}}$>3550 K) whose parameter values most likely suffer from large uncertainties.

In our future work, we aim to expand our spectroscopic sample using a large SDSS/SEGUE catalog. We will apply the pipeline to those spectra with good quality and examine their distribution in some important photometric, chemical and kinematic diagrams. The larger number of metal-poor M subdwarfs in the SDSS/SEGUE sample may disclose some substructure in the Galactic halo.

We greatly thank the anonymous referee for their careful reading and helpful comments and  suggestions that significantly improved our manuscript. We would also like to acknowledge Ian Czekala for his useful discussions and suggestions. We extend our thanks to Douglas Gies, Russel White, Philip Muirhead, and Derek Homeier for their extremely insightful comments and suggestions. We would like to appreciate Ilija Medan for his help with the Gaia data. We finally wish to thank Justin Cantrell and Jeremy Simmons for their technical support. This material is based on work supported by the National Science Foundation under grant No. AST 09-08419, the National Aeronautics and Space Administration under grant Nos. NNX15AV65G, NNX16AI63G, and NNX16AI62G issued through the SMD/Astrophysics Division as part of the K2 Guest Observer Program. Parts of this research were supported  by the Australian Research Council Centre of Excellence for All Sky Astrophysics in 3 Dimensions (ASTRO 3D), through project number CE170100013.

This work has made use of data from the European Space Agency (ESA) mission Gaia (https://www.cosmos.esa.int/gaia), processed by the Gaia Data Processing and Analysis Consortium (DPAC, https://www.cosmos.esa.int/web/gaia/dpac/consortium). Funding for the DPAC has been provided by national institutions, in particular the institutions participating in the Gaia Multilateral Agreement.

\section{References}

\noindent
\footnotesize{Adibekyan, V. Z., Figueira, P., Santos, N. C., et al. 2013, A\&A, 554, A44 (DOI: 10.1051/0004-6361/201321520)}

\noindent
\footnotesize{Adibekyan, V. Z., Sousa, S. G., Santos, N. C., et al. 2012, A\&A, 545, A32 (DOI: 10.1051/0004-6361/201219401)}

\noindent
\footnotesize{Allard, F., \& Hauschildt, P. H. 1995, ApJ, 445, 433 (DOI: 10.1086/175708)}

\noindent
\footnotesize{An, D. 2019. ApJL, 878, L31 (DOI: 10.3847/2041-8213/ab2467)}

\noindent
\footnotesize{Baraffe, I., Homeier, D., Allard, F., \& Chabrier, G. 2015, A\&A, 577, A42 (DOI: 10.1051/0004-6361/201425481)}

\noindent
\footnotesize{Belokurov, V.,  Penoyre, Z.,  Oh, S. et al. 2020, MNRAS, 496, 1922 (DOI: 10.1093/mnras/staa1522)}

\noindent
\footnotesize{Bergemann, M., \& Nordlander, T. 2014, Determination of Atmospheric Parameters of B-, A-, F- and G-Type (Basel: Springer), 169 (DOI: 10.1007/978-3-319-06956-2$\_$16)}

\noindent
\footnotesize{Bessell, M. S. 1991, AJ, 101, 662 (DOI: 10.1086/115714)}

\noindent
\footnotesize{Bochanski, J. J., Hawley, S. L., Covey, K. R., et al. 2010, AJ, 139, 2679 DOI: 10.1088/0004-6256/139/6/2679}

\noindent
\footnotesize{Buder, S., Sharma, S., Kos, J., et al. 2021, MNRAS, 506, 150 (DOI: 10.1093/mnras/stab1242)}

\noindent
\footnotesize{Chavez, J., \& Lambert, D. L. 2009, ApJ, 699, 1906 (DOI: 10.1088/0004-637X/699/2/1906)}

\noindent
\footnotesize{Chiappini, C. 2001, AmSci, 89, 506 (DOI: 10.1511/2001.6.506)}

\noindent
\footnotesize{Chiappini, C., Matteucci, F., Beers, T., \& Nomoto, K. 1999, ApJ, 515, 226 (DOI: 10.1086/307006)}

\noindent
\footnotesize{Chiappini, C., Matteucci, F., \& Romano, D. 2001, ApJ, 554, 1044 (DOI: 10.1086/321427)}

\noindent
\footnotesize{Choi, J.,  Dotter, A.,  Conroy, C., et al. 2016, ApJ, 823, 102 (DOI: 10.3847/0004-637X/823/2/102)}

\noindent
\footnotesize{Croswell, K. 1995, The Alchemy of the Heavens: Searching for Meaning in the Milky Way (New York: Anchor Books)}

\noindent
\footnotesize{Czekala, I., Andrews, S. M., Mandel, K. S., et al.  2015, ApJ, 812, 128 (DOI: 10.1088/0004-637X/812/2/128)}

\noindent
\footnotesize{De Simone, R. A., Wu, X., \& Tremaine, S. 2004, MNRAS, 350, 627 (DOI:  10.1111/j.1365-2966.2004.07675.x)}

\noindent
\footnotesize{Dotter, A., Chaboyer, B., Jevremovic, D. et al. 2008, ApJS, 178, 89, (DOI: 10.1086/589654)} 

\noindent
\footnotesize{Dotter, A. 2016, ApJS, 222, 8 (DOI:10.3847/0067-0049/222/1/8)}

\noindent
\footnotesize{Du, C-H., Zhou, X., Ma, J., et al. 2004, AJ, 128, 2265 (DOI: 10.1086/424858)}

\noindent
\footnotesize{Fux, R. 2001, A\&A, 373, 511 (DOI: 10.1051/0004-6361:20010561)}

\noindent
\footnotesize{Gaia Collaboration, Brown, A. G. A., Vallenari, A., et al. 2018, A\&A, 616, A1  (DOI: 10.1051/0004-6361/201833051)}

\noindent
\footnotesize{Gaia Collaboration, Brown, A. G. A., Vallenari, A., Prusti, T., et al. 2021, A\&A, 649, A1 (DOI: 10.1051/0004-6361/202039657)}

\noindent
\footnotesize{Gao, X., Lind, K., Amarsi, A. M., et al. 2018, MNRAS, 481, 2666 (DOI: 10.1093/mnras/sty2414)}

\noindent
\footnotesize{Gustafsson, B., Edvardsson, B., Eriksson, K., et al. 2008, A\&A, 486, 951 (DOI: 10.1051/0004-6361:200809724)}

\noindent
\footnotesize{Hawkins, K.,  Lucey, M.,  Ting, Y.-S., et al. 2020, MNRAS, 492, 1164  (DOI: 10.1093/mnras/stz3132)}

\noindent
\footnotesize{Hawkins, K.,  Jofr\'e, P.,  Masseron, T., \&   Gilmore, G. 2015, MNRAS, 2015, 453, 758 (DOI (10.1093/mnras/stv1586)}

\noindent
\footnotesize{Hayden, M. R., Bovy, J., Holtzman, J. A., et al. 2015, ApJ, 808, 132 (DOI: 10.1088/0004-637X/808/2/132)}

\noindent
\footnotesize{Haywood, M. 2001, MNRAS, 325, 1365 (DOI: 10.1046/j.1365-8711.2001.04510.x)}

\noindent
\footnotesize{Hejazi, N., De Robertis, M. M., \& Dawson, P. C. 2015, AJ, 149, 140 (DOI: 10.1088/0004-6256/149/4/140)}

\noindent
\footnotesize{Hejazi, N., Lepine, S., Homeier, D., Rich, R. M., \& Shara, M. M. 2020, AJ,  159, 30 (DOI: 10.3847/1538-3881/ab563c)}

\noindent
\footnotesize{Heringer, E., Pritchet, C., \& van Kerkwijk, M. H. 2019, ApJ, 882, 52 (DOI: 10.3847/1538-4357/ab32dd)}

\noindent
\footnotesize{Hinkle, K. H.,  Wallace, L., \&  Livingston, W. 2003. AAS, Meeting No. 203, 35,1260}

\noindent
\footnotesize{Kervella, P.,  Arenou, F., \&  Thevenin, F. 2022, A\&A, 657, A7 (DOI: 10.1051/0004-6361/202142146)}

\noindent
\footnotesize{Kesseli, A. Y., Kirkpatrick, J. D., Sergio, B. F., et al. 2019, AJ, 157, 63 (DOI: 10.3847/1538-3881/aae982)}

\noindent
\footnotesize{Kippenhahn, R., \& Weigert, A. 1990, Stellar Structure and Evolution (Springer-Verlag Berlin Heidelberg New York)}

\noindent
\footnotesize{Kouwenhoven, M. B. N., Goodwin, S. P., Parker, R. J., et al.  2010, MNRAS, 404, 1835 (DOI: 10.1111/j.1365-2966.2010.16399.x)}

\noindent
\footnotesize{Kuznetsov, M. K., del Burgo, C., Pavlenko, Ya. V., \& Frith, J. 2019, ApJ, 878, 134 (DOI: 10.3847/1538-4357/ab1fe9)}

\noindent
\footnotesize{Lee, Y. S., Beers, T. C, Prieto, C. A., et al. 2011a, AJ,141, 90  (DOI: 10.1088/0004-6256/141/3/90)}

\noindent
\footnotesize{Lee, Y. S., Beers, T. C., An, D., et al. 2011b, ApJ, 738, 187 (DOI: 10.1088/0004-637X/738/2/187)}

\noindent
\footnotesize{Lee, J.-E., Lee, S., Dunham, M. M., et al. 2017, NatAs, 1, 0172  (DOI: 10.1038/s41550-017-0172)}

\noindent
\footnotesize{L\'epine, S., \& Shara, M. M. 2005, AJ, 129, 1483 (DOI: 10.1086/427854)}

\noindent
\footnotesize{L\'epine, S. 2005, AJ, 130, 1680 (DOI:  10.1086/432792)}

\noindent
\footnotesize{Lépine, S., Rich, R. M., \& Shara, M. M. 2007, ApJ, 669, 1235 (DOI: 10.1086/521614)}

\noindent
\footnotesize{L\'epine, S., Hilton, E. J., Mann, A. W., et al. 2013, AJ, 145, 102 (DOI: 10.1088/0004-6256/145/4/102)}

\noindent
\footnotesize{Mann, A. W., Brewer, J. M., Gaidos, E., Lépine, S., \& Hilton, E. J. 2013, AJ, 145, 52 (DOI: 10.1088/0004-6256/145/2/52)}

\noindent
\footnotesize{Mann, A. W.,  Feiden, G. A., Gaidos, E., et al. 2015, ApJ, 804, 64 (DOI: 10.1088/0004-637X/804/1/64)}

\noindent
\footnotesize{Mann, A. W.,  Dupuy, T., Kraus, A. L., et al. 2019, ApJ, 871, 63  (DOI: 10.3847/1538-4357/aaf3bc)}

\noindent
\footnotesize{Moeckel, N. \& Clarke C. J.  2011, MNRAS, 415, 1179  (DOI: 10.1111/j.1365-2966.2011.18731.x)}

\noindent
\footnotesize{Nordlander, T., Amarsi, A. M., Lind, K., et al. 2017, A\&A, 597, A6}

 \noindent
\footnotesize{Nordstr\"om, B.,  Mayor, M.,  Andersen, J., et al. 2004,  A\&A, 418, 989 (DOI: 10.1051/0004-6361:20035959)}

\noindent
\footnotesize{Offner, S. S. R., Kratter, K. M., Matzner, C. D., et al.  2010, ApJ, 725, 1485  (DOI: 10.1088/0004-637X/725/2/1485)}

\noindent
\footnotesize{Pagel, B. E. 1997, Nucleosynthesis and Chemical Evolution of Galaxies (Cambridge: Cambridge Univ. Press)}

\noindent
\footnotesize{Passegger, V. M., Wende-von Berg, S., \& Reiners, A. 2016, A\&A, 587, A19  (DOI: 10.1051/0004-6361/201322261)}

\noindent
\footnotesize{Passegger, V. M., Reiners, A., Jeffers, S. V., et al.  2018, A\&A, 615, A6 (DOI: 10.1051/0004-6361/201732312)}

\noindent
\footnotesize{Pipino, A., \&  Matteucci, F. 2009, The Ages of Stars, Proceedings of the International Astronomical Union, IAU Symposium,  258, 39 (DOI: 10.1017/S174392130903169X)}

\noindent
\footnotesize{Prsa, A,  Harmanec, P.,  Torres, G., et al. 2016, AJ, 152, 41 (DOI: 10.3847/0004-6256/152/2/41)}

\noindent
\footnotesize{Rains, A. D., Zerjal, M., Ireland, M. J., et al. 2021, MNRAS, 504, 5788 (DOI: 10.1093/mnras/stab1167)}

\noindent
\footnotesize{Rajpurohit, A. S., Reyl\'e, C., Allard, F., et al. 2014, A\&A, 564, A90 (DOI: 10.1051/0004-6361/201322881)}

\noindent
\footnotesize{Rajpurohit, A. S., Allard, F., Rajpurohit, S., et al. 2018, A\&A, 620, A180 (DOI: 10.1051/0004-6361/201833500)}

\noindent
\footnotesize{Recio-Blanco, A., de Laverny, P., Kordopatis, G., et al. 2014, A\&A, 567, A5 (DOI: 10.1051/0004-6361/201322944)}

\noindent
\footnotesize{Reid, I. N., \& Gizis, J. E. 1997, AJ, 114, 1992 (DOI: 10.1086/118620)}

\noindent
\footnotesize{Reipurth B. \& Mikkola S. 2012, Nature, 492, 221  (DOI: 10.1038/nature11662 )}

\noindent
\footnotesize{Rocha-Pinto H.J., \&  Maciel, W.J. 1998, A\&A, 339, 79}

\noindent
\footnotesize{Sch\"onrich, R., Binney, J. \& Dehnen, W. 2010, MNRAS, 403, 1829 (DOI: 10.1111/j.1365-2966.2010.16253.x)}

\noindent
\footnotesize{Simpson, J. D., Martell, S. L., Da Costa, G., et al. 2019, MNRAS, 482, 5302  (DOI: 10.1093/mnras/sty3042)}

\noindent
\footnotesize{Tokovinin A. 2017, MNRAS, 468, 3461  (DOI: 10.1093/mnras/stx707)}

\noindent
\footnotesize{Valenti, J. A., Piskunov, N. E., \& Johns-Krull, C. M. 1998, ApJ, 498, 851 (DOI: 10.1086/305587)}

\noindent
\footnotesize{Veyette, M. J., Muirhead, P. S., Mann, A. W., et al. 2017, ApJ, 851, 26 (DOI: 10.3847/1538-4357/aa96aa)}

\noindent
\footnotesize{West, A. A., Morgan, D. P., Bochanski, J. J., et al. 2011, AJ, 141, 97 (DOI: 10.1088/0004-6256/141/3/97)}

\noindent
\footnotesize{Wood, M. L.,  Mann, A. W., \& Kraus, A. L.  2021, AJ, 162,128 (DOI: 10.3847/1538-3881/ac0ae9)}

\noindent
\footnotesize{Woolf, V. M. \&  Wallerstein, G. 2020, MNRAS, 494, 2718 (DOI: 10.1093/mnras/staa878)}

\noindent
\footnotesize{Yan, Y.,  Du, C.,  Liu, S. et al. 2019, ApJ,  880, 36 (DOI:  10.3847/1538-4357/ab287d)}

\noindent
\footnotesize{Zhang, S., Luo, A-Li., Comte, G., et al. 2021, ApJ, 908, 131 (DOI: 10.3847/1538-4357/abcfc5)}

\noindent
\footnotesize{Ziegler, C., Tokovinin, A., Briceno, C., et al. 2020, AJ, 159, 19 (DOI: 10.3847/1538-3881/ab55e9)}

 \end{document}